\documentclass[12pt,a4paper]{article}
\usepackage{cite}
\usepackage{latexsym}
\usepackage{amsbsy}
\usepackage{amssymb}
\usepackage{epsfig}
\usepackage{rotate}
\newcounter{figures}
\newcounter{tables}

\newlength{\templength}
\newenvironment{equationarray}
{ \setlength{\templength}{\arraycolsep} \setlength{\arraycolsep}{2pt} 
\begin{eqnarray} }
{ \setlength{\arraycolsep}{\templength} \end{eqnarray} }
\newenvironment{equationarrayzero}
{ \setlength{\templength}{\arraycolsep} \setlength{\arraycolsep}{0pt} 
\begin{eqnarray} }
{ \setlength{\arraycolsep}{\templength} \end{eqnarray} }
\newcommand{\vet}[1]{\ensuremath{\hskip-1pt\vec{\hskip1pt#1}}}
\newcommand{\Nmass}{N}
\newcommand{\nmass}{\nu}
\begin{document}

\null

\begin{flushright}
\begin{tabular}{r}
UWThPh-1998-61\\
DFTT 69/98\\
KIAS-P98045\\
SFB 375-310\\
TUM-HEP 340/98\\
hep-ph/9812360
\end{tabular}
\end{flushright}

\vspace{0.5cm}

%\vspace{3.5cm}

\begin{center}
\Large \bfseries
Phenomenology of \\Neutrino Oscillations
\\[0.5cm]
\large \mdseries \upshape
S.M. Bilenky
\\[0.25cm]
\normalsize \itshape
Joint Institute for Nuclear Research, Dubna, Russia, and
\\
Institut f\"ur Theoretische Physik,
Technische Universit\"at M\"unchen,
D--85748 Garching, Germany
\\[0.5cm]
\large \mdseries \upshape
C. Giunti
\\[0.25cm]
\normalsize \itshape
INFN, Sezione di Torino, 
and
Dipartimento di Fisica Teorica,
\\
Universit\`a di Torino,
Via P. Giuria 1, I--10125 Torino, Italy, and
\\
School of Physics, Korea Institute for Advanced Study,
Seoul 130-012, Korea  
\\[0.5cm]
\large \mdseries \upshape
W. Grimus
\\[0.25cm]
\normalsize \itshape
Institute for Theoretical Physics, University of Vienna,
\\
Boltzmanngasse 5, A--1090 Vienna, Austria
\\
\vspace{0.5cm}
\large \upshape
Abstract
\\[0.25cm]
\normalsize
\begin{minipage}[t]{0.9\textwidth}
This review is focused on neutrino mixing and neutrino
oscillations in the light of the recent experimental developments. 
After discussing possible types of neutrino mixing for Dirac
and Majorana neutrinos and considering in detail
the phenomenology of neutrino oscillations in vacuum and matter,
we review all existing evidence and indications
in favour of neutrino oscillations that have been obtained in the
atmospheric, solar and LSND experiments. We present the results of 
the analyses
of the neutrino oscillation data in the framework of mixing of three
and four massive neutrinos and investigate possibilities to test the
different neutrino mass and mixing schemes obtained in this way. 
We also discuss briefly future neutrino oscillation experiments.
\end{minipage}
\end{center}

\newpage
\tableofcontents
\newpage

\section{Introduction}
\label{Introduction}
\setcounter{equation}{0}
\setcounter{figures}{0}
\setcounter{tables}{0}

The strong evidence in favour of oscillations of atmospheric neutrinos
found by the Super-Kamiokande Collaboration
\cite{SK-atm-nu98,SK-atm-98}
opened a new era in particle physics. 
There is no doubt that new experiments are necessary to understand
the nature
of neutrino masses and mixing which are intimately connected with
neutrino oscillations,
but the first decisive step has been done:
massive and mixed neutrinos can now be considered as real physical objects.

The problem of neutrino mass has a long history.
Originally, Pauli considered
the neutrino as a particle with a small but non-zero mass
(smaller than the electron mass) \cite{Pauli}
and
the method for the measurement of the neutrino mass
through the investigation of
the $\beta$-spectrum near the end point was proposed
in the first theoretical
papers on $\beta$-decay of Fermi \cite{Fermi33,Fermi34}
and Perrin \cite{Perrin34}.

The first experiments on the measurement of the neutrino mass,
based on the Fermi-Perrin method,
yielded the upper bound
$ m_{\nu} \lesssim 500 \, \mathrm{eV} $ \cite{pont-massbound} 
which was improved in the fifties to
$ m_{\nu} \lesssim 250 \, \mathrm{eV} $ \cite{LM52}.
Therefore, it became evident that
the neutrino mass (if non-zero at all) 
is much smaller than the electron mass.
This was the main reason that in 1957,
after the discovery of parity violation in $\beta$-decay,
the authors of the two-component theory of the neutrino
(Landau \cite{Landau57},
Lee and Yang \cite{Lee-Yang57},
Salam \cite{Salam57})
assumed that
the neutrino is a massless particle,
the field of which is either
a left-handed field $\nu_{L}$ or a right-handed field $\nu_{R}$.

In 1958, Goldhaber \textit{et al.} \cite{Goldhaber58}
measured the helicity of the neutrino.
The result of this experiment was in agreement with the
two-component neutrino theory and
it was established that the neutrino field is
$\nu_{L}$.\footnote{Notice that in the Goldhaber \textit{et al.}
experiment the helicity of the electron neutrino was measured. The
helicity of the muon neutrino was measured in several experiments (for
the references see the review of V.L. Telegdi \cite{telegdi}). The
best accuracy in the measurement of the muon neutrino was achieved in
the experiment by Gr\'enacs \textit{et al.} \cite{grenacs}.}
The results of the experiment of Goldhaber \textit{et al.}
could not exclude, however, the
possibility of a small neutrino mass.
In the V$-$A theory
(Feynman and Gell-Mann \cite{Feynman-Gell-Mann58},
Sudarshan and Marshak \cite{Sudarshan-Marshak58})
the Hamiltonian of weak interactions
contains the left-handed component of the neutrino field
$\nu_{L}$, and also the left-handed components of \emph{all massive fields}.
Therefore the possibility for the neutrino to be nevertheless a
massive particle became more natural \cite{private} after the 
confirmation of the V$-$A theory.

In 1957, B. Pontecorvo \cite{Pontecorvo57,Pontecorvo58}
proposed the idea that 
the state of neutrinos
produced in weak interaction processes
is a superposition of states of two
Majorana neutrinos \cite{majorana} with definite masses
(analogous to the states 
$|K^0\rangle$ and $|\bar{K}^0\rangle$
which are the superposition of
$|K_1\rangle$ and  $|K_2\rangle$,
the states of particles with definite masses and widths).
In this way, B. Pontecorvo arrived at the hypothesis of neutrino oscillations
(analogous to $K^0 \leftrightarrows \bar{K}^0$ oscillations).
At that time only one type of neutrino was known.
The possibility of mixing of the two species of neutrinos 
$\nu_e$ and $\nu_{\mu}$ was considered in
Ref.~\cite{Maki62}. All possible types of neutrino oscillations for
this case were investigated by Pontecorvo
in 1967 \cite{Pontecorvo67}.

Gribov and Pontecorvo proposed in 1969 \cite{Pontecorvo69}
the first phenomenological theory of neutrino mixing
and oscillations.
In this theory, the two left-handed neutrino fields 
$\nu_{{e}L}$ and $\nu_{{\mu}L}$ are linear combinations of
the left-handed
components of the fields of Majorana neutrinos with definite masses
and
the neutrino mass term contains
only the left-handed fields $\nu_{{e}L}$ and $\nu_{{\mu}L}$.

In 1976,
neutrino oscillations were considered in the scheme of mixing of two Dirac
neutrinos based on the analogy between quarks and leptons
\cite{Bilenky-Pontecorvo76b,fritzsch}
and in the same year in the general Dirac--Majorana scheme 
\cite{Bilenky-Pontecorvo76a} (for later works see
Refs.~\cite{barger80a,BHP80,Kobzarev80,Schechter-Valle80b}).

The theoretical arguments in favour of non-zero neutrino masses and mixing
are based on the models beyond the Standard Model
(see, for example, Ref.~\cite{Mohapatra-Pal91}).
In such models 
the fields of quarks, charged leptons and neutrinos
are grouped in the same multiplets
and the generation of the masses of quarks and charged leptons
with the Higgs
mechanism as a rule provides also non-zero neutrino masses.

In 1979,
the see-saw mechanism for the generation
of neutrino masses was proposed
\cite{GRS79,Yanagida79,MS80}.
This mechanism
connects the smallness of neutrino masses with the possible violation of
lepton number conservation at a very large energy scale.

At present the effects of neutrino masses and mixing are investigated
in many different experiments. There are three types of experiments in
which the effects of small neutrino masses (say, of the order of 1 eV
or smaller) and mixing can be revealed (for a review and references
see Ref.~\cite{PDG98}):

\begin{enumerate}

\item
Neutrino oscillations experiments.

\item
Experiments on the search for neutrinoless double $\beta$-decay.

\item
Experiments on the measurement of the electron neutrino
mass 
with the precise
investigation of the high energy part of
the $\beta$-spectrum of $^3$H.

\end{enumerate}

Three indications in favour of neutrino masses and mixing have been found
so far. These indications were obtained in the following experiments:

\begin{enumerate}

\item
Solar neutrino experiments
(Homestake \cite{Homestake68,Homestake94,Homestake98},
Kamiokande \cite{Kam-sun-88,Kam-sun-91-PRD,Kam-sun-96},
GALLEX \cite{GALLEX92a,GALLEX96},
SAGE \cite{SAGE91,SAGE96},
Super-Kamiokande \cite{SK-sun-97-Inoue,SK-sun-98-PRL,SK-sun-nu98}).

\item
Atmospheric neutrino experiments
(Super-Kamiokande \cite{SK-atm-nu98,SK-atm-98},
Kamiokande \cite{Kam-atm-94},
IMB \cite{Bec95},
Soudan \cite{All97}, MACRO \cite{Amb98}).

\item
The accelerator LSND experiment \cite{LSND96,LSND98}.

\end{enumerate}

Many other neutrino oscillation experiments with neutrinos 
from reactor and accelerators
did not find any evidence for neutrino oscillations. In
the experiments on the search for neutrinoless double $\beta$-decay 
no indications for non-zero neutrino
masses were found (see Section 6.2).
The present upper bound for the electron neutrino mass
obtained in the Troitsk experiment 
\cite{lobashov} is 2.7 eV (see also the Mainz experiment \cite{Mainz-nu98}). 
The upper limits on the masses of $\nu_\mu$ and
$\nu_\tau$ are 170 keV (90\% CL) and 18.2 MeV (95\% CL), 
respectively \cite{PDG98}.
Neutrinos play an important role in cosmology and astrophysics
and many bounds on neutrino properties can be derived in this
context. For reviews see, \textit{e.g.}, Refs.~\cite{Mohapatra-Pal91,turner}.

In this review we discuss the phenomenological theory of
neutrino mixing (Section 2),
neutrino oscillations in vacuum (Section 3),
neutrino oscillations and transitions in matter (Section 4) and
experimental data and results of analyses of the data (Sections 5 and 6).
We also consider the implications
of the existing experimental results on neutrino oscillations
for experiments in preparation. After the conclusions (Section 7) we
discuss some properties of Majorana neutrinos and fields in Appendix A.

We hope that this review will be useful not only for the physicists
that are working
in the field but also for those who are interested in this exciting field of
physics. 
In many cases we present not only results but also
derivations of the results. 
For those who start to study the subject we refer to the books
Refs.~\cite{Boehm-Vogel87,Kayser89-book,Bahcall89-book,Mohapatra-Pal91,Kim93}
and the reviews
Refs.~\cite{Bilenky-Pontecorvo78,Frampton-Vogel82,Vergados86,Bilenky-Petcov87,%
Valle91,Oberauer-vonFeilitzsch92,Gelmini-Roulet95,Brunner97,Zuber98-NEUTRINOS,%
Conrad-98-Vancouver,Valle98,Raffelt99}.

\section{Neutrino Mixing}
\label{Neutrino Mixing}
\setcounter{equation}{0}
\setcounter{figures}{0}
\setcounter{tables}{0}

All the numerous data on \emph{weak interaction processes}
are perfectly well described by 
the Standard Model
\cite{Glashow61,Weinberg67,Salam68}.
The standard weak interactions
are due to the coupling of quarks and leptons
with the gauge $W$ and $Z$ vector bosons,
described by the charged-current (CC)
and neutral-current (NC)
interaction Lagrangians
\begin{equationarrayzero}
&&
\mathcal{L}_{I}^{\mathrm{CC}}
=
- \frac{g}{2\sqrt{2}} \,
j^{\mathrm{CC}}_{\rho} \, W^{\rho}
+
\mathrm{h.c.}
\,,
\label{CC}
\\
&&
\mathcal{L}_{I}^{\mathrm{NC}}
=
- \frac{g}{2\cos\theta_{W}} \,
j^{\mathrm{NC}}_{\rho} \, Z^{\rho}
\,.
\label{NC}
\end{equationarrayzero}%
Here $g$ is the SU(2)$_L$
gauge coupling constant,
$\theta_{W}$ is the weak angle and
the charged and neutral currents
$j^{\mathrm{CC}}_{\rho}$ and  $j^{\mathrm{NC}}_{\rho}$
are given by the expressions
\begin{equationarrayzero}
&&
j^{\mathrm{CC}}_{\rho}
=
2
\sum_{\ell=e,\mu,\tau} \overline{\nu_{{\ell}L}} \, \gamma_{\rho}
\, \ell_L
+
\ldots
\,,
\label{jCC}
\\
&&
j^{\mathrm{NC}}_{\rho}
=
\sum_{\ell=e,\mu,\tau} \overline{\nu_{{\ell}L}} \, 
\gamma_{\rho} \, \nu_{{\ell}L}
+
\ldots
\,,
\label{jNC}
\end{equationarrayzero}%
where the $\ell$ are the physical charged lepton fields with
masses $m_\ell$ and 
we have written explicitly only the terms containing the neutrino fields.
The flavour neutrinos 
$\nu_e$,
$\nu_{\mu}, \ldots$ 
are determined by CC weak interactions: 
for example, the
$\nu_{\mu}$ is the particle produced in the decay
$ \pi^+ \to \mu^+ + \nu_\mu $
and so on. 
The number of light flavour neutrinos is given by the
invisible width of the $Z$ boson \cite{bertlmann} in the Standard Model. 
It was shown in the famous LEP experiments on the
measurement of the invisible width of the $Z$ boson that the number of the
light flavour neutrinos is equal to 3. The most recent experimental 
value of the number of neutrino flavours is $2.994 \pm 0.012$
\cite{PDG98},
showing that are no other neutrino flavours than the well-known
$\nu_e$, $\nu_\mu$, $\nu_\tau$.

The CC and NC interactions conserve the
electron $L_e$,
muon $L_{\mu}$
and tau $L_{\tau}$
lepton numbers,
which are assigned as shown in Table~\ref{lepton numbers}.
 
\begin{table}[t]
\begin{center}
\renewcommand{\arraystretch}{1.45}
\setlength{\tabcolsep}{0.5cm}
\begin{tabular}{|cccc|}
\hline
&
$L_{e}$
&
$L_{\mu}$
&
$L_{\tau}$
\\
\hline
$\left( \nu_{e} \, , \, e^{-} \right)$
&
$+1$
&
0
&
0
\\
$\left( \nu_{\mu} \, , \, \mu^{-} \right)$
&
0
&
$+1$
&
0
\\
$\left( \nu_{\tau} \, , \, \tau^{-} \right)$
&
0
&
0
&
$+1$
\\
\hline
\end{tabular}
\end{center}
\refstepcounter{tables}
\label{lepton numbers}
\footnotesize
Table \ref{lepton numbers}.
Assignment of lepton numbers.
The corresponding antiparticles have opposite lepton numbers.
\end{table}

There are no indications in favour of violation  
of the law of conservation of lepton numbers in weak processes
and
very strong bounds on the probabilities of the lepton number violating
processes have been obtained from the experimental data.
The most stringent limits (90\% CL) are
(see Ref.~\cite{PDG98}):
\begin{equationarrayzero}
&&
\Gamma( \mu \to e \, \gamma ) / \Gamma( \mu \to \mathrm{all} )
<
4.9 \times 10^{-11}
\,,
\label{lim01}
\\
&&
\Gamma( \mu \to 3 \, e ) / \Gamma( \mu \to \mathrm{all} )
<
1.0 \times 10^{-12}
\,,
\label{lim02}
\\
&&
\sigma( \mu^- \, \mathrm{Ti} \to e^- \, \mathrm{Ti} )
 / \sigma( \mu^- \, \mathrm{Ti} \to \mathrm{capture} )
<
4.3 \times 10^{-12}
\,,
\label{lim03}
\\
&&
\Gamma( K_L \to e \, \mu ) / \Gamma( K_L \to \mathrm{all} )
<
3.3 \times 10^{-11}
\,,
\label{lim04}
\\
&&
\Gamma( K^+ \to \pi^+ \, e^- \, \mu^+ ) / \Gamma( K^+ \to \mathrm{all} )
<
2.1 \times 10^{-10}
\,.
\label{lim05}
\end{equationarrayzero}% 

According to the \emph{neutrino mixing hypothesis}
\cite{Pontecorvo57,Pontecorvo58,Maki62},
the conservation of the lepton numbers is only approximate.
It is violated because of
non-zero neutrino masses and neutrino mixing.
In the case of neutrino mixing,
the left-handed flavour neutrino
fields
$\nu_{{\ell}L}$
are superpositions
of
the left-handed components
$\nu_{kL}$
of the fields of neutrinos with definite masses
$m_k$:
\begin{equation}
\nu_{{\ell}L}
=
\sum_{k=1}^{n}
U_{{\ell}k}
\,
\nu_{kL}
\qquad
(\ell=e,\mu,\tau)
\,,
\label{mixing}
\end{equation}
where $U$
is a unitary mixing matrix. The number of massive neutrino fields $n$
could be equal or more than 3. If $n$ is larger than 3 there are
\emph{sterile neutrinos} that do not take part in the 
standard weak interactions (\ref{jCC}) and (\ref{jNC}).

It is well-established that
quarks take part in CC weak interactions in mixed form with the V$-$A
current 
\begin{equation}
\sum_{q'=u,c,t} \, \sum_{q=d,s,b} \, \bar{q}'_L\, \gamma_\rho 
V_{q'q} \, q_L
\,,
\label{quark-mix}
\end{equation}
where $V$ is the Cabibbo-Kobayashi-Maskawa mixing matrix
\cite{Cabibbo63,Kobayashi-Maskawa73}.
The relation (\ref{mixing})
is analogous to the mixing in Eq.(\ref{quark-mix}).
However, between the mixing of neutrinos and quarks
there can be a fundamental difference.
In fact,
quarks are four-component Dirac particles: quarks and antiquarks have
opposite electric and baryonic charges.
Instead,
neutrinos are electrically neutral particles.
If the total lepton charge
\begin{equation}
L = L_e + L_\mu + L_\tau
\label{L}
\end{equation}
is conserved,
neutrinos with definite masses
are four-component Dirac particles
like quarks
(in this case a neutrino differs from
an antineutrino by the opposite value of $L$). 
If the total lepton number (\ref{L})
is not conserved,
massive neutrinos are truly neutral two-component Majorana particles.
These possibilities are realized in different
models and correspond to different \emph{neutrino mass terms}.

\subsection{Dirac mass term}
\label{Dirac mass term}

A Dirac neutrino mass term can be generated by the Higgs mechanism 
with the
standard Higgs doublet which is responsible for the generation of the masses of
quarks and charged leptons.\footnote{One needs
to assume the existence of right-handed SU(2)$_L$ singlet fields
$\nu_{{\ell}R}$.
If these fields do not exist,
it is not possible to construct
a Dirac mass term for the neutrinos,
\textit{i.e.}, neutrinos are massless particles.
This model is sometimes called Minimal Standard Model
\cite{Glashow61,Weinberg67,Salam68}.}
In this case
the neutrino mass term is given 
by\footnote{Note that in this review we use four-component spinors. For
the equivalent alternative of two-component spinors see, \textit{e.g.},
Refs.~\cite{Schechter-Valle80b,Schechter-Valle81}.}
\begin{equation}
\mathcal{L}^{\mathrm{D}}
=
-
\sum_{\ell,\ell'}
\overline{\nu_{{\ell}R}}
\,
M^{\mathrm{D}}_{\ell\ell'}
\,
\nu_{{\ell'}L}
+
\mathrm{h.c.}
\qquad
(\ell=e,\mu,\tau)
\,,
\label{Dirac}
\end{equation}
and $M^{\mathrm{D}}$ is a complex $3 \times 3$ matrix. 

The Dirac mass term (\ref{Dirac}) can be diagonalized
taking into account that the complex matrix $M^{\mathrm{D}}$
can be written as
\begin{equation}
M^{\mathrm{D}}
=
V \, \hat{m} \, U^\dagger
\,,
\label{complex matrix}
\end{equation}
where 
$V$ and $U$ are unitary matrices and 
$\hat{m}$ is a positive definite diagonal matrix,
$ \hat{m}_{kj} = m_k \delta_{kj} $,
with
$ m_k \geq 0 $.
Hence,
the Dirac mass term (\ref{Dirac})
can be written in the diagonal form
\begin{equation}
\mathcal{L}^{\mathrm{D}}
=
- \sum_{k=1}^{3} m_k \, \overline{\nu_k} \, \nu_k
\,,
\label{Dirac diagonal}
\end{equation}
with
\begin{equation}
\nu_{{\ell}L}
=
\sum_{k=1}^{3} U_{{\ell}k} \, \nu_{kL}
\qquad
(\ell=e,\mu,\tau)
\,.
\label{Dirac mixing}
\end{equation}
Therefore,
in the case of the Dirac mass term (\ref{Dirac})
the three flavour fields 
$\nu_{{\ell}L}$ ($\ell=e,\mu,\tau$)
are linear unitary combinations of
the left-handed components $\nu_{kL}$ of three fields
of neutrinos with masses
$m_k$ ($k=1,2,3$). On the other hand, we also have
\begin{equation}
\nu_{\ell R}
=
\sum_{k=1}^{3} V_{\ell k} \, \nu_{kR}
\qquad
(\ell=e,\mu,\tau)
\,,
\label{Dirac mixing R}
\end{equation}
but these right-handed fields do not occur in the standard weak
interaction Lagrangian. Therefore, right-handed singlets are
sterile and are not mixed in this scheme with the active
neutrinos.\footnote{In the Dirac case it is possible to conceive
also the existence of left-handed sterile singlet fields mixed with the
active flavour neutrino fields, in contrast to the right-handed
singlets which cannot be mixed with the active neutrinos because
of lepton number conservation.}

The field $\nu_k$ is a Dirac field if not only the mass term
(\ref{Dirac}) but also the total Lagrangian 
is invariant under the global U(1) transformation
\begin{equation}
\nu_\ell \to e^{i\varphi} \, \nu_\ell
\,,
\qquad
\ell \to e^{i\varphi} \, \ell
\qquad
(\ell=e,\mu,\tau)
\,,
\label{global gauge transformations}
\end{equation}
where the phase $\varphi$ is the same for all neutrino 
and charged lepton fields. Then,
using Noether's theorem,
one can see that
the invariance of the Lagrangian under the transformation
(\ref{global gauge transformations}) implies that
the total lepton charge $L$ is conserved. Therefore, $L$ is the 
quantum number that distinguishes a neutrino from an antineutrino.
 
The Dirac mass term (\ref{Dirac})
allows processes like
$ \mu \to e + \gamma $, $ \mu^- \to e^- + e^+ + e^- $.
However,
the contribution of neutrino mixing to the probabilities of such
processes is negligibly small
\cite{Petcov77,Cheng-Li77,Lee-Shrock77}.

The unitary $3\times3$ mixing matrix $U$
can be written in terms of
3 mixing angles and 6 phases.
However, only one phase is measurable
\cite{Kobayashi-Maskawa73}.
This is due to the fact that in the Standard Model the only term in the Lagrangian
where the mixing matrix $U$ enters
is the CC interaction Lagrangian (\ref{CC}).
With neutrino mixing the lepton charged-current
is given by
\begin{equation}
{j_{\rho}^{\mathrm{CC}}}^\dagger
=
2
\sum_{\ell=e,\mu,\tau}
\overline{\ell_L}
\,
\gamma_{\rho}
\,
\nu_{{\ell}L}
=
2
\sum_{\ell=e,\mu,\tau}
\sum_{k=1}^{3}
\overline{\ell_L}
\,
\gamma_{\rho}
\,
U_{{\ell}k}
\,
\nu_{kL}
\,.
\label{U01}
\end{equation}
Because the phases of Dirac fields are arbitrary one can eliminate from
the mixing matrix $U$ five phases
by a redefinition of the
phases of the charged lepton and neutrino fields, with
only one physical phase remaining in $U$.
The presence of this phase causes the violation of CP invariance
in the lepton sector.
A convenient parameterization of the mixing matrix $U$
is the one proposed in Ref.~\cite{Chau-Keung84}:
\begin{equation}
U
=
\left(
\begin{array}{ccc}
c_{12}
c_{13}
&
s_{12}
c_{13}
&
s_{13}
e^{i\delta_{13}}
\\
-
s_{12}
c_{23}
-
c_{12}
s_{23}
s_{13}
e^{i\delta_{13}}
&
c_{12}
c_{23}
-
s_{12}
s_{23}
s_{13}
e^{i\delta_{13}}
&
s_{23}
c_{13}
\\
s_{12}
s_{23}
-
c_{12}
c_{23}
s_{13}
e^{i\delta_{13}}
&
-
c_{12}
s_{23}
-
s_{12}
c_{23}
s_{13}
e^{i\delta_{13}}
&
c_{23}
c_{13}
\end{array}
\right)
\,,
\label{par3}
\end{equation}
where
$ c_{ij} \equiv \cos\vartheta_{ij} $
and
$ s_{ij} \equiv \sin\vartheta_{ij} $
and
$\delta_{13}$ is the CP-violating phase.

Since in the parameterization (\ref{par3})
of the mixing matrix
the CP-violating phase $\delta_{13}$ is associated with $s_{13}$,
it is clear that CP violation
is negligible in the lepton sector if the mixing angle $\vartheta_{13}$
is small.
More generally,
it is possible to show that if any of the elements
of the mixing matrix is zero,
the CP-violating phase can be rotated away by a suitable
rephasing of the charged lepton and neutrino
fields.\footnote{\label{jarlskog}This can also be seen
by noticing that the Jarlskog rephasing-invariant parameter
\cite{Jarlskog,DGW,Dunietz}
vanishes if one of the elements of the mixing matrix is zero.}

\subsection{Dirac--Majorana mass term}
\label{Dirac-Majorana mass term}

If none of the lepton numbers is conserved and both, 
left-handed flavour fields $\nu_{\ell L}$ ($\ell=e,\mu,\tau$) and
sterile right-handed gauge singlet fields $\nu_{sR}$ ($s=s_1,s_2,\ldots$)
enter into the mass term
we have the so-called Dirac--Majorana mass term
\begin{equation}
\mathcal{L}^{\mathrm{D+M}}
=
\mathcal{L}^{\mathrm{M}}_L
+
\mathcal{L}^{\mathrm{D}}
+
\mathcal{L}^{\mathrm{M}}_R
\,,
\label{Dirac-Majorana}
\end{equation}
with
\begin{equationarrayzero}
&&
\mathcal{L}^{\mathrm{D}}
=
-
\sum_{s,\ell}
\overline{\nu_{sR}}
\,
M^{\mathrm{D}}_{s\ell}
\,
\nu_{{\ell}L}
+
\mathrm{h.c.}
\,,
\label{LMD}
\\
&&
\mathcal{L}^{\mathrm{M}}_L
=
- \frac{1}{2}
\sum_{\ell,\ell'}
\overline{(\nu_{{\ell}L})^c}
\,
M^{L}_{\ell\ell'}
\,
\nu_{{\ell'}L}
+
\mathrm{h.c.}
\,,
\label{LML}
\\
&&
\mathcal{L}^{\mathrm{M}}_R
=
- \frac{1}{2}
\sum_{s,s'}
\overline{\nu_{sR}}
\,
M^{R}_{ss'}
\,
(\nu_{s'R})^c
+
\mathrm{h.c.}
\label{LMR}
\end{equationarrayzero}%
Here $M^{\mathrm{D}}$, 
$M^L$ and $M^R$
are complex matrices and the indices $s,s'$ run over $n_R$ values.
Let us stress that
the number of right-handed singlet fields
could be different from the number of neutrino flavours
(see \cite{Schechter-Valle80a,Schechter-Valle80b}).

The charge-conjugate fields are defined by
\begin{equation}
(\nu_{{\ell}L})^c \equiv \mathcal{C} \, \overline{\nu_{{\ell}L}}^T
\,,
\qquad
(\nu_{sR})^c \equiv \mathcal{C} \, \overline{\nu_{sR}}^T
\,,
\label{charge-conjugated}
\end{equation}
where $\mathcal{C}$ is the charge conjugation matrix (see Appendix A).
Notice that $(\nu_{{\ell}L})^c$ and 
$(\nu_{sR})^c$ are right-handed and left-handed components, respectively.
Indeed, with $\mathcal{C}^{-1} \gamma_5 \mathcal{C} = \gamma_5^T$ one
finds
\begin{equation}
\frac{1+\gamma_5}{2} (\nu_{{\ell}L})^c = 
\mathcal{C} \left( \overline{\nu_{{\ell}L}} \frac{1+\gamma_5}{2} \right)^T =
\mathcal{C} \overline{\nu_{{\ell}L}}^T =
(\nu_{{\ell}L})^c
\end{equation}
and the analogous reasoning holds for $(\nu_{sR})^c$.
Furthermore, we have
\begin{equation}
\overline{(\nu_{{\ell}L})^c} = - \nu_{{\ell}L}^T \, \mathcal{C}^{-1}
\,,
\qquad
\overline{(\nu_{sR})^c} = - \nu_{sR}^T \, \mathcal{C}^{-1}
\,.
\label{hermitian of charge-conjugated}
\end{equation}

The matrices $M^{L}$ and $M^{R}$
are symmetric. This can be shown with the help of the relation
$\nu_{\ell L}^T \mathcal{C}^{-1} \nu_{\ell' L} =
\nu_{\ell' L}^T \mathcal{C}^{-1} \nu_{\ell L}$,
which follows from the fact that $\mathcal{C}$ is an antisymmetric
matrix and from the
anticommutation property of fermion fields. An analogous relation holds for
the right-handed fields.
Then, using Eq.(\ref{hermitian of charge-conjugated}),
one obtains
\begin{equation}
\sum_{\ell,\ell'}
\overline{(\nu_{{\ell}L})^c}
\,
M^{L}_{\ell\ell'}
\,
\nu_{{\ell'}L}
=
-
\sum_{\ell,\ell'}
\nu_{{\ell'}L}^T
\,
\mathcal{C}^{-1}
\,
M^{L}_{\ell\ell'}
\,
\nu_{{\ell}L}
=
\sum_{\ell,\ell'}
\overline{(\nu_{{\ell}L})^c}
\,
M^{L}_{\ell'\ell}
\,
\nu_{{\ell'}L}
\,.
\label{transposing the mass terms}
\end{equation}

Let us introduce the left-handed column vector
\begin{equation}
n_L
\equiv
\left(
\begin{array}{c} \displaystyle
\nu_L
\\ \displaystyle
(\nu_R)^c
\end{array}
\right)
\quad \mbox{with} \quad
\nu_L
\equiv
\left(
\begin{array}{c} \displaystyle
\nu_{eL}
\\ \displaystyle
\nu_{{\mu}L}
\\ \displaystyle
\nu_{{\tau}L}
\end{array}
\right)
\,,
\label{nL}
\end{equation}
where $\nu_R$ is the column vector of right-handed fields.
With the relation
\begin{equation}
\sum_{s,\ell}
\overline{\nu_{sR}}
\,
M^{\mathrm{D}}_{s\ell}
\,
\nu_{{\ell}L}
=
-
\sum_{s,\ell}
\nu_{{\ell}L}^T
\,
M^{\mathrm{D}}_{s\ell}
\,
\overline{\nu_{sR}}^T
=
\sum_{s,\ell}
\overline{(\nu_{{\ell}L})^c}
\,
(M^{\mathrm{D}})^T_{\ell s}
\,
(\nu_{sR})^c \,,
\end{equation}
the Dirac--Majorana mass term $\mathcal{L}^{\mathrm{D+M}}$
can be written in the form
\begin{equation}
\mathcal{L}^{\mathrm{D+M}}
=
- \frac{1}{2}
\,
\overline{(n_L)^c}
\,
M^{\mathrm{D+M}}
\,
n_L
+
\mathrm{h.c.}
\,,
\label{LDM}
\end{equation}
with the symmetric $(3+n_R) \times (3+n_R)$ matrix
\begin{equation}
M^{\mathrm{D+M}}
\equiv
\left(
\begin{array}{cc} \displaystyle
M^L & (M^{\mathrm{D}})^T
\\ \displaystyle
M^{\mathrm{D}} & M^R
\end{array}
\right)
\,.
\label{MDM}
\end{equation}

The complex symmetric matrix
$M^{\mathrm{D+M}}$
can be diagonalized with the help of a unitary matrix $U$ \cite{schur,zumino}:
\begin{equation}
M^{\mathrm{D+M}}
=
(U^\dagger)^T \, \hat{m} \, U^\dagger
\,,
\label{MDMdiag}
\end{equation}
with
$ \hat{m}_{kj} = m_{k} \delta_{kj} $ and $m_{k} \geq 0$.
Using this relation the mass term (\ref{LDM})
can be written as
\begin{equation}
\mathcal{L}^{\mathrm{D+M}}
=
- \frac{1}{2} \, \overline{\Nmass} \, \hat{m} \, \Nmass
=
- \frac{1}{2}
\sum_{k=1}^{3+n_R}
m_k \, \overline{\nu_k} \, \nu_k
\,,
\label{LDMdiag}
\end{equation}
with
\begin{equation}
\Nmass
\equiv
\left(
\begin{array}{c} \displaystyle
\nu_1
\\ \displaystyle 
\nu_2
\\ \displaystyle 
\vdots
\end{array}
\right)
=
U^\dagger \, n_L + (U^\dagger n_L)^c
\,.
\label{Nmass}
\end{equation}
The fields $\nu_k$ satisfy the Majorana condition
\begin{equation}
(\nu_k)^c = \nu_k
\qquad
(k=1,\ldots,3+n_R)
\,.
\label{Majorana}
\end{equation}
Therefore,
in the general case of a Dirac--Majorana mass term
the fields of particles with definite masses
are \emph{Majorana fields}. 
Majorana particles are particles with spin $1/2$ and all charges
equal to zero (particle $=$ antiparticle). 
Some properties of
Majorana fields are discussed in Appendix~A.

We want to emphasize that it is natural that the diagonalization of the mass
term (\ref{LDM}) leads to fields of Majorana particles with definite
masses: the mass
term (\ref{LDM}) for the case of a general matrix $M^{\mathrm{D+M}}$
is not invariant
under any global phase transformation.
In other words,
in the general case of the Dirac and Majorana mass term
there are no conserved quantum numbers
that allow to
distinguish a particle from its antiparticle.

From Eq.(\ref{Nmass}), for the left-handed components
of the neutrino fields we have
the mixing relations
\begin{equationarrayzero}
&&
\nu_{{\ell}L}
=
\sum_{k=1}^{3+n_R}
U_{{\ell}k} \, \nmass_{kL}
\qquad
(\ell=e,\mu,\tau)
\,,
\label{mixL}
\\
&&
(\nu_{sR})^c
=
\sum_{k=1}^{3+n_R}
U_{sk} \, \nmass_{kL}
\,.
\label{mixR}
\end{equationarrayzero}%
Thus,
in the case of a Dirac--Majorana mass term the flavour fields
$\nu_{\ell L}$
are linear unitary combinations of the left-handed components 
of the Majorana fields
$\nmass_k$
of neutrinos with definite masses.
These components are also connected with
the sterile fields $(\nu_{sR})^c$
through the relation (\ref{mixR}).
If the masses $m_k$ are small,
the mixing relations (\ref{mixL}) and (\ref{mixR}) imply that
flavour neutrinos $\nu_e$, $\nu_{\mu}$ and $\nu_{\tau}$ can
oscillate into sterile states that are quanta of
the right-handed fields
$\nu_{sR}$.
We will discuss such transitions in Section 6.

Let us stress that in order to have all three terms in the
Dirac--Majorana mass term (\ref{LDM}) not only a Higgs doublet but
also a Higgs triplet 
\cite{Gelmini-Roncadelli81,Georgi83,JEKim87}
and a Higgs singlet are necessary.
Thus,
it can be generated only in
the framework of models beyond the Standard Model.
A typical example is the SO(10) model
(see, for example, \cite{Mohapatra-Pal91}). For a discussion of
radiative corrections to the Dirac--Majorana mass term see
Ref.~\cite{grimus89}. 

Up to now we did not make any special assumption about the
Dirac--Majorana mass term (\ref{LDM}).
Let us now consider the possibility
that CP invariance in the lepton sector holds.
In this case we have
\begin{equation}
U_{\mathrm{CP}}
\,
\mathcal{L}^{\mathrm{D+M}}(x)
\,
U_{\mathrm{CP}}^{-1}
=
\mathcal{L}^{\mathrm{D+M}}(x_{\mathrm{P}})
\,.
\label{LCP}
\end{equation}
where $U_{CP}$ is the operator of CP conjugation,
$x\equiv(x^0,\vec{x})$
and
$x_{\mathrm{P}}\equiv(x^0,-\vec{x})$.
For the fields $n_{L}(x)$ we take the usual CP transformation
\begin{equation}
U_{\mathrm{CP}}
\,
n_{L}(x)
\,
U_{\mathrm{CP}}^{-1}
=
\eta
\,
\gamma^0
\,
\mathcal{C}
\,
\overline{n_{L}}^T(x_{\mathrm{P}})
\,,
\label{nCP}
\end{equation}
where $\eta$ is a diagonal matrix of phase factors.
The mass term $\mathcal{L}^{\mathrm{D+M}}$
can be written in the form
\begin{equation}
\mathcal{L}^{\mathrm{D+M}}(x)
=
\frac{1}{2}
n_L^T(x)
\,
\mathcal{C}^{-1}
\,
M^{\mathrm{D+M}}
\,
n_L(x)
-
\frac{1}{2}
\overline{n_L}(x)
\,
(M^{\mathrm{D+M}})^\dagger
\,
\mathcal{C}
\,
\overline{n_L}^T(x)
\,.
\label{LDM2}
\end{equation}
Using Eq.(\ref{nCP}) we obtain
\begin{equation}
U_{\mathrm{CP}}
\,
n_L^T(x)
\,
\mathcal{C}^{-1}
\,
M^{\mathrm{D+M}}
\,
n_L(x)
\
U_{\mathrm{CP}}^{-1}
=
\overline{n_L}(x_{\mathrm{P}})
\,
\eta M^{\mathrm{D+M}} \eta
\,
\mathcal{C}
\,
\overline{n_L}^T(x_{\mathrm{P}})
\,.
\label{CP0}
\end{equation}
Hence,
the CP invariance condition (\ref{LCP})
is satisfied if
\begin{equation}
\eta M^{\mathrm{D+M}} \eta
=
-
(M^{\mathrm{D+M}})^\dagger
\,.
\label{CP1}
\end{equation}
Up to now $\eta$ was an arbitrary diagonal matrix of phase factors.
If we choose $\eta = i$,
in this case
CP invariance in the lepton sector implies that
\begin{equation}
M^{\mathrm{D+M}}
=
(M^{\mathrm{D+M}})^\dagger
\,.
\label{CP2}
\end{equation}
Taking into account that $M^{\mathrm{D+M}}$
is a symmetric matrix,
the condition (\ref{CP2}) is equivalent to
\begin{equation}
M^{\mathrm{D+M}}
=
(M^{\mathrm{D+M}})^*
\,.
\label{CP3}
\end{equation}
The real symmetric matrix $M^{\mathrm{D+M}}$ can be diagonalized with
the transformation
\begin{equation}
M^{\mathrm{D+M}}
=
\mathcal{O} \, m' \, \mathcal{O}^T
\,,
\label{orth}
\end{equation}
where $\mathcal{O}$
is an orthogonal matrix ($\mathcal{O}^T=\mathcal{O}^{-1}$)
and $m'$ is a diagonal matrix,
$ m'_{kj} = m'_k \delta_{kj} $.
The eigenvalues $m'_k$
can be positive or negative and the neutrino masses are thus given by
$ m_k = |m'_k| $. Therefore, we write the eigenvalues as
\begin{equation}
m'_k = m_k \, \rho_k
\,,
\label{mass-eigenvalues}
\end{equation}
where
$ \rho_k = \pm 1 $
is the sign of the $k^{\mathrm{th}}$
eigenvalue of the matrix $M^{\mathrm{D+M}}$.
Then the relation (\ref{orth}) can be rewritten in the form
\begin{equation}
M^{\mathrm{D+M}}
=
(U^\dagger)^T \, \hat{m} \, U^\dagger
\label{orth2}
\end{equation}
with
\begin{equation}
U^\dagger = \sqrt{\rho} \, \mathcal{O}^T
\label{orth3}
\end{equation}
and $ \hat{m}_{kj} = m_k \, \delta_{kj} $.
Here
$\rho$
is the diagonal matrix with elements
$ \rho_{kj} = \rho_k \delta_{kj} $.
Therefore,
if CP is conserved in the lepton sector,
the neutrino mixing matrix has the simple structure
shown in Eq.(\ref{orth3}), which implies that
\begin{equation}
U^* = U \, \rho
\,.
\label{orth4}
\end{equation}

The CP transformation of the Majorana field $\nmass_k$
is given by (see Appendix A)
\begin{equation}
U_{\mathrm{CP}}
\,
\nmass_k(x)
\,
U_{\mathrm{CP}}^{-1}
=
\eta^{\mathrm{CP}}_k \, \gamma_0 \, \nmass_k(x_\mathrm{P})
\,,
\label{NmassCP1}
\end{equation}
where
$\eta^{\mathrm{CP}}_k$
is the CP parity of the Majorana field $\nmass_k$
which is
$ \eta^{\mathrm{CP}}_k = \pm i $
(see Ref.~\cite{Bilenky-Petcov87}).
From Eq.(\ref{NmassCP1}) we have
\begin{equation}
U_{\mathrm{CP}}
\,
\nmass_{kL}(x)
\,
U_{\mathrm{CP}}^{-1}
=
\eta^{\mathrm{CP}}_k \, \gamma_0 \, \nmass_{kR}(x_\mathrm{P})
\,.
\label{NmassCP2}
\end{equation}
This relation and Eqs.(\ref{Nmass}), (\ref{nCP}) and (\ref{orth4}) imply
that the CP parity of the Majorana field $\nmass_k$ is 
\cite{Wolfenstein81-PL,Bilenky-Nedelcheva-Petcov84,Kayser84}
\begin{equation}
\eta^{\mathrm{CP}}_k = i \, \rho_k
\,.
\label{etaCPk}
\end{equation}
Indeed, we have
\begin{equation}
U_{\mathrm{CP}}
\,
\Nmass_{L}(x)
\,
U_{\mathrm{CP}}^{-1}
=
U^\dagger \, i \, \gamma_0 \, \mathcal{C} \, \overline{n_L}^T(x_{\mathrm{P}})
=
U^\dagger \, U^* \, i \, \gamma_0 \, \mathcal{C} \, \overline{\Nmass_L}^T(x_{\mathrm{P}})
=
i \, \rho \, \gamma_0 \, \Nmass_R(x_{\mathrm{P}})
\,.
\label{NmassCP3}
\end{equation}

\subsection{Majorana mass term}
\label{Majorana mass term}

If only left-handed neutrino flavour fields
$\nu_{{\ell}L}$ ($\ell=e,\mu,\tau$) enter into the Lagrangian 
we can write down the mass term \cite{Pontecorvo69,BP87}
\begin{equation}
\mathcal{L}^{\mathrm{M}}
=
- \frac{1}{2}
\sum_{\ell,\ell'}
\overline{(\nu_{{\ell}L})^c}
\,
M^{L}_{\ell\ell'}
\,
\nu_{{\ell'}L}
\,,
\label{Majorana mass}
\end{equation}
where $M^{L}$ is a symmetric complex matrix (see previous section).
Then we have for the mixing 
\begin{equation}
\nu_{{\ell}L}
=
\sum_{k=1}^{3} U_{{\ell}k} \, \nu_{kL}
\,,
\label{Majorana mixing}
\end{equation}
where
$\nu_k$
is the field of Majorana neutrinos with mass $m_k$.
In this case
the number of massive Majorana neutrinos is equal
to the number of lepton flavours (three).
Note that the generation
of the Majorana mass term (\ref{Majorana mass}) requires an
enlargement of the scalar sector of the Minimal Standard Model
\cite{konetschny}:  
with a Higgs triplet like in the Majoron models 
Ref.~\cite{Gelmini-Roncadelli81,Georgi83,JEKim87} Majorana masses are
obtained at the tree level, whereas 
radiative generation is possible at the 1-loop level with a singly
charged Higgs singlet (plus an additional Higgs doublet)
\cite{Zee80,Zee85,Zee86,Voloshin88,Babu-Mohapatra91,Mohapatra-Pal91})
or at the 2-loop level with a doubly charged Higgs singlet (plus an
additional singly charged scalar) \cite{babu88}.

Since the Majorana condition (\ref{Majorana})
does not allow the rephasing
of the neutrino fields,
only three of the six phases in the
$3\times3$ mixing matrix $U$
can be absorbed into
the charged lepton fields
in the charged current (\ref{U01}).
Therefore,
in the Majorana case the mixing matrix contains three CP-violating phases
\cite{BHP80,Kobzarev80,Doi81} in contrast
to the single CP-violating phase of the Dirac case discussed in
Section~\ref{Dirac mass term}.
However, the additional CP-violating phases in the Majorana case
have no effect
on neutrino oscillations
in vacuum
\cite{BHP80,Kobzarev80,Doi81}
(see Section~\ref{Neutrino oscillations in vacuum})
as well as in matter
\cite{Langacker87}.

\subsection{The one-generation case}
\label{A simple example}

Let us consider the Dirac--Majorana mass term in the simplest case of one
generation.
We have
\begin{equationarray}
\mathcal{L}^{\mathrm{D+M}}
\null & \null = \null & \null
- \frac{1}{2} \, m_{L}
\,
\overline{(\nu_L)^c}
\,
\nu_L
-
m_{\mathrm{D}}
\,
\overline{\nu_R}
\,
\nu_L
- \frac{1}{2} \, m_{R}
\,
\overline{\nu_R}
\,
(\nu_R)^c
+
\mathrm{h.c.}
\nonumber
\\
\null & \null = \null & \null
- \frac{1}{2} \, m_{R}
\,
\overline{(n_L)^c}
\,
M
\,
n_L
+
\mathrm{h.c.}
\,,
\label{two01}
\end{equationarray}%
with
\begin{equation}
n_L
\equiv
\left(
\begin{array}{c} \displaystyle
\nu_L
\\ \displaystyle
(\nu_R)^c
\end{array}
\right)
\,,
\qquad
M
\equiv
\left(
\begin{array}{cc} \displaystyle
m_{L}
\null & \null \displaystyle
m_{\mathrm{D}}
\\ \displaystyle
m_{\mathrm{D}}
\null & \null \displaystyle
m_{R}
\end{array}
\right)
\,.
\label{two02}
\end{equation}

For simplicity we assume CP invariance 
in the lepton sector (see Subsection 2.2).
In this case $m_{L}$, $m_{\mathrm{D}}$ and $m_{R}$
are real parameters.
In order to diagonalize the matrix $M$,
let us write it in the form
\begin{equation}
M
=
\frac{1}{2}
\,
\mathrm{Tr}M
+
M^0
=
\frac{1}{2} \, ( m_L + m_R )
+
M^0
\,,
\label{two03}
\end{equation}
with
\begin{equation}
M^0
=
\left(
\begin{array}{cc} \displaystyle
- \frac{1}{2} \, ( m_R - m_L )
\null & \null \displaystyle
m_{\mathrm{D}}
\\ \displaystyle
m_{\mathrm{D}}
\null & \null \displaystyle
\frac{1}{2} \, ( m_R - m_L )
\end{array}
\right)
\,.
\label{two04}
\end{equation}
The eigenvalues of the matrix $M^0$ are
\begin{equation}
m^0_{1,2}
=
\mp \frac{1}{2} \,
\sqrt{ ( m_R - m_L )^2 + 4 \, m_{\mathrm{D}}^2 }
\,.
\label{two05}
\end{equation}
Furthermore, we have
\begin{equation}
M^0 = \mathcal{O} \, m^0 \mathcal{O}^T
\,,
\label{two06}
\end{equation}
where $ m^0 = \mathrm{diag}(m^0_1,m^0_2) $
and
$\mathcal{O}$
is the orthogonal matrix
\begin{equation}
\mathcal{O}
=
\left(
\begin{array}{rr} \displaystyle
\cos\vartheta
\null & \null \displaystyle
\sin\vartheta
\\ \displaystyle
- \sin\vartheta
\null & \null \displaystyle
\cos\vartheta
\end{array}
\right)
\label{two07}
\end{equation}
with the mixing angle $\vartheta$ given by
\begin{equation}
\cos 2\vartheta = 
\frac{m_R-m_L}{\sqrt{ ( m_R - m_L )^2 + 4 \, m_{\mathrm{D}}^2 }} 
\,, \quad
\sin 2\vartheta = 
\frac{2m_D}{\sqrt{ ( m_R - m_L )^2 + 4 \, m_{\mathrm{D}}^2 }} \,.
\label{two08}
\end{equation}
Hence, the matrix $M$ can be written as
\begin{equation}
M = \mathcal{O} \, m' \, \mathcal{O}^T
\,,
\label{two09}
\end{equation}
where 
$ m' = \mathrm{diag}(m'_1,m'_2) $
and
\begin{equation}
m'_{1,2}
=
\frac{1}{2}
\left[
( m_R + m_L )
\mp \sqrt{ ( m_R - m_L )^2 + 4 \, m_{\mathrm{D}}^2 }
\right]
\,.
\label{two10}
\end{equation}
The eigenvalues of the matrix $M$ are real but can have positive or negative sign.
Let us write them as
\begin{equation}
m'_k = m_k \, \rho_k
\,,
\label{two11}
\end{equation}
where
$ m_k = |m'_k| $
and
$\rho_k$
is the sign of the
$k^{\mathrm{th}}$ eigenvalue of the matrix $M$.
As shown in Eq.(\ref{etaCPk}),
the CP parity of the Majorana field $\nu_k$ with definite mass $m_k$
is $ \eta^{\mathrm{CP}}_k = i \rho_k $.
The relation (\ref{two09})
can be written in the form
\begin{equation}
M
=
(U^\dagger)^T \, \hat{m} \, U^\dagger
\,,
\label{two12}
\end{equation}
with
$ \hat{m} = \mathrm{diag}(m_1,m_2) $
and
\begin{equation}
U^\dagger
\equiv
\sqrt{\rho} \, \mathcal{O}^T
\,,
\label{two13}
\end{equation}
with
$ \rho = \mathrm{diag}(\rho_1,\rho_2) $.
Using now the general formulas obtained in
Section~\ref{Dirac-Majorana mass term},
we have the mixing relation
\begin{equation}
\left( \begin{array}{c} \nu_L \\ (\nu_R)^c \end{array} \right) = 
U \, 
\left( \begin{array}{c} \nu_{1L} \\ \nu_{2L} \end{array} \right)
\,,
\label{two141}
\end{equation}
with
\begin{equation}
U
=
\left(
\begin{array}{rr} \displaystyle
(\sqrt{\rho_1})^* \, \cos\vartheta
& \displaystyle
(\sqrt{\rho_2})^* \, \sin\vartheta
\\ \displaystyle
- (\sqrt{\rho_1})^* \, \sin\vartheta
& \displaystyle
(\sqrt{\rho_2})^* \, \cos\vartheta
\end{array}
\right)
\,.
\label{two14}
\end{equation}

Therefore,
the three parameters $m_{L}$, $m_{\mathrm{D}}$, $m_{R}$
are related with
the mixing angle $\vartheta$
and the neutrino masses $m_k$
by the relations (\ref{two08}) and (\ref{two10}), (\ref{two11}).
The signs of the eigenvalues of $M$ determine
the CP parities of the massive Majorana fields $\nu_k$.

In the framework of CP conservation, the relations obtained so far are general.
In the following part of this section
we consider some particular cases
with special physical significance.

\subsection{The see-saw mechanism}
\label{The see-saw mechanism}

Let us consider
the Dirac--Majorana mass term (\ref{two01})
and assume \cite{GRS79,Yanagida79,MS80} that
$ m_{L} = 0 $,
$ m_{\mathrm{D}} \simeq m^f$,
where $m^{f}$
is the mass of a quark or a charged lepton
of the same generation,
and $ m_R \simeq \mathcal{M} \gg m^f $.
In this case,
from the relations (\ref{two08}), (\ref{two10}) and (\ref{two11})
we have
\begin{equationarrayzero}
&&
\vartheta \simeq \frac{m^f}{\mathcal{M}} \ll 1
\,,
\label{ss03}
\\
&&
m_1 \simeq \frac{(m^f)^2}{\mathcal{M}} \ll m^f
\,,
\qquad
\rho_1 = -1
\,,
\label{ss01}
\\
&&
m_2 \simeq \mathcal{M}
\,,
\qquad \qquad
\rho_2 = 1
\,.
\label{ss02}
\end{equationarrayzero}%
These relations imply that approximately
\begin{equation}
\nu_L = - i \, \nu_{1L}
\,,
\qquad
(\nu_R)^c = \nu_{2L}
\,,
\label{ss04}
\end{equation}
and the Majorana fields $\nu_1$, $\nu_2$
are connected with the fields $\nu_{L}$
and $\nu_{R}$ by 
\begin{equation}
\nu_1 = i \, [ \nu_L - (\nu_L)^c ]
\,,
\qquad
\nu_2 = \nu_R + (\nu_R)^c
\,.
\label{ss06}
\end{equation}

The mechanism which we consider here is called see-saw mechanism 
\cite{GRS79,Yanagida79,MS80}.
It is based on the assumption that the conservation of
the total lepton number $L$
is violated by the right-handed Majorana mass term at the scale
$\mathcal{M}$ that is
much larger than the scale of the electroweak
symmetry breaking.
Several models which implement the see-saw mechanism are possible
(see, for example, \cite{Mohapatra-Pal91,Valle91,Gelmini-Roulet95}
and references therein)
and the scale $\mathcal{M}$
depends on the model.
This scale could be as low as the TeV scale
(for example, in left-right symmetric models
\cite{pati,MS81,Mohapatra-Pal91})
or an intermediate scale,
or as high as the grand unification scale
$ \sim 10^{15} \, \mathrm{GeV} $
or even the Plank scale
$ \sim 10^{19} \, \mathrm{GeV} $.
The great attractiveness of the see-saw model
lies in the fact that,
through the relation (\ref{ss01}),
it gives an explanation
of the smallness of neutrino masses with respect to the masses of other
fundamental fermions.\footnote{Note that the see-saw mechanism can also be
realized for Dirac neutrinos \cite{ecker87,branco89}.}

In the case of three generations
the see-saw mechanism leads to
a spectrum of masses of Majorana particles
with three light neutrino masses $m_k$
and three very heavy masses $M_k$ ($k=1,2,3$)
of the order of the scale of violation of the lepton numbers. 
This is realized if the mass matrix
(\ref{MDM}) 
has $M^L=0$,
\begin{equation}
M^{\mathrm{D+M}}
\equiv
\left(
\begin{array}{cc} \displaystyle
0 & \displaystyle (M^{\mathrm{D}})^T
\\ \displaystyle
M^{\mathrm{D}} & \displaystyle M^R
\end{array}
\right)
\,,
\label{ss08}
\end{equation}
and $M^R$
is such that all its eigenvalues
are much bigger than the elements of
$M^{\mathrm{D}}$.
In this case
the mass matrix is block-diagonalized
(up to corrections of order $(M^R)^{-1}M^{\mathrm{D}}$)
by the unitary transformation
\begin{equation}
W^T
\,
M^{\mathrm{D+M}}
\,
W
\simeq
\left(
\begin{array}{cc} \displaystyle
M_{\mathrm{light}} & \displaystyle0
\\ \displaystyle
0 & \displaystyle M_{\mathrm{heavy}}
\end{array}
\right)
\,,
\label{ss09}
\end{equation}
with
\begin{equation}
W
\simeq
\left(
\begin{array}{cc} \displaystyle
1 - \frac{1}{2} \, (M^{\mathrm{D}})^\dagger \, (M^R
{(M^R)}^\dagger)^{-1} \, 
M^{\mathrm{D}}
& \displaystyle
(M^{\mathrm{D}})^\dagger \, {(M^R)^\dagger}^{-1}
\\ \displaystyle
- (M^R)^{-1} \, M^{\mathrm{D}}
& \displaystyle
1 - \frac{1}{2} \, (M^R)^{-1} \, M^{\mathrm{D}} \,
(M^{\mathrm{D}})^\dagger \,
{(M^R)^\dagger}^{-1} 
\end{array}
\right)
\,.
\label{ss10}
\end{equation}
The matrices for the light and heavy masses are given by
\cite{Kanaya80,Schechter-Valle82}
\begin{equation}
M_{\mathrm{light}}
\simeq
-(M^{\mathrm{D}})^T \, (M^R)^{-1} \, M^{\mathrm{D}}
\,,
\qquad
M_{\mathrm{heavy}}
\simeq
M^R
\,.
\label{ss11}
\end{equation}
The mass eigenvalues of the light neutrinos
are determined by the specific form
of $M^{\mathrm{D}}$ and $M^R$. 
Note that in left-right symmetric models and in SO(10) models the
matrix $M^L$ in the big matrix $M^\mathrm{D+M}$ (\ref{MDM})
can be important (see, \textit{e.g.},
Ref.~\cite{ml}). 

Two simple possibilities are discussed in the literature
(see Refs.~\cite{Bludman-Kennedy-Langacker92a,Bludman-Kennedy-Langacker92b}):

\begin{enumerate}

\item
If
$ M^R = \mathcal{M} \, I $,
where $I$ is the identity matrix,
one obtains the \emph{quadratic see-saw},
\begin{equation}
M_{\mathrm{light}}
\simeq -
\frac{ (M^{\mathrm{D}})^T \, M^{\mathrm{D}} }{ \mathcal{M} }
\,,
\label{ss12}
\end{equation}
and the light neutrino masses are given by
\begin{equation}
m_k
=
\frac{ (m^f_k)^2 }{ \mathcal{M} }
\qquad
(k=1,2,3)
\,,
\label{ss13}
\end{equation}
where
$m^f_k$
is the mass of a quark or a charged lepton of the
$k^{\mathrm{th}}$ generation.
In this case the neutrino masses $m_k$ scale as the squares of the
masses $m^f_k$:
\begin{equation}
m_1 : m_2 : m_3
=
(m^f_1)^2 : (m^f_2)^2 : (m^f_3)^2
\,.
\label{ss14}
\end{equation}

\item
If
$ M^R = \frac{ \mathcal{M} }{ \mathcal{M}_{\mathrm{D}} } \, M_{\mathrm{D}} $,
where
$\mathcal{M}_{\mathrm{D}}$
characterizes the scale of
$M_{\mathrm{D}}$,
one obtains the \emph{linear see-saw}
(see, for example, Ref.~\cite{Witten80}),
\begin{equation}
M_{\mathrm{light}}
\simeq -
\frac{ \mathcal{M}_{\mathrm{D}} }{ \mathcal{M} }
\, M^{\mathrm{D}}
\,,
\label{ss15}
\end{equation}
and the light neutrino masses are given by
\begin{equation}
m_k
=
\frac{ \mathcal{M}_{\mathrm{D}} }{ \mathcal{M} } \, m^f_k
\qquad
(k=1,2,3)
\,.
\label{ss16}
\end{equation}
In this case the neutrino masses $m_k$ scale as the
masses $m^f_k$:
\begin{equation}
m_1 : m_2 : m_3
=
m^f_1 : m^f_2 : m^f_3
\,.
\label{ss17}
\end{equation}

\end{enumerate}

Let us stress that in any case
the see-saw mechanism implies
the hierarchical relation
\begin{equation}
m_1 \ll m_2 \ll m_3
\label{ss18}
\end{equation}
for the three light Majorana neutrino masses.

\subsection{Effective Lagrangians}
\label{Effective operators}

In the Standard Model
without right-handed singlet neutrino fields
there are no renormalizable interactions that
give masses to the neutrinos
after the spontaneous breaking of the
$ \mathrm{SU(2)}_L \times \mathrm{U(1)}_Y $
symmetry with the Higgs doublet mechanism.
However,
there is a general belief that the Standard Model
is the low-energy manifestation of a more complete theory
\cite{Pati-Salam73,Georgi-Glashow74}
(for reviews see Refs.~\cite{Langacker81,Mohapatra-Pal91}).
The effect of this new theory is to induce in the Lagrangian
of the Standard Model
non-renormalizable interactions
which preserve the
$ \mathrm{SU(2)}_L \times \mathrm{U(1)}_Y $
symmetry
above the electroweak symmetry breaking scale,
but violate
the conservation of lepton and baryon numbers
(see Ref.~\cite{Wilczek-Zee79} and references therein).
These non-renormalizable interactions are operators of dimension
$ d > 4 $
and must be multiplied by coupling constants that have dimension
$ \mathcal{M}^{4-d} $,
where $ \mathcal{M} $ is a mass scale characteristic of the new theory.
It is clear that the dominant effects at low energies
are produced by the operators with lowest dimension.

In the Standard Model the lepton number non-conserving operator
with minimum dimension that can generate a neutrino mass is\footnote{This
operator can be written also as
$$
\frac{ 1 }{ \mathcal{M} }
\sum_{\ell,\ell'}
g_{\ell\ell'}
\,
( L_\ell^T \, \sigma_2 \, \phi )
\, \mathcal{C}^{-1} \,
( \phi^T \, \sigma_2 \, L_{\ell'} )
+
\mathrm{h.c.}
$$}
\cite{Weinberg79,Weinberg80,Weldon-Zee80,Weinberg87,ABS92,ABST93}
\begin{equation}
\frac{ 1 }{ \mathcal{M} }
\sum_{\ell,\ell'}
\frac{ g_{\ell\ell'} }{ 2 }
\,
( L_\ell^T \, \mathcal{C}^{-1} \, \sigma_2 \, \vet{\sigma} \, L_{\ell'} )
\,
( \phi^T \, \sigma_2 \, \vet{\sigma} \, \phi )
+
\mathrm{h.c.}
\,,
\label{eo01}
\end{equation}
where $g$ is a $3\times3$ matrix of coupling constants,
$\vet{\sigma}$
are the Pauli matrices,
$L_\ell$ are the standard leptonic doublets
and $\phi$ is the standard Higgs doublet:
\begin{equation}
L_\ell
\equiv
\left(
\begin{array}{c} \displaystyle
\nu_{{\ell}L}
\\ \displaystyle
\ell_L
\end{array}
\right)
\qquad
(\ell=e,\mu,\tau)
\,,
\qquad
\phi
\equiv
\left(
\begin{array}{c} \displaystyle
\varphi^+
\\ \displaystyle
\varphi^0
\end{array}
\right)
\,.
\label{eo02}
\end{equation}
The effective operator in Eq.(\ref{eo01})
has dimension five
and the coupling constant is proportional to $\mathcal{M}^{-1}$.

When
$ \mathrm{SU(2)}_L \times \mathrm{U(1)}_Y $
is broken by the vacuum expectation value
$ v / \sqrt{2} $
of
$\varphi^0$,
the effective interaction (\ref{eo01})
generates the Majorana mass term
\begin{equation}
\mathcal{L}^{\mathrm{M}}
=
\frac{1}{2}
\,
\frac{ v^2 }{ \mathcal{M} }
\sum_{\ell,\ell'}
g_{\ell\ell'}
\,
\nu_{{\ell}L}^T \,
\mathcal{C}^{-1}
\nu_{{\ell'}L}
+
\mathrm{h.c.}
=
-
\frac{1}{2}
\,
\frac{ v^2 }{ \mathcal{M} }
\sum_{\ell,\ell'}
\overline{(\nu_{{\ell}L})^c}
\,
g_{\ell\ell'}
\,
\nu_{{\ell'}L}
+
\mathrm{h.c.}
\,,
\label{eo03}
\end{equation}
where $v \simeq 246$ GeV. In this mass term there is a suppression
factor $v/\mathcal{M}$ which is responsible for the smallness of the
neutrino masses \cite{Weinberg79,Weinberg80}. 

\subsection{Maximal mixing}
\label{Maximal mixing}

The expression (\ref{two08}) for the mixing angle $\vartheta$
implies that the mixing is maximal,
\textit{i.e.},
$ \vartheta = \pi/4 $,
if $m_{R}=m_{L}$.
In this case,
assuming that $ m_{L} \geq |m_{\mathrm{D}}| $, 
the Majorana neutrino masses are given by
\begin{equation}
m_{1,2} = m_L \mp m_\mathrm{D}
\,.
\label{mm01}
\end{equation}
The fields $\nu_{L}$ and $(\nu_{R})^c$
are connected with the fields $\nu_1$ and $\nu_2$
by the relations
\begin{equation}
\nu_L
=
\frac{1}{\sqrt{2}}
\left( \nu_{1L} + \nu_{2L} \right)
\,,
\qquad
(\nu_R)^c
=
\frac{1}{\sqrt{2}}
\left( - \nu_{1L} + \nu_{2L} \right)
\,.
\label{mm02}
\end{equation}
The massive Majorana fields are given by
\begin{equation}
\nu_{1,2}
=
\frac{1}{\sqrt{2}}
\left\{
\left[ \nu_L + (\nu_L)^c \right]
\mp
\left[ \nu_R + (\nu_R)^c \right]
\right\}
\,.
\label{mm04}
\end{equation}

If 
$m_{R}=m_{L}=0$ and $m_{\mathrm{D}}>0$,
the mass term (\ref{two01}) is simply a Dirac mass term.
Applying Eqs.(\ref{two08}) and (\ref{two10}), we obtain
\begin{equation}
\vartheta = \frac{\pi}{4}
\,,
\qquad
m'_{1,2} = \mp m_{\mathrm{D}}
\,.
\label{mm05}
\end{equation}
Using Eq.(\ref{two14}) for the mixing matrix $U$,
one can see that
the fields $\nu_{L}$ and $(\nu_{R})^c$
are connected to the Majorana fields
$\nu_1$ and $\nu_2$
by the relations
\begin{equation}
\nu_L
=
\frac{1}{\sqrt{2}}
\left( - i \nu_{1L} + \nu_{2L} \right)
\,,
\qquad
(\nu_R)^c
=
\frac{1}{\sqrt{2}}
\left( i \nu_{1L} + \nu_{2L} \right)
\,,
\label{mm06}
\end{equation}
where $\nu_{1,2}$ are Majorana fields with the same mass 
$ m_1 = m_2 = m_{\mathrm{D}} $ and with CP parities
$ \eta^{\mathrm{CP}}_1 = - i $
and
$ \eta^{\mathrm{CP}}_2 = + i $ (see Eq.(\ref{etaCPk})).
Eq.(\ref{mm04})
implies for a Dirac field 
\begin{equation}
\nu
=
\frac{1}{\sqrt{2}}
\left( - i \nu_{1} + \nu_{2} \right)
\,.
\label{mm08}
\end{equation}
Thus we arrive at the well-known result that a Dirac field can always be
represented as an equal mixture of two Majorana fields with the same mass and
opposite CP parities.

One can
see this result also directly:
\begin{equation}
\nu
=
- \frac{i}{\sqrt{2}}
\left( i \, \frac{ \nu - \nu^c }{ \sqrt{2} } \right)
+
\frac{1}{\sqrt{2}}
\,
\frac{ \nu + \nu^c }{ \sqrt{2} }
=
- \frac{i}{\sqrt{2}} \, \nu_1
+ \frac{1}{\sqrt{2}} \, \nu_2
\,,
\end{equation}
where 
\begin{equation}
\nu_1 = i \, \frac{ \nu - \nu^c }{ \sqrt{2} }
\,,
\qquad
\nu_2 = \frac{ \nu + \nu^c }{ \sqrt{2} }
\end{equation}
are Majorana fields.

Finally, there is 
the possibility that
$ |m_L| , |m_R| \ll m_{\mathrm{D}} $ but at least one of the
parameters $m_{L,R}$ is non-zero.
In this case Eqs.(\ref{mm06}) and (\ref{mm08}) are approximately valid
and $\nu_{1,2}$ are
two Majorana neutrinos with opposite CP parities and almost degenerate
masses given by
\begin{equation}
m_{1,2}
\simeq
m_{\mathrm{D}} \mp \frac{1}{2} \left( m_L + m_R \right)
\,.
\label{mm10}
\end{equation}
The field $\nu$ is called in this case a \emph{pseudo-Dirac neutrino} field
\cite{Wolfenstein81-NP,Bilenky-Pontecorvo78,Bilenky-Petcov87,%
KLM91,GKL92-PRD46}.

\section{Neutrino oscillations in vacuum}
\label{Neutrino oscillations in vacuum}
\setcounter{equation}{0}
\setcounter{figures}{0}
\setcounter{tables}{0}

\subsection{The general formalism}
\label{The general formalism}

If there is neutrino mixing,
the left-handed components of the neutrino fields
$\nu_{{\alpha}L}$
($\alpha=e,\mu,\tau,s_1,s_2,\ldots$)
are unitary linear combinations of the left-handed components of the
$n$
(Dirac or Majorana)
neutrino fields $\nu_k$ ($k=1,\ldots,n$) with masses $m_k$:
\begin{equation}
\nu_{{\alpha}L}
=
\sum_{k=1}^{n} U_{{\alpha}k} \, \nu_{kL}
\,.
\label{mix01}
\end{equation}
The number $n$ of massive neutrinos
is 3 for the Dirac mass term discussed in
Section~\ref{Dirac mass term}
and for the Majorana mass term discussed in
Section~\ref{Majorana mass term}, in which cases there are only the three
active flavour neutrinos.
The number $n$ of massive neutrinos is more than three in the case of a 
Dirac--Majorana mass term discussed in
Section~\ref{Dirac-Majorana mass term} with a
mixing of both, active and sterile neutrinos. In general, the
number of light massive neutrinos can be more than three.
We enumerate the neutrino masses in such a way that
\begin{equation}
m_1 \leq m_2 \leq m_3 \leq \ldots \leq m_n
\,.
\label{enumerate}
\end{equation}
In this section we will consider in detail the phenomenon of neutrino
oscillations in vacuum which is implied by
the mixing relation (\ref{mix01}). 

If all neutrino mass differences are small,
a state of a flavour neutrino
$\nu_\alpha$
produced in a weak process
(as the $ \pi^+ \to \mu^+ \nu_\mu $ decay, nuclear beta-decays, etc.)
with momentum $p \gg m_k$
is described by the coherent superposition of mass eigenstates
(for a discussion of the quantum mechanical problems
of neutrino oscillations
see Refs.~\cite{Nussinov76,Bilenky-Pontecorvo78,Kayser81,
GKL91-PRD44,GKL92-PRD45,GKL92-PLB274,GKLL93-PRD48,Rich93,Kim93,
Kiers-Nussinov-Weiss96,Grimus-Stockinger96,Campagne97,
Grossman-Lipkin97-PRD55,
Kiers-Weiss98,GKL97-PLB421,GK98-PRD58,
Grimus-Stockinger-Mohanty98,Stodolsky98})
\begin{equation}
|\nu_\alpha\rangle
=
\sum_{k=1}^{n} U_{{\alpha}k}^* \, |\nu_k\rangle
\,.
\label{state}
\end{equation}
Here
$|\nu_k\rangle$
is the state of a neutrino with negative helicity, mass $m_k$
and energy
\begin{equation}
E_k
=
\sqrt{ p^2 + m_k^2 }
\simeq
p + \frac{ m_k^2 }{ 2 p }
\,.
\label{energy}
\end{equation}

Let us assume that at the production point and at time $t=0$
the state of a neutrino is described by Eq.(\ref{state}).
According to the Schr{\"o}dinger equation the mass eigenstates
$|\nu_k\rangle$
evolve in time with the phase factors $\exp (-i E_k t)$
and at the time $t$ at the detection point
we have
\begin{equation}
|\nu_\alpha\rangle_t
=
\sum_{k=1}^{n} U_{{\alpha}k}^* \, e^{ - i E_k t } \, |\nu_k\rangle
\,.
\label{state-t}
\end{equation}
Neutrinos are detected by observing weak interaction processes.
Expanding the state (\ref{state-t})
in the basis of flavour neutrino states
$|\nu_\beta\rangle$,
we obtain
\begin{equation}
|\nu_\alpha\rangle_t
=
\sum_\beta
\mathcal{A}_{\nu_\alpha\to\nu_\beta}(t) \, |\nu_\beta\rangle
\,,
\label{state-t1}
\end{equation}
where
\begin{equation}
\mathcal{A}_{\nu_\alpha\to\nu_\beta}(t)
=
\sum_{k=1}^{n} U_{{\beta}k} \, e^{ - i E_k t } \, U_{{\alpha}k}^*
\label{ampli}
\end{equation}
is the amplitude of
$\nu_\alpha\to\nu_\beta$
transitions at the time $t$
at a distance $L \simeq t$.
Consequently, the probability of this transition is given by
\begin{equation}
P_{\nu_\alpha\to\nu_\beta}
=
|\mathcal{A}_{\nu_\alpha\to\nu_\beta}(t)|^2
=
\left|
\sum_{k=1}^{n} U_{{\beta}k} \, e^{ - i E_k t } \, U_{{\alpha}k}^*
\right|^2
\,.
\label{prob1}
\end{equation}
This formula has a very simple interpretation. $U^*_{\alpha k}$
is the amplitude to find the neutrino mass eigenstate $| \nu_k \rangle$ 
with energy $E_k$ in the state of the flavour neutrino $| \nu_\alpha \rangle$, 
the factor $\exp ( - i E_k t )$ gives the time evolution of the
mass eigenstate and, finally, the term $U_{{\beta}k}$ gives the
amplitude to find the flavour neutrino state $| \nu_\beta \rangle$ 
in the mass eigenstate $| \nu_k \rangle$.
We want to remark that weak interaction processes responsible for
neutrino production and detection involve active
neutrinos. Therefore, strictly speaking, 
in the derivation of the Eqs.(\ref{ampli}) and 
(\ref{prob1}) the indices $\alpha$ and $\beta$ take only active flavour
indices. However, transitions into sterile neutrinos can be
revealed through neutral current neutrino experiments (disappearance
of active neutrinos). In this sense,
the formulas (\ref{ampli}) and (\ref{prob1}) have a meaning
also for transitions between active and sterile states. 

Notice
that in order to have a non-negligible
active-sterile transition probability
the sterile fields must have a mixing with the
light neutrino mass eigenfields the number of which must be more than three.
Such a possibility is phenomenologically given by the Dirac--Majorana
mass term (\ref{Dirac-Majorana}), but it is not realized
in the simple see-saw scheme discussed
in Section~\ref{The see-saw mechanism}, where the scale of the
right-handed Majorana mass term is large.
However, the see-saw scenario can be modified by
additional assumptions to include light sterile neutrinos 
(``singular see-saw'' \cite{CKL98},
``universal see-saw \cite{Koide98,Koide-Fusaoka98}).

At this point a remark concerning the unitarity of the mixing matrix
$U$ is at order. If some of the mass eigenstates
are so heavy that they are not produced in the standard weak processes
then these mass eigenstates will not occur in the flavour state 
$| \nu_\alpha \rangle$ (\ref{state}). Let us assume that the
first $n'$ mass eigenstates are light ($n' < n$). Consequently,
only that part of $U$ plays a role in neutrino oscillations
where $k \leq n'$. In the following we will always assume that
in the situation described here we can confine ourselves to an
$n' \times n'$ submatrix of $U$ which is unitary to a good
approximation (see, \textit{e.g.}, Ref.~\cite{BG93}). This is realized in
the see-saw mechanism with a sufficiently large right-handed scale.
In the further discussion we will drop the distinction between
$n$ and $n'$.

From the relation (\ref{mix01}) it follows that the state describing a flavour 
antineutrino $\bar\nu_\alpha$ is given by
\begin{equation}
|\bar\nu_\alpha\rangle
=
\sum_{k=1}^{n} U_{{\alpha}k} \, |\bar\nu_k\rangle
\,.
\label{anti-state}
\end{equation}
Thus the amplitude of
$\bar\nu_\alpha\to\bar\nu_\beta$
transitions is given by
\begin{equation}
\mathcal{A}_{\bar\nu_\alpha\to\bar\nu_\beta}(t)
=
\sum_{k=1}^{n} U_{{\beta}k}^* \, e^{ - i E_k t } \, U_{{\alpha}k}
\,.
\label{aampli}
\end{equation}
Notice that the amplitude for antineutrino transitions
differs from the corresponding amplitude (\ref{ampli})
for neutrinos only by the exchange $U \to U^*$.

Using the unitarity relation
\begin{equation}
\sum_{k=1}^{n} U_{{\beta}k} \, U_{{\alpha}k}^* =
\delta_{\alpha\beta}\,,
\label{uni1}
\end{equation}
the probability (\ref{prob1})
can be written as
\begin{equation}
P_{\nu_\alpha\to\nu_\beta}
=
\left|
\delta_{\alpha\beta}
+
\sum_{k=2}^{n} U_{{\beta}k} \, U_{{\alpha}k}^*
\left[ \exp\left( - i \, \frac{ \Delta{m}^2_{k1} L }{ 2 E } \right) - 1 \right]
\right|^2
\,,
\label{prob2}
\end{equation}
where $\Delta m^2_{kj} \equiv m^2_k - m^2_j$ and the ultrarelativistic
approximation (\ref{energy}) has been used. Thus the probability
of $\nu_\alpha\to\nu_\beta$ transitions depends on the
elements of the mixing matrix, on $n-1$ independent mass-squared
differences and on the parameter $L/E$, whose range is determined
by the experimental setup.

If there is no mixing ($U=I$) or/and 
$ \Delta{m}^2_{k1} L / E \ll 1 $
for all $k=2,\ldots,n$,
there are no transitions
($P_{\nu_\alpha\to\nu_\beta} = \delta_{\alpha\beta}$).
Neutrino transitions can be observed only if
neutrino mixing takes place
and at least one\footnote{We use the symbol $\Delta{m}^2$
for a generic mass-squared difference.
Hence $\Delta{m}^2$
can be any of the $\Delta{m}^2_{kj}$'s.}
$\Delta{m}^2$
satisfies the condition
\begin{equation}
\Delta{m}^2
\gtrsim
\frac{E}{L}
\,.
\label{cond}
\end{equation}
In this inequality, 
$\Delta{m}^2$ is the neutrino mass-squared in eV$^2$,
$L$ is the distance between neutrino source and
detector in m (km)
and
$E$ is neutrino energy in MeV (GeV).
Thus, the larger the value of the parameter $L/E$,
the smaller are values of
$\Delta{m}^2$
which can be probed in an experiment.
The inequality (\ref{cond}) allows to estimate (for large mixing
angles) the sensitivity to
the parameter $\Delta m^2$ of
different types of neutrino oscillation experiments.
These estimates are presented in Table~\ref{estimates}.
Let us stress that this table gives only a very rough idea of the
sensitivity to $\Delta m^2$. 
For example, in the LSND short-baseline (SBL) experiment 
antineutrino energies are between 20 and 60 MeV, the distance is
approximately 30 m and the minimal value of $\Delta m^2$ probed 
in this experiment is $4 \times 10^{-2}$ eV$^2$.

\begin{table}[t]
\begin{center}
\renewcommand{\arraystretch}{1.45}
\setlength{\tabcolsep}{0.5cm}
\begin{tabular}{|cccc|}
\hline
Experiment
&
$L$ (m)
&
$E$ (MeV)
&
$\Delta{m}^2$ (eV$^2$)
\\
\hline
Reactor SBL
&
$10^2$
&
$1$
&
$10^{-2}$
\\
Reactor LBL
&
$10^3$
&
$1$
&
$10^{-3}$
\\
Accelerator SBL
&
$10^3$
&
$10^3$
&
$1$
\\
Accelerator LBL
&
$10^6$
&
$10^3$
&
$10^{-3}$
\\
Atmospheric
&
$10^7$
&
$10^3$
&
$10^{-4}$
\\
Solar
&
$10^{11}$
&
$1$
&
$10^{-11}$
\\
\hline
\end{tabular}
\end{center}
\refstepcounter{tables}
\label{estimates}
\footnotesize
Table \ref{estimates}.
Order of magnitude estimates of the values of
$\Delta{m}^2$
which can be probed in
reactor short-baseline (SBL) and long-baseline (LBL),
accelerator SBL and LBL,
atmospheric and solar
neutrino oscillation experiments.
Note,
however,
that the energies and distances of the various types of experiments
can vary in a wide range and only some representative values are given in
this table.
\end{table}

The probability (\ref{prob1}) and the corresponding one for antineutrinos
are invariant under the phase transformation
\begin{equation}
U_{{\alpha}k} \to e^{-i\varphi_\alpha} \, U_{{\alpha}k} \, e^{i\psi_k}
\label{phasetrafo}
\end{equation}
Therefore, it is clear that the probabilities
of
$\nu_\alpha\to\nu_\beta$
and
$\bar\nu_\alpha\to\bar\nu_\beta$
transitions do not depend
\cite{BHP80,Kobzarev80,Doi81}
on the Majorana CP-violating phases
discussed at the end of Section~\ref{Majorana mass term} and it is not
possible to distinguish the Dirac and Majorana cases by the
observation of neutrino oscillations.

Comparing the expressions
(\ref{ampli}) and (\ref{aampli})
for the transition amplitudes
of neutrinos and antineutrinos we see that
$
\mathcal{A}_{\nu_\alpha\to\nu_\beta}(t)
=
\mathcal{A}_{\bar\nu_\beta\to\bar\nu_\alpha}(t)
$.
Therefore,
for the transition probabilities we have
\begin{equation}
P_{\nu_\alpha\to\nu_\beta}
=
P_{\bar\nu_\beta\to\bar\nu_\alpha}
\,.
\label{pna}
\end{equation}
This relation is a consequence of CPT invariance 
inherent in any local field theory \cite{CPT}.
From the equality (\ref{pna})
it follows that the neutrino and antineutrino survival
probabilities are equal:
\begin{equation}
P_{\nu_\alpha\to\nu_\alpha}
=
P_{\bar\nu_\alpha\to\bar\nu_\alpha}
\,.
\label{spna}
\end{equation}
On the other hand,
the transition probabilities of neutrinos and antineutrinos in general are
different for $\alpha \neq \beta$.
They are equal only if there is CP invariance in the lepton sector.
In fact,
in the case of massive Dirac neutrinos the phases of
the neutrino fields and of the charged lepton fields
can be chosen in such a way that
the mixing matrix $U$ is real. 
In the case of Majorana neutrinos,
CP invariance implies that
(see Eq.(\ref{orth4}))
\begin{equation}
U_{{\alpha}k}^* = U_{{\alpha}k} \, \rho_k
\,,
\label{orth41}
\end{equation}
with
$ \rho_k = - i \, \eta^{\mathrm{CP}}_k = \pm 1 $,
where
$\eta^{\mathrm{CP}}_k$
is the CP parity of the Majorana neutrinos with mass $m_k$
(see Eq.(\ref{etaCPk})).
Thus, 
in the Dirac as well as in the Majorana case,
CP invariance implies that
\begin{equation}
P_{\nu_\alpha\to\nu_\beta}
=
P_{\bar\nu_\alpha\to\bar\nu_\beta}
\,.
\label{pCP}
\end{equation}

\subsection{Oscillations in the two-neutrino case}

The results of neutrino oscillation experiments are usually analysed
under the simplest assumption of oscillations between two neutrino
types. In this case, for the transition
probability (\ref{prob2}) we get
\begin{equation}
P_{\nu_\alpha\to\nu_\beta}
=
\left|
\delta_{\alpha\beta}
+
U_{\beta 2} U^*_{\alpha 2}
\left[ \exp\left( - i \frac{ \Delta{m}^2 L }{ 2 E } \right) - 1 \right]
\right|^2
\,,
\label{prob3}
\end{equation}
where $\Delta m^2 = m_2^2 - m_1^2$ and $\alpha$, $\beta$ are
$e$, $\mu$,
or
$\mu$, $\tau$,
etc$\ldots$
Thus in the simplest case of
transitions between two neutrino types the probability is determined
only by the elements of $U$ which connect flavour neutrinos with $\nu_2$
(or $\nu_1$). It is obvious that phases drop out in the expression
(\ref{prob3}). This is an illustration of Eq.(\ref{phasetrafo})
and of the fact that no information about CP violation can be obtained
in the case of transitions between only two neutrino types. 
If we put $U_{\alpha 2} = \sin \vartheta$ then we have 
$U_{\beta 2} = \cos \vartheta$ and the transition and survival
probabilities are given by the following standard expressions
valid for both, neutrinos and antineutrinos:
\begin{equationarrayzero}
&&
P_{\stackrel{\makebox[0pt][l]
{$\hskip-3pt\scriptscriptstyle(-)$}}{\nu_{\alpha}}
\to\stackrel{\makebox[0pt][l]
{$\hskip-3pt\scriptscriptstyle(-)$}}{\nu_{\beta}}}
=
\frac{1}{2}
\,
\sin^22\vartheta
\left( 1 - \cos \frac{ \Delta{m}^2 L }{ 2 E } \right)
\qquad
(\alpha\neq\beta)
\,,
\label{trans2}
\\
&&
P_{\stackrel{\makebox[0pt][l]
{$\hskip-3pt\scriptscriptstyle(-)$}}{\nu_{\alpha}}
\to\stackrel{\makebox[0pt][l]
{$\hskip-3pt\scriptscriptstyle(-)$}}{\nu_{\alpha}}}
=
P_{\stackrel{\makebox[0pt][l]
{$\hskip-3pt\scriptscriptstyle(-)$}}{\nu_{\beta}}
\to\stackrel{\makebox[0pt][l]
{$\hskip-3pt\scriptscriptstyle(-)$}}{\nu_{\beta}}}
=
1
-
P_{\stackrel{\makebox[0pt][l]
{$\hskip-3pt\scriptscriptstyle(-)$}}{\nu_{\alpha}}
\to\stackrel{\makebox[0pt][l]
{$\hskip-3pt\scriptscriptstyle(-)$}}{\nu_{\beta}}}
\,.
\label{surv2}
\end{equationarrayzero}%
The transition probability (\ref{trans2})
can be written in the form
\begin{equation}
P_{\stackrel{\makebox[0pt][l]
{$\hskip-3pt\scriptscriptstyle(-)$}}{\nu_{\alpha}}
\to\stackrel{\makebox[0pt][l]
{$\hskip-3pt\scriptscriptstyle(-)$}}{\nu_{\beta}}}
=
\frac{1}{2}
\,
\sin^22\vartheta
\left( 1 - \cos 2.53 \frac{ \Delta{m}^2 L }{ E } \right)
\qquad
(\alpha\neq\beta)
\,.
\label{trans22}
\end{equation}
where $L$ is the source -- detector distance expressed in m (km),
$E$ is the neutrino energy in MeV (GeV) and 
$\Delta{m}^2$ is the neutrino mass-squared difference in eV$^2$.
Thus, the transition probability is a periodic function of $L/E$. This
phenomenon is called \emph{neutrino oscillations}.
The amplitude of oscillations is $\sin^22\vartheta$
and the oscillation length is given by
\begin{equation}
L^{\mathrm{osc}}
=
\frac{ 4 \, \pi \, E }{ \Delta{m}^2 }
\simeq
2.48
\,
\frac{ E \ (\mathrm{MeV}) }{ \Delta{m}^2 \ (\mathrm{eV}^2) }
\:
\mbox{m}
\,.
\label{length}
\end{equation}
The condition (\ref{cond}) can be rewritten in the form
\begin{equation}
L^{\mathrm{osc}} \lesssim L
\,.
\label{cond2}
\end{equation}
Therefore,
neutrino oscillations can be observed if the oscillation length
is not much larger than the source -- detector distance $L$.

\begin{figure}[t]
\begin{center}
\mbox{\epsfig{file=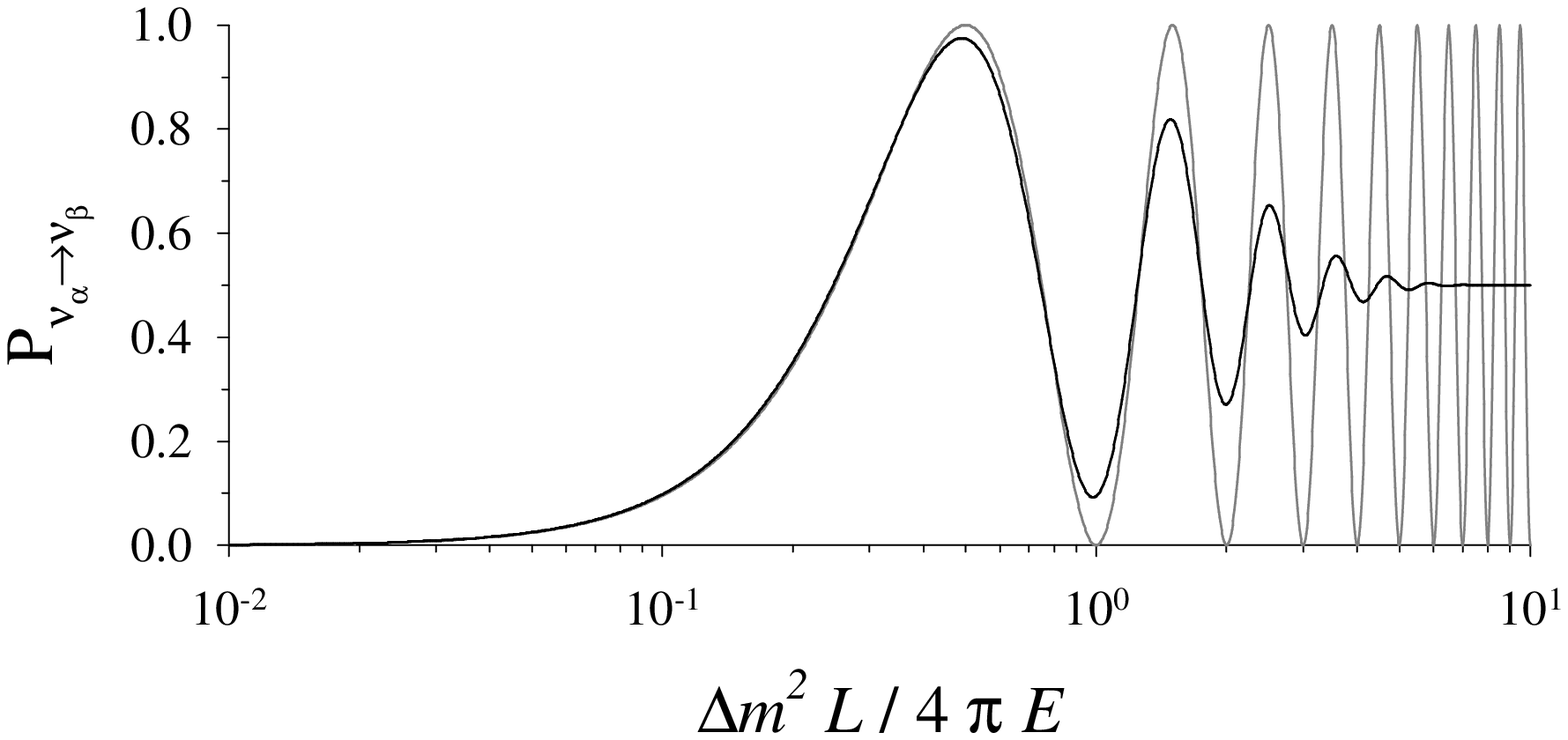,width=0.95\linewidth}}
\end{center}
\refstepcounter{figures}
\label{prob}
\footnotesize
Figure \ref{prob}.
Transition probability for $\sin^22\vartheta=1$
as a function of
$ \Delta{m}^2 L / 4 \pi E = L / L^{\mathrm{osc}} $,
where
$L^{\mathrm{osc}}$
is the oscillation length.
The grey line represents the transition probability (\ref{trans2})
and the black line represents the same transition probability
averaged over a Gaussian energy spectrum with mean value $E$ and
standard deviation $\sigma = E/10$.
\end{figure}

The oscillatory behaviour of the transition probability (\ref{trans2})
with $\sin^22\vartheta=1$
is shown
in Fig.~\ref{prob},
where we have plotted it as a function of
$ \Delta{m}^2 L / 4 \pi E = L / L^{\mathrm{osc}} $.
The grey line represents the transition probability (\ref{trans2}),
whereas, in order to demonstrate the effect of energy averaging,
the black line represents the transition probability (\ref{trans2})
averaged over a Gaussian energy distribution with mean value $E$ and
standard deviation $\sigma = E/10$.
The averaged probability is
the measurable quantity in neutrino oscillation experiments.
One can see that the averaging over the energy spectrum practically
reduces the probability to the constant
$ 1 - \sin^22\vartheta / 2 $
for $ L \gg L^{\mathrm{osc}} $.

\begin{figure}[t!]
\begin{tabular*}{\linewidth}{@{\extracolsep{\fill}}cc}
\begin{minipage}{0.47\linewidth}
\begin{center}
\mbox{\epsfig{file=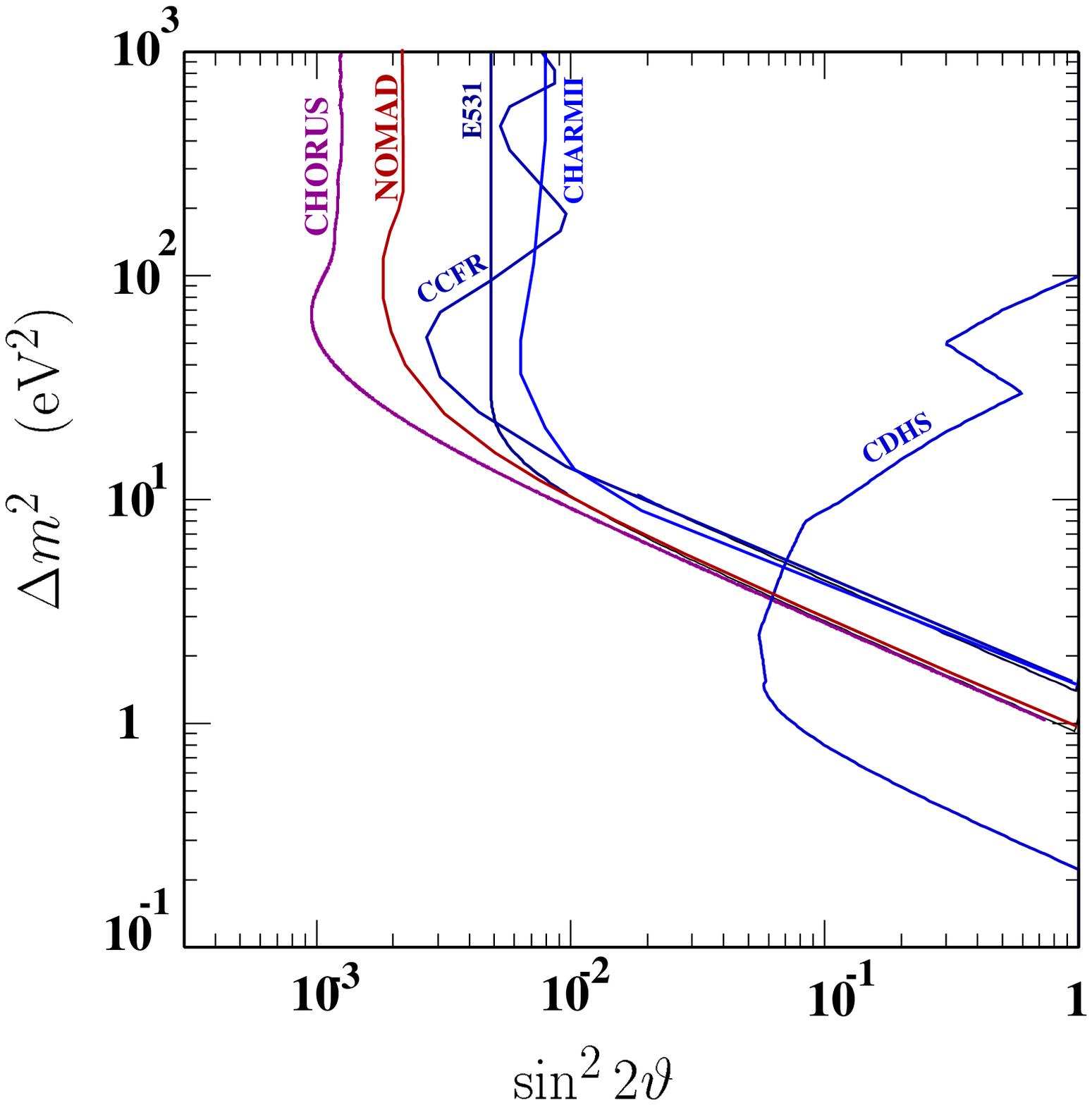,width=0.95\linewidth}}
\end{center}
\end{minipage}
&
\begin{minipage}{0.47\linewidth}
\begin{center}
\mbox{\epsfig{file=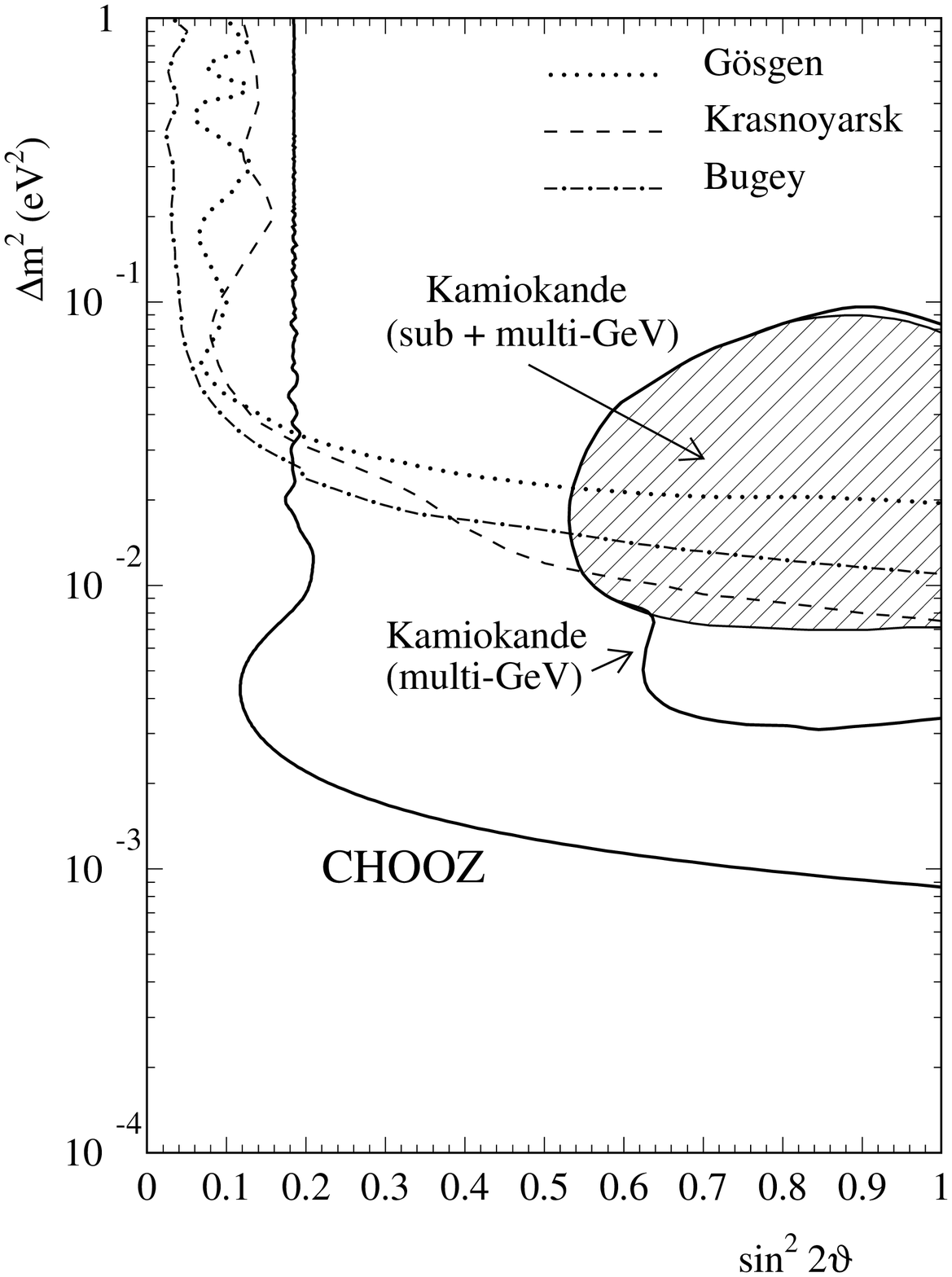,width=0.95\linewidth}}
\end{center}
\end{minipage}
\\
\begin{minipage}{0.47\linewidth}
\refstepcounter{figures}
\label{nomad}
\footnotesize
Figure \ref{nomad}.
Exclusion curves (90\% CL) in the
$\nu_\mu\to\nu_\tau$
channel
obtained in the
CDHS \protect\cite{CDHS84},
FNAL E531 \protect\cite{FNALE531},
CHARM II \cite{CHARMII94},
CCFR \protect\cite{CCFR95},
CHORUS \protect\cite{CHORUS98}
and
NOMAD \protect\cite{NOMAD98}
experiments.
\end{minipage}
&
\begin{minipage}{0.47\linewidth}
\refstepcounter{figures}
\label{chooz}
\footnotesize
Figure \ref{chooz}.
Exclusion curves (90\% CL) in the
$\bar\nu_e\to\bar\nu_e$
channel
obtained in the
CHOOZ \protect\cite{CHOOZ98},
G\"osgen \protect\cite{Gosgen86},
Krasnoyarsk \protect\cite{Krasnoyarsk94}
and
Bugey \protect\cite{Bugey95}
experiments.
The shadowed region is allowed by the results of the Kamiokande
atmospheric neutrino experiment \protect\cite{Kam-atm-94}.
\end{minipage}
\end{tabular*}
\end{figure}

The expressions (\ref{trans2}) and (\ref{surv2})
are usually employed in analyses of the data of 
neutrino oscillation experiments.
In many SBL experiments with neutrinos from reactors
and accelerators, no indication in favour of neutrino oscillations was found.
The data of these experiments give an upper bound 
for the transition probability which implies
an excluded region in the space of the parameters
$\Delta{m}^2$ and $\sin^2{2\vartheta}$.
A typical exclusion plot is presented in Fig.~\ref{nomad} \cite{nomad98b}.
This plot shows the exclusion curves in the
$\nu_\mu\to\nu_\tau$
channel
obtained in the
CDHS \cite{CDHS84},
FNAL E531 \cite{FNALE531},
CHARM II \cite{CHARMII94},
CCFR \cite{CCFR95},
CHORUS \cite{CHORUS98}
and
NOMAD \cite{NOMAD98}
experiments.
The excluded region lies on the right of the curves.

The two most stringent exclusion curves
in Fig.~\ref{nomad} have been obtained in the
CHORUS \cite{CHORUS98}
and
NOMAD \cite{NOMAD98}
experiments, which are operating at CERN using the
neutrino beam from the SPS (with an average energy of about 30 GeV).
800 kg of emulsions are used in the CHORUS experiment as target.
The production and decay of $\tau$'s 
in the emulsion is searched for.
In the NOMAD experiment a magnetic detector is used and
the production of $\tau$'s is identified with kinematical criteria.

Figure~\ref{chooz}
shows the exclusion curves obtained in the
CHOOZ \cite{CHOOZ98},
G\"osgen \cite{Gosgen86},
Krasnoyarsk \cite{Krasnoyarsk94}
and
Bugey \cite{Bugey95}
reactor $\bar\nu_e\to\bar\nu_e$ experiments.
The region allowed by the results of the Kamiokande
atmospheric neutrino experiment \cite{Kam-atm-94}
(see Section \ref{Indications: atmospheric neutrino experiments})
is also depicted in this figure.
The G\"osgen, Krasnoyarsk and Bugey experiments
are SBL reactor experiments,
whereas
the recent CHOOZ experiment is the first LBL reactor experiment.
In this experiment the detector
(5 tons of liquid scintillator loaded with Gd)
is at the distance of about 1 km from the CHOOZ power station, which has
two water reactors with a total thermal power of 8.5 GW.
The antineutrinos are detected through the observation of
the reaction
\begin{equation}
\bar\nu_e + p \to e^+ + n
\label{nuebar-p}
\end{equation}
(the photons from annihilation of the positron and
the delayed photons
from the capture of neutron by Gd are detected).
No indications in favour of neutrino oscillations were found in the CHOOZ
experiment. The ratio $R$ of the numbers of measured antineutrino
events and of expected antineutrino
events in the case of absence of neutrino oscillations 
in the CHOOZ detector is \cite{CHOOZ98}
\begin{equation}
R = 0.98 \pm 0.04 \pm 0.04
\,.
\label{R-CHOOZ}
\end{equation}
Since the average value of $L/E$ in the CHOOZ experiment is
approximately 300
($ \langle{E}\rangle \simeq 3 \, \mathrm{MeV} $,
$ L \simeq 1000 \, \mathrm{m} $),
in this
experiment it is possible to probe the value of
the relevant
neutrino mass-squared difference down to $10^{-3} \, \mathrm{eV}^2$.

With the help of the expression (\ref{trans22}),
it is possible to understand
qualitatively the general features of exclusion curves.
In the region of large
$\Delta{m}^2$ such that
the oscillation length is much smaller than the
source -- detector distance $L$,
the cosine in the expression (\ref{trans22})
oscillates very rapidly as a function of the neutrino energy $E$.
Since in practice all neutrino beams have an energy spectrum
and
the neutrino sources and detectors are extended in space,
only the average transition probability
\begin{equation}
\langle
P_{\stackrel{\makebox[0pt][l]
{$\hskip-3pt\scriptscriptstyle(-)$}}{\nu_{\alpha}}
\to\stackrel{\makebox[0pt][l]
{$\hskip-3pt\scriptscriptstyle(-)$}}{\nu_{\beta}}}
\rangle
=
\frac{1}{2} \, \sin^22\vartheta
\label{ave}
\end{equation}
can be determined in the region of large $\Delta m^2$.
The average probability is independent from $\Delta{m}^2$
and, therefore, from
an experimental upper bound
$
\langle
P_{\stackrel{\makebox[0pt][l]
{$\hskip-3pt\scriptscriptstyle(-)$}}{\nu_{\alpha}}
\to\stackrel{\makebox[0pt][l]
{$\hskip-3pt\scriptscriptstyle(-)$}}{\nu_{\beta}}}
\rangle_0
$
on
$
\langle
P_{\stackrel{\makebox[0pt][l]
{$\hskip-3pt\scriptscriptstyle(-)$}}{\nu_{\alpha}}
\to\stackrel{\makebox[0pt][l]
{$\hskip-3pt\scriptscriptstyle(-)$}}{\nu_{\beta}}}
\rangle
$
one obtains the vertical-line part of the exclusion curve.

At
\begin{equation}
\Delta{m}^2
\simeq
\frac{\pi}{2}
\,
\frac{\langle E \rangle}{1.27 \, \langle L \rangle}
\,,
\label{minimum}
\end{equation}
where $\langle E \rangle$ is the average energy and $\langle L
\rangle$ is the average distance, 
the parameter $\sin^2 2\vartheta$ has the minimal value
\begin{equation}
(\sin^22\vartheta)_{\mathrm{min}} = 
\langle
P_{\stackrel{\makebox[0pt][l]
{$\hskip-3pt\scriptscriptstyle(-)$}}{\nu_{\alpha}}
\to\stackrel{\makebox[0pt][l]
{$\hskip-3pt\scriptscriptstyle(-)$}}{\nu_{\beta}}}
\rangle_0
\end{equation}
on the boundary curve.
Note that we are using here and in the rest of the section the same
units as in Eq.(\ref{trans22}).

Typically, the upper bound
$
\langle
P_{\stackrel{\makebox[0pt][l]
{$\hskip-3pt\scriptscriptstyle(-)$}}{\nu_{\alpha}}
\to\stackrel{\makebox[0pt][l]
{$\hskip-3pt\scriptscriptstyle(-)$}}{\nu_{\beta}}}
\rangle_0
$
is much less than one. Then, in the region where $\sin^22\vartheta$ is
large
the expression (\ref{trans22}) for the transition probability
can be approximated by
\begin{equation}
P_{\stackrel{\makebox[0pt][l]
{$\hskip-3pt\scriptscriptstyle(-)$}}{\nu_{\alpha}}
\to\stackrel{\makebox[0pt][l]
{$\hskip-3pt\scriptscriptstyle(-)$}}{\nu_{\beta}}}
\simeq
\,
\sin^22\vartheta
\left( 1.27\, \frac{ \Delta{m}^2 L }{ E } \right)^2
\label{trans23}
\end{equation}
and in this region the boundary curve in the exclusion plot is given by
\begin{equation}
\Delta{m}^2
\simeq
\frac
{ \sqrt{
\langle
P_{\stackrel{\makebox[0pt][l]
{$\hskip-3pt\scriptscriptstyle(-)$}}{\nu_{\alpha}}
\to\stackrel{\makebox[0pt][l]
{$\hskip-3pt\scriptscriptstyle(-)$}}{\nu_{\beta}}}
\rangle_0
}}
{ 1.27 \, \sqrt{ \sin^22\vartheta \, 
\langle L^2 \rangle \langle E^{-2} \rangle }}
\,.
\label{up1}
\end{equation}
Therefore, this part of the exclusion curve 
is a straight line in the
$\log\sin^22\vartheta$ -- $\log\Delta{m}^2$ plot
as can be seen from Fig.~\ref{nomad}.
From Eq.(\ref{up1}) it follows that the minimal
value of the parameter $\Delta{m}^2$ that can be probed by
an experiment 
is
\begin{equation}
\Delta{m}^2
\simeq
\frac
{ \sqrt{
\langle
P_{\stackrel{\makebox[0pt][l]
{$\hskip-3pt\scriptscriptstyle(-)$}}{\nu_{\alpha}}
\to\stackrel{\makebox[0pt][l]
{$\hskip-3pt\scriptscriptstyle(-)$}}{\nu_{\beta}}}
\rangle_0
}}
{ 1.27 \, \sqrt{\langle L^2 \rangle \langle E^{-2} \rangle }} 
\,.
\label{up3}
\end{equation}
It corresponds to
$ \sin^22\vartheta = 1 $.

\section{Neutrino oscillations and transitions in matter}
\label{Neutrino transitions in matter}
\setcounter{equation}{0}
\setcounter{figures}{0}
\setcounter{tables}{0}

\subsection{The effective Hamiltonian for neutrinos in matter}

It has been pointed out by Wolfenstein \cite{wol} and by
Mikheyev and Smirnov \cite{MS} that the neutrino
oscillation pattern in vacuum can get significantly modified by the
passage of neutrinos through matter because of the effect of coherent
forward scattering. This effect can be described by an effective
Hamiltonian. Starting with neutrino oscillations in vacuum, 
one can easily check that the transition
probability (\ref{prob2}) can be obtained by considering the evolution of
the state vector $\psi$ of the neutrino types with the
``Schr\"odinger equation''
\begin{equation}\label{vac}
i \frac{\mathrm{d}\psi(x)}{\mathrm{d}x} = H^{\mathrm{vac}}_\nu \psi(x) = 
\frac{1}{2E} \, U \hat{m}^2 U^\dagger \psi(x) \,,
\end{equation}
where $H^{\mathrm{vac}}_\nu$ is the effective Hamiltonian for
neutrino oscillations in vacuum
with $U$ being the mixing matrix, $E$ the neutrino energy and
$\hat{m}$ the diagonal neutrino mass matrix. The presence of matter will give a
correction to Eq.(\ref{vac}). Note that the variable in
Eq.(\ref{vac}) is not time but space and the components of $\psi(x)$ are
given by the amplitudes denoted by $a_\alpha(x)$ for the neutrino
types $\alpha = e,\mu,\tau,s$. The
corresponding effective Hamiltonian for antineutrinos
$H^{\mathrm{vac}}_{\bar \nu}$ is obtained from $H^{\mathrm{vac}}_\nu$
by the exchange $U \to U^*$. After the seminal work of Wolfenstein
about neutrino oscillations in matter (see also Ref.~\cite{barger80})
and the discovery of the 
importance of $H^{\mathrm{mat}}_\nu$,
the analogue of $H^{\mathrm{vac}}_\nu$ in matter, 
for the solar neutrino problem
\cite{MS} many rederivations of the effective matter Hamiltonian
\cite{rosen,bethe} have been presented using a coupled
system of Dirac equations \cite{dirac,gri93}, field theory \cite{pal-pham}
or studying coherent forward scattering in more detail \cite{liu}.

We want to sketch a derivation using the Dirac equation and
following Ref.~\cite{gri93}. The starting point in most
derivations is the expectation value of the currents for 
isotropic non-relativistic
matter given by \cite{wol,MS,pal}
\begin{equation}\label{nf}
\langle \bar f_L \gamma_\mu f_L \rangle_{\mathrm{matter}} = 
\frac{1}{2} N_f \delta_{\mu 0} \,,
\end{equation}
where $N_f$ is the number density of the particles represented
by the field $f$. Note that the $\gamma_5$ term does not
contribute for non-aligned spins \cite{GKL91-PRD43}.
Eq.(\ref{nf}) is a good approximation even for electrons in the
core of the sun with a temperature $T \simeq 16 \times 10^6$ K 
\cite{Bahcall-Ulrich88,Bahcall-Pinsonneault95} 
and thus an average velocity of the electrons of
around 10\% of the velocity of light and corrections of order 
$kT/m_e \simeq 2.6 \times 10^{-3}$. 
Starting from the weak interaction Lagrangians
(\ref{CC}) and (\ref{NC}) one gets for low-energy neutrino
interactions of flavour $\ell$ with the background matter
\begin{equation}\label{matterL}
- \mathcal{L}^{\mathrm{mat}}_{\nu_\ell} = 
\frac{G_F}{\sqrt{2}} \nu_\ell^\dagger
(1 - \gamma_5) \nu_\ell \sum_f N_f(\delta_{\ell f} + T_{3f_L} -
2 \sin^2 \theta_W Q_f) \,,
\end{equation}
where $G_F$ is the Fermi coupling constant, $\theta_W$ the weak
mixing angle, $T_{3f_L}$ the eigenvalue of the field $f_L$ of the third
component of the weak isospin ($T_{3f_R} = 0$ in the Standard
Model) and $Q_f$ is the charge of $f$. In the matter Lagrangian (\ref{matterL})
the charged current interaction is represented by the
Kronecker symbol $\delta_{\ell f}$ saying that for neutrinos
of flavour $\ell$ the charged current only contributes when
background matter containing charged leptons of the same flavour
is present. Concentrating now on realistic matter with
electrons, protons and neutrons which is electrically neutral,
\textit{i.e.}, $N_e = N_p$, and taking into account that we 
have $T_{3e_L} = -T_{3p_L} = T_{3n_L} = -1/2$ and 
$Q_e = -Q_p = -1$, $Q_n = 0$ for electrons, protons and
neutrons, respectively,\footnote{For protons and neutrons 
this is ensured by CVC (conserved vector current)
for the $T_3$ part and conservation of the electromagnetic current.}
we get an effective Hamiltonian
\begin{equation}\label{DH}
H^D_\nu = -i \vec{\alpha} \cdot \vec{\nabla} + 
\beta (M P_L+M^\dagger P_R) +
\sqrt{2}\, G_F\, \mbox{diag}\, \left( N_e - \frac{1}{2} N_n,
- \frac{1}{2} N_n, - \frac{1}{2} N_n \right) P_L
\end{equation}
acting on the vector $\Psi(t,x)$ of the flavour neutrino wave functions,
where we have defined $P_{L,R} = ( \mathbf{1} \mp \gamma_5 )/2$,
$\alpha_j = \gamma^0 \gamma^j$ and $\beta = \gamma^0$. $M$ is
the non-diagonal mass matrix. The last
term in Eq.(\ref{DH}) is called the matter potential term.
Several remarks concerning the Hamiltonian (\ref{DH}), which is
the point of departure for deriving the effective matter
Hamiltonian in Refs.~\cite{dirac,gri93}, are at order:
\begin{enumerate}
\item $H^D_\nu$ has spinor and flavour indices, it leads thus to a
system of Dirac equations coupled via the mass term.
\item The neutral current contributions of electrons and protons
cancel for realistic matter
because of opposite $T_{3f_L}$ quantum numbers and electric
charges (see discussion above before Eq.(\ref{DH})).
\item \label{Dcase}
The Hamiltonian (\ref{DH}) is valid for Dirac neutrinos. For
antineutrinos, $M$ is replaced by $M^T$ and 
the matter potential term of $H^D_{\bar \nu}$
has the opposite sign and the projector $P_R$ instead of $P_L$. This is a
consequence of proceeding along the same lines as before but
using the charge-conjugate fields to describe antineutrinos 
($\bar \nu_\ell \gamma_\rho P_L \nu_\ell = 
- \overline{\nu^c_\ell} \gamma_\rho P_R \nu^c_\ell$).
\item \label{Mcase}
In the Majorana case, in order to obtain the Hamiltonian $H^M_\nu$ 
the antineutrino matter potential has
to be added to $H^D_\nu$ and the neutrino field vector $\Psi$ is
subject to the Majorana condition $\Psi = \mathcal{C} \gamma_0^T \Psi^*$.
This condition is compatible with the time evolution governed by
$H^M_\nu$ \cite{gri93}.
\item Sterile neutrinos can easily be incorporated into
$H^{D}_{\stackrel{\scriptscriptstyle (-)}{\nu}}$ and $H^M_\nu$
by simply adding the entry 0 to the matter potential term. Note that
the probability of active -- active transitions does not depend on the
neutron density $N_n$, which is, however, important for active --
sterile transitions (see Table~\ref{Alist}).
\end{enumerate}

The essence of deriving an effective Hamiltonian for neutrino
oscillations in matter is to get rid of the spinor indices in $H^D_\nu$
and to obtain an equation involving only the indices $\alpha$ for the different
neutrino types as in the
vacuum case (\ref{vac}). To this end,
let us now assume that the neutrino propagation proceeds along
the $x^3 \equiv x$ axis, that the neutrino momentum $p>0$
corresponding to propagation in vacuum is much larger than the
matter potentials and that the neutrinos are ultrarelativistic. 
Thus we consider a one-dimensional problem
from now on. Defining a wave function $\phi$ via
\begin{equation}
\Psi(t,x) = e^{ipx} \phi(x,t) \quad \mbox{with} \quad
i \frac{\partial \phi}{\partial t} = 
\left( p \alpha_3 + H^D_\nu \right) \phi \,,
\end{equation}
this wave function changes little over distances of the order of
the de Broglie wave length of the neutrino. With respect to
$\alpha_3$ we can decompose the Hamiltonian (\ref{DH}) into
$H^D_\nu = H_{\mathrm{even}} + H_{\mathrm{odd}}$ which are the
parts commuting or anticommuting with $\alpha_3$.
$H_{\mathrm{odd}}$ consists solely of the mass term and
$H_{\mathrm{even}}$ contains the rest of $H^D_\nu$. With 
$p$ being a large parameter we can perform a Foldy -- Wouthuysen
transformation where we truncate the series at $1/p$ leading to
the Hamiltonian \cite{FW,gri93}
\begin{eqnarray}
H^{D}_{\mathrm{FW},\nu} & = &  
p \alpha_3 + H_{\mathrm{even}} +
\frac{1}{2p} \alpha_3 H_{\mathrm{odd}}^2  \nonumber \\
& = & \alpha_3 \left( p -i \frac{\partial}{\partial x} \right) 
+
\sqrt{2}\, G_F\, \mbox{diag} \left( N_e - \frac{1}{2} N_n,
- \frac{1}{2} N_n, - \frac{1}{2} N_n \right) P_L \nonumber \\
& + &
\frac{1}{2p}\, \alpha_3 ( M^\dagger M P_L + M M^\dagger P_R ) \,,
\label{FW}
\end{eqnarray}
such that all terms in $H^D_{\mathrm{FW},\nu}$ commute with $\alpha_3$
and, therefore, all matrices in $H^D_{\mathrm{FW},\nu}$ with spinor indices 
can be diagonalized at the same time in order to separate
positive and negative energy states. Denoting this unitary
diagonalization matrix by $U_0$ we can achieve 
$U_0 \alpha_3 U_0^\dagger = \mbox{diag} (1,1,-1,-1)$,
$U_0 \frac{\mathbf{1}-\gamma_5}{2} U_0^\dagger = \mbox{diag} (0,1,1,0)$
and
$U_0 \frac{\mathbf{1}+\gamma_5}{2} U_0^\dagger = \mbox{diag} (1,0,0,1)$.
Note that there are no
transitions between left and right-handed states through 
$H^{D}_{\mathrm{FW},\nu}$. For neutrinos  with magnetic moments
(electric dipole moments) the method described above would also
yield the appropriate left-right transitions \cite{gri93}. 

Going back to the Foldy -- Wouthuysen transform of the wave
function $\Psi$ instead of $\phi$ removes the $p$ from the
Hamiltonian (\ref{FW}). The final step in deriving the effective
matter Hamiltonian consists of considering stationary states and
splitting off the plane wave part by
\begin{equation} 
\Psi_{\mathrm{FW}}(t,x) = e^{-iE(t-x)} \psi(x) \,.
\end{equation}
Then the wave function $\psi(x)$, taken for positive energies
and left-handed neutrinos, fulfills the first order differential
equation (\ref{vac}) with $H^{\mathrm{vac}}_\nu$ replaced by \cite{wol}
\begin{equation}\label{mat}
H^{\mathrm{mat}}_\nu = \frac{1}{2E}
\left( M^\dagger M + 
2\sqrt{2}\,E G_F\, \mbox{diag} \left( N_e - \frac{1}{2} N_n,
- \frac{1}{2} N_n, - \frac{1}{2} N_n \right) \right) \,,
\end{equation}
correct up to order $1/E$. This effective Hamiltonian is valid
for left-handed Dirac neutrinos or Majorana neutrinos, whereas
in the case of right-handed (Dirac) antineutrinos or
right-handed Majorana neutrinos we have
\begin{equation}\label{mat-anti}
H^{\mathrm{mat}}_{\bar \nu} = \frac{1}{2E}
\left( M^T M^* - 
2\sqrt{2}\,E G_F\, \mbox{diag} \left( N_e - \frac{1}{2} N_n,
- \frac{1}{2} N_n, - \frac{1}{2} N_n \right) \right) \,,
\end{equation}
because for (Dirac) antineutrinos one has to replace $M$ by
$M^T$ in the Hamiltonian (\ref{FW}) and for Majorana neutrinos
$M=M^T$ holds (see also points \ref{Dcase} and \ref{Mcase} after
Eq.(\ref{DH})). With
\begin{equation}
M = U_R \hat{m} U_L^\dagger \quad \mbox{and} \quad U_L \equiv U
\end{equation}
only the left-handed mixing matrix $U$ enters and connection
between Eqs.(\ref{mat}) and (\ref{vac}) is made. 
Thus in the final results (\ref{mat}) and (\ref{mat-anti})
there is no distinction between the
Dirac and Majorana case for ultrarelativistic neutrinos.
This is an illustration of the fact that with neutrino
oscillation experiments the Dirac or Majorana nature cannot be
distinguished \cite{BP87,lang-pet87}. 

The Hamiltonians (\ref{mat}) and (\ref{mat-anti}) have been used to
investigate neutrino oscillations in the sun, in the earth and 
in supernovae. In the following we will only be concerned
with the first two subjects. Application limits of the neutrino
evolution equations in matter have been discussed in
Ref.~\cite{MS-rev}. Elastic and inelastic neutrino scattering
introduces quantum damping into the evolution equations which
is proportional to the neutrino interaction rate \cite{stodolsky}. In
the sun and the earth this effect is negligible, in particular, for
low energies, whereas in the early universe it is of crucial
importance \cite{stodolsky}. Density fluctuations have been found to
influence considerably neutrino propagation in the sun
\cite{sun-fluct}, however, more realistic considerations with
helioseismic waves as density fluctuations show no observable effect
on the solar neutrino problem discussed in terms of neutrino
oscillations \cite{bamert}. Solar neutrinos are also influenced by
their passage through the earth \cite{baltz}.

\subsection{The two-neutrino case and adiabatic transitions}

Let us now concentrate on left-handed neutrinos and specify the effective
Hamiltonian (\ref{mat}) to two neutrino types. Thus for 
two-neutrino oscillations in matter with the definitions
\begin{equation}
N(\nu_\alpha) \equiv \delta_{\alpha e} N_e - \frac{1}{2} N_n \quad 
(\alpha = e, \mu, \tau) \,, \quad N(\nu_s) \equiv 0
\end{equation}
and
\begin{equation}\label{Adef}
A \equiv 2\sqrt{2} G_F E \left( N(\nu_\alpha) - N(\nu_\beta) \right) \,,
\end{equation}
the differential equation
\begin{equationarray}
i \frac{\mathrm{d}}{\mathrm{d}x} 
\left( \begin{array}{c} a_\alpha \\ a_\beta \end{array} \right) & = &
H^{\mathrm{mat}}_\nu 
\left( \begin{array}{c} a_\alpha \\ a_\beta \end{array} \right)
\nonumber \\
& = & \frac{1}{4E} 
\left\{ 
\left[ m_1^2+m_2^2 + 
2\sqrt{2} G_F \left( N(\nu_\alpha) + N(\nu_\beta) \right)
\right]
\left( \begin{array}{cc} 1 & 0 \\ 0 & 1 \end{array} \right) 
\right. \nonumber \\ & & +
\left. \left( \begin{array}{cc} 
A - \Delta m^2 \cos 2\vartheta & \Delta m^2 \sin 2\vartheta \\
\Delta m^2 \sin 2\vartheta & -A + \Delta m^2 \cos 2\vartheta 
\end{array} \right) 
\right\}
\left( \begin{array}{c} a_\alpha \\ a_\beta \end{array} \right) 
\label{master2}
\end{equationarray}%
has to be studied \cite{wol,MS,MS-rev,Kuo-Pantaleone89,pal,haxton95},
where $a_\alpha$, $a_\beta$ are the amplitudes for the neutrino types
$\alpha$, $\beta$ ($\alpha, \beta = e,\mu,\tau,s$), respectively.
The evolution equation for antineutrinos is obtained from
Eq.(\ref{master2}) by $A \to -A$. In the two-neutrino case
there is one $\Delta m^2 = m_2^2-m_1^2$ and the mixing matrix
\begin{equation}\label{vacangle}
U = \left( \begin{array}{rr} \cos \vartheta & \sin \vartheta \\
-\sin \vartheta & \cos \vartheta \end{array} \right)
\end{equation}
is real without loss of generality.
\begin{table}
\renewcommand{\arraystretch}{1.5}
\begin{center}
\begin{tabular}{|c|cccc|} \hline
& $\nu_e\to\nu_{\mu,\tau}$ & $\nu_e\to\nu_s$ &
$\nu_\mu\to\nu_\tau$ & $\nu_{\mu,\tau}\to\nu_s$ \\ \hline\hline
\rule[-0.4cm]{0pt}{0.5cm}
$\frac{A}{2\sqrt{2}EG_F}$ & $N_e$ & $N_e-\frac{1}{2}N_n$ & $0$ & 
$-\frac{1}{2}N_n$\\ \hline \end{tabular}
\end{center}
\refstepcounter{tables}
\label{Alist}
\footnotesize
Table \ref{Alist}.
The list of matter densities 
relevant for two-neutrino oscillations (for the definition of $A$
see Eq.(\ref{Adef})).
\end{table}
The list of all possible matter densities which determine $A$ and
occur in the
different oscillation channels is given in Table~\ref{Alist}.
Evidently, $\nu_\mu\leftrightarrow\nu_\tau$ oscillations proceed
as in vacuum because $A=0$ and the term proportional to the unit
matrix in Eq.(\ref{master2}) has no effect on 
transitions.
As mentioned before, in the rest of this section we will have in mind that
Eq.(\ref{master2}) is applied to neutrino propagation in the sun
and the earth.

Let us first define the eigenfunctions of the effective
Hamiltonian as
\begin{equation}\label{eigen}
H^{\mathrm{mat}}_\nu \psi_{mj} = E_j \psi_{mj} \quad \mbox{with} \quad
\psi_{m1} = \left( \begin{array}{r} \cos \vartheta_m \\ -\sin
\vartheta_m \end{array} \right) \,, \;
\psi_{m2} = \left( \begin{array}{r} \sin \vartheta_m \\ 
\cos \vartheta_m \end{array} \right) \,,
\end{equation}
where the matter angle $\vartheta_m$ and the energy eigenvalues 
$E_j$ are functions of $x$. In
the limit of vanishing matter density $\vartheta_m$ is identical with 
the vacuum angle $\vartheta$ (\ref{vacangle}). The eigenvalues
$E_j$ and the matter angle are given by
\begin{equationarray}
E_{1,2}
\null & \null = \null & \null
\frac{1}{4E} \left\{ m_1^2 + m_2^2 + 
2\sqrt{2} G_F \left( N(\nu_\alpha) + N(\nu_\beta) \right)
\right.
\nonumber
\\
\null & \null & \null
\left.
\hspace{2cm}
\mp
\sqrt{(A-\Delta m^2 \cos 2 \vartheta)^2 + (\Delta m^2 \sin 2\vartheta)^2}
\right\}
\label{eigenvalues}
\end{equationarray}%
% \begin{equation}\label{eigenvalues}
% E_{1,2} = \frac{1}{4E} \left\{ m_1^2 + m_2^2 + 
% 2\sqrt{2} G_F \left( N(\nu_\alpha) + N(\nu_\beta) \right) \mp
% \sqrt{(A-\Delta m^2 \cos 2 \vartheta)^2 + (\Delta m^2 \sin 2\vartheta)^2}
% \right\}
% \end{equation}
and
\begin{equation}\label{matterangle}
\tan 2\vartheta_m = \frac{\tan 2\vartheta}{\displaystyle 1 - 
\frac{A}{\Delta m^2 \cos 2\vartheta}} \,,
\end{equation}
respectively.
The general solution of the differential equation
(\ref{master2}) can be represented as 
\begin{equation}\label{solution}
\psi(x) = \sum_{j=1}^2 a_j(x) \psi_{mj}(x) \exp \left( -i
\int_{x_0}^x \mathrm{d}x' E_j(x') \right) 
\end{equation}
with the coefficients $a_{1,2}$ fulfilling
\begin{equation}\label{coeff}
\frac{\mathrm{d}}{\mathrm{d}x} 
\left( \begin{array}{c} a_1 \\ a_2 \end{array} \right) =
\left( \begin{array}{cc} 0 & -\frac{\mathrm{d}\vartheta_m}{\mathrm{d}x}
\exp \left( i \int_{x_0}^x \mathrm{d}x'\Delta E(x') \right) \\
\frac{\mathrm{d} \vartheta_m}{\mathrm{d}x} 
\exp \left( -i \int_{x_0}^x \mathrm{d}x' \Delta E(x') \right) & 0 
\end{array} \right)
\left( \begin{array}{c} a_1 \\ a_2 \end{array} \right)
\end{equation}
and $\Delta E \equiv E_2-E_1$. 

The \emph{adiabatic solution} is defined as an approximate
solution where $\dot{a_j} \simeq 0$ ($j=1,2$).\footnote{We use
the dot above a symbol as abbreviation for $\frac{\mathrm{d}}{\mathrm{d}x}$.}
Postponing 
the discussion of the question under which
condition adiabaticity is fulfilled, we consider temporarily the
case of an arbitrary number $n$ of neutrino flavours or types and 
define the mixing matrix in matter $U_m(x)$ via
\begin{equation}\label{eigenfunctions}
U_m(x)^\dagger H^{\mathrm{mat}}_\nu U_m(x) = 
\mbox{diag} (E_1(x), \ldots, E_{n}(x)) \quad \mbox{with} \quad
U_m(x) = (\psi_{m1}, \ldots, \psi_{m n}) \,,
\end{equation} 
where the eigenvectors $\psi_{mj}$ generalize Eq.(\ref{eigen}). One
readily obtains the generalization of the vacuum oscillation
amplitude Eq.(\ref{prob2}) for neutrino oscillations in matter if the
evolution of the neutrino wave function is adiabatic:
\begin{equation}\label{adiab}
\mathcal{A}^{\mathrm{adiab}}_{\nu_\alpha\to\nu_\beta} = 
\sum_j U_m(x_1)_{\beta j} 
\exp \left( -i \left\{ \delta_j + \int_{x_0}^{x_1} \mathrm{d}x' E_j(x') \right\} \right)
U^*_m(x_0)_{\alpha j} \,.
\end{equation}
In this formula neutrino production and detection happen at
$x_0$ and $x_1$, respectively. The adiabatic phases $\delta_j$,
which are defined by 
\begin{equation}\label{dphase}
\delta_j \equiv -i \int_{x_0}^{x_1} \mathrm{d}x' 
\psi_{mj}(x')^\dagger \dot{\psi}_{mj}(x') \,,
\end{equation}
are necessary for the correct evolution of $\psi(x)$ in the
adiabatic limit. Usually, they can be absorbed into the eigenvectors
$\psi_{mj}$
\cite{schiff}, except for special matter densities, where these phases
acquire a topological meaning, together with the presence of CP
violation \cite{Naumov-phases}. Note that for real vectors $\psi_{mj}$ one has
$\psi_{mj}^\dagger \dot{\psi}_{mj} = 0$ and therefore
$\delta_j = 0$.\footnote{The scalar product in Eq.(\ref{dphase}) is
imaginary for arbitrary vectors $\psi_{mj}(x)$ normalized to one 
$\forall x$, therefore, for real vectors it must be zero.} 
This is so in the two-dimensional case Eq.(\ref{eigen}).

Evaluating Eq.(\ref{adiab}) for two neutrino types and assuming that an 
averaging over neutrino energies takes place such that
\begin{equation}
\left\langle \exp \left( -i \int_{x_0}^{x_1} \mathrm{d}x' \Delta E(x') \right)
\right\rangle_{\mathrm{av}} = 0 \,,
\end{equation}
it is easy to show that the averaged two-neutrino survival probability can be
written in the adiabatic case as 
\begin{equation}\label{avprob}
\bar P_{\nu_\alpha\to\nu_\alpha} = 
\frac{1}{2} (1 + \cos 2\vartheta_m(x_0) \cos 2\vartheta_m(x_1)) \,.
\end{equation}

Let us now derive a condition for adiabaticity in the two-neutrino case.
The evolution of the neutrino state in matter is adiabatic if
the right-hand side of Eq.(\ref{coeff}) can be neglected. 
A formal solution of Eq.(\ref{coeff}) is given by
\begin{equation}\label{sum}
\left( \begin{array}{c} a_1(x) \\ a_2(x) \end{array} \right) =
\sum_{k=0}^\infty \int_{x_0}^x \mathrm{d}x' J(x') \int_{x_0}^{x'}
\mathrm{d}x^{\prime\prime} J(x^{\prime\prime}) \cdots \int_{x_0}^{x^{(k-1)}}
\mathrm{d}x^{(k)} J(x^{(k)})
\left( \begin{array}{c} a_1(x_0) \\ a_2(x_0) \end{array} \right) \,,
\end{equation}
where $J$ denotes the matrix on the right-hand side of
Eq.(\ref{coeff}). Defining the variable 
\begin{equation}
\alpha \equiv \frac{1}{2} \int_{x_0}^x \mathrm{d}x' \Delta E(x')
\end{equation}
and the \emph{adiabaticity parameter} (as a function of $x$) 
\cite{parke,haxton,kuo,haxton95}
\begin{equation}\label{adpar}
\gamma (x) \equiv \frac{\Delta E}{2 |\dot{\vartheta}_m|}
\end{equation}
one can determine an upper bound to the typical integral which occurs in all
the terms of the sum (\ref{sum}) on the right end. In an interval
$[y_0,y_1]$ where $\gamma$ is monotonous there is a value $y_c$
such that \cite{born}
\begin{equation}
\pm \int_{y_0}^{y_1} dy\, \dot{\vartheta}_m \cos 2\alpha =
\int_{\alpha(y_0)}^{\alpha(y_1)} d\alpha \,\frac{1}{\gamma}\, \cos
2\alpha = \frac{1}{\gamma(y_0)} \int_{y_0}^{y_c} d\alpha \cos 2\alpha
+
\frac{1}{\gamma(y_1)} \int_{y_c}^{y_1} d\alpha \cos 2\alpha
\end{equation}
according to the second mean value theorem of integral calculus 
($y_0 < y_c < y_1$). This consideration allows to put the bound
\begin{equation}
\left| \int_{y_0}^{y_1} dy\, \dot{\vartheta}_m e^{2i\alpha} \right| \leq
\frac{2\sqrt{2}}{\gamma_\mathrm{min}} \quad \mbox{with} \quad
\gamma_\mathrm{min} \equiv \min_{y \in [y_0,y_1]} \gamma(y) \,.
\end{equation}
Having found this bound, in the rest of the integrals in the terms in
Eq.(\ref{sum}) one can simply take the absolute values of the
integrands and one is left with integrations of the type
$\mathrm{d}x |\dot{\vartheta}_m| =  |\mathrm{d} \vartheta_m|$. With
the refinement that the interval $[x_0,x]$ possibly has to be divided into
several parts labelled by $a$ such that in each part interval
$\gamma(y)$ is monotonous we can define 
\begin{equation}
{\bar\gamma}^{-1} \equiv \sum_a \gamma_{\mathrm{min},a}^{-1}
\end{equation}
and obtain the exact bound
\begin{equation}\label{adco}
| a_j(x) - a_j(x_0) | \leq \frac{2\sqrt{2}}{\bar\gamma}
\left. \left( \begin{array}{cc} 
\cosh \Delta \Theta & \sinh \Delta \Theta \\ 
\sinh \Delta \Theta & \cosh \Delta \Theta 
\end{array} \right)
\left( \begin{array}{c} |a_1(x_0)| \\ |a_2(x_0)| \end{array} \right)
\right|_j \,,
\end{equation}
where the symbol $|_j$ denotes the $j$-th element of the vector on the
right-hand side of this inequality and $\Delta \Theta$ has a
contribution $|\Delta \vartheta_m|$ for every interval where
$\dot{\vartheta}_m$ has a definite sign. ($\Delta \vartheta_m$ is the
difference between the angle $\vartheta_m$ in the initial and the 
final point of such an interval.)
Thus with the exact bound (\ref{adco}) we have found the following 
sufficient condition:
\begin{equation}
\bar\gamma \gg 1 \quad \Rightarrow \quad
\mbox{\textbf{adiabatic evolution}} \,. 
\end{equation}
It is fulfilled if on the scale of the oscillation length in matter
given by $1/\Delta E$ the matter angle $\vartheta_m$ changes
very little, \textit{i.e.}, $|\dot{\vartheta}_m| \ll \Delta E$. 
Eq.(\ref{adco}) allows to get an upper bound on the crossing or
jumping probability from $\psi_{m1}$ to $\psi_{m2}$. Assuming the
initial conditions $a_1(x_0)=1$ and $a_2(x_0)=0$ we get
\begin{equation}
|a_2(x)| \leq \frac{2\sqrt{2}}{\bar\gamma} \sinh \Delta \Theta
\end{equation}
which illustrates once more the importance of $\bar\gamma$ for
adiabaticity. 

\subsection{The resonance}

In the context of the solar neutrino problem the discovery that
a \emph{resonance} in the passage of $\nu_e$ through
the sun is possible \cite{MS} gave a major boost to the
investigation of the propagation of neutrinos in matter. The
possibility of a resonance is most easily understood in the
adiabatic approximation by looking at Eqs.(\ref{matterangle}) and
(\ref{avprob}). For different neutrino masses we can always label them
in such a way that $\Delta m^2 > 0$. If on the way from the creation 
point $x_0$ in matter of $\nu_e$ to a 
point $x_1$ in vacuum the neutrino passes 
through a point $x_{\mathrm{res}}$ where the \emph{resonance condition}
\begin{equation}\label{rescond}
A(x_{\mathrm{res}}) = \Delta m^2 \cos 2\vartheta
\end{equation}
is fulfilled then $\vartheta_m(x_1) \equiv \vartheta$
has to be between $0^\circ$ and $45^\circ$ 
(for $\nu_e$ the quantity $A$ is positive!) whereas
according to Eq.(\ref{matterangle}) the matter angle
$\vartheta_m(x_0)$ is found between $45^\circ$ and
$90^\circ$.\footnote{Without loss of generality we consider 
$0^\circ < \vartheta < 90^\circ$.}
Consequently, the product of cosines in the $\nu_e$ survival
probability (\ref{avprob}) is negative and the probability to
find a $\nu_e$ after the passage through the sun is less than
1/2. The interesting phenomenon is that this can happen even for
relatively small mixing angles, provided the matter potential at
the production point is large enough such that there is a point
$x_{\mathrm{res}}$ where the condition (\ref{rescond}) is fulfilled.
Indeed, for $\vartheta$ close to $0^\circ$, one can even have
$\vartheta_m(x_1)$ close to $90^\circ$ and thus 
$\bar P_{\nu_e\to\nu_e} \simeq 0$ (\ref{avprob}).
It has been shown that the resonance is also effective in a
certain area in the $\Delta m^2$--$\sin^2 2\vartheta$ plane
where the evolution of the neutrino state is non-adiabatic. For
antineutrinos a resonance is possible if
$45^\circ < \vartheta < 90^\circ$.

If there is a resonance, then for reasonable matter densities the
adiabaticity parameter (\ref{adpar}) is smallest at the resonance
and thus adiabaticity is most likely violated there. Therefore,
considering  $\gamma$ at the resonance,
from Eq.(\ref{matterangle}) one easily computes
$\left. \dot{\vartheta}_m \right|_{\mathrm{res}} =
\dot{A}_{\mathrm{res}}/(2 \sin 2\vartheta\, \Delta m^2)$. A
suitable measure of adiabaticity in the case of a resonance
is thus given by \cite{parke,haxton}
\begin{equation}
\gamma_\mathrm{res} \equiv 
\left. \frac{\Delta E}{2 |\dot{\vartheta}_m|} \right|_{\mathrm{res}} =
\frac{\Delta m^2 \sin^2 2\vartheta}{2E \cos 2\vartheta\,
(|\dot{A}|/A)_{\mathrm{res}}} \,.
\end{equation}

In the solar interior the electron density is maximal in
the center with $\left. N_e \right|_\mathrm{max} \simeq 100
\times N_\mathrm{avo}$ where $N_\mathrm{avo} = 6.022 \times
10^{23} \; \mbox{cm}^{-3}$ and $\left. N_n \right|_\mathrm{max}
\simeq \frac{1}{2} \left. N_e \right|_\mathrm{max}$ 
\cite{Bahcall-Ulrich88}.
This leads to
\begin{equation}
2\sqrt{2} G_F E \left. N_e \right|_\mathrm{max} \simeq 1.5 \times
10^{-5} \left( \frac{E}{1 \; \mbox{MeV}} \right) \: \mbox{eV}^2 \,.
\end{equation}
Considering the resonance condition (\ref{rescond}) one can read
off from this equation for which $\Delta m^2$ a resonance is possible.
Whereas the density variation of $N_e$ and $N_n$ in the sun is smooth,
in the earth due to its layered structure (lithosphere -- mantle --
core) the densities can be approximated as step functions
\cite{lisi97,GKM98-atm}. Furthermore, approximately $N_n \simeq N_e$
is valid in all layers. In the lithosphere, which is relevant for LBL neutrino
oscillation experiments, the average mass density $\rho$ 
is around 3 g/cm$^3$. It increases to around 13 g/cm$^3$ in the core. 
Thus we can conveniently write
\begin{equation}\label{litho}
2\sqrt{2} G_F E\, N_e \simeq 2.3 \times 10^{-4} \: \mbox{eV}^2 \: 
\left( \frac{\rho}{3\, \mathrm{g}\, \mathrm{cm}^{-3}} \right)
\left( \frac{E}{1 \; \mbox{GeV}} \right) .
\end{equation}

\subsection{Non-adiabatic neutrino oscillations in matter and
crossing probabilities}

The analogue of the adiabatic formula (\ref{adiab}) in terms of
probabilities is given by
\begin{equation}\label{nonadiab}
P_{\nu_\alpha\to\nu_\beta} = 
\left| \sum_{j,k} U_m(x_1)_{\beta j} B(x_1,x_0)_{jk}
U^*_m(x_0)_{\alpha k} \right|^2
\end{equation}
with
\begin{eqnarray}
B(x_1,x_0) & = &
U_m^\dagger(x_1)\, P \exp \left\{ -i \int_{x_0}^{x_1}
\mathrm{d}x\, H_\nu^\mathrm{mat}(x) \right\} U_m(x_0) \nonumber\\
& = &
P \exp \left\{ -i \int_{x_0}^{x_1} \mathrm{d}x
\left( \hat{E} -i\, U_m^\dagger(x) \dot{U}_m(x) \right) \right\} \,,
\label{B}
\end{eqnarray}
where $\hat{E}$ is the diagonal matrix of energy eigenvalues present in
Eq.(\ref{eigenfunctions}) (see Ref.~\cite{Kuo-Pantaleone89}). 
$B(x_1,x_0)$ is a unitary $n \times n$ matrix which is
diagonal in the adiabatic limit corresponding to 
$-i\, U_m^\dagger(x) \dot{U}_m(x) \simeq 
\mathrm{diag}\, (\delta_1, \ldots, \delta_n)$ (see Eq.(\ref{dphase})). 

Confining ourselves now to two neutrino types
and having in mind neutrino production in matter and detection in
vacuum we use the notation $\vartheta_m(x_0) \equiv \vartheta_m^0$ 
and $\vartheta_m(x_1) = \vartheta$. With the crucial assumption that
averaging over neutrino energies and the neutrino production region
all terms other than probabilities can be dropped we get
\begin{equation}\label{BPc}
\langle B_{jk} B^*_{j'k'} \rangle_\mathrm{av} = \delta_{jj'} \delta_{kk'}
|B_{jk}|^2 \quad \mbox{and} \quad
|B_{12}|^2 = |B_{21}|^2 = P_c \,,\; |B_{11}|^2 = |B_{22}|^2 = 1-P_c \,,
\end{equation}
where $P_c$ is the crossing probability from $\psi_{m1}$ to
$\psi_{m2}$ and in the second part of Eq.(\ref{BPc}) unitarity has been
used. Such averaging procedures have been shown to be
effective in the context of solar neutrinos, \textit{e.g.}, in
Refs.~\cite{pet88,pet-rich}. Inserting the relations (\ref{BPc}) into
Eq.(\ref{nonadiab}) specialized to the survival probability of $\nu_e$
one obtains after some algebra with trigonometric functions the
Parke formula 
\begin{equation}\label{surviv}
\bar P_{\nu_e\to\nu_e} = \frac{1}{2} + 
\left( \frac{1}{2} - P_c \right) \cos 2\vartheta \cos 2\vartheta_m^0 \,,
\end{equation}
which is the
generalization of Eq.(\ref{avprob}) \cite{kuo,Kuo-Pantaleone89}.
This formula was first derived in the context of an exact solution of
Eq.(\ref{master2}) for matter densities linear in $x$
\cite{haxton,pet87} (see also Refs.~\cite{zener,parke}).

The probability $P_c$ can be estimated with the Landau -- Zener method
\cite{landau,zener}. Following the derivation of Landau, the idea 
is to make an analytic
continuation  of the matter densities and therefore of the effective
matter Hamiltonian (\ref{mat}) into the complex plane by $x \to z$
where $z$ is a complex variable. Then also the 
neutrino state $\psi$ (\ref{solution}) is analytic. Considering the
energy eigenvalues $E_{1,2}(x)$ (\ref{eigenvalues}) as functions of
$z$ we find two branching points $z_0$ and $z^*_0$ 
($\mbox{Im}\, z_0 > 0$) defined by the equation
\begin{equation}\label{z0}
E_1(z_0) = E_2(z_0) \quad \Leftrightarrow
A(z_0) = \Delta m^2 e^{\pm 2i\vartheta} \,.
\end{equation}
We consider $E_1(z)$ and $\psi_{m1}(z)$
for definiteness and follow its evolution along a
curve $C(z)$ which starts at a value $z$ in the vicinity of the real
axis and goes along a negatively oriented loop around $z_0$ such that we end
at the same point $z$ but now located on the second sheet.\footnote{It
is convenient to imagine cuts from $z_0$ to $z_0 + i\infty$ and 
$z^*_0$ to $z^*_0 - i\infty$ in the complex plane. Crossing these
cuts, one moves from one sheet to the other.}
Along this path
$E_1(z)$ changes into $E_2(z)$ \cite{landau} and $\psi_{m1}(z)$ into
$\psi_{m2}(z)$.\footnote{In general this change is accompanied by a
factor of geometrical origin. It has no effect in our case for the
jumping probability \cite{joye}.} The crucial observation
is that, though $E_j(z)$ and $\psi_{mj}(z)$ ($j=1,2$) are two-valued
analytic objects, the analytic continuation $\psi(z)$ of the solution
(\ref{solution}) of the differential equation (\ref{master2}) is
single-valued \cite{joye}. From this fact it can easily be shown 
with Eq.(\ref{solution}) that \cite{joye}
\begin{equation}\label{ancont}
\tilde{a}_1(z) = \exp \left( i\int_{C(z)} E_1(z)dz \right) a_2(z) \,,
\end{equation}
where $\tilde{a}_1(z)$ is the analytic continuation of $a_1(x)$ in the
vicinity of the real axis and then along $C(z)$. The Landau -- Zener
crossing probability is derived from the following
consideration. Starting on the real axis at $x_- \ll x_\mathrm{res}$,
where the evolution of $\psi$ is adiabatic and where the boundary
conditions $a_1(x_-) = 1$, $a_2(x_-) = 0$ hold, and going along a path in
the complex plane which passes above $z_0$ and returning to the real
axis at a point $x_+ \gg x_\mathrm{res}$ on the second sheet 
this path can be chosen such
that the evolution in the complex variable $z$ is adiabatic because we
have passed the resonance in safe distance and we have thus
$|\tilde{a}_1(x_+)| \simeq 1$. On the other hand, we can go from $x_+$
on the first sheet to the
point $x_+$ on the second sheet via $C(x_+)$ and 
with Eq.(\ref{ancont}) we thus obtain
\begin{equation}\label{Pcc}
P_c = |a_2(x_+)|^2 = 
\left| \exp \left( -i\int_{C(x_+)} E_1(z)dz \right) \right|^2 \,,
\end{equation}
neglecting deviations of $|\tilde{a}_1(x_+)|$ from 1.
From this formula we can read off that integrations along the real axis do
not contribute to $P_c$. Because of analyticity, the path
$C(x_+)$ can be deformed such that it goes along the real axis from
$x_+$ to $x_\mathrm{res}$ from where it leads to
$z_0$, circles around the branching point with infinitely small radius
and then goes back to $x_\mathrm{res}$ and $x_+$ on the second
sheet. In this way, we obtain from Eq.(\ref{Pcc})
the final result for the Landau -- Zener crossing probability
\begin{equation}\label{Pc}
\ln P_c = -\frac{1}{E}\, \mbox{Im} \int_{x_\mathrm{res}}^{z_0} \! dz \,
[(A-\Delta m^2 \cos 2\vartheta)^2 + (\Delta m^2 \sin 2\vartheta)^2 ]^{1/2}
\,.
\end{equation}
To get this formula we have used the explicit form of the energies of
the adiabatic states (\ref{eigenvalues}). 

With the variable
transformation $A = A(z)$ and $dz = dA/ \frac{dA}{dz}$ it is easy to
evaluate Eq.(\ref{Pc}) for a linear density
\cite{parke,haxton,kuo,pal} and is also possible for an exponential
density \cite{pizz}. These calculations can be summarized by
\cite{kuo,Kuo-Pantaleone89} 
\begin{equation}
P_c = \exp \left(-\frac{\pi}{2} \gamma_\mathrm{res}F \right)
\end{equation}
with $F=1$ for the linear and $F=1-\tan^2 \vartheta$ for the
exponential case. For an exponentially varying matter density 
the dependence of $P_c$ on $\vartheta$ is particularly simple because
$\gamma_\mathrm{res}F = 4\delta \sin^2 \vartheta$
with $\delta \equiv \gamma_\mathrm{res} \cos 2\vartheta/\sin^2
2\vartheta$ which is independent of the vacuum mixing angle.
For a more general discussion of the crossing probability (\ref{Pc}) 
see Ref.~\cite{beacom}.

Exact solutions of the differential equation (\ref{master2}) for
neutrino oscillations in matter exist not only for the linear case
\cite{zener,parke,haxton,pet87} in terms of Weber functions but also for the
exponentially varying matter density in terms of Whittaker functions
\cite{exponential,pet88}. This case is of particular importance
for solar neutrinos because it approximates the real density variation
in the sun. Further exact solutions are known for $A$ varying with
$\tanh x$ \cite{noetzold} and $1/x$ \cite{kuo}. 
In Refs.~\cite{kuo,Kuo-Pantaleone89}
a list of the factors $F$ for all these cases is given, calculated with
the Landau -- Zener formula (\ref{Pc}).

The survival probability (\ref{surviv}) with $P_c$ in the Landau --
Zener approximation does not reproduce well the exact survival
probability in the extremely non-adiabatic region \cite{pet87,pet88} 
(see also Refs.~\cite{kuo,Kuo-Pantaleone89,pal}). In these references the
example of an extremely dense medium ($A \to \infty$ and thus $\vartheta^0_m
= \pi/2$) with a sharp boundary to the vacuum is given. In this case
the neutrino does not oscillate in the medium and therefore
$\bar P_{\nu_e\to\nu_e} = 1 - \frac{1}{2} \sin^2 2\vartheta$ stems purely
from vacuum oscillations. This has to be compared with $P_c = 1$,
because of the jump in density one gets $\gamma_\mathrm{res}=0$, inserted into
Eq.(\ref{surviv}). Obviously, the resulting expression
$\bar P_{\nu_e\to\nu_e} = \cos^2 \vartheta$ does not agree 
with the previous one. 

A remedy of
this deficiency was found in the framework of the exact solution for
an exponentially varying matter density leading to the following 
modification \cite{pet88} of the Landau -- Zener crossing probability
$P_c$ (\ref{Pc}):
\begin{equation}\label{Pc1}
P_c = \frac{\exp \left(-{\displaystyle \frac{\pi}{2}} \gamma_\mathrm{res} F 
\vphantom{-{\displaystyle \frac{\pi}{2}} \gamma 
{\displaystyle \frac{F}{\sin^2 \vartheta}}}
\right) -
\exp \left(-{\displaystyle \frac{\pi}{2}} \gamma_\mathrm{res}
{\displaystyle \frac{F}{\sin^2 \vartheta}} \right)}%
{1 - \exp \left(-{\displaystyle \frac{\pi}{2}} \gamma_\mathrm{res}
{\displaystyle \frac{F}{\sin^2 \vartheta}} \right)} \,.
\end{equation} 
Numerical calculations for solar neutrinos with an exponential
electron density using this formula agree
very well with numerical integrations of the differential equation
(\ref{master2}), typically the agreement is in the percent range for
relevant mixing parameters \cite{krapet88}. It has been conjectured
in Ref.~\cite{kuo} that the form (\ref{Pc1}) of $P_c$ holds for a wide
class of matter density profiles. In any case, in the limit
$\gamma_\mathrm{res} \to 0$ one gets $P_c \to \cos^2 \vartheta$ which,
when inserted into formula (\ref{surviv}),
correctly describes the survival probability in the above example of
an extremely non-adiabatic evolution.

Concluding this section, we consider again an arbitrary number of neutrino
flavours or types and envisage the interesting case 
where one of the eigenfunctions
(\ref{eigenfunctions}) of $H^\mathrm{mat}_\nu$ labelled by the index $\ell_0$ 
has an adiabatic evolution whereas the other part of $\psi$ has an
arbitrary evolution. We assume
exact adiabaticity for $\psi_{m \ell_0}$ which amounts to 
$B_{j\ell_0} = B_{\ell_0 k} = 0$ for $j,k \neq \ell_0$. This
allows to write
\begin{equation}
P_{\nu_\alpha\to\nu_\beta} = 
(1-|U_m(x_1)_{\beta \ell_0}|^2) (1-|U_m(x_0)_{\alpha \ell_0}|^2)
P^{(\ell_0)}_{\nu_\alpha\to\nu_\beta} + 
|U_m(x_1)_{\beta \ell_0}|^2 |U_m(x_0)_{\alpha \ell_0}|^2 \,,
\label{aaa}
\end{equation}
where we have defined
\begin{equationarray}
P^{(\ell_0)}_{\nu_\alpha\to\nu_\beta}
\null & \null = \null & \null
[(1-|U_m(x_1)_{\beta \ell_0}|^2) (1-|U_m(x_0)_{\alpha \ell_0}|^2)]^{-1}
\left| \sum_{j,k \neq \ell_0} U_m(x_1)_{\beta j} B(x_1,x_0)_{jk}
U^*_m(x_0)_{\alpha k} \right|^2
\nonumber
\\
\null & \null \leq \null & \null
1
\,.
\label{p0}
\end{equationarray}%
% \begin{equation}\label{p0}
% P^{(\ell_0)}_{\nu_\alpha\to\nu_\beta} =
% [(1-|U_m(x_1)_{\beta \ell_0}|^2) (1-|U_m(x_0)_{\alpha \ell_0}|^2)]^{-1}
% \left| \sum_{j,k \neq \ell_0} U_m(x_1)_{\beta j} B(x_1,x_0)_{jk}
% U^*_m(x_0)_{\alpha k} \right|^2 \leq 1 \,.
% \end{equation}
That $P^{(\ell_0)}_{\nu_\alpha\to\nu_\beta}$ is smaller than 1
follows from the Cauchy -- Schwarz inequality and from the
above assumption for $B$ because after dropping the row and 
the column labelled 
by $\ell_0$ in $B$ the remaining matrix is again unitary.
Eq.(\ref{p0}) and similar cases are useful if neutrino masses
differing by orders of magnitude occur.

\section{Indications of neutrino oscillations}
\label{Indications of neutrino oscillations}
\setcounter{equation}{0}
\setcounter{figures}{0}
\setcounter{tables}{0}

\subsection{Atmospheric neutrino experiments}
\label{Indications: atmospheric neutrino experiments}

\subsubsection{The atmospheric neutrino flux}

In 1912 it was discovered by V.F. Hess \cite{hess} in a manned balloon flight 
that the intensity of the ionizing radiation in the atmosphere 
as a function of the altitude did not conform with the idea 
that this ionization was caused by radioactive elements in the surface
of the earth but rather
pointed to an extraterrestrial origin. In the following decades,
before the advent of accelerator physics, this
radiation, which was called first ``ultraradiation'' and later
baptized ``cosmic rays'' by R.A. Millikan, proved to be one of the most
fruitful means for doing particle physics experiments. At the end of
the first half of the 20th century such experiments had lead to the
discovery of the positron, the pion and the muon 
and also the first particles with strangeness were found with
cosmic rays \cite{Pai94}. Eventually, 
in the beginning of the fifties, proton beams from accelerators
replaced the cosmic proton flux as an experimental tool. 
However, after many years
where accelerator physics was dominating in particle physics, at
the end of the 20th century cosmic rays play again a major role
through atmospheric neutrinos which allow to use the whole
globe as a neutrino physics laboratory and to probe neutrino
mass-squared differences down to a few $10^{-4}$ eV$^2$. In this way 
convincing evidence for the existence of neutrino oscillations and thus
for non-zero neutrino masses has been obtained \cite{SK-atm-98}.

In a simplified picture, the production of atmospheric neutrinos
\cite{Vol80,Mit86,Bugaev-86,Bugaev-87,But89,Bar89,Lee89,Bug89,%
Hon90,Hon95,Agr96,Gai96,Lip98b} 
proceeds in three steps \cite{Gai90}. 
In the first step the primary cosmic rays \cite{Gai90,Berezinsky90}
hit the nuclei in the
atmosphere, thereby producing charged pions and kaons, either
directly or via intermediate particles. In the second
step, the decay of these particles gives rise to part of the
atmospheric $\nu_\mu$ and $\bar{\nu}_\mu$ neutrino fluxes:
\begin{equation}
\pi^+ \to \mu^+ + \nu_\mu \,, \; \pi^- \to \mu^- + \bar{\nu}_\mu
\quad \mbox{and} \quad
K^+ \to \mu^+ + \nu_\mu \,, \; K^- \to \mu^- + \bar{\nu}_\mu \,.
\label{pik}
\end{equation}
In the third step, the $\nu_e$ and $\bar{\nu}_e$ fluxes and further
$\nu_\mu$ and $\bar{\nu}_\mu$ fluxes are produced by
\begin{equation}
\mu^+ \to e^+ + \nu_e + \bar{\nu}_\mu 
\quad \mbox{and} \quad
\mu^- \to e^- + \bar{\nu}_e + \nu_\mu \,.
\label{mu}
\end{equation}
There is also a contribution to the neutrino fluxes from the decays 
\begin{equation}
K_L \to \pi^+ + \ell^- + \bar{\nu}_\ell
\quad \mbox{and} \quad
K^+ \to \pi^0 + \ell^+ + \nu_\ell
\label{kl} 
\end{equation}
with $\ell = e, \mu$ and the charge conjugate processes 
\cite{Gai90,Lip93,Hon95} which does not conform with the simple 3-step
picture. At low energies the most important process in
Eq.(\ref{pik}) is the pion decay. The contribution of 
$K_{\ell 3}$ decays (\ref{kl}) to the neutrino fluxes is small
\cite{Naumov-97,Naumov-98}. Let us now describe
some useful details of the production of the atmospheric neutrino flux
and its properties.

\paragraph{Cosmic rays:}
Galactic cosmic rays enter the solar system as an isotropic flux of
particles. There is convincing evidence that the bulk of the radiation with
energies less than $10^6$ GeV comes from our galaxy. Though there is
no conclusive proof of the origin of cosmic rays yet, plausible
mechanisms range from material ejected by supernovae to interstellar
medium accelerated in supernova shock waves \cite{Sim83}.
Solar cosmic rays are emitted irregularly by major solar flares on the sun.
Apart from electrons, primary cosmic rays consist of protons and bare
nuclei (mostly He). Roughly speaking, 
the chemical composition of the galactic cosmic ray particles
is given approximately by 90 \% H, 9 \% He and less than 1 \% heavier
nuclei. However, the chemical composition varies with energy. At
energies of around 100 MeV per nucleon the particle number ratio H/He
is less than 5, it increases to 10 at 1 GeV and is around 30 at 100
GeV \cite{Web74,Seo91}. 
The absolute flux of cosmic protons is not very large: it is of the order of 
$1000 / \mbox{m}^2 \times \mbox{sec} \times \mbox{sr}$  
for energies of a few GeV and above the atmosphere. As a function of
the energy the proton flux above a few GeV is well described
by the power law $E^{-2.7}$ until $10^6$ GeV \cite{Web74}.
Above an energy of 100 GeV per nucleon the cosmic ray
fluxes are less precisely known \cite{Web74,Seo91}. 

\paragraph{Solar modulation:}
The solar cosmic rays (solar wind)
are important in so far as they weaken and modulate the flux of galactic 
cosmic ray particles with the solar activity. The stronger the solar wind
the more difficult it is for the low energy galactic cosmic rays to
enter the solar sphere of influence. This modulation is noticeable for
kinetic energies of around 10 GeV per nucleon or less \cite{Sim83,Hon95}.
This variation of the primary cosmic ray flux with the approximate
11-year cycle of solar activity can be
parameterized as a function of the neutron monitor at Mt.\ Washington
\cite{Hon95}. It induces a corresponding modulation of the
low energy atmospheric neutrino flux but its influence is small. The effect
of the solar wind on cosmic ray particles
depends on the rigidity of the nuclei which is defined by
momentum/charge. It is strongest for
cosmic ray particles with small rigidity. For
neutrino energies $E_\nu \sim 2$ GeV it is practically negligible, for
energies around 1 GeV only appreciable at high geomagnetic latitudes
because the geomagnetic cut-off admits cosmic ray particles with
smaller rigidity there.

\paragraph{Geomagnetic cut-off:}
The geomagnetic field prevents primary cosmic particles with low
rigidity from entering the atmosphere 
\cite{Bugaev-88,Hon95,Lip95b,Lip98b}. In the atmospheric
neutrino flux calculations it is assumed that the momenta of
the neutrinos have the same direction as the primary cosmic ray
particles responsible for their production \cite{Hon95,Gai96}. This is
a good approximation for $E_\nu > 200$ MeV as shown in
Ref.~\cite{Lee88}. In a first
approximation, for a given direction, there is a cut-off in
rigidity for the primary cosmic ray particles below which such a
particle cannot reach the top of the atmosphere in a vertical
altitude of around 20 km. The value of the geomagnetic cut-off
can be obtained by a computer simulation with the help of the
``backtracking technique''. In this technique, to establish if a
particle with a given charge and momentum coming from a
certain direction can reach a point on top of the atmosphere, one
integrates the equations of motion for a particle with opposite
charge and reflected momentum starting from its final position.
If the backtracked particle reaches infinity it is assumed that
the trajectory is allowed. As infinity one can take a distance
of around 10 times the radius of the earth where the geomagnetic
field has decreased to the level of the interstellar magnetic
field with a strength $\sim 3 \times 10^{-8}$ Tesla
\cite{Hon95}. More refined calculations replace the rigidity
cut-off by a probability distribution \cite{Bugaev-88,Lip95b,Lip98b}. In
the rigidity cut-off approximation one can give for every
experimental location a contour map depicting curves with
constant cut-off in a plot with the coordinate axes representing
azimuth and zenith angles of the primary cosmic ray particle
direction. For Kamioka such a plot is found in Ref.~\cite{Hon95}. 
At the geomagnetic latitude of $90^\circ$ the
geomagnetic cut-off disappears, therefore experiments situated at
high geomagnetic latitudes have a higher flux of low energy
atmospheric neutrinos. The magnetic field of the earth is not
exactly of dipole form but also has a considerable multipole
contribution which affects the local rigidity cut-off maps and
thus the atmospheric neutrino flux in particular for down-going
neutrinos \cite{Hon95}.

\paragraph{Hadronic interactions:}
After entering the atmosphere the cosmic ray particles collide
with the nuclei in the air.
Though the fractions of other particles than protons in the cosmic
rays are small, their importance for the atmospheric neutrino
flux gets
enhanced because the predecessors of the neutrinos, pions and kaons,
are created through the hadronic interactions of the cosmic rays with
the air nuclei. These processes depend rather on the number of
nucleons than on the number of nuclei. The ratio of the 
charged kaon to the charged pion flux depends sensitively on the
description of hadronic interactions. In this context the quantities 
\begin{equation}
Z_{p \mathcal{M}} \equiv \int_0^1 \mathrm{d}x \, x^\gamma \frac{\mathrm{d}N_{p
\mathcal{M}}}{\mathrm{d}x} \,,
\end{equation}
where $\mathcal{M} = \pi, K$ and $\gamma \simeq 1.7$ comes from the
power law of the primary cosmic protons, $x$ is the fraction of the proton
momentum carried by the meson and $\mathrm{d}N/\mathrm{d}x$ is the distribution of the
charged mesons produced by collisions of protons with nuclei in the
atmosphere, are important indicators for the fraction of the muon neutrino flux
originating from the $\mathcal{M}$ meson flux.
The ratio $Z_{pK^\pm}/Z_{p\pi^\pm}$ ranges from 0.10 to
0.15 in different calculations \cite{Gai96}. As a consequence, at 
neutrino energies below 100 GeV the pions dominate as neutrino sources
\cite{Lip93,Agr96}. 

\paragraph{Characteristics of the atmospheric neutrino fluxes:}
Some of the characteristic properties of the atmospheric
neutrino fluxes are simple consequences of the production
mechanisms (\ref{pik}) and (\ref{mu}), whereas others follow from
a Monte Carlo or analytic calculation \cite{Lip93} of the neutrino fluxes. 
As mentioned above, the flux calculations are
one-dimensional and energy losses in the air, in particular, for the
muons, have to be taken into account \cite{Gai90}.
\begin{enumerate}
\renewcommand{\labelenumi}{\alph{enumi}.}
\item Approximately, the 
$\stackrel{\scriptscriptstyle (-)}{\nu}_{\hskip-3pt \mu}$ 
fluxes have a power law
energy spectrum $E_\nu^{-3}$ for $1 \lesssim E_\nu \lesssim 10^3$ GeV
whereas the $\stackrel{\scriptscriptstyle (-)}{\nu}_{\hskip-3pt e}$ 
fluxes decrease like
$E_\nu^{-3.5}$ \cite{Hon95}.
\item Fixing a neutrino energy $E_\nu$, the largest contribution to
the flux of this energy originates from cosmic ray particles
with energy $E_{\mathrm{cr}} \sim 10 \times E_\nu$ \cite{Hon95,Agr96}.
\item Denoting neutrino fluxes by $(\nu)$ then it follows
immediately from Eqs.(\ref{pik}) and (\ref{mu}) that
\begin{equation}
\frac{(\nu_\mu) + (\bar{\nu}_\mu)}{(\nu_e) + (\bar{\nu}_e)}
\simeq 2 \,.
\label{emratio}
\end{equation}
However, it can easily be estimated that, at neutrino energies
larger than 1 GeV, muons from the reactions (\ref{pik})
start to reach the surface of the earth before they decay. 
Therefore the neutrino fluxes from muon decay decrease, in particular,
the $\stackrel{\scriptscriptstyle (-)}{\nu}_{\hskip-3pt e}$ flux,
and the ratio (\ref{emratio}) begins to rise \cite{Gai90,Hon95}.
\item For the same reason muons cease to be the dominant source of 
$(\nu_e) + (\bar{\nu}_e)$ above $E_\nu \simeq 100$ GeV and the
reactions (\ref{kl}) take over \cite{Gai90,Lip93}.
\item Pions cease to be the dominant source of the 
$\stackrel{\scriptscriptstyle (-)}{\nu}_{\hskip-3pt \mu}$ flux for neutrino
energies above 100 GeV \cite{Lip93}. The reason is given by the different
masses and lifetimes of the parent mesons $\mathcal{M} = \pi, K$ 
in the decays $\mathcal{M} \to \mu \nu$.
For relativistic parent mesons and for a given energy $E_\nu$,
the lower bound on the energy of the parent meson in such a
decay is given by
$E_{\mathcal{M}} \geq E_\nu / (1-m_\mu^2/m^2_{\mathcal{M}})$. 
Thus, neutrinos with energy $E_\nu$ from the decay of the heavier
parent meson need in
average a lower parent meson energy where the corresponding meson flux
is higher. This together with the shorter kaon lifetime leads to
the prevailing of kaons as the source of muon neutrinos at
sufficiently high neutrino energies \cite{Osb65,Lip93,Hon95,Agr96}.
\item The ratio $(\bar{\nu}_e) / (\nu_e)$ is smaller than 1
because among the primary cosmic ray nucleons the protons by far
dominate which
leads to an enhancement of positively charged pions over
negative ones in the hadronic interactions and,
consequently,\footnote{In general, we denote particle fluxes by putting
the particle symbol within parentheses.}
$(\pi^+)/(\pi^-) \simeq (\mu^+)/(\mu^-) > 1$ \cite{Lip93,Hon95}.
\item On the other hand, for $E_\nu \lesssim 1$ GeV, where the muons
have time to decay in the atmosphere, one has
$(\bar{\nu}_\mu)/(\nu_\mu) \simeq 1$ because the muons supply the
antineutrinos (neutrinos) to the neutrinos (antineutrinos) from pions
and kaons. However, for neutrino energies above a few GeV, when part
of the muons reaches the surface of the earth before they decay, this
ratio becomes smaller than one because the $\pi^+$ are more numerous
than the $\pi^-$ (see above).
\item For $E_\nu \gtrsim 5$ GeV the geomagnetic effects become negligible.
\end{enumerate}
The largest error in the atmospheric neutrino flux calculations arises
from an overall uncertainty of primary cosmic ray measurements of
the order of $\pm 15 \%$ \cite{Hon95,Gai96}. Therefore, it is
mandatory to use quantities which are ratios such that this
uncertainty cancels or, in the case of a fit to the data, to include 
the overall normalization of the primary cosmic ray flux in the
set of parameters to be determined by the fit. Among the flux calculations of
the different groups the major source of differences comes from
the treatment of pion production in collisions of protons with light nuclei 
\cite{Gai96,Gai97}. Thus, the calculated absolute fluxes have
uncertainties of the order of $\pm 20 \%$ and even larger
uncertainties at
neutrino energies below 1 GeV, whereas flux ratios
differ in general only by a few percent (see the comparison
between different flux calculations in Refs.~\cite{Gai96,Gai97}). 

\paragraph{The angular dependence of the atmospheric neutrino flux:}
The direction of the atmospheric neutrino flux at the location
of an experiment is described by the zenith angle $\theta$ and
the azimuth angle $\phi$. Neutrinos going vertically downward
have $\theta = 0$ whereas those coming vertically upward through the earth
have $\theta = \pi$. The azimuth angle indicates the direction
of the flux in the horizontal plane. There are two natural causes for an 
angular dependence \cite{Lip98b} of the atmospheric neutrino flux:
\begin{enumerate}
\renewcommand{\labelenumi}{\alph{enumi}.}
\item The development of cosmic ray showers in the atmosphere
depends on the density of the air. At zenith angles 
$\theta \sim \pi/2$, where along the line of flight the increase
in density is less steep, pion and kaon decay is enhanced
compared to vertical directions. It is easily checked that the
situation is symmetric with respect to 
$\theta \to \pi - \theta$. Therefore, the generation of pions and
kaons in the atmosphere caused by primary cosmic rays induces a
dependence of the atmospheric neutrino flux on $|\cos \theta|$.
In other words, this mechanism does not introduce an up-down
asymmetry in the flux.
\item The geomagnetic field, which acts on the primary cosmic
ray flux, is the cause for a $\phi$ and $\theta$ dependence of
the atmospheric neutrino flux. Because of the positive charge of
the primary cosmic rays the neutrino flux is highest (lowest) for
directions coming from the west (east). The %local 
nature of the
geomagnetic effects also generates an up-down asymmetry for
low energy neutrinos \cite{Hon95,Lip98b}. 
%For an exact dipole
%field and using St\"ormer's formula for the geomagnetic cut-off
%which neglects the shadow of the earth the atmospheric neutrino
%flux would be again up-down symmetric \cite{Lip98b}.
\end{enumerate}

A third cause for a dependence on the zenith angle is possibly
given by neutrino oscillations which arises because varying
$\theta$ from 0 to $\pi$ the neutrino path length varies from
around 10 km \cite{Gai98} to around 13000 km. Clearly, this dependence
is not up-down symmetric. If it is disentangled from the
up-down asymmetry caused by the geomagnetic effects, \textit{e.g.}, by using
the high energy component of the neutrino flux, it provides us
with valuable information on neutrino masses and mixing.

\subsubsection{Experiments with atmospheric neutrinos}

Early efforts for detecting atmospheric neutrinos (see Ref. \cite{Gai94}
for a summary) concentrated
on neutrino-induced upward muons, \textit{i.e.}, muons with zenith angles
$90^\circ \leq \theta \leq 180^\circ$, or horizontal muons,
\textit{i.e.}, muons with a zenith angle around $90^\circ$, using the process
\begin{equation}
\stackrel{\scriptscriptstyle (-)}{\nu}_{\hskip-3pt \mu} + N \to \mu^\pm + X
\,,
\end{equation}
where $N$ is a nucleon in the rock surrounding the detector
located underground \cite{Mar60,Gre60}. This method was proposed
to distinguish muons generated by atmospheric neutrinos from
muons originating in the meson decays (\ref{pik}) in the
atmosphere. There were two experiments, the Kolar Gold Field
experiment in India \cite{Ach65} and an experiment in South
Africa \cite{Rei65}, reporting the first evidence for
atmospheric muon neutrinos. Since these experiments could not
distinguish up and down directions they used horizontal muons
crossing the detector
which lay, due to the great depth of the experimental sites, in
the large zenith angle intervals $60^\circ \leq \theta \leq 120^\circ$ and
$45^\circ \leq \theta \leq 135^\circ$ for the Indian and the South
African experiment, respectively. These early experiments were
accompanied already by atmospheric neutrino flux
calculations \cite{Mar61,Zat62,Osb65,Cow66}.

Recent detectors \cite{Bei94} are divided into two classes: water Cherenkov
detectors where the neutrino target is a large volume of water
surveyed by a huge array of photomultiplier tubes sitting on the
surface of the volume (the Kamiokande \cite{Hir88,Hir92,Kam-atm-94},
Super-Kamiokande \cite{SK-atm-98} and 
IMB \cite{Bec92,Bec95,Cla97} collaborations)
and iron plate calorimeters where
neutrino-induced charged particles ionize the gas between the
plates and the particle paths are reconstructed electronically
(the Fr\'ejus \cite{Dau95}, NUSEX \cite{Agl89} and Soudan-2
\cite{All97} collaborations). 
In contrast to the early detectors
the recent detectors are sensitive to the direction of tracks and can thus
distinguish between up and down through-going tracks. However,
they cannot measure the charge of the leptons $\ell^\pm$ and thus
cannot distinguish between $\nu_\ell$ and $\bar \nu_\ell$.
Therefore, at low energies they are approximately sensitive the flux
combination $(\nu_\ell)+\frac{1}{3}(\bar{\nu}_\ell)$ ($\ell = e, \mu$)
because at low energies quasi-elastic scattering is predominant
and the ratio of quasi-elastic antineutrino to neutrino
cross sections is approximately 1/3 \cite{Gai94,Gai95,Gai96}. In
atmospheric neutrino physics it is important to distinguish
$e$-like and $\mu$-like events which is accomplished by
distinguishing between showers or diffuse Cherenkov rings 
($e^+, e^-, \gamma$) and
tracks or sharp Cherenkov rings ($\mu^+, \mu^-$ and charged
pions, kaons, protons etc.). The separation between $e$-like and
$\mu$-like events is very good, \textit{e.g.}, for Super-Kamiokande its
efficiency is estimated to be 98\% or better \cite{SK-atm-98}.

For the deep underground detectors two different event classes are
defined \cite{Bei94}. Events in which the neutrino interacts with the material
inside the detector and where all particles from the neutrino interaction 
deposit their energies inside the detector are called 
\emph{contained events}. The second class refers to events where muon
neutrinos interact with the material surrounding the detector via
charged current interactions such that the high energy muons enter the
detector \cite{Fra93}. In this way one distinguishes 
\emph{through-going muons} and
\emph{stopping muons}. Recent measurements of the upward muon
flux were performed by the Baksan \cite{Baksan-up-mu}, Kamiokande
\cite{Tot93,Hat98}, IMB \cite{Bec95}, Fr\'ejus \cite{Dau95}, MACRO
\cite{Amb98} and Super-Kamiokande \cite{SK-atm-nu98} collaborations.

Super-Kamiokande -- and before also Kamiokande
-- has an inner detector surrounded by an outer detector. This allows
to further subdivided the contained events into \emph{fully contained
events} (FC) with all energy of an event deposited in the inner
detector and \emph{partially contained events} (PC) which have exiting
tracks detected also in the outer detector. 
In Kamiokande and Super-Kamiokande, FC
events are separated into those having a visible energy
$E_{\mathrm{vis}} < 1.33$ GeV, the \emph{sub-GeV} events, and those with 
$E_{\mathrm{vis}} > 1.33$ GeV, the \emph{multi-GeV} events. 
Among the contained events in Cherenkov detectors the single-ring
events are well understood, they are predominantly produced by
quasi-elastic scattering of electron and muon neutrinos. In
Kamiokande and Super-Kamiokande for the analyses of FC events only
single-ring events are used with the additional criteria $p_e
\geq 100$ MeV and $p_\mu \geq 200$ MeV for electron and muon
momenta, respectively, in the case of sub-GeV events. 
To quote the numbers of Super-Kamiokande 
\cite{SK-atm-98}, a Monte Carlo simulation has shown that 88\% (96\%) of
the sub-GeV $e$-like ($\mu$-like) events are charged current
interactions whereas for multi-GeV events the number is 84\%
(99\%). The remainder is given by neutral current events.
The PC events were estimated to be 98\% 
$\stackrel{\scriptscriptstyle (-)}{\nu}_{\hskip-3pt \mu}$-induced 
events for single
and multi-ring configurations and, therefore, all PC events are used for
the analyses. 

For the analyses of atmospheric neutrino data it is very important
to have a good understanding of the neutrino interactions in the
detector. The deep-inelastic scattering (DIS) formulas are only valid
for sufficiently high momentum transfer $Q^2$ from the leptons
to the hadrons. For neutrinos with a few GeV interacting with
the detector material, the lowest multiplicity exclusive channels
represent an important fraction of the cross section. Therefore
the following decomposition of the neutrino cross section
\cite{Lip95a} has been proposed:
\begin{equation}
\sigma_\nu^{\mathrm{CC}} = 
\sigma_{\mathrm{QEL}} + \sigma_{1\pi} + \sigma_{\mathrm{DIS}} \,,
\end{equation}
where QEL indicates the quasi-elastic $\nu_\ell + n \to \ell^- + p$
and $\bar{\nu}_\ell + p \to \ell^+ + n$ cross sections and
$1\pi$ single pion production. In the case of the latter it is
assumed that $\Delta(1232)$ dominates. To avoid double counting in
$\sigma_{1\pi}+\sigma_{\mathrm{DIS}}$, the
maximal mass of the pion -- nucleon system in the final state is
taken to be $W_c = 1.4$ GeV for single pion production whereas the
kinematical region for DIS is bounded by $W \geq W_c$,
where $W$ is the mass of the hadronic final state.

\subsubsection{The atmospheric neutrino anomaly}

The first observable to be measured in recent atmospheric neutrino
experiments was the ratio of $\mu$-like to $e$-like events
denoted by $(\mu/e)_{\mathrm{data}}$. As discussed earlier, in
flux ratios the large uncertainty in the overall normalization
of the primary cosmic ray flux cancels and there is also
some cancellation of errors in the theoretical calculation.
However, the above ratio is only a limited reflection of the
corresponding ratio of atmospheric neutrino fluxes because of 
detector efficiencies and event selection criteria. Thus for the
expected ratio one has to fold the theoretical flux calculations
with the cross sections of the neutrino interactions in the
detector and the detection efficiencies and apply the event
selection criteria. Quoting the result of the Monte Carlo
calculation of Super-Kamiokande as an example, this
collaboration obtains $(\mu/e)_{\mathrm{MC}} = 1.50$ and $2.83$
for the sub-GeV ratio of FC events and the ratio considering multi-GeV
FC and PC events, respectively \cite{SK-atm-98}, using the neutrino flux
calculations of Ref.~\cite{Hon95}. Note that these numbers 
significantly deviate from the naive expectation 2. Therefore,
the actual physically relevant quantity is given by the double ratio
\begin{equation}\label{R}
R = \frac{(\mu/e)_{\mathrm{data}}}{(\mu/e)_{\mathrm{MC}}} \,.
\end{equation}

The first indication that this ratio is smaller than 1 was
reported more than ten years ago \cite{Hai86}. In the meantime
the most impressive measurements of $R$ are represented by the
Kamiokande and Super-Kamiokande results:\footnote{For recent results on
atmospheric neutrinos see \emph{Note added}.}
\begin{equation}
R = \left\{
\begin{array}{rr}
\left. 
\begin{array}{crr}
0.60 \!\begin{array}{c} +0.07 \\[-3pt] 
                      -0.06 \end{array} \!\pm 0.05 &
\; \mbox{sub-GeV} \;   & \mbox{\protect\cite{Hir92}} \\
0.57 \!\begin{array}{c} +0.08 \\[-3pt] -0.07 \end{array} \!\pm 0.07 &
\; \mbox{multi-GeV} \; & \mbox{\protect\cite{Kam-atm-94}}
\end{array} 
\right\} & 
\quad \mbox{Kamiokande} \\[8mm]
\left. \!
\begin{array}{crr}
0.63 \pm 0.03 \pm 0.05 & \; \mbox{sub-GeV}   \; & \mbox{\protect\cite{SK-atm-98}} \\
0.65 \pm 0.05 \pm 0.08 & \; \mbox{multi-GeV} \; & \mbox{\protect\cite{SK-atm-98}}
\end{array} 
\right\} & 
\quad \mbox{Super-Kamiokande}
\end{array}
\right.
\end{equation}
The PC events have been added to the multi-GeV data. The
results $R = 0.54 \pm 0.05 \pm 0.11$ of
the IMB Collaboration \cite{Bec95,kearns} and $R = 0.61 \pm 0.15 \pm 0.05$
of Soudan-2 \cite{kafka} are in agreement with those of the
experiments in Kamioka. However, the early iron calorimeter
experiments found values of $R$ compatible with 1, namely
$R = 1.00 \pm 0.15 \pm 0.08$ (Fr\'ejus Coll. \cite{Dau95}) and 
$R = 0.96 \begin{array}{l} +0.32 \\[-3pt] -0.28 \end{array}$
(NUSEX Coll. \cite{Agl89}).
Apart from the latter two experiments, all others hint at a reduction
of $R$ compared to the expectation $R=1$. 

Kamiokande and, in particular,
Super-Kamiokande have enough statistics to study the zenith angle
dependence of the measured 
$\stackrel{\scriptscriptstyle(-)}{\nu}_{\hskip-3pt e}$
and $\stackrel{\scriptscriptstyle(-)}{\nu}_{\hskip-3pt \mu}$ fluxes. 
To this end,
the $\cos \theta$ interval $[-1, 1]$ is divided into five bins of
length 0.4. Kamiokande \cite{Kam-atm-94} has observed a zenith angle 
variation of $R$
for the FC multi-GeV + PC events with indications that the zenith
angle variation rather comes from $\mu$-like events than from the
$e$-like events. All this is amply confirmed by Super-Kamiokande with
much more statistics and with a significant zenith angle variation in
the $\mu$-like events for both, sub-GeV and
multi-GeV. 

\begin{figure}[t!]
\begin{tabular*}{\linewidth}{@{\extracolsep{\fill}}cc}
\begin{minipage}{0.47\linewidth}
\begin{center}
\mbox{\epsfig{file=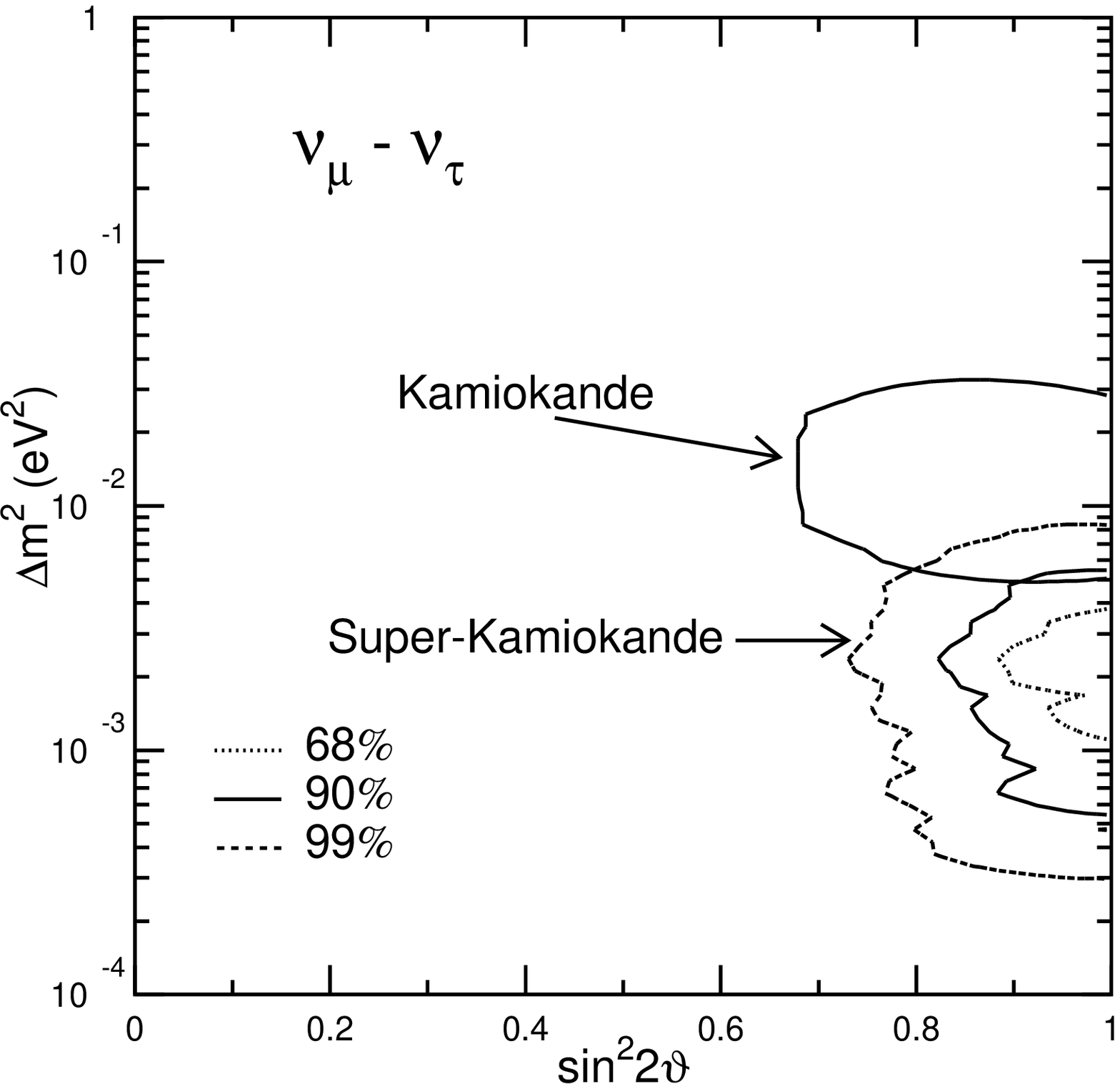,width=0.95\linewidth}}
\end{center}
\end{minipage}
&
\begin{minipage}{0.47\linewidth}
\begin{center}
\mbox{\epsfig{file=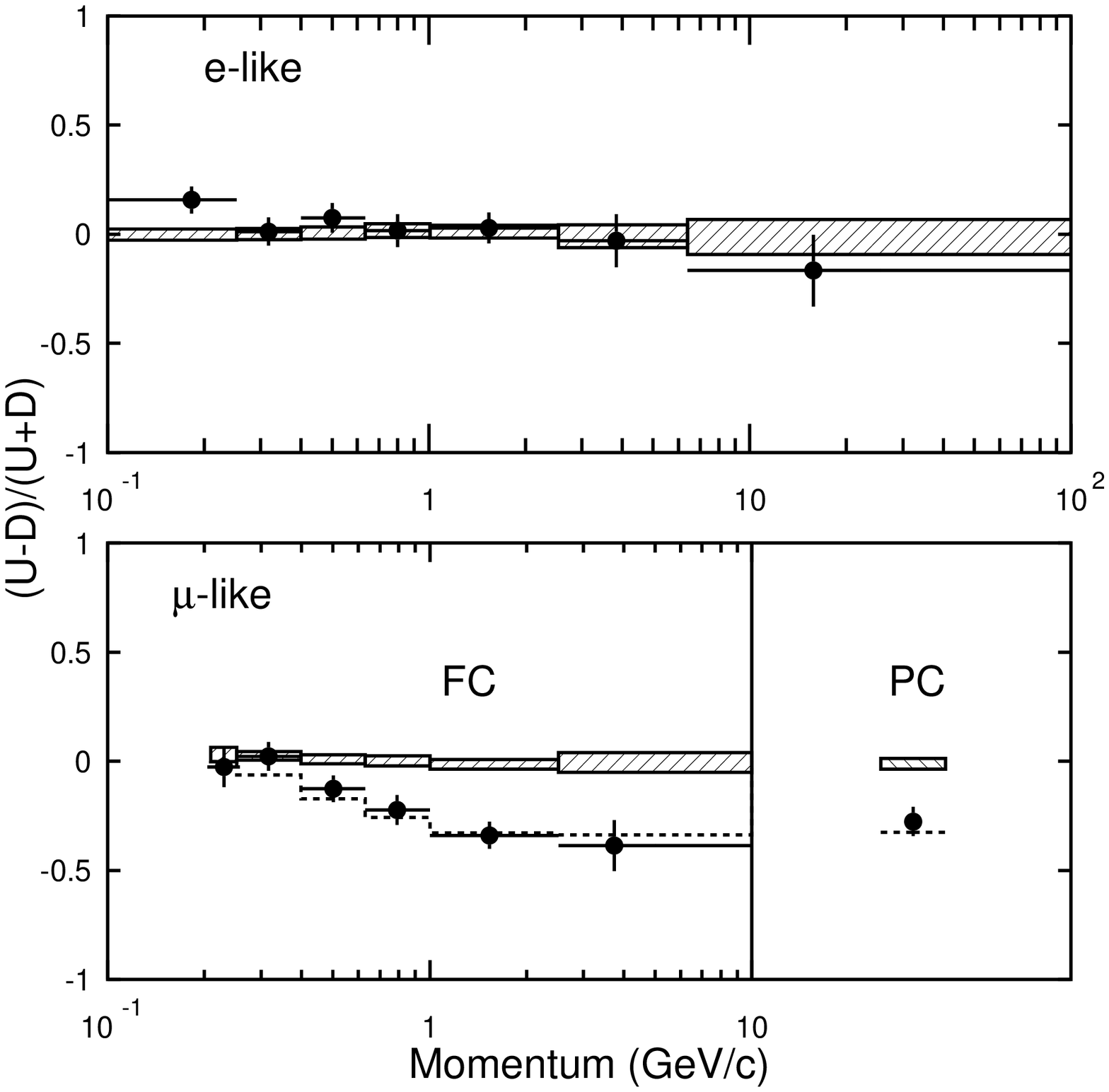,width=0.95\linewidth}}
\end{center}
\end{minipage}
\\
\begin{minipage}{0.47\linewidth}
\refstepcounter{figures}
\label{kam-sk}
\footnotesize
Figure \ref{kam-sk}.
The 68\%, 90\% and 99\%
confidence regions in the
$\sin^22\vartheta$--$\Delta{m}^2$
plane for
$\nu_\mu\to\nu_\tau$ oscillations
obtained from 33.0 kton year of Super-Kamiokande data
\protect\cite{SK-atm-98}.
The 90\% confidence region obtained by the Kamiokande experiment is also shown.
\end{minipage}
&
\begin{minipage}{0.47\linewidth}
\refstepcounter{figures}
\label{sk-asym}
\footnotesize
Figure \ref{sk-asym}.
Up-down asymmetry measured in the Super-Kamiokande
atmospheric neutrino experiment
as a function of the momentum of $e$-like and $\mu$-like events
\protect\cite{SK-atm-98}.
The hatched region shows the theoretical expectation
without neutrino oscillations,
with statistical and systematic errors added in quadrature.
The dashed line for $\mu$-like events
is the theoretical expectation in the case of two-generation
$\nu_\mu\to\nu_\tau$ oscillations
with
$ \Delta{m}^2 = 2.2 \times 10^{-3} \, \mathrm{eV}^2 $
and
$ \sin^22\vartheta = 1.0 $.
\end{minipage}
\end{tabular*}
\end{figure}

It is important to study the zenith angle variation independent
of the double ratio $R$ in order to disentangle the zenith angle
dependencies of the electron and muon neutrino fluxes. To this end,
in addition to the oscillation parameters,
also the normalization of primary cosmic ray flux has to be fitted.
Super-Kamiokande has performed a statistical analysis of
the data under the assumption of 
$\stackrel{\scriptscriptstyle (-)}{\nu}_{\hskip-3pt \mu} \to
\stackrel{\scriptscriptstyle (-)}{\nu}_{\hskip-3pt \tau}$ oscillations. With
$e$-like and $\mu$-like events in five $\cos \theta$ bins and seven
momentum bins there are altogether 70 data points and three
quantities to be determined: the mixing angle $\vartheta$, the
neutrino mass-squared difference $\Delta m^2$ and the overall neutrino
flux normalization. The best fit gives $\sin^2 2\vartheta =1$ and
$\Delta m^2 = 2.2 \times 10^{-3}$ eV$^2$ with
$\chi^2_{\mathrm{min}} = 65.2$ for 67 DOF.
The regions in the $\sin^22\vartheta$--$\Delta{m}^2$
plane allowed at
68\%, 90\% and 99\%
confidence level in the case of
$\stackrel{\scriptscriptstyle (-)}{\nu}_{\hskip-3pt \mu} \to
\stackrel{\scriptscriptstyle (-)}{\nu}_{\hskip-3pt \tau}$
oscillations
are shown in Fig.~\ref{kam-sk}
\cite{SK-atm-98}.
At 90\% CL the mass-squared
difference lies in the interval 
$5 \times 10^{-4} \: \mbox{eV}^2 < \Delta m^2 < 6 \times 10^{-3}
\: \mbox{eV}^2$. 
For the simulation of neutrino oscillations the profiles for the
neutrino production heights of Ref.~\cite{Gai98} were used.
An analogous procedure with the hypothesis of
$\stackrel{\scriptscriptstyle (-)}{\nu}_{\hskip-3pt \mu} \leftrightarrow
\stackrel{\scriptscriptstyle (-)}{\nu}_{\hskip-3pt e}$ oscillations, taking
into account matter effects in the earth, gives a
poor fit with $\chi^2_{\mathrm{min}} = 87.8$ for 67 DOF.

It is interesting to note that,
as shown in Fig.~\ref{kam-sk},
a fit for Kamiokande with 
$\stackrel{\scriptscriptstyle (-)}{\nu}_{\hskip-3pt \mu} \to
\stackrel{\scriptscriptstyle (-)}{\nu}_{\hskip-3pt \tau}$ oscillations gives 
$5 \times 10^{-3} \: \mbox{eV}^2 < \Delta m^2 < 3 \times 10^{-2} 
\: \mbox{eV}^2$ with 90\% CL \cite{Kam-atm-94} for the multi-GeV data
whereas the sub-GeV data show no indication for a zenith angle
dependence of the number of events.
However, Kamiokande has a lower statistics
than Super-Kamiokande. 
In Refs.~\cite{Gon98a,Gon98b,Foot-Volkas-Yasuda98-PRD58,Yasuda98,FLMS98}
analyses of all available experiments were performed using the
three possible oscillation hypotheses 
$\stackrel{\scriptscriptstyle (-)}{\nu}_{\hskip-3pt \mu} \leftrightarrow
\stackrel{\scriptscriptstyle (-)}{\nu}_{\hskip-3pt e,\tau,s}$ with similar
results as described above and showing in addition that also the
hypothesis of oscillations into sterile neutrinos gives a satisfactory fit.
For analyses with three neutrinos and including all neutrino
oscillation experiments see \cite{FLM94,FLM95,FL95,FLMS97-PRD55,FLMM98,FLMS98}.
In Ref.~\cite{Cla97} (IMB) no zenith angle variation of $R$ was seen though
with rather small statistics.

In the Super-Kamiokande experiment a significant up-down
asymmetry for the $\mu$-like events was found.
The measured value of the asymmetry
$(U-D)/(U+D)$
as a function of momentum
for $e$-like and $\mu$-like events is shown in Fig.~\ref{sk-asym}
\cite{SK-atm-98}.
Here
$U$ is the number of upward-going events with zenith
angles in the range $-1 < \cos \theta < -0.2$ and $D$ the number
of downward-going events with $0.2 < \cos \theta < 1$.
The value of the asymmetry for
FC and PC multi-GeV $\mu$-like events is
\cite{SK-atm-98}
\begin{equation}\label{asymm}
A_\mu \equiv \left(\frac{U-D}{U+D}\right)_\mu = -0.296 \pm 0.048 \pm 0.01
\,.
\end{equation}
The
geomagnetic effect which is one of the causes of an up-down
asymmetry contributes less than $\pm 0.01$ for multi-GeV events
\cite{SK-atm-98}. For the PC events with a mean neutrino energy of
15 GeV \cite{SK-atm-98} this effect is even less important. Note
that for multi-GeV events
the average angle between the incoming neutrino direction
and the charged lepton seen in the single-ring events is of the
order of $20^\circ$ or less. Since $\cos \theta = \pm 0.2$
corresponds to $90^\circ \pm 11.5^\circ$, the asymmetry $A$
of the $\mu$-like events is in effect an up-down asymmetry of
the $\stackrel{\scriptscriptstyle (-)}{\nu}_{\hskip-3pt \mu}$ flux. The
result (\ref{asymm}) constitutes the best indication in favour
of neutrino oscillations found so far with $A_\mu$ being 6 standard
deviations away from 0. This is corroborated by 
the plots of $A_{e,\mu}(p)$ as a function of the lepton momentum $p$
in the case of
the single-ring events
shown in Fig.~\ref{sk-asym}
\cite{SK-atm-98}. For $e$-like events, $A_e(p)$ is a flat function
consistent with no 
$
\stackrel{\scriptscriptstyle (-)}{\nu}_{\hskip-3pt e} 
\leftrightarrow
\stackrel{\scriptscriptstyle (-)}{\nu}_{\hskip-3pt \mu}
$
oscillations. The asymmetry for $e$-like events analogous to
Eq.(\ref{asymm}) is compatible with zero: $A_e = -0.036 \pm 0.067 \pm 0.02$.
For the $\mu$-like events, $A_\mu(p)$ starts with
zero and decreases, thus indicating that at small energies both
up and down-going neutrinos oscillate with averaged
$\stackrel{\scriptscriptstyle (-)}{\nu}_{\hskip-3pt \mu} \to
\stackrel{\scriptscriptstyle (-)}{\nu}_{\hskip-3pt \mu}$
probabilities 1/2 whereas at muon energies in the GeV range the
probability of $\stackrel{\scriptscriptstyle (-)}{\nu}_{\hskip-3pt \mu} \to
\stackrel{\scriptscriptstyle (-)}{\nu}_{\hskip-3pt \mu}$ survival
for the down-going neutrinos approaches 1. This behaviour is
in agreement with the $L/E_\nu$ dependence of the oscillation
probabilities. 
It is interesting to note that the experimental values of the 
asymmetries $A_{e,\mu}$ can be used to discriminate between
different oscillation scenarios
\cite{Flanagan-Learned-Pakvasa98,Bun97,Foo98a,Foo98b}.

The importance of the results of the atmospheric neutrino
oscillation experiments requires further scrutiny to test
the interpretation in terms of neutrino oscillations. It has
been proposed for Super-Kamiokande to use ratios of charged
current events (CC) to neutral current (NC) events \cite{K2K,Vis98a} in 
the spirit of the SNO experiment \cite{SNO-nu98}
in the context of solar neutrinos. 
The basic idea is that in Super-Kamiokande NC events could be
seen through
\begin{equation}\label{pi0}
\stackrel{\scriptscriptstyle (-)}{\nu}_{\hskip-3pt \ell} + N \to 
\stackrel{\scriptscriptstyle (-)}{\nu}_{\hskip-3pt \ell} + \pi^0 + N 
\quad \mbox{with} \quad N = p,n \, \quad \ell = e, \mu, \tau \,,
\end{equation}
whereas CC reactions are likely to produce a single charged pion via
\begin{equation}\label{pi+-}
\stackrel{\scriptscriptstyle (-)}{\nu}_{\hskip-3pt \ell} + N \to
\ell^\mp + \pi^\pm + N \quad \mbox{with} \quad \ell = e, \mu \,.
\end{equation}
To produce a test
with the reactions (\ref{pi0}) and (\ref{pi+-}) it is necessary
to discuss their experimental signatures and the contaminations
of these signature with other processes.
Neutral pions are selected by taking events with two diffuse rings from the two
decay photons with an invariant mass between 90 and 180 MeV and
momenta smaller 400 MeV in order to separate the two rings. Such events
were already observed in Kamiokande \cite{Fuk96}. 
Because of the selection criteria for $\pi^0$'s,
single pion production is important only for neutrino energies
of order 1 GeV, therefore it involves the sub-GeV events and no
tau lepton will produced in the CC reaction. Defining
$N_{\pi^0}$ via this procedure then the two-ring
ratios \cite{Vis98a}
\begin{equation}
\mathcal{R}_e   \equiv \frac{N_{DS}}{N_{\pi^0}} \quad \mbox{and} \quad
\mathcal{R}_\mu \equiv \frac{N_{SS}}{N_{\pi^0}} 
\end{equation}
act as measures of the CC to NC event ratios because in the
ideal situation one can make the identification
$N_{\pi^0} = N^{NC}_{\pi^0}$, $N_{DS} = N_{e^\mp \pi^\pm}^{CC}$ and
$N_{SS} = N_{\mu^\mp \pi^\pm}^{CC}$ since electrons produce
diffuse and muons and charged pions sharp rings. This picture is
blurred \cite{Vis98a,Bal98} by detector efficiencies (\textit{e.g.}, 
it is 0.77 for the
identification of neutral pions via two diffuse rings),
misidentifications (like misidentifying a $\pi^0$ whose two
photons cannot be resolved with an electron) and contaminations
(most notably, NC events with $\pi^+ \pi^-$ in the final state
contribute to $N_{SS}$, electronic CC events with a $\pi^0$ whose photons
cannot be resolved add to $N_{\pi^0}$ and muonic CC events with
the same $\pi^0$ configuration add to $N_{DS}$). Taking all this into
account \cite{Vis98a,Bal98}, one has nevertheless reason the expect that
with increasing statistics in Super-Kamiokande the two-ring events can
be used to obtain information on distinguishing between oscillation of
the muon neutrino into tau or sterile neutrinos and to discriminate
between different regions of the parameter space of neutrino mixing
\cite{Vis98a,Bal98,Vis98b}. Furthermore, it was suggested to use
the asymmetry
$A_N \equiv (U_{\pi^0}-D_{\pi^0})/(U_{\pi^0}+D_{\pi^0})$
for up and down-going $\pi^0$'s coming from the NC reaction
(\ref{pi0}) as an observable to distinguish muon neutrino
oscillations into active neutrinos from those into sterile
neutrinos \cite{Learned-Pakvasa-Stone98}.
One has to assume in this case that
up and down $\pi^0$'s originate from up and down neutrinos,
respectively, with high probability. This is not so obvious, however, 
since for the identification of the $\pi^0$ the events should rather have low
energy as explained above.

The further tests discussed here concern the stopping and through-going
muon events where special efforts have been made to calculate the
fluxes \cite{Fra93,Akh93,Lip95a}. Whereas the FC events have neutrino
energies of around 1 GeV the stopping muon events correspond to a mean
neutrino energy of 10 GeV and the through-going muons to 100 GeV
\cite{Fra93,Gai97}. Thus we are now discussing a different range of energy
compared to the discussion above (with the exception of the PC sample).

Obviously, also the zenith angle distribution of upward stopping or
through-going muons can be used to test the neutrino oscillation
hypothesis \cite{Fog98}. Among other experiments (see above) 
upward through-going muons have been studied by Kamiokande \cite{Hat98} 
and upward muons by MACRO \cite{Amb98}. Kamiokande
has 372 such events above an energy threshold of 1.6 GeV. Fitting the
data to the $\stackrel{\scriptscriptstyle (-)}{\nu}_{\hskip-3pt \mu} \to
\stackrel{\scriptscriptstyle (-)}{\nu}_{\hskip-3pt \tau}$ 
oscillation hypothesis yields a best fit with 
$\Delta m^2 = 3.2 \times 10^{-3}$ eV$^2$ agreeing
rather well with the Super-Kamiokande result. The analysis of the
MACRO Coll. based on 479 events gives a similar result for $\Delta m^2$, 
however, the zenith angle distribution does not fit very well with the
oscillation hypothesis into tau neutrinos. An attempt has been made to
explain the zenith angle distribution of the MACRO experiment with
$\stackrel{\scriptscriptstyle (-)}{\nu}_{\hskip-3pt \mu} \to
\stackrel{\scriptscriptstyle (-)}{\nu}_{\hskip-3pt s}$ 
oscillations where matter
effects in the earth play a crucial role \cite{Liu98}.

In the earth, the density profile can approximately be represented 
by constant densities in the mantle and the core, respectively 
(see end of Section 4.3). Such a profile can lead to an enhancement of
neutrino transitions due to
the matter effect\footnote{Note that for the case of a periodical
matter density the effect of enhancement was considered 
in Refs.~\cite{akhmedov88}.}
in the earth if atmospheric 
neutrinos cross the core\footnote{Neutrinos with $\cos
\theta \leq -0.837$} such that the phase picked up by
a neutrino wave function traversing the mantle for the first time and
the phase acquired by traversing the core are each approximately equal
to $\pi$. Such an effect for atmospheric neutrinos was recently 
considered in detail in Refs.~\cite{Liu98,LMS98,petcov98,akhmedov98,chizhov}. 
%periodically varying matter potential \cite{akhmedov88}.  
%Thus, instead of infinitely many periods, neutrinos crossing the
%core undergo ``1.5 periods'' \cite{ADLS98}, which is nevertheless
%sufficient to produce an effect for those neutrinos which fulfill the
%above-mentioned phase conditions. 
It has also been proposed to exploit this
effect to discriminate between 
$\stackrel{\scriptscriptstyle (-)}{\nu}_{\hskip-3pt \mu} \to
\stackrel{\scriptscriptstyle (-)}{\nu}_{\hskip-3pt \tau}$
(no matter effects) and 
$\stackrel{\scriptscriptstyle (-)}{\nu}_{\hskip-3pt \mu} \to
\stackrel{\scriptscriptstyle (-)}{\nu}_{\hskip-3pt s}$
transitions of atmospheric neutrinos
\cite{Liu98} and to explain an excess of $e$-like events \cite{ADLS98}
possibly seen in the Super-Kamiokande experiment \cite{SK-atm-98}.

It was suggested in Ref.~\cite{Bec92} to use the observable
\begin{equation}\label{r}
r = \frac{N_{\mathrm{stop}}}{N_{\mathrm{thru}}}
\end{equation}
where $N_{\mathrm{stop}}$ and $N_{\mathrm{thru}}$ are the numbers of
stopping and though-going muons, respectively, as an indicator for
neutrino oscillations because this ratio is reduced for neutrino
oscillations with respect to the no-oscillation hypothesis
\cite{Lip95a,Lip98a}. Also the ratios of ``horizontal'' to ``vertical''
muon events for stopping and through-going muons defined by
\begin{equation}\label{S}
S_{\mathrm{stop,thru}} \equiv
\left. \frac{N_{\mathrm{hor}}}{N_{\mathrm{vert}}}
\right|_{\mathrm{stop, thru}} \,,
\end{equation}
with
\begin{equation}
N_{\mathrm{hor}} = \int^0_{\cos \theta_c} d \cos \theta
\frac{dN}{d \cos \theta}
\quad \mbox{and} \quad
N_{\mathrm{vert}} = \int^{\cos \theta_c}_{-1} d \cos \theta
\frac{dN}{d \cos \theta}
\end{equation}
are useful observables as shown in Ref.~\cite{Lip98a}. In this work a
study was made for Super-Kamiokande, taking $\cos \theta_c = -0.5$ as
the boundary between ``horizontal'' and ``vertical'', showing that the
prospects are good for confirming atmospheric neutrino oscillations found with
the FC single-ring and PC events by using stopping and through-going
muons and the observables $r$ (\ref{r}), $S_{\mathrm{stop}}$ and
$S_{\mathrm{thru}}$ (\ref{S}). However, for a precise determination of 
$\Delta m^2$ these variables are not suitable.

\subsubsection{Long-baseline experiments and tests of the atmospheric
neutrino oscillation parameters}

The explanation of the atmospheric neutrino anomaly in terms of
neutrino oscillations 
can be checked with long-baseline neutrino oscillation experiments.
The first long-baseline reactor experiment CHOOZ \cite{CHOOZ98}
(see Section 3.2 and Fig.~3.3)
has already excluded atmospheric $\nu_\mu\leftrightarrows\nu_e$
oscillations with a large mixing angle for
$ \Delta{m}^2_{\mathrm{atm}} \gtrsim 10^{-3} \, \mathrm{eV}^2 $.
Two other long-baseline reactor experiments are under construction:
Palo Verde \cite{PaloVerde,Palo-Verde-TAUP97}
and
Kam-Land \cite{Kam-Land,Suzuki-now98}.
The Palo Verde experiment has a setup and a sensitivity similar to 
the CHOOZ experiment, whereas the Kam-Land experiment,
which is the result of the conversion of the old Kamiokande detector
to a liquid scintillator detector,
will detect $\bar\nu_e$'s produced by Japanese reactors
150--200 km away
and will be sensitive to
$ \Delta{m}^2 \gtrsim 10^{-5} \, \mathrm{eV}^2 $
and a large mixing angle.
The Borexino experiment
(see the end of the Section 5.2 and
\cite{Schonert-TAUP97,Suzuki-now98})
will allow to perform a similar measurement.

Accelerator long-baseline experiments will study the oscillation
channels $\nu_\mu\to\nu_{e,\mu,\tau}$. 
The K2K \cite{K2K}
experiment,
with a baseline of about 235 km from KEK to Super-Kamiokande
and a neutrino beam with 1.4 GeV average energy,
will be sensitive to
$\nu_\mu$ disappearance and $\nu_\mu\to\nu_e$ transitions
with
$ \Delta{m}^2 \gtrsim 2 \times 10^{-3} \, \mathrm{eV}^2 $.
A near 1 kton water-Cherenkov detector
will be placed at a distance of about 1 km from the beam dump and
will allow to measure the initial flux and energy spectrum of $\nu_\mu$'s.
This experiment is under construction and
is planned to begin taking data in the year 1999.

Also the MINOS \cite{MINOS} experiment is under construction.
This experiment will have a near detector at Fermilab
and a baseline of about 730 km from Fermilab
to the Soudan mine,
where the far detector will be placed.
The neutrino beam will be produced by protons from the
new Main Injector at Fermilab
and will have an average energy of about 10 GeV.
The far detector is an 8 kton sampling calorimeter made of
magnetized iron and scintillators.
This experiment will be sensitive to
$\nu_\mu$ disappearance
and
$\nu_\mu\to\nu_e$,
$\nu_\mu\to\nu_\tau$,
$\nu_\mu\to\nu_s$
transitions,
with the possibility to distinguish the different channels,
for $ \Delta{m}^2 \gtrsim 10^{-3} \, \mathrm{eV}^2 $
(the possibility to extend the sensitivity to
$ \Delta{m}^2 \gtrsim 5 \times 10^{-5} \, \mathrm{eV}^2 $
lowering the neutrino energy is under study).
In particular
$\nu_\mu\to\nu_s$
transitions can be revealed through the measurement of a deficit in the
NC/CC ratio.
The MINOS experiment is scheduled to start data-taking around the year 2003.

The ICARUS experiment \cite{ICARUS} in Gran Sasso,
constituted of a 0.6 kton liquid argon detector is
scheduled to start in the year 2000.
In the future three new modules with a  total mass of 2.4 kton will be installed.
This detector will be sensitive to atmospheric and solar neutrinos
and will allow to reveal long-baseline
$\nu_\mu\to\nu_\tau$
oscillations using a neutrino beam
produced at CERN about 730 km away.
Since the average energy of the neutrino beam is rather high,
about 25 GeV,
in order to allow the detection of $\nu_\tau$ through the CC production of
a $\tau$,
this experiment will be sensitive to
$ \Delta{m}^2 \gtrsim 10^{-3} \, \mathrm{eV}^2 $.
Four other detectors for future LBL CERN--Gran Sasso experiments,
OPERA \cite{OPERA},
NOE \cite{NOE},
AQUA-RICH \cite{AQUA-RICH}
and
NICE \cite{NICE},
have been proposed and
are under consideration
(see Ref.~\cite{CERN-LNGS}),
together with the
possibility of a new atmospheric neutrino detector
consisting of a large-mass and high-density
tracking calorimeter
\cite{Aglietta-CERN-SPSC-98-28}.

In Ref.~\cite{Bat98} a comparison is made between the possibilities using
atmospheric neutrinos and LBL neutrino experiments for the
determination of the oscillation parameters. 
The feasibility to distinguish between atmospheric $\nu_\mu\to\nu_\tau$
and $\nu_\mu\to\nu_s$ oscillations using a combination of the results of
future atmospheric, LBL and SBL experiments is discussed in
Ref.~\cite{Geiser-now98}.    

\subsection{Solar neutrino experiments}
\label{Indications: solar neutrino experiments}

The earliest indication in favour of neutrino oscillations
was obtained about 30 years ago in the radiochemical 
solar neutrino experiment by R. Davies 
\textit{et al.} \cite{Homestake68}. The flux of solar electron
neutrinos measured in this experiment was significantly less than the
predicted one. This phenomenon was called \emph{solar neutrino problem}.
The existence of this problem was confirmed in 
all five solar neutrino experiments
(Homestake \cite{Homestake68,Homestake94,Homestake98},
Kamiokande \cite{Kam-sun-88,Kam-sun-91-PRD,Kam-sun-96},
GALLEX \cite{GALLEX92a,GALLEX96},
SAGE \cite{SAGE91,SAGE96}
and
Super-Kamiokande \cite{SK-sun-97-Inoue,SK-sun-98-PRL,SK-sun-nu98})
which measure a flux of electron neutrinos
significantly smaller than the one
predicted by the Standard Solar Model (SSM)
\cite{Bahcall-Ulrich88,Bahcall89-book,Bahcall-Pinsonneault92,%
Bahcall-Pinsonneault95,BP98,Turck88,Turck-Lopes93,Turck93,%
Castellani93-AA,Castellani94,Ciacio97}.
The solar neutrino problem
(see, for example,
\cite{Bahcall89-book,Koshiba92,Castellani97,Berezinsky97-Durban,
Bahcall97-SLAC,Bahcall-nu98,Smirnov-nu98})
arose in the Homestake experiment by the low counting rate showing 
that the flux of
the high-energy $^8\mathrm{B}$ neutrinos
($ E_{^8\mathrm{B}} \lesssim 15 \, \mathrm{MeV} $)
and of
the medium-energy  $^7\mathrm{Be}$ neutrinos
($ E_{^7\mathrm{Be}} = 0.862 \, \mathrm{MeV} $)
is suppressed by a factor of about 1/3
with respect to the SSM prediction.
In 1988 the solar neutrino problem was confirmed by the
results of the real-time
water-Cherenkov Kamiokande experiment \cite{Kam-sun-88}
which measured a flux of $^8\mathrm{B}$ neutrinos of
about half of the SSM flux.
The measurements of the Kamiokande experiment
proved that the observed neutrinos arrive
at the detector from the direction of the sun.
In 1992 the radiochemical GALLEX \cite{GALLEX92a}
and SAGE experiments \cite{SAGE96}
succeeded in measuring the neutrino flux
with a low energy threshold
$ E_{\mathrm{th}} = 233 \, \mathrm{keV} $,
which allowed to detect low-energy $pp$ neutrinos
produced by the fundamental reaction of the $pp$ cycle.
Also these experiments measured a neutrino flux of about half
of the one predicted by the SSM.
Finally,
the Super-Kamiokande experiment
has recently confirmed \cite{SK-sun-97-Inoue,SK-sun-98-PRL,SK-sun-nu98}
with high statistics
the suppression of the
$^8\mathrm{B}$ neutrino flux with respect to the SSM one
by a factor of about 1/2.

\begin{figure}[p!]
\setlength{\templength}{\arraycolsep} \setlength{\arraycolsep}{0pt}
\newlength{\newtabularwidth}
\setlength{\newtabularwidth}{0.98\textheight}
\rotate[l]{
\begin{tabular*}{\newtabularwidth}{@{\extracolsep{\fill}}cc}
\begin{minipage}{0.48\newtabularwidth}
\begin{center}
\null \vspace{2cm} \null
\mbox{\epsfig{file=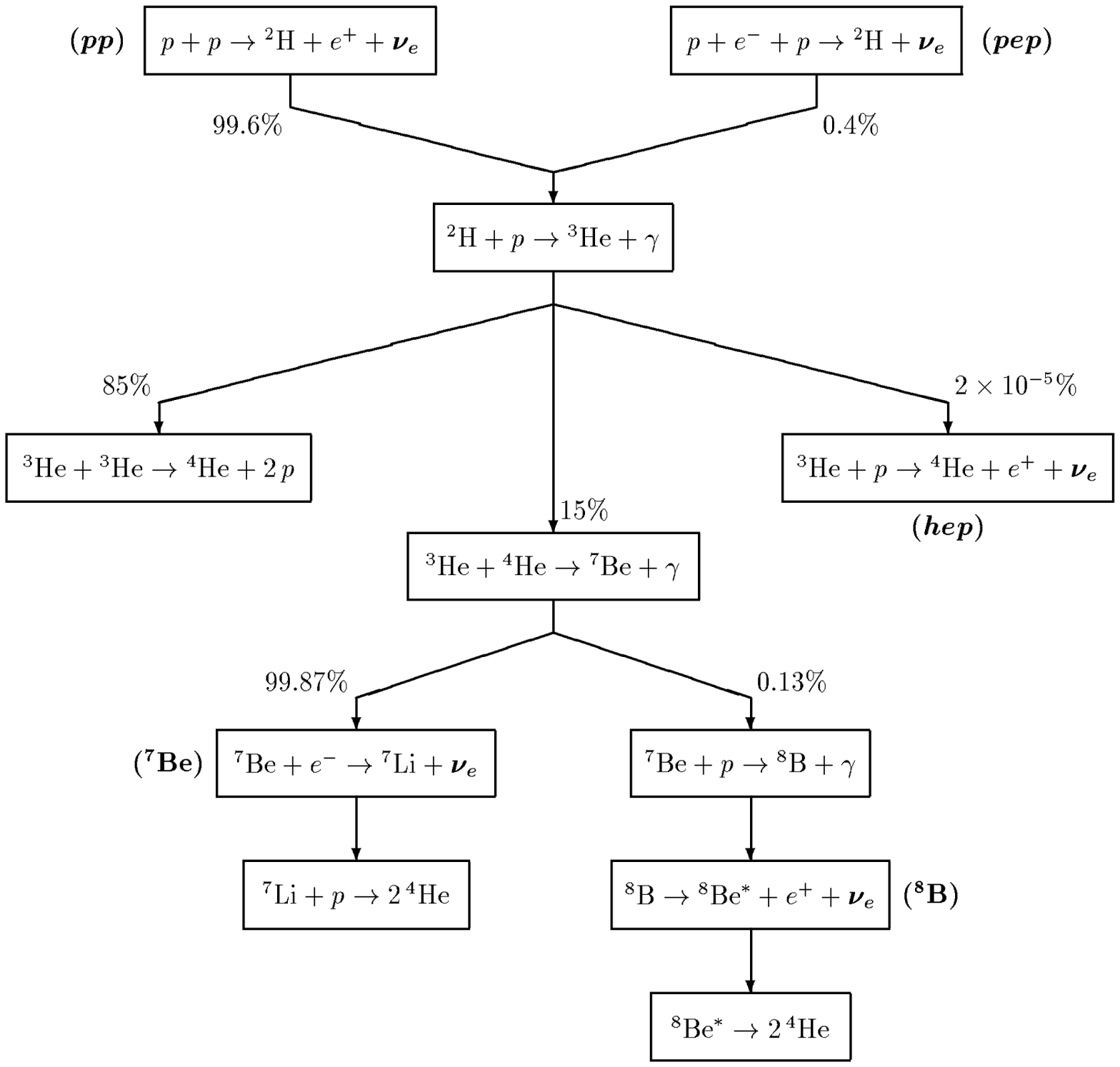,width=0.46\newtabularwidth}}
\end{center}
\end{minipage}
&
\begin{minipage}{0.48\newtabularwidth}
\begin{center}
\null \vspace{2cm} \null
\mbox{\epsfig{file=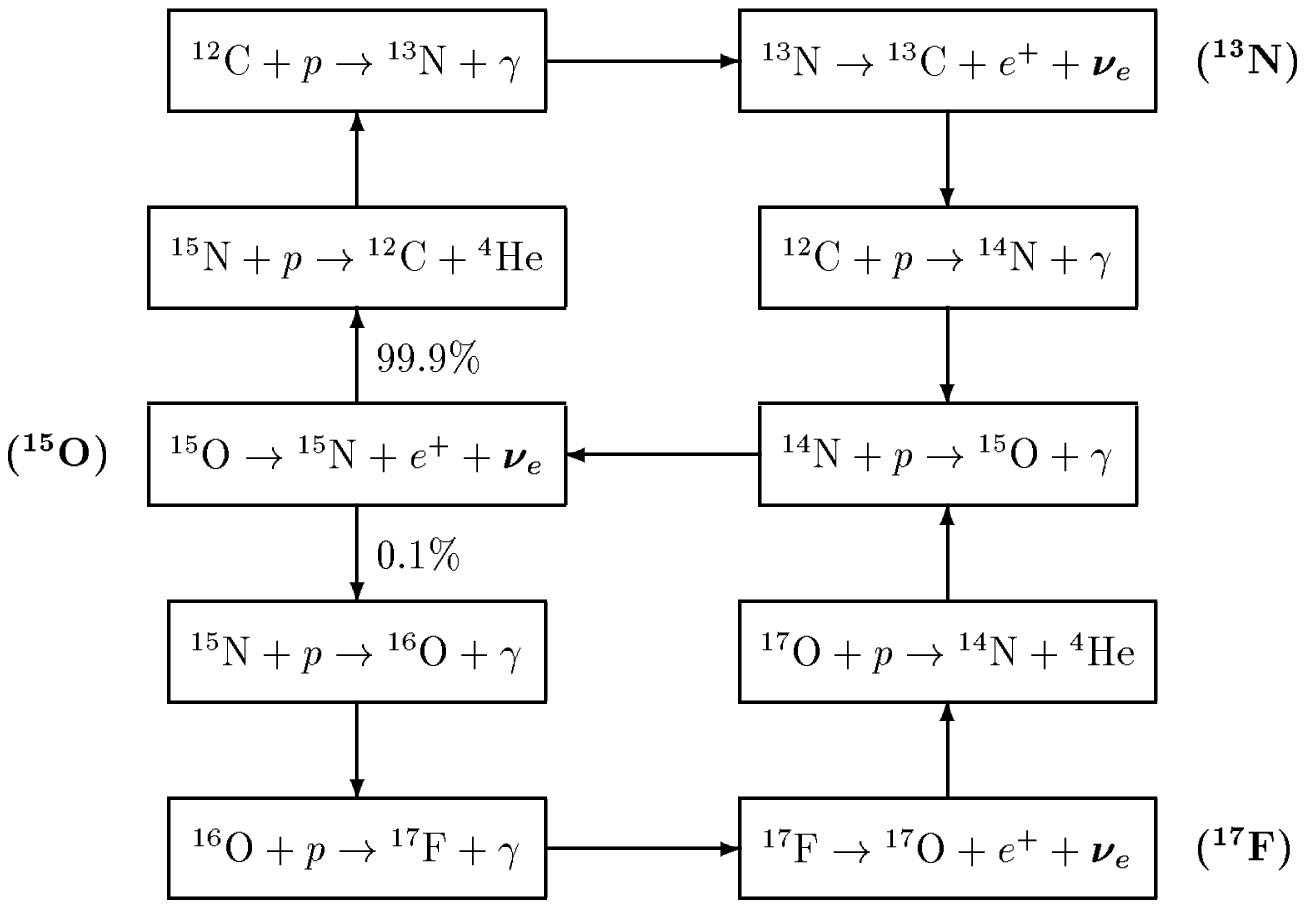,width=0.46\newtabularwidth}}
\end{center}
\end{minipage}
\\
\begin{minipage}{0.48\newtabularwidth}
\refstepcounter{figures}
\label{pp_cycle}
\footnotesize
\begin{center}
\null \vspace{1cm} \null
Figure \ref{pp_cycle}.
The $pp$ cycle.
\end{center}
\end{minipage}
&
\begin{minipage}{0.48\newtabularwidth}
\refstepcounter{figures}
\label{cnocycle}
\footnotesize
\begin{center}
\null \vspace{1cm} \null
Figure \ref{cnocycle}.
The CNO cycle.
\end{center}
\end{minipage}
\end{tabular*}
}
\setlength{\arraycolsep}{\templength}
\end{figure}

The energy of the sun is produced in
the reactions of the thermonuclear $pp$ and CNO cycles
shown in Figs.~\ref{pp_cycle} and \ref{cnocycle} (see, \textit{e.g.},
Ref.~\cite{Bahcall89-book}). 
The overall result of both cycles is the transition 
\begin{equation}
4 \, p + 2 \, e^-
\to
{}^4\mathrm{He} + 2 \, \nu_e + Q
\,,
\label{601}
\end{equation}
where
$
Q
=
4 \, m_p + 2 \, m_e - m_{^4\mathrm{He}}
=
26.73 \, \mathrm{MeV}
$
is the energy release.\footnote{Here
$ m_p = 938.272 \, \mathrm{MeV} $
is the proton mass,
$ m_e = 0.511 \, \mathrm{MeV} $
is the electron mass
and
$ m_{^4\mathrm{He}} = 2 m_p + 2 m_n - E^B_{^4\mathrm{He}} $
is the mass of the $^4\mathrm{He}$-nucleus,
where
$ m_n = 939.566 \, \mathrm{MeV} $
is the neutron mass
and
$ E^B_{^4\mathrm{He}} = 28.296 \, \mathrm{MeV} $
is the binding energy of the $^4\mathrm{He}$-nucleus.}
Hence,
\emph{the production of energy in the sun is accompanied
by the emission of electron neutrinos}.
The main part of the solar energy
is radiated through photons
and
a small part (about 2\%)
is emitted through neutrinos.
The sources of solar neutrinos are
listed in Table \ref{Sources of solar neutrinos}.
The $pp$, $pep$, $^7$Be, $^8$B and $hep$
reactions belong to the $pp$ cycle
(see Fig.~\ref{pp_cycle}),
whereas the
$^{13}$N,  $^{15}$O and $^{17}$F
reactions
belong to the CNO cycle
(see Fig.~\ref{cnocycle}),
which produces only about 2\% of the solar energy.
The average and maximum neutrino energies
listed in Table \ref{Sources of solar neutrinos}
are taken from Refs.~\cite{Bahcall96-PRC,Bahcall97-PRC}.
The neutrino fluxes and predictions
for the neutrino capture rates in the chlorine Homestake experiment
and in the gallium GALLEX and SAGE experiments
given by the
Bahcall-Pinsonneault 1998 (BP98) \cite{BP98} SSM
are listed in Table~\ref{solar neutrinos fluxes}.

\begin{table}[t]
\begin{center}
\begin{tabular}{cccc}
Source $r$
&
Reaction
&
\begin{tabular}{c}
Average Neutrino
\\
Energy $\langle{E}\rangle_r$ (MeV)
\end{tabular}
&
\begin{tabular}{c}
Maximum Neutrino
\\
Energy (MeV)
\end{tabular}
\\
\hline
\hline
$pp$
&
$ p + p \to d + e^+ + \nu_e $
&
$ 0.2668 $
&
$ 0.423 \pm 0.03 $
\vphantom{\bigg|}
\\
\hline
$pep$
&
$ p + e^- + p \to d + \nu_e $
&
$ 1.445 $
&
$ 1.445 $
\vphantom{\bigg|}
\\
\hline
$^7$Be
&
$ e^- + {}^7\mathrm{Be} \to {}^7\mathrm{Li} + \nu_e $
&
\begin{tabular}{c}
$ 0.3855 $
\\
$ 0.8631 $
\end{tabular}
&
\begin{tabular}{c}
$ 0.3855 $
\\
$ 0.8631 $
\end{tabular}
\vphantom{\bigg|}
\\
\hline
$^8$B
&
$ {}^8\mathrm{B} \to {}^8\mathrm{Be}^* + e^+ + \nu_e $
&
$ 6.735 \pm 0.036 $
&
$ \sim 15 $
\vphantom{\bigg|}
\\
\hline
$hep$
&
$ {}^3\mathrm{He} + p \to {}^4\mathrm{He} + e^+ + \nu_e $
&
$ 9.628 $
&
$ 18.778 $
\vphantom{\bigg|}
\\
\hline
$^{13}$N
&
$ {}^{13}\mathrm{N} \to {}^{13}\mathrm{C} + e^+ + \nu_e $
&
$ 0.7063 $
&
$ 1.1982 \pm 0.0003 $
\vphantom{\bigg|}
\\
\hline
$^{15}$O
&
$ {}^{15}\mathrm{O} \to {}^{15}\mathrm{N} + e^+ + \nu_e $
&
$ 0.9964 $
&
$ 1.7317 \pm 0.0005 $
\vphantom{\bigg|}
\\
\hline
$^{17}$F
&
$ {}^{17}\mathrm{F} \to {}^{17}\mathrm{O} + e^+ + \nu_e $
&
$ 0.9977 $
&
$ 1.7364 \pm 0.0003 $
\vphantom{\bigg|}
\\
\hline
\hline
\end{tabular}
\end{center}
\refstepcounter{tables}
\label{Sources of solar neutrinos}
\footnotesize
Table \ref{Sources of solar neutrinos}.
Sources of solar neutrinos \cite{Bahcall96-PRC,Bahcall97-PRC,Bahcall94-PRD49}.
\end{table}

\begin{table}[t]
\begin{center}
\begin{tabular}{cccccc}
Source $r$
&
\begin{tabular}{c}
Flux $\Phi_r$
\\
($ \mathrm{cm}^{-2} \, \mathrm{s}^{-1} $)
\end{tabular}
&
\begin{tabular}{c}
$\langle\sigma_{\mathrm{Cl}}\rangle_r$
\\
($ 10^{-44} \, \mathrm{cm}^2 $)
\end{tabular}
&
\begin{tabular}{c}
$S_{\mathrm{Cl}}^{(r)}$
\\
(SNU)
\end{tabular}
&
\begin{tabular}{c}
$\langle\sigma_{\mathrm{Ga}}\rangle_r$
\\
($ 10^{-44} \, \mathrm{cm}^2 $)
\end{tabular}
&
\begin{tabular}{c}
$S_{\mathrm{Ga}}^{(r)}$
\\
(SNU)
\end{tabular}
\\
\hline
\hline
$pp$
&
$ ( 5.94 \pm 0.06 ) \times 10^{10} $
&
--
&
--
&
$ 0.117 \pm 0.003 $
&
$ 69.6 \pm 0.7 $
\vphantom{\bigg|}
\\
\hline
$pep$
&
$ ( 1.39 \pm 0.01 ) \times 10^{8} $
&
$ 0.16 $
&
$ 0.2 $
&
$ 2.04 \, {}^{+0.35}_{-0.14} $
&
$ 2.8 $
\vphantom{\bigg|}
\\
\hline
$^7$Be
&
$ ( 4.80 \pm 0.43 ) \times 10^{9} $
&
$ 0.024 $
&
$ 1.15 \pm 0.1 $
&
$ 0.717 \, {}^{+0.050}_{-0.0.021} $
&
$ 34.4 \pm 3.1 $
\vphantom{\bigg|}
\\
\hline
$^8$B
&
$ ( 5.15 \, {}^{+0.98}_{-0.72} ) \times 10^{6} $
&
$ 114 \pm 11 $
&
$ 5.9 \, {}^{+1.1}_{-0.8} $
&
$ 240 \, {}^{+77}_{-36} $
&
$ 12.4 \, {}^{+2.4}_{-1.7} $
\vphantom{\bigg|}
\\
\hline
$hep$
&
$ 2.10 \times 10^{3} $
&
$ 390 $
&
$ 0.0 $
&
$ 714 \, {}^{+228}_{-114} $
&
$ 0.0 $
\vphantom{\bigg|}
\\
\hline
$^{13}$N
&
$ ( 6.05 \, {}^{+1.15}_{-0.77} ) \times 10^{8} $
&
$ 0.017 $
&
$ 0.1 $
&
$ 0.604 \, {}^{+0.036}_{-0.018} $
&
$ 3.7 \, {}^{+0.7}_{-0.5} $
\vphantom{\bigg|}
\\
\hline
$^{15}$O
&
$ ( 5.32 \, {}^{+1.17}_{-0.80} ) \times 10^{8} $
&
$ 0.068 \pm 0.001 $
&
$ 0.4 \pm 0.1 $
&
$ 1.137 \, {}^{+0.136}_{-0.057} $
&
$ 6.0 \, {}^{+1.3}_{-0.9} $
\vphantom{\bigg|}
\\
\hline
$^{17}$F
&
$ ( 6.33 \, {}^{+0.76}_{-0.70} ) \times 10^{6} $
&
$ 0.069 $
&
$ 0.0 $
&
$ 1.139 \, {}^{+0.137}_{-0.057} $
&
$ 0.1 $
\vphantom{\bigg|}
\\
\hline
Total
&
&
&
$ 7.7 \, {}^{+1.2}_{-1.0} $
&
&
$ 129 \, {}^{+8}_{-6} $
\vphantom{\bigg|}
\\
\hline
\hline
\end{tabular}
\end{center}
\refstepcounter{tables}
\label{solar neutrinos fluxes}
\footnotesize
Table \ref{solar neutrinos fluxes}.
Standard Solar Model \protect\cite{BP98}
neutrino fluxes,
average neutrino cross sections \cite{Bahcall89-book,Bahcall96-PRC,Bahcall97-PRC}
and SSM predictions
for the neutrino capture rates \protect\cite{BP98}
in the chlorine (Cl) Homestake experiment
and in the gallium (Ga) GALLEX and SAGE experiments.
\end{table}

As it is seen from the Tables \ref{Sources of solar neutrinos} and
\ref{solar neutrinos fluxes},
the major part of solar neutrinos are low energy neutrinos
coming from the $pp$ reaction.
Monoenergetic neutrinos with
intermediate energy
are produced in the capture of electrons by $^7$Be
and in the $pep$ reaction.
High energy neutrinos 
are produced in the decay of $^8$B
(the flux of $hep$ neutrinos is so small that its contribution
to the event rates of solar neutrino experiments
is negligible).
The flux of $^8$B neutrinos is much smaller
than the fluxes
of $pp$, $^7$Be and $pep$ neutrinos.
However,
as we will see later, these neutrinos
give the major contribution to the event rates of experiments
with a high energy detection threshold.
The CNO $^{13}\mathrm{N}$, $^{15}\mathrm{O}$, $^{17}\mathrm{F}$
reactions are sources of intermediate energy neutrinos
with a spectrum that extends up to about
$ 1.7 \, \mathrm{MeV} $.
Their contribution to the event rates of solar neutrino experiment
is small but not negligible.

The neutrino flux
coming from each source
as a function of the neutrino energy $E$
can be written as
\begin{equation}
\phi_r(E)=\Phi_r\,X_r(E)
\qquad \qquad
(r=pp,pep,{}^7\mathrm{Be},{}^8\mathrm{B},hep,
{}^{13}\mathrm{N},{}^{15}\mathrm{O},{}^{17}\mathrm{F})
\,,
\label{609}
\end{equation}
where
$\Phi_r$ is the total flux
and
$X_r(E)$
is the energy spectrum
($ \int \mathrm{d}E \, X_r(E) = 1 $).
The energy spectrum $X_r(E)$ for each source $r$
is known with negligible uncertainties
\cite{Bahcall91-PRD,Bahcall96-PRC}
because it is determined by the weak interactions
and it is practically independent from solar physics.
On the other hand,
the total flux
$\Phi_r$ of each source $r$
must be calculated with a solar model
and the resulting uncertainties
represent one of the main problem for the interpretation
of the experimental results.
However,
there are some relations
that allow to extract model-independent information
on the neutrino fluxes from the experimental data.
The main one is the \emph{luminosity constraint},
which is based on the assumption that the sun is in a stable state
(the energy is produced in the central region
of the sun and for its electromagnetic part
it takes more than $10^4$ years to reach the surface,
whereas neutrinos escape the sun in about two seconds).
Let us consider a solar neutrino with energy $E$.
The luminous energy released together with this neutrino
is
$ Q/2 - E $.
Multiplying this quantity with the total flux of neutrinos
$ \sum_r \phi_r(E) $
and
integrating over the neutrino energy one obtains the luminosity
constraint\footnote{The corrections to this relation
due to the generation of gravitational energy
\cite{Bahcall-Ulrich88}
and due to the fact that the abundance of $^3\mathrm{He}$
nuclei is out of equilibrium in the outer regions of the solar core
\cite{Bahcall-Krastev96}
are estimated to be less than 1\%
and thus negligible at the present level of accuracy.}
\begin{equation}
\sum_r
\left( \frac{ Q }{ 2 } - \langle{E}\rangle_r \right)
\Phi_r
=
\frac{ \mathcal{L}_\odot }{ 4 \pi R^2 }
\,,
\label{luminosity1}
\end{equation}
where
$ \langle{E}\rangle_r = \int E \, X_r(E) \, \mathrm{d}E $
is the average energy of the neutrinos from the source $r$
(see Table~\ref{Sources of solar neutrinos}),
$ \mathcal{L}_\odot = 2.40 \times 10^{39} \, \mathrm{MeV} \, \mathrm{s}^{-1} $
\cite{PDG98}
is the luminosity of the sun,
$ R = 1.496 \times 10^{13} \, \mathrm{cm} $
\cite{PDG98}
is the sun -- earth distance.
The luminosity constraint can be rewritten in the compact form
\begin{equation}
\sum_r Q_r \, \Phi_r
=
K_\odot
\,,
\label{luminosity2}
\end{equation}
where
$ Q_r \equiv Q/2 - \langle{E}\rangle_r $
and
$
K_\odot
\equiv
\mathcal{L}_\odot / 4 \pi R^2
=
8.54 \times 10^{11} \, \mathrm{MeV} \, \mathrm{cm}^{-2} \, \mathrm{s}^{-1}
$
is the solar constant.
Let us emphasize that the luminosity relation
is valid under the assumption that solar $\nu_e$'s
on their way to the earth do not transform into other states.
Neglecting the small fraction of energy carried away by neutrinos,
the luminosity constraint gives the approximate value of the
total solar neutrino flux $ \Phi = \sum_r \Phi_r $:
\begin{equation}
\Phi
\simeq
\frac{ 2 \, K_\odot }{ Q }
=
6.4 \times 10^{10} \, \mathrm{cm}^{-2} \, \mathrm{s}^{-1}
\,.
\label{total flux}
\end{equation}

Since the $^3\mathrm{He}$ nuclei necessary for the
formation of $^7\mathrm{Be}$ and $^8\mathrm{B}$
are created by the $pp$ or $pep$ reactions,
there is another model-independent constraint for
the solar neutrino fluxes of the $pp$ cycle
(see \cite{Bahcall-Krastev96}):
\begin{equation}
\phi_{^7\mathrm{Be}}
+
\phi_{^8\mathrm{B}}
\leq
\phi_{pp}
+
\phi_{pep}
\,.
\label{sun101}
\end{equation}

Let us now consider the experimental data.
The results of \emph{five solar neutrino experiments}
are available at present
and are listed in Table \ref{results of solar neutrino experiments}.

\begin{table}
\begin{center}
\begin{tabular}{cccc}
Experiment
&
Result
&
Theory
&
$ \frac{\mbox{Result}}{\mbox{Theory\rule[-0.2cm]{0pt}{0.5cm}}}$
\vphantom{\bigg|}
\\
\hline
\hline
Homestake \protect\cite{Homestake98}
&
\begin{tabular}{c}
$ 2.56 \pm 0.16 \pm 0.16 $
\rule{0pt}{0.4cm}
\\
($ 2.56 \pm 0.23 $)
\rule[-0.2cm]{0pt}{0.5cm}
\end{tabular}
&
$ 7.7 \, {}^{+1.2}_{-1.0} $
&
$ 0.33 \, {}^{+0.06}_{-0.05} $
\\
\hline
GALLEX \protect\cite{GNO-nu98}
&
\begin{tabular}{c}
$ 77.5 \pm 6.2 \, {}^{+4.3}_{-4.7} $
\rule{0pt}{0.5cm}
\\
($ 78 \pm 8 $)
\rule[-0.2cm]{0pt}{0.5cm}
\end{tabular}
&
$ 129 \, {}^{+8}_{-6} $
&
$ 0.60 \pm 0.07 $
\\
\hline
SAGE \protect\cite{SAGE-nu98}
&
\begin{tabular}{c}
$ 66.6 \, {}^{+6.8}_{-7.1} \, {}^{+3.8}_{-4.0} $
\rule{0pt}{0.5cm}
\\
($ 67 \pm 8 $)
\rule[-0.2cm]{0pt}{0.5cm}
\end{tabular}
&
$ 129 \, {}^{+8}_{-6} $
&
$ 0.52 \pm 0.07 $
\\
\hline
Kamiokande \protect\cite{Kam-sun-96}
&
\begin{tabular}{c}
$ 2.80 \pm 0.19 \pm 0.33 $
\rule{0pt}{0.4cm}
\\
($ 2.80 \pm 0.38 $)
\rule[-0.2cm]{0pt}{0.5cm}
\end{tabular}
&
$ 5.15 \, {}^{+1.0}_{-0.7} $
&
$ 0.54 \pm 0.07 $
\\
\hline
Super-Kamiokande \protect\cite{SK-sun-nu98}
&
\begin{tabular}{c}
$ 2.44 \pm 0.05 \, {}^{+0.09}_{-0.07} $
\rule{0pt}{0.5cm}
\\
($ 2.44 \, {}^{+0.10}_{-0.09} $)
\rule[-0.2cm]{0pt}{0.5cm}
\end{tabular}
&
$ 5.15 \, {}^{+1.0}_{-0.7} $
&
$ 0.47 \, {}^{+0.07}_{-0.09} $
\\
\hline
\hline
\end{tabular}
\end{center}
\refstepcounter{tables}
\label{results of solar neutrino experiments}
\footnotesize
Table \ref{results of solar neutrino experiments}.
The results of solar neutrino experiments
confronted with the corresponding theoretical predictions \protect\cite{BP98}.
The results of the Homestake, GALLEX and SAGE experiments are expressed in
terms of event rates in SNU units
($ 1 \:\mathrm{SNU} \equiv 10^{-36} \:
\mathrm{events} \: \mathrm{atom}^{-1} \, \mathrm{s}^{-1} $),
whereas the results of the Kamiokande and Super-Kamiokande experiments
are expressed in terms of the $^8\mathrm{B}$ neutrino flux in units of
$ 10^6 \, \mathrm{cm}^{-2} \mathrm{s}^{-1} $.
The first experimental error is statistical and the second is systematic.
The experimental values in parenthesis have the
statistical and systematic errors added in quadrature.
\end{table}

Homestake \cite{Homestake68,Homestake94,Homestake98},
GALLEX \cite{GALLEX92a,GALLEX96}
and
SAGE \cite{SAGE91,SAGE96}
are radiochemical experiments.
In the pioneering chlorine Homestake experiment 
of R. Davis \textit{et. al.},
which started in 1967,
the detector is a tank with a volume of
$6 \times 10^5$ liters filled with C$_2$Cl$_4$.
Radioactive atoms of $^{37}$Ar
are produced by solar electron neutrinos through the reaction
\cite{Pontecorvo46,Alvarez49}
\begin{equation}
\nu_e + {}^{37}\mathrm{Cl} \to e^- + {}^{37}\mathrm{Ar}
\,,
\label{604}
\end{equation}
which has an energy threshold
$ E_{\mathrm{th}} = 0.81 \, \mathrm{MeV} $.
The radioactive $^{37}$Ar atoms that
are created  during the time of exposition of each run
(about two months) are extracted from the detector
by purging it with $^4$He
and counted in small proportional counters
which detect the Auger electron produced in the
electron-capture of the $^{37}$Ar nuclei.
About 0.5 atoms of $^{37}$Ar are produced every day by solar neutrinos
and about 16 atoms are extracted in each run
(this number is smaller than 30 because of the $^{37}$Ar lifetime of about 35 days
and because of the extraction efficiency of about 90\%).
Since the energy threshold is above the end-point of the $pp$ neutrino
spectrum
and the cross section of the detecting process (\ref{604})
grows with the neutrino energy,
the main contribution
to the counting rate in the
Homestake experiment comes from $^8$B and $^7$Be neutrinos.
According to the BP98 SSM \cite{BP98}
the event rate in the Homestake experiment should
be\footnote{One Solar Neutrino Unit (SNU) is defined as
$ 10^{-36} \, \mathrm{events} \, \mathrm{atom}^{-1} \, \mathrm{s}^{-1} $.}
$ 7.7 \, {}^{+1.2}_{-1.0} \, \mathrm{SNU} $,
of which
5.9 SNU come from $^8$B neutrinos,
1.15 SNU are produced by $^7$Be neutrinos
and
0.7 SNU are due to $pep$ and CNO neutrinos
(see Table \ref{solar neutrinos fluxes}).
The Homestake event rate presented in 
Table \ref{results of solar neutrino experiments}
is the event rate averaged over 108 runs \cite{Homestake98}.

In the radiochemical gallium experiments GALLEX and SAGE 
electron neutrinos from the sun are detected through the observation
of radioactive $^{71}$Ge that is produced in the process
\begin{equation}
\nu_e + {}^{71}\mathrm{Ga} \to e^- + {}^{71}\mathrm{Ge}
\,.
\label{605}
\end{equation}
The GALLEX detector
is a tank containing 30.3 tons of $^{71}$Ga
(100 tons of a water solution of gallium chloride),
whereas the SAGE experiment
uses about 57 tons of $^{71}$Ga in metallic form.
The event rate measured in the GALLEX experiment is
$ 0.699 \pm 0.069 $ events per day.
(Since the $^{71}$Ge lifetime is about 16.5 days, only around 
7 atoms of $^{71}$Ge are extracted
in each 3-weeks run.)

Since the threshold of the process (\ref{605}) is
$ E_{\mathrm{th}}^{^{71}\mathrm{Ga}} = 0.233 \, \mathrm{MeV} $,
neutrinos from all sources are detected
in gallium experiments. 
According to the SSM,
the contributions to the total predicted event rate
from $pp$, $^7$Be and $^8$B neutrinos are \cite{BP98}
54\%, 27\% and 10\%,
respectively.

As can be seen from Table~\ref{results of solar neutrino experiments},
both gallium detectors measure an event rate
that is about one half of the SSM prediction.
The weighted average of the GALLEX and SAGE results
yields the event rate
\begin{equation}
S_{\mathrm{Ga}}^{\mathrm{exp}}
=
72.5 \pm 6 \: \mathrm{SNU}
\,.
\label{611}
\end{equation}
Taking into account the experimental and theoretical errors,
this rate differs from the SSM prediction by about seven standard deviations!

Both the GALLEX and SAGE detectors have been calibrated using an intense
$^{51}\mathrm{Cr}$ neutrino source.
The ratio of observed and expected events is
$ 0.93 \pm 0.08 $ for GALLEX \cite{GALLEX98-Cr}
and
$ 0.95 \pm 0.12 $ for SAGE \cite{SAGE96}.
These results demonstrate the absence of unexpected systematic
errors at the 10\% level
in both experiments.
In addition,
the GALLEX Collaboration calibrated the detector
by introducing a known number of
radioactive $^{71}\mathrm{As}$ atoms
in the target solution
\cite{GALLEX98-Cr,GALLEX98-As}.
The atoms of $^{71}\mathrm{Ge}$ resulting from $^{71}\mathrm{As}$
decay have been extracted in the usual way,
and the As tests prove, at the 1\% level, the reliability of the technique
\cite{GALLEX-TAUP97} (the number of $^{71}\mathrm{Ge}$ atoms 
produced in these tests is of the order of $10^5$,
whereas the number produced in the $^{51}\mathrm{Cr}$
tests is of the order of ten per day).
As emphasized by the GALLEX Collaboration \cite{GALLEX98-Cr,GALLEX98-As},
these results rule out the presence of unexpected radiochemical effects
that could explain the deficit of solar $\nu_e$'s
measured by the gallium experiments.

It is necessary to emphasize that
the gallium experiments are not only very important for the assessment of
the solar neutrino problem but also for the theory of
thermonuclear energy production in the sun: 
\emph{They have provided the 
first observation of low-energy solar neutrinos
produced in the $pp$ reaction
that is the basic reaction of the $pp$ cycle and, therefore, 
the first direct experimental confirmation of the theory of
thermonuclear origin of solar energy production.}

In the
Kamiokande \cite{Kam-sun-88,Kam-sun-91-PRD,Kam-sun-96}
and
Super-Kamiokande \cite{SK-sun-97-Inoue,SK-sun-98-PRL,SK-sun-nu98}
experiments
water-Cherenkov detectors are used for the detection of
solar neutrinos
through the observation of the Cherenkov light emitted by
the recoil electrons in the elastic-scattering process
\begin{equation}
\nu + e^- \to \nu + e^-
\,.
\label{606}
\end{equation}
Since the direction of the recoil electron is peaked in the direction
of the incoming neutrino,
water-Cherenkov detectors measure the direction of the neutrino flux
and the results of
Kamiokande
and
Super-Kamiokande
have proven that there is a flux of
high-energy neutrinos coming from the sun
with an intensity about half of that predicted by the SSM.

The energy threshold of water-Cherenkov detectors is given by the threshold
for the detection of the recoil electron in the reaction (\ref{606}).
It is higher than in other solar neutrino experiments
because of the large background at low energies:
$ E_{\mathrm{th}}^{\mathrm{Kam}} \simeq 7 \, \mathrm{MeV} $
in the Kamiokande experiment and
$ E_{\mathrm{th}}^{\mathrm{SK}} \simeq 6.5 \, \mathrm{MeV} $
in the Super-Kamiokande experiment.
Therefore,
only $^8\mathrm{B}$ neutrinos can be detected in water-Cherenkov experiments.
If nothing happens to electron neutrinos during their trip from
the core of the sun to the earth
(\textit{i.e.}, no neutrino oscillations or other transitions),
the results of the Kamiokande and Super-Kamiokande experiments provide
a measurement of the total flux
$\Phi_{^8\mathrm{B}}$ of $^8\mathrm{B}$ neutrinos.
Therefore, the results of the Kamiokande and Super-Kamiokande experiments
are usually presented in terms of the measured 
flux of $^8\mathrm{B}$ neutrinos. In the Super-Kamiokande experiment
it was found
\begin{equation}
\Phi_{^8\mathrm{B}}^{\mathrm{SK}}
=
( 2.44 \, {}^{+0.10}_{-0.09} ) \times 10^6 \, \mathrm{cm}^{-2} \, \mathrm{s}^{-1}
\,,
\label{6061}
\end{equation}
which is about one half of the flux predicted by the SSM
(see Table~\ref{results of solar neutrino experiments}).\footnote{For
recent Super-Kamiokande results on solar neutrinos see \emph{Note added}.}

From the experimental results and
the most updated theoretical predictions \cite{BP98}
listed in Table \ref{results of solar neutrino experiments}
one can see that the observed event rates in all solar neutrino experiments
are significantly smaller than the predicted rates
(see also \cite{Bahcall-Krastev-Smirnov98}
where the comparison with other theoretical predictions is discussed).
This discrepancy constitutes the \emph{solar neutrino problem}.
The SSM is robust and it has been recently
tested in a convincing way by comparing its predicted
value for the sound speed in the interior of the sun
with precise helioseismological measurements
(see \cite{Bahcall-Pinsonneault97-PRL78,Ricci97-PLB407,
BP98,Bahcall-nu98,Gough-nu98,Turck98}).
However,
it is clear that a model-independent proof of the existence 
of the solar neutrino problem
would be more convincing.
Such a model-independent approach has been discussed in several papers
\cite{Spiro-Vignaud90,Castellani93-AA,Castellani93-PL,
Hata-Bludman-Langacker94,Berezinsky94,Bahcall94,
Fiorentini95,Parke95,Heeger-Robertson96,Bahcall-Krastev96,
Castellani97,Hata-Langacker97,Bahcall-Krastev-Smirnov98,Minakata-Nunokawa98}
using the model-independent luminosity constraint (\ref{luminosity2})
and the fact that the energy spectra $X_r(E)$
Eq.(\ref{609})
of the various neutrino sources are practically independent
from solar physics
\cite{Bahcall91-PRD,Bahcall96-PRC}.

From the luminosity constraint it is possible to obtain a model-independent
lower bound on the gallium event rate.
Indeed,
since $pp$ neutrinos have the smallest average energy
and
the $\nu_e$--$^{71}\mathrm{Ga}$ cross section $\sigma_{\mathrm{Ga}}(E)$
increases with the neutrino energy $E$,
we have
\begin{equation}
S_{\mathrm{Ga}}
=
\int \sigma_{\mathrm{Ga}}(E) \sum_r \phi_r(E) \, \mathrm{d}E
=
\sum_r \langle\sigma_{\mathrm{Ga}}\rangle_r \, \Phi_r
\geq
\langle\sigma_{\mathrm{Ga}}\rangle_{pp} \Phi
\,,
\label{607}
\end{equation}
where
$
\langle\sigma_{\mathrm{Ga}}\rangle_r
=
\int \sigma_{\mathrm{Ga}}(E) \, X_r(E) \, \mathrm{d}E
$
is the average cross section of neutrinos from the source $r$
(see Table~\ref{solar neutrinos fluxes}).
Since the luminosity constraint (\ref{luminosity2}) implies that
\begin{equation}
Q_{pp} \, \Phi
\geq
K_\odot
\,,
\label{608}
\end{equation}
we obtain
\begin{equation}
S_{\mathrm{Ga}}
\geq
\frac{ \langle\sigma_{\mathrm{Ga}}\rangle_{pp} \, K_\odot }{ Q_{pp} }
=
76 \pm 2 \: \mathrm{SNU}
\,,
\label{610}
\end{equation}
where we used the value of
$ \langle\sigma_{\mathrm{Ga}}\rangle_{pp} = 11.7 \times 10^{-46} 
\, \mathrm{cm}^2 $
given in Table~\ref{solar neutrinos fluxes}
and
$ Q_{pp} = 13.1 \, \mathrm{MeV} $.

The lower bound (\ref{610})
is just compatible with the
combined result (\ref{611}) of the gallium experiments.
This means that the results of the GALLEX and SAGE experiments
can be explained if practically only $pp$ neutrinos are emitted by the sun.
This possibility is incompatible with any solar model
constrained by the helioseismological data
\cite{BP98,Bahcall-nu98}.

More stringent model-independent conclusions
on the existence of the solar neutrino problem
can be obtained by comparing the results of different solar 
neutrino experiments.
If the survival probability of solar $\nu_e$'s is equal to one,
the result of the Super-Kamiokande experiment gives
the value (\ref{6061}) for the flux of $^8\mathrm{B}$ neutrinos,
whose contributions to the event rates of the chlorine and 
gallium experiments are
\begin{equationarrayzero}
&&
S_{\mathrm{Cl}}^{^8\mathrm{B},\mathrm{SK}}
=
\langle\sigma_{\mathrm{Cl}}\rangle_{^8\mathrm{B}}
\,
\Phi_{^8\mathrm{B}}^{\mathrm{SK}}
=
2.78 \pm 0.27 \, \mathrm{SNU}
\,,
\label{621}
\\
&&
S_{\mathrm{Ga}}^{^8\mathrm{B},\mathrm{SK}}
=
\langle\sigma_{\mathrm{Ga}}\rangle_{^8\mathrm{B}}
\,
\Phi_{^8\mathrm{B}}^{\mathrm{SK}}
=
5.9 \, {}^{+1.9}_{-0.9} \, \mathrm{SNU}
\,.
\label{622}
\end{equationarrayzero}%

Subtracting the contribution (\ref{621}) of $^8\mathrm{B}$
$\nu_e$'s from the event rate measured in the Homestake experiment
(see Table~\ref{results of solar neutrino experiments}),
one obtains
\begin{equation}
S_{\mathrm{Cl}}^{\mathrm{exp}}
-
S_{\mathrm{Cl}}^{^8\mathrm{B},\mathrm{SK}}
=
- 0.22 \pm 0.35
\,.
\label{623}
\end{equation}
Since the quantity
$
S_{\mathrm{Cl}}
-
S_{\mathrm{Cl}}^{^8\mathrm{B}}
$
represents the contribution to the chlorine event rate due to
$pep$, $^7\mathrm{Be}$, $hep$ and CNO neutrinos,
it must be positive (or zero)
but the result in Eq.(\ref{623})
is only marginally compatible with a positive value
(the probability is less than 26\%).
Furthermore,
the result Eq.(\ref{623})
shows that the fluxes of intermediate energy neutrinos
($pep$, $^7\mathrm{Be}$ and CNO neutrinos)
are strongly suppressed with respect to the SSM prediction
(see Table~\ref{solar neutrinos fluxes}).
This is incompatible with any solar model
constrained by the helioseismological data
\cite{BP98,Bahcall-nu98}
and with the fact that,
since both the reactions that produce
$^7\mathrm{Be}$
and
$^8\mathrm{B}$ neutrinos
originate from $^7\mathrm{Be}$ nuclei
(see Fig.~\ref{pp_cycle}),
it is very difficult
to have a suppression of the $^7\mathrm{Be}$ neutrino flux
(with respect to the SSM value)
that is stronger than the suppression of the
$^8\mathrm{B}$ neutrino flux.\footnote{Non-standard solar models
with low central temperature yield a suppression of the
$^8\mathrm{B}$ neutrino flux that is stronger than that
of the $^7\mathrm{Be}$ neutrino flux
(see \cite{Castellani97} and references therein).}

Let us consider now the gallium experiments.
Subtracting the contribution of $^8\mathrm{B}$
neutrinos from the luminosity constraint (\ref{luminosity2}),
we have
\begin{equation}
K_\odot - Q_{^8\mathrm{B}} \, \Phi_{^8\mathrm{B}}
=
\sum_{r\neq{}^8\mathrm{B}} Q_r \, \Phi_r
\leq
Q_{pp} \sum_{r\neq{}^8\mathrm{B}} \Phi_r
\,.
\label{626}
\end{equation}
Hence,
following the same reasoning as in Eq.(\ref{607})
we obtain
\begin{equation}
S_{\mathrm{Ga}}^{\mathrm{min}}
=
S_{\mathrm{Ga}}^{^8\mathrm{B},\mathrm{SK}}
+
\langle\sigma_{\mathrm{Ga}}\rangle_{pp}
\,
\frac
{ K_\odot - Q_{^8\mathrm{B}} \, \Phi_{^8\mathrm{B}}^{\mathrm{SK}} }
{ Q_{pp} }
=
82 \, {}^{+3}_{-2} \, \mathrm{SNU}
\,,
\label{627}
\end{equation}
and the contribution of
$pep$, $^7\mathrm{Be}$, $hep$ and CNO neutrinos
to the gallium event rate cannot be bigger than
\begin{equation}
S_{\mathrm{Ga}}^{\mathrm{exp}}
-
S_{\mathrm{Ga}}^{\mathrm{min}}
=
- 9.5 \pm 6
\,.
\label{628}
\end{equation}
This result is compatible with a positive value
with a probability smaller than 6\%.
Again,
the result (\ref{628}) implies that
the fluxes of intermediate energy neutrinos
($pep$, $^7\mathrm{Be}$ and CNO neutrinos)
are strongly suppressed with respect to the SSM prediction,
in contradiction with any solar model
constrained by the helioseismological data
and with the moderate suppression of the $^8\mathrm{B}$ neutrino flux.

Further model-independent methods for proving
the existence of the solar neutrino problem
on the basis of the data of solar neutrino experiments
are nicely discussed in Ref.~\cite{Castellani97}.
In the following
we will assume that there is a solar neutrino problem
that is caused by neutrino oscillations.

The solar neutrino data have been analysed in many papers
under the assumption of two-neutrino mixing
(see \cite{Hata-Langacker97,FLM98-sun-AP,Bahcall-Krastev-Smirnov98}
and references therein)
and in a few papers
under the assumption of three-neutrino mixing
(see \cite{NMRS96-PRD53,Goswami-Kar-Raychaudhuri97,FLM96-PRD54,Osland-Vigdel98}
and references therein).
Here we will briefly review the results
of the most updated two-generation analysis
\cite{Bahcall-Krastev-Smirnov98}
of all the solar
neutrino data,
including those obtained in the first 504 days of operation
of the Super-Kamiokande experiment
(see Table~\ref{results of solar neutrino experiments}).

The deficit of solar $\nu_e$'s can be explained
in terms of two-generation neutrino mixing
either through vacuum oscillations
or through MSW resonant transitions in matter \cite{MS,wol}.
Furthermore,
the two mixed neutrinos can be the electron neutrino
and another active ($\nu_\mu$ or $\nu_\tau$) neutrino
or
the electron neutrino and a sterile neutrino.
There are two differences between transitions of
solar $\nu_e$'s into active and sterile neutrinos:
\begin{enumerate}

\item
The probability of MSW transitions
is different in the two cases,
because sterile neutrinos do not have the weak-interaction
potential due to elastic forward scattering
that is present for active neutrinos.

\item
The active neutrinos
($\nu_\mu$'s or $\nu_\tau$'s)
that are produced by the oscillation mechanism
(either in vacuum or in matter)
contribute to the event rates measured in the Kamiokande and Super-Kamiokande
water-Cherenkov experiments
(notice, however,
that the cross section of $\nu_\mu$--$e^-$ and $\nu_\tau$--$e^-$
scattering
is about six times smaller than the $\nu_e$--$e^-$
cross section).

\end{enumerate}

\begin{figure}[t!]
\begin{tabular*}{\linewidth}{@{\extracolsep{\fill}}cc}
\begin{minipage}{0.47\linewidth}
\begin{center}
\mbox{\epsfig{file=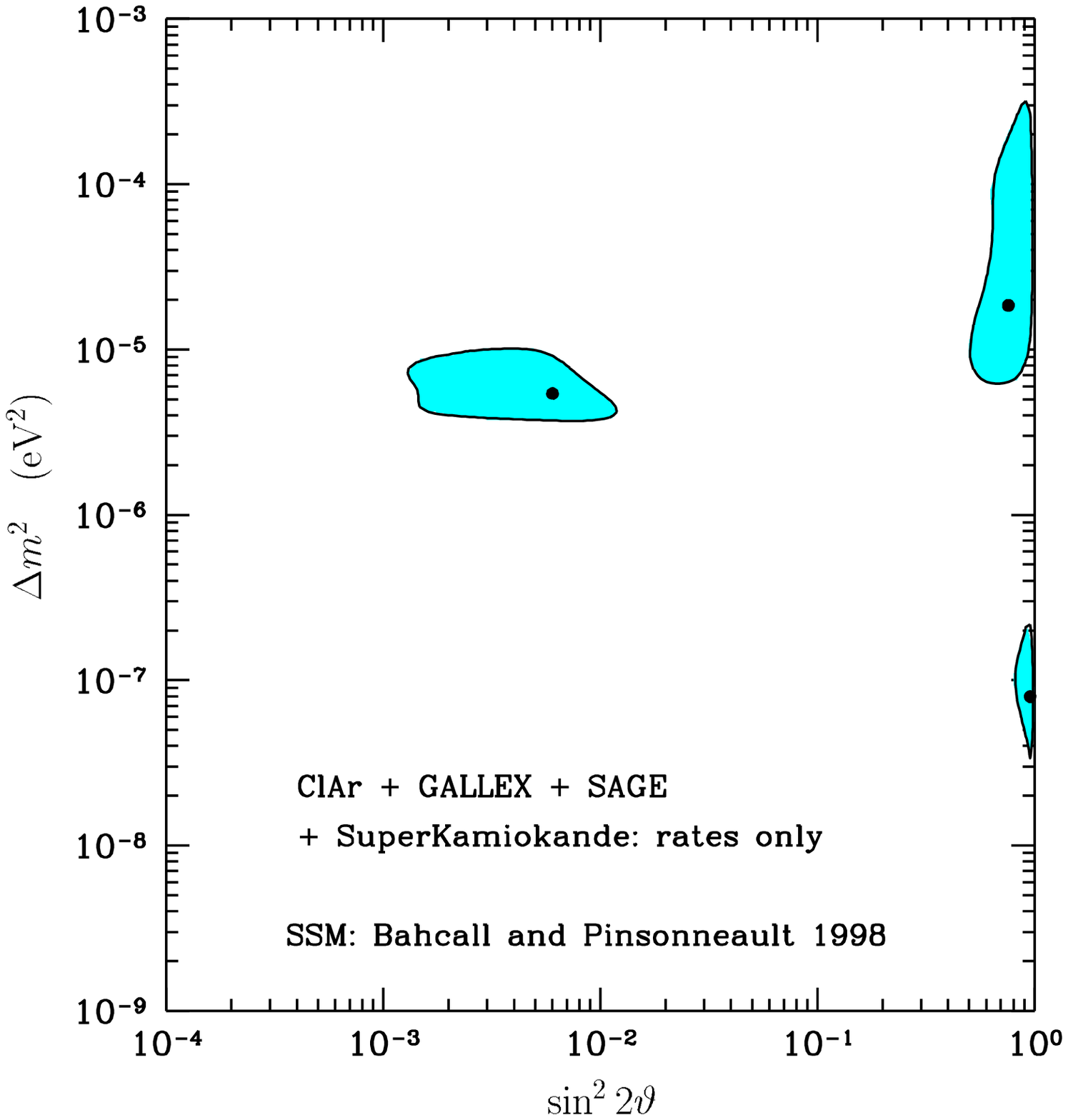,width=0.95\linewidth}}
\end{center}
\end{minipage}
&
\begin{minipage}{0.47\linewidth}
\begin{center}
\mbox{\epsfig{file=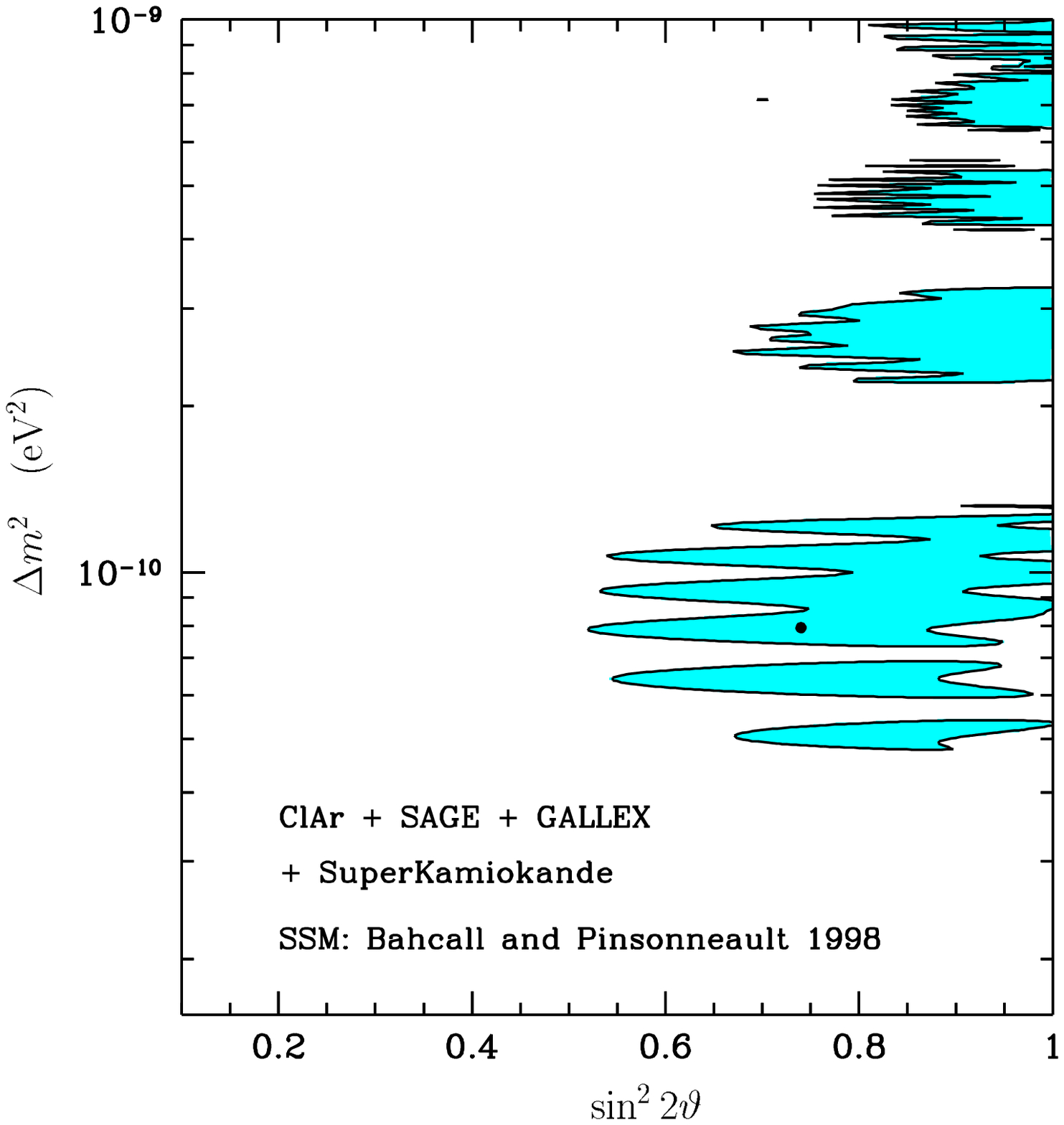,width=0.95\linewidth}}
\end{center}
\end{minipage}
\\
\begin{minipage}{0.47\linewidth}
\refstepcounter{figures}
\label{bks-f2}
\footnotesize
Figure \ref{bks-f2}.
MSW $\nu_e\to\nu_\mu$ or $\nu_e\to\nu_\tau$ transitions:
result of the fit of the experimental event rates listed in
Table~\protect\ref{results of solar neutrino experiments}.
The shadowed regions are allowed at 99\% CL
\protect\cite{Bahcall-Krastev-Smirnov98}.
The dots indicate the best-fit points in each allowed region.
\end{minipage}
&
\begin{minipage}{0.47\linewidth}
\refstepcounter{figures}
\label{bks-f5}
\footnotesize
Figure \ref{bks-f5}.
Vacuum $\nu_e\to\nu_\mu$ or $\nu_e\to\nu_\tau$ oscillations:
result of the fit of the experimental event rates listed in
Table~\protect\ref{results of solar neutrino experiments}.
The shadowed regions are allowed at 99\% CL
\protect\cite{Bahcall-Krastev-Smirnov98}.
The dot indicates the best-fit point.
\end{minipage}
\end{tabular*}
\end{figure}

The formulas for the survival probability of solar $\nu_e$'s
in the case of vacuum oscillations and MSW transitions
are given in Eq.(\ref{surv2}) (with $\alpha=e$)
and
in Eq.(\ref{surviv}) together with Eq.(\ref{Pc1}) \cite{pet88},
respectively.
These formulas depend on the two mixing parameters
$\Delta{m}^2$ and $\sin^22\vartheta$.
The allowed regions in the
$\sin^22\vartheta$--$\Delta{m}^2$ plane,
obtained in Ref.~\cite{Bahcall-Krastev-Smirnov98}
from the fit
of the measured event rates listed in
Table~\ref{results of solar neutrino experiments}
by using the BP98 SSM \cite{BP98} and the analytic formulas in
Ref.~\cite{pet88} for the MSW case,
are shown in Figs.~\ref{bks-f2}--\ref{bks-f5}.

Figure~\ref{bks-f2}
\cite{Bahcall-Krastev-Smirnov98}
shows the three allowed regions
in the case of
MSW $\nu_e\to\nu_\mu$ or $\nu_e\to\nu_\tau$ transitions.
They are the small mixing angle (SMA-active) region at
\begin{equation}
\Delta{m}^2 \simeq 5 \times 10^{-6} \, \mathrm{eV}^2
\,,
\qquad
\sin^22\vartheta \simeq 6 \times 10^{-3}
\qquad \qquad
\mbox{(SMA-active)}
\,,
\label{SMA-active}
\end{equation}
and the large mixing angle (LMA) region at
\begin{equation}
\Delta{m}^2 \simeq 2 \times 10^{-5} \, \mathrm{eV}^2
\,,
\qquad
\sin^22\vartheta \simeq 0.76
\qquad \qquad
\mbox{(LMA)}
\,.
\label{LMA}
\end{equation}
The best fit, with a confidence level of 19\%,
is obtained in the SMA region,
whereas the LMA region has a confidence level of 4\%.
In Ref.~\cite{Bahcall-Krastev-Smirnov98} there is also
the so-called low mass (LOW) region at
$\Delta{m}^2 \simeq 8 \times 10^{-8} \: \mathrm{eV}^2$
and
$\sin^22\vartheta \simeq 0.96$,
however, the LOW region is only marginally acceptable,
with a confidence level of 0.7\%.

The allowed regions
in the $\sin^22\vartheta$--$\Delta{m}^2$ plane
in the case of
$\nu_e\to\nu_\mu$ or $\nu_e\to\nu_\tau$
vacuum oscillations are shown in Fig.~\ref{bks-f2}
\cite{Bahcall-Krastev-Smirnov98}.
These regions extend over large ranges of
$\sin^22\vartheta$ and $\Delta{m}^2$ around the best fit values
\begin{equation}
\Delta{m}^2 \simeq 8 \times 10^{-11} \, \mathrm{eV}^2
\,,
\qquad
\sin^22\vartheta \simeq 0.75
\qquad \qquad
\mbox{(Vac. Osc.)}
\,.
\label{Vac. Osc.}
\end{equation}
The confidence level of the fit is 3.8\%. 
There is no allowed region in the case of
$\nu_e\to\nu_s$
vacuum oscillations
(the fit has a confidence level of 0.05\%)
\cite{Bahcall-Krastev-Smirnov98,pet-solar-vac}.

\begin{figure}[t!]
\begin{tabular*}{\linewidth}{@{\extracolsep{\fill}}cc}
\begin{minipage}{0.47\linewidth}
\begin{center}
\mbox{\epsfig{file=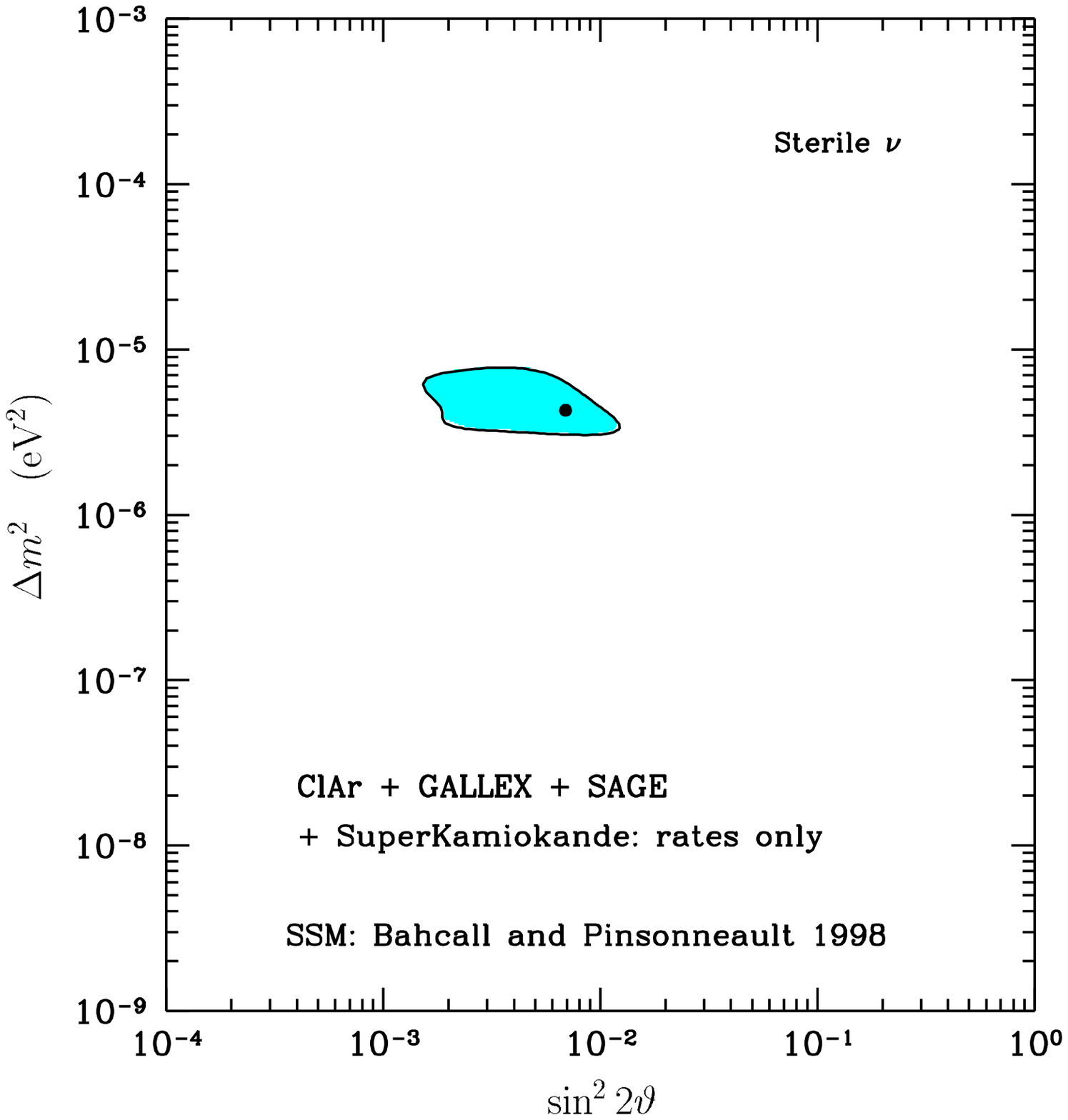,width=0.95\linewidth}}
\end{center}
\end{minipage}
&
\begin{minipage}{0.47\linewidth}
\begin{center}
\mbox{\epsfig{file=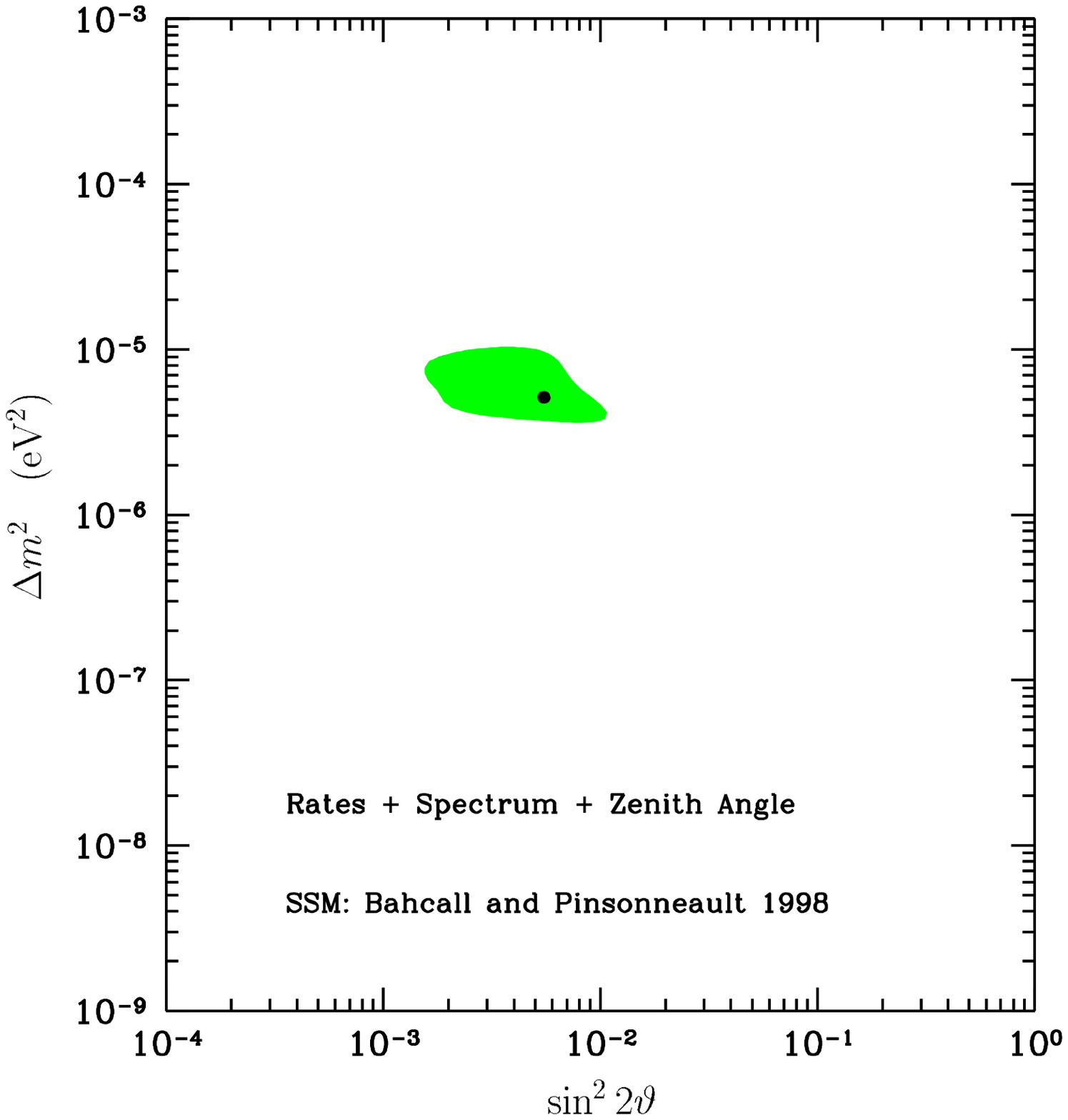,width=0.95\linewidth}}
\end{center}
\end{minipage}
\\
\begin{minipage}{0.47\linewidth}
\refstepcounter{figures}
\label{bks-f4}
\footnotesize
Figure \ref{bks-f4}.
MSW $\nu_e\to\nu_s$ transitions:
result of the fit of the experimental event rates listed in
Table~\protect\ref{results of solar neutrino experiments}.
The shadowed region is allowed at 99\% CL
\protect\cite{Bahcall-Krastev-Smirnov98}.
The dot indicates the best-fit point.
\end{minipage}
&
\begin{minipage}{0.47\linewidth}
\refstepcounter{figures}
\label{bks-f15b}
\footnotesize
Figure \ref{bks-f15b}.
MSW $\nu_e\to\nu_\mu$ or $\nu_e\to\nu_\tau$ transitions:
result of the global fit of the experimental event rates listed in
Table~\protect\ref{results of solar neutrino experiments}
and of the energy spectrum and zenith-angle distribution
measured in the Super-Kamiokande experiment
\cite{SK-sun-nu98}.
The shadowed region is allowed at 99\% CL
\protect\cite{Bahcall-Krastev-Smirnov98}.
The dot indicates the best-fit point.
\end{minipage}
\end{tabular*}
\end{figure}

As shown in Fig.~\ref{bks-f4}
\cite{Bahcall-Krastev-Smirnov98},
only the small mixing angle (SMA-sterile) region at
\begin{equation}
\Delta{m}^2 \simeq 4 \times 10^{-6} \, \mathrm{eV}^2
\,,
\qquad
\sin^22\vartheta \simeq 7 \times 10^{-3}
\qquad \qquad
\mbox{(SMA-sterile)}
\,,
\label{SMA-sterile}
\end{equation}
is allowed in the case of
MSW $\nu_e\to\nu_s$ transitions,
with a confidence level of 19\%
(the large mixing angle and low mass solutions
have a confidence level of 0.001\%
and 0.003\%,
respectively) \cite{Bahcall-Krastev-Smirnov98,pet-solar-msw}.

The authors of Ref.~\cite{Bahcall-Krastev-Smirnov98}
performed also global fits of the solar neutrino data
including the energy spectrum and zenith-angle
distribution
of the recoil electrons
measured in the Super-Kamiokande experiment
\cite{SK-sun-nu98}.
As noted in Ref.~\cite{Bahcall-Krastev-Smirnov98},
since these data are still preliminary,
the results of this analysis are less robust than those
obtained fitting only the global rates.
However,
it is interesting to note that
in the case of
MSW $\nu_e\to\nu_\mu$ or $\nu_e\to\nu_\tau$ transitions
only the SMA-active region remains allowed,
as shown in Fig.~\ref{bks-f15b}
\cite{Bahcall-Krastev-Smirnov98},
with a confidence level of 7\%.

During the night, solar neutrinos
pass through the earth and the matter effect can
cause a regeneration of $\nu_e$'s
(see Ref.~\cite{Smirnov-nu98} and references therein) and, therefore, a
zenith angle dependence of the solar neutrino flux.
The size of this effect depends on the values of
$\Delta{m}^2$ and $\sin^22\vartheta$
and is sizable only for large values of the mixing angle.
The preliminary value of the day-night asymmetry
of solar neutrino events measured in the Super-Kamiokande experiment is
\cite{SK-sun-nu98}
\begin{equation}
\frac{D-N}{D+N}
=
- 0.023 \pm 0.020 \pm 0.014
\,,
\end{equation}
compatible with zero.
Hence this experimental result
tends to exclude solutions of the solar neutrino problem
with a large mixing angle. 

As was shown in Refs.~\cite{petcov98,maris} the
step-like profile of the matter density of the earth could lead to an
enhancement of neutrino transitions not only for atmospheric neutrinos
(see Section 5.1.3) but also for solar neutrinos.

In conclusion of this section,
we would like to emphasize that the results of solar neutrino experiments
provide a rather strong indication on favour of neutrino mixing
and several experiments are under construction
\cite{SNO-nu98,Borexino-nu98,GNO-nu98}
or
in project
(see \cite{future-sun,Lanou-nu98} and references therein).
In particular,
the measurement of the electron neutrino spectrum
at SNO may allow to obtain model-independent information
on the neutrino oscillation probability and on the flux
$\Phi_{^8\mathrm{B}}$ of $^8\mathrm{B}$ neutrinos
\cite{BG93-towards,BG-sterile-sun,Kwong-Rosen95-PRD51,%
Kwong-Rosen96-PRD54,FLV98},
the measurement of the flux of $^7\mathrm{Be}$ neutrinos
on the earth in the Borexino experiment
and
the results of Super-Kamiokande, SNO, Borexino and GNO
will allow to distinguish the different
possible solutions of the solar neutrino problem
(see
Refs.~\cite{pet-solar-msw,Bahcall-Krastev97-PRC55,Bahcall-Krastev97-PRC56,% 
FLM98-PLB434,FFLM98} and references therein).

The Sudbury Neutrino Observatory (SNO) \cite{SNO-nu98,SNO-www}
is located 6800 feet below ground in the Creighton mine,
near Sudbury in Ontario (Canada). In the SNO experiment
a Cherenkov detector
with 1 kton of heavy water (D$_2$O) contained in an
acrylic vessel of 12 m diameter will be used.
The Cherenkov light is detected with a geodesic array 
of $10^4$ photomultiplier tubes
surrounding the heavy water vessel.
The heavy water detector is immersed in normal water in order to reduce the
background.
Solar neutrinos will be observed in real-time through the
charged-current (CC), neutral-current (NC) and
elastic scattering (ES) reactions
\begin{eqnarray}
\nu_e + d \to e^- + p + p
\null & \null
\qquad \qquad
\null & \null
\mbox{(CC)}
\,,
\label{SNO-CC}
\\
\nu_\ell + d \to \nu_\ell + p + n
\null & \null
\,,
\qquad
\ell=e,\mu,\tau
\,,
\qquad
\null & \null
\mbox{(NC)}
\,,
\label{SNO-NC}
\\
\nu_\ell + e^- \to \nu_\ell + e^-
\null & \null
\,,
\qquad
\ell=e,\mu,\tau
\,,
\qquad
\null & \null
\mbox{(ES)}
\,.
\label{SNO-ES}
\end{eqnarray}
Since the energy threshold for the observation of the recoil 
electrons in the CC and ES
processes is about 5 MeV
and the neutrino energy threshold for the NC reaction is 2.2 MeV,
only $^8$B neutrinos can be observed.
The event rates predicted by the SSM are around
23 per day for the CC reaction,
7 per day for the NC reaction and
3 per day for the ES reaction.
From the hit pattern of the photomultipliers, the direction and 
energy of the neutrino
in the CC reaction can be reconstructed,
allowing a direct measurement of the spectrum of the
solar electron neutrino flux on the earth.
The observation of a distortion of this spectrum
with respect to the one calculated without neutrino oscillations
will represent a model-independent
proof of the occurrence of neutrino oscillations. 
The NC reaction will allow to detect all active neutrinos
with the same cross section
(because of the $e$--$\mu$--$\tau$
universality of weak interactions),
whereas the cross sections of $\nu_\mu$ and $\nu_\tau$ scattering
in the ES reaction is about six times smaller than that of $\nu_e$.
The measurement of the neutral current reaction will
allow to determine the total flux of active neutrinos from the sun
\cite{BG93-towards,FLV98},
which coincides with the produced flux if there are only
active-active neutrino oscillations,
which can be revealed trough a measurement of
a deficit in the NC/CC event rate.
On the other hand,
if the NC/CC event rate will not show any deficit and
a distortion of the electron neutrino spectrum will be observed,
it will mean that there are transitions of solar $\nu_e$'s into sterile states
\cite{BG-sterile-sun}
(of course, complicated scenarios with transitions
solar $\nu_e$'s into active and sterile states are not excluded
and must be taken into account).
The SNO experiment
is expected to release the first results in the year 1999. 

The Borexino experiment \cite{Borexino-nu98,Borexino-www}
is designed to detect low-energy
solar neutrinos
in real-time
through the observation of the elastic scattering
process (\ref{SNO-ES})
with an energy threshold for the recoil
electron of 250 keV.
It will use 300 tons of liquid scintillator
in an unsegmented detector with 2000 photomultiplier tubes.
The event rate predicted by the SSM for a fiducial volume of about 100 tons
is about 50 events per day,
mostly generated by the 0.86 MeV monoenergetic line of $^7$Be solar neutrinos.
Since this line gives a characteristic spectral signature in the ES process,
the flux on the earth of $^7$Be solar neutrinos will be determined
and it will be possible to check if
it is suppressed with respect to the one predicted by the SSM
as suggested by the results of current experiments.
The neutrino oscillation solutions predict a measurable flux:
$\sim 10$ events per day for the SMA MSW solution,
$\sim 30$ events per day for the LMA MSW solution
and
$\sim 35$ events per day for the vacuum oscillation solution
(see \cite{Bahcall-Krastev97-PRC55}).
The Borexino experiment is
under construction in the Laboratori Nazionali del Gran Sasso in Italy
and is scheduled to start data taking around the year 2001.

\subsection{The LSND experiment}

The Liquid Scintillator Neutrino Detector (LSND) experiment
\cite{LSND95,LSND96,LSND96-PRC54,LSND97-NIM388,LSND98-PRC,%
LSND98,LSND-nu98,LSND-www}
at the Los Alamos Meson Physics Facility (LAMPF) is a SBL neutrino
oscillation experiment with a baseline of about 30 meters. 
It utilizes the LAMPF proton beam with a kinetic energy of 800 MeV which 
produces pions by hitting a 30 cm long water target
located about 1 m upstream from a copper beam stop.
Most of the $\pi^+$'s
are stopped in the target
and decay into muons which come to
rest and decay in the target as well.
The decay at rest of the positively charged muons allows
to investigate $\bar\nu_\mu\to\bar\nu_e$ oscillations
\cite{LSND95,LSND96,LSND96-PRC54,LSND-nu98,LSND-www}.
A small fraction of the positively charged pions ($\sim3.4\%$) decays in 
flight in the 1 m long space between the target
and the beam stop
and is used for the study of $\nu_\mu\to\nu_e$ 
oscillations
\cite{LSND98,LSND98-PRC,LSND-nu98,LSND-www}
(only $\sim0.001\%$ of the $\mu^+$'s
decay in flight and produce a small contamination of
$\nu_e$'s in the $\nu_\mu$ beam).
The detector is an approximately cylindrical tank
8.3 m long by 5.7 m in diameter,
filled with a scintillator medium
consisting mainly of mineral oil (CH$_2$).
In the following we review
the results obtained by the LSND Collaboration
from the analyses of the data from
$\mu^+$ decay at rest
and
$\pi^+$ decay in flight.

\paragraph{$\mu^+$ decay at rest}
\cite{LSND95,LSND96,LSND96-PRC54,LSND-nu98,LSND-www}.
In this case one has the decay chain
\begin{equation}
\setlength{\arraycolsep}{0pt}
\begin{array}{rll} \displaystyle
\pi^+ \to
\null & \null \displaystyle
\mu^+
\null & \null \displaystyle
+ \nu_\mu
\\ \displaystyle
\null & \null \displaystyle
\downarrow
\null & \null \displaystyle
\\ \displaystyle
\null & \null \displaystyle
e^+
\null & \null \displaystyle
+ \nu_e + \bar\nu_\mu
\,.
\end{array}
\label{chain}
\end{equation}
Note that in these decays no $\bar\nu_e$ appears,
which makes the
study of $\bar\nu_\mu \to \bar\nu_e$ transitions possible.
In $\mu^+$
decay at rest the upper bound on the $\bar\nu_\mu$ energy is 
$m_\mu/2 = 52.8$ MeV ($m_\mu$ is the muon mass)
and the energy spectrum of the $\bar\nu_\mu$ flux is very well known,
being determined by the kinematics of muon decay.
The electron antineutrinos are detected by 
\begin{equation}\label{signal}
\bar \nu_e + p \to e^+ + n \,.
\end{equation}
The $\bar\nu_e$ signature is a coincidence between $e^+$ and a delayed
$\gamma$ from the capture
$n + p \to d + \gamma\,(2.2\:\mbox{MeV})$.

There is a background of $\bar\nu_e$ from
the decay chain symmetrical to Eq.(\ref{chain})
starting with $\pi^-$. This $\nu_e$ flux
is suppressed relative to the $\bar\nu_\mu$ flux from the
$\pi^+$ chain by a factor of $7.8 \times 10^{-4}$ for three
reasons: $\pi^-$ production in the beam stop is a factor of 8
smaller than $\pi^+$ production, only 5 \% of the $\pi^-$ decay
in flight (all stopped $\pi^-$ are absorbed) and of the
resulting $\mu^-$ only 12 \% decay before they are captured by
nuclei. In the latter case they give rise to $\nu_\mu$ 
but not to $\bar\nu_e$.
The LSND apparatus does not distinguish between $e^+$ and $e^-$.
Therefore, the reaction 
$\nu_e + {}^{12}C \to e^- + n + {}^{11}N$ gives a signal like a
$\bar\nu_e$ but can only occur for $E_e < 20$ MeV.

\begin{figure}[t!]
\begin{tabular*}{\linewidth}{@{\extracolsep{\fill}}cc}
\begin{minipage}{0.47\linewidth}
\begin{center}
\mbox{\epsfig{file=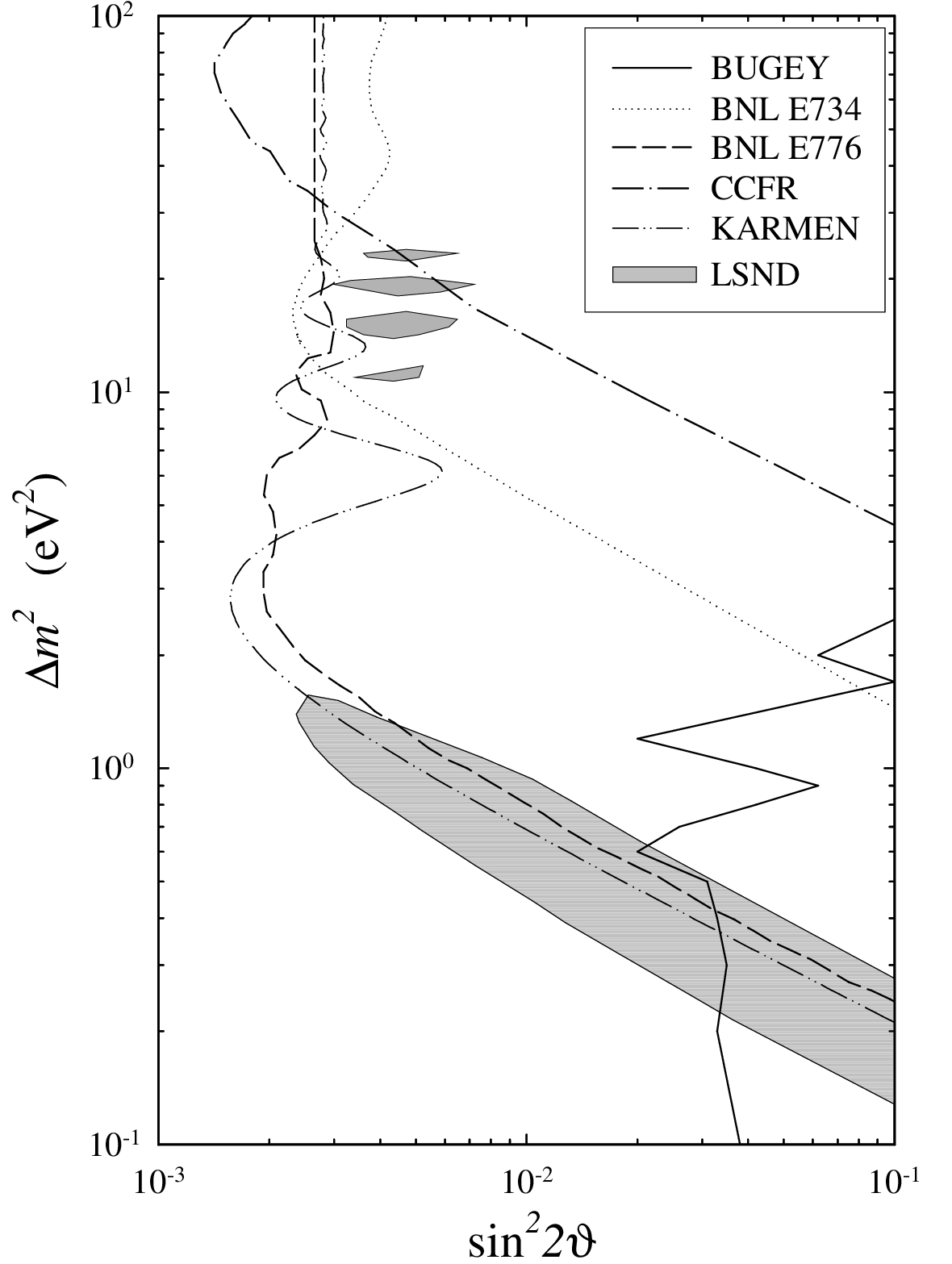,width=0.95\linewidth,clip=}}
\end{center}
\end{minipage}
&
\begin{minipage}{0.47\linewidth}
\begin{center}
\mbox{\epsfig{file=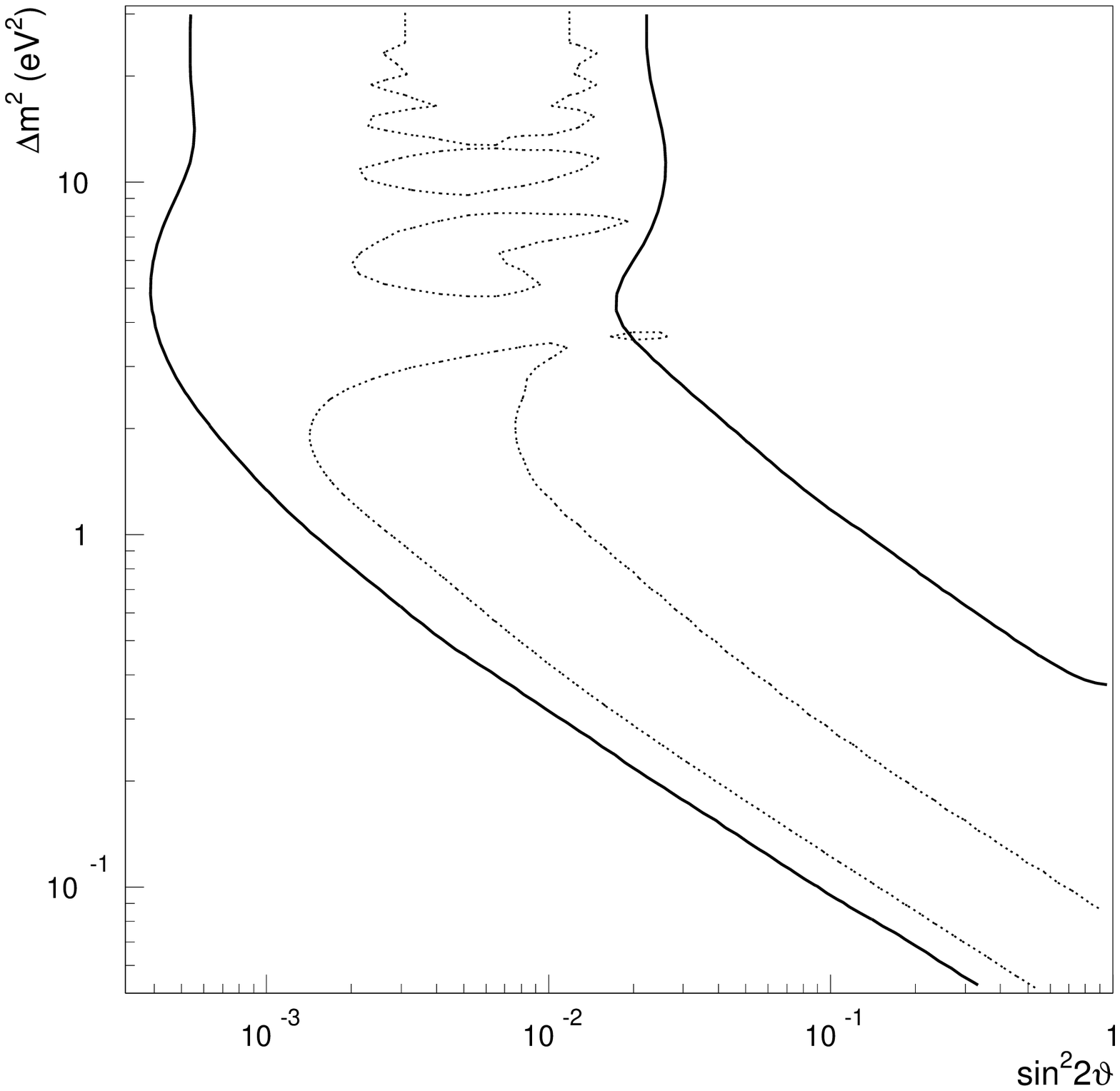,width=0.95\linewidth}}
\end{center}
\end{minipage}
\\
\begin{minipage}{0.47\linewidth}
\refstepcounter{figures}
\label{lsnd-dar}
\footnotesize
Figure \ref{lsnd-dar}.
The shadowed regions
are allowed at 90\% CL
by the $\mu^+$ decay at rest analysis of the LSND data
\protect\cite{LSND-nu98,LSND-www}.
Also shown are the exclusion curves of the
Bugey \protect\cite{Bugey95},
BNL E734 \protect\cite{BNLE734},
BNL E776 \protect\cite{BNLE776},
CCFR \protect\cite{CCFR97}
and
KARMEN (Bayesian analysis) \protect\cite{KARMEN98-nu98}
experiments.
\end{minipage}
&
\begin{minipage}{0.47\linewidth}
\refstepcounter{figures}
\label{lsnd-dif}
\footnotesize
Figure \ref{lsnd-dif}.
The region between the two solid lines is allowed at 95\% CL
by the $\pi^+$ decay in flight analysis of the LSND data
\protect\cite{LSND98,LSND98-PRC}.
The regions enclosed by dashed lines
are the 99\% CL LSND favoured regions obtained with the 1996
muon decay at rest analysis \cite{LSND96,LSND96-PRC54}.
\end{minipage}
\end{tabular*}
\end{figure}

LSND has seen an excess of $e^+$ events with energies between 20
and 60 MeV of 
$51.0 \, {}^{+20.2}_{-19.5} \pm 8.0$
corresponding to a $\bar\nu_\mu\to\bar\nu_e$ transition
probability of $(3.1 \pm 1.2 \pm 0.5) \times 10^{-3}$
\cite{LSND96}.
In Refs.~\cite{LSND-nu98,LSND-www} preliminary results can be
found for the data taken from 1993 to 1997.
The value for the transition probability
agrees with the previous one,
except for a reduced statistical error of
$0.9 \times 10^{-3}$ instead of $1.2 \times 10^{-3}$.
The plot of the favoured region
in the $\sin^22\vartheta$--$\Delta{m}^2$
plane obtained by the LSND collaboration
is shown in Fig.~\ref{lsnd-dar}. In the positron energy range between
36 and 60 MeV the background is reduced because in this
energy range there are no electrons from 
$\nu_e + {}^{12}C \to e^- + {}^{12}N$ induced by electron neutrinos from
$\mu^+$ decay at rest \cite{LSND96}. LSND has
observed 22 events with such positron energies compared to a background
of only $4.6 \pm 0.6$ events \cite{LSND96}.

\paragraph{$\pi^+$ decay in flight}
\cite{LSND98,LSND98-PRC,LSND-nu98,LSND-www}.
The $\pi^+ \to \mu^+ + \nu_\mu$ decay in flight generates a $\nu_\mu$ 
beam with which $\nu_\mu \to \nu_e$ oscillations are studied. The electron
neutrinos are detected by electrons from the inclusive reaction
\begin{equation}
\nu_e + {}^{12}C \to e^- + X
\,,
\end{equation}
where the electrons have an energy $E_e$ between 60 and 200 MeV.
In this case the signal is a single electron. The
electron neutrinos from $\mu^+ \to e^+ + \nu_e + \bar\nu_\mu$
decay in flight are suppressed by the longer muon lifetime and
the kinematics of the three-body decay, whereas those from
$\pi^+ \to e^+ + \nu_e$ by the small branching ratio of 
$1.23 \times 10^{-4}$. The lower bound of 60 MeV on $E_e$ lies
well above $m_\mu/2 = 52.8$ MeV, the endpoint of the electron
spectrum from muon decay at rest.
Such muons can be induced by beam-related 
$\stackrel{\scriptscriptstyle (-)}{\nu}_{\hskip-3pt \mu}$ events. Also
$\nu_e$ from $\mu^+$ decay at rest (\ref{chain}) generate
electrons below this energy. LSND has seen 40 electron events
compared to a background of $21.9 \pm 2.1$. This corresponds to
a $\nu_\mu\to\nu_e$ transition probability of 
$(2.6 \pm 1.0 \pm 0.5) \times 10^{-3}$ \cite{LSND98},
consistent with the one
determined by muon decay at rest.
The favoured region
in the $\sin^22\vartheta$--$\Delta{m}^2$
plane obtained by the LSND collaboration
is shown in Fig.~\ref{lsnd-dif} as the area between the two solid lines
\cite{LSND98}.
The regions in Fig.~\ref{lsnd-dif} enclosed by dashed lines
are the LSND favoured regions obtained with the 1996
muon decay at rest analysis \cite{LSND96,LSND96-PRC54}.
One can see that the two results are perfectly compatible.

Recently, the LSND collaboration has reported \cite{now98}
new evidence in favour of $\nu_\mu\to\nu_e$
oscillations using the $\nu_\mu$ beam produced by $\pi^+$ decay in flight
and detecting the $\nu_e$'s through the observation of the reaction
\begin{equation}
\nu_e + {}^{12}\mathrm{C} \to {}^{12}\mathrm{N}_{\mathrm{g.s.}} + e^-
\,,
\label{Ngs}
\end{equation}
where $^{12}\mathrm{N}_{\mathrm{g.s.}}$ is the ground state of
$^{12}\mathrm{N}$,
which decays with a lifetime of 15 ms into
$ {}^{12}\mathrm{C} + e^+ + \nu_e $.
A temporal ($\lesssim15\,\mathrm{ms}$)
and
spatial ($\lesssim30\,\mathrm{cm}$)
correlation between the observations of the $e^-$ and $e^+$
allows an almost background-free measurement.
The LSND collaboration reported \cite{now98}
the observation of 5 events,
with an expected background of 0.5 events.
These numbers are in agreement with the results
of the other two analyses of the LSND data
summarized above.

The region of neutrino mixing parameters indicated by the results
of the LSND experiments is presently explored
also by the KARMEN experiment
\cite{KARMEN90,KARMEN94,KARMEN98-Moriond,KARMEN98-nu98,
KARMEN98-nu98-proc,KARMEN98-now98,KARMEN-www}.
The KARMEN experiment is located at
the ISIS spallation neutron facility of the Rutherford Laboratories (UK).
At ISIS a pulsed proton beam is used:
two 100 ns wide proton pulses separated by 330 ns
are produced every 20 ms.
The proton pulses are directed on a target where they produce
positive pions
(the negative pions are absorbed in the source before they decay)
which decay at rest according to the chain (\ref{chain})
producing an equal number of
$\nu_\mu$, $\nu_e$ and $\bar\nu_\mu$.
The KARMEN Collaboration searches for
$\bar\nu_e$
%produced by $\bar\nu_\mu\to\bar\nu_e$ oscillations
at a distance of about 18 m. %17.75 m.
The time structure of the neutrino beam
is important for the identification of the neutrino induced reactions
in the KARMEN detector
and for the effective suppression of the cosmic ray background.
The KARMEN experiment started in 1990 and ran until 1995 as KARMEN~1
\cite{KARMEN94}.
In 1996 the experiment was upgraded to KARMEN~2,
eliminating the main cosmic ray induced background component
in the search for
$\bar\nu_\mu\to\bar\nu_e$
oscillations
\cite{KARMEN98-Moriond}.
So far, the KARMEN~2 experiment
measured no events
\cite{KARMEN98-nu98,KARMEN98-nu98-proc,KARMEN98-now98,KARMEN-www},
also no background events, of which $ 2.88 \pm 0.13 $ events were expected.
The resulting exclusion curve obtained with a Bayesian analysis
\cite{KARMEN98-nu98}
is shown in Fig.~\ref{lsnd-dar}.
One can see that the null result of the KARMEN~2 experiment
excludes part of LSND-allowed region in favour of $\bar\nu_\mu\to\bar\nu_e$
oscillations, but at present
the sensitivity of the KARMEN experiment is not enough to
check all the LSND-allowed region.\footnote{The KARMEN~2 result
has also been analysed using the Unified Approach \cite{Feldman-Cousins98},
which is a frequentist method with proper coverage
(see Ref.~\cite{Cousins95}).
The resulting exclusion curve seems to forbid almost all of the 
LSND-allowed region \cite{KARMEN98-nu98-proc,KARMEN-www},
but this result depends on the non-observation of the expected 
background events.
The analysis of the KARMEN~2 null result with an alternative
frequentist method with proper coverage \cite{Giunti98-poisson}
leads to an exclusion curve compatible
with the Bayesian exclusion curve shown in Fig.~\ref{lsnd-dar}
\cite{Giunti98-poisson,Giunti-now98-PWG}.}

The KARMEN experiment is expected to reach a sensitivity 
that will allow to cover the LSND-allowed region in one or two years
\cite{KARMEN98-Moriond}.
If the KARMEN experiment will find an oscillation signal
compatible with the one observed in the LSND experiment,
the LSND evidence in favour of 
$\stackrel{\scriptscriptstyle (-)}{\nu}_{\hskip-3pt \mu}
\to
\stackrel{\scriptscriptstyle (-)}{\nu}_{\hskip-3pt e}$
oscillations will be confirmed.
On the other hand,
if the KARMEN experiment will not find any neutrino oscillation signal,
then the KARMEN exclusion curve will lie on the left of the
LSND allowed region in the $\sin^22\vartheta$--$\Delta{m}^2$ plane, 
but not far away from it \cite{KARMEN98-Moriond}.
Hence,
it is important to test the LSND result with
a new experiment with much higher sensitivity.
Four such experiments have been proposed and are under study:
BooNE \cite{BooNE} at Fermilab,
I-216 \cite{I-216} at CERN,
ORLaND \cite{ORLaND} at Oak Ridge
and
NESS at the European Spallation Source \cite{NESS}.
Among these proposals BooNE is approved and will start in 2001.

We would like to finish this section by emphasizing the importance
of the results obtained in the LSND experiment for
neutrino physics.
As we will discuss in the next section,
the LSND evidence in favour of SBL 
$\stackrel{\scriptscriptstyle (-)}{\nu}_{\hskip-3pt \mu}
\to
\stackrel{\scriptscriptstyle (-)}{\nu}_{\hskip-3pt e}$
oscillations
taken together with the indications in favour of
solar and atmospheric neutrino oscillations
require the existence of at least one sterile neutrino,
which represents a manifestation of new physics beyond the Standard Model.
Hence it is important that the
LSND indication is checked as soon as possible by another experiment
with high sensitivity. 

\section{Analysis of neutrino oscillation data}
\label{Analysis of neutrino oscillation data}
\setcounter{equation}{0}
\setcounter{figures}{0}
\setcounter{tables}{0}

As we have discussed in the previous section,
at present there are three
indications in favour of non-zero neutrino masses and mixing.
The analysis of the experimental data
(based on the simplest assumption of two neutrino mixing)
indicate that there must be at least three
\emph{different} $\Delta{m}^2$,
each one relevant for the oscillations of neutrinos in one type of
experiments:
$ \Delta{m}^2 \sim 10^{-5} \, \mathrm{eV}^2 $
(MSW)
or
$ \Delta{m}^2 \sim 10^{-10} \, \mathrm{eV}^2 $
(Vac. Osc.)
is relevant for oscillations of solar neutrinos,
$ \Delta{m}^2 \sim 10^{-3} \, \mathrm{eV}^2 $
is relevant for oscillations of atmospheric neutrinos,
and
$ \Delta{m}^2 \sim 1 \, \mathrm{eV}^2 $
is relevant
for SBL neutrino oscillations according to the LSND experiment.
Three different 
$\Delta m^2$ require the mixing of at least four massive
neutrinos.
Here we consider two possible scenarios:

\begin{enumerate}

\item
Only the indications in favour of neutrino mixing obtained in
atmospheric and solar neutrino experiments are confirmed by future
experiments.
In this case the mixing of
three massive neutrinos can accommodate all data.

\item
All the existing indications in favour of neutrino mixing,
including that obtained in the LSND experiment,
are confirmed by future experiments.
In this case,
the minimal mixing scheme that allows to accommodate all data
is one with mixing of four neutrinos.\footnote{From time to time
special schemes
with three neutrinos have been proposed
in order to accommodate all neutrino oscillation data
\cite{Cardall-Fuller96,Kang-Kim-Ko96,Acker-Pakvasa97,Teshima98,
Thun-McKee98,Barenboim98,Conforto98}
However,
none of these schemes can fit all data in a satisfactory way
\cite{FLMS97-PRD56,Fogli-now98-PWG}.
Schemes with more than four neutrinos have been considered in the papers in
Refs.~\cite{GKL92-PRD46,more-than-four}
(see the review Ref.~\cite{Geiser-now98-PWG}).}.

\end{enumerate}

\subsection{Mixing of three massive neutrinos}
\label{Mixing of three massive neutrinos}

Let us consider the mixing of three massive neutrinos,
\textit{i.e.},
$n=3$
in Eqs.(\ref{state}) and (\ref{anti-state}),
\begin{equation}
|\nu_\alpha\rangle
=
\sum_{k=1}^{3} U_{{\alpha}k}^* \, |\nu_k\rangle
\,,
\qquad
|\bar\nu_\alpha\rangle
=
\sum_{k=1}^{3} U_{{\alpha}k} \, |\bar\nu_k\rangle
\qquad
(\alpha=e,\mu,\tau)
\,,
\label{3mix}
\end{equation}
and assume the neutrino mass hierarchy\footnote{Actually,
the phenomenology of neutrino oscillations discussed in this section
depends only on the hierarchy
$ \Delta{m}^2_{21} \ll \Delta{m}^2_{32} $
of mass-squared differences.
Hence,
the discussion of neutrino oscillations presented in this section
applies, for example, also to the case of three almost degenerate
neutrinos with masses in the eV range
that give a contribution to the dark matter in the universe.
Furthermore,
with a cyclic permutation of the indices 1, 2, 3,
the discussion of neutrino oscillations presented in this section
can be applied to all cases in which
$ \Delta{m}^2_{32} \ll \Delta{m}^2_{21} $,
as, for example,
the case of an inverted mass hierarchy
$ m_1 \ll m_2 \simeq m_3 $.}
\begin{equation}
m_1 \ll m_2 \ll m_3
\label{hierarchy}
\end{equation}
that is indicated by the see-saw mechanism
(see Section~\ref{The see-saw mechanism})
and by the models beyond the Standard Model
(see Section~\ref{Effective operators}).
In this case,
the only available possibility for the explanation 
of the solar and atmospheric neutrino 
anomalies through neutrino oscillations is that
$\Delta{m}^2_{21} \equiv m_2^2 - m_1^2$ is relevant for the transitions of
solar neutrinos and $\Delta{m}^2_{31} \equiv m_3^2 -  m_1^2$ is
relevant for the transitions of atmospheric neutrinos:
\begin{equationarrayzero}
&&
\Delta{m}^2_{21}
\sim
10^{-5} \, \mathrm{eV}^2
\
\mbox{(MSW)}
\quad
\mbox{or}
\quad
\Delta{m}^2_{21}
\sim
10^{-10} \, \mathrm{eV}^2
\
\mbox{(Vac. Osc.)}
\,,
\label{dm2sun}
\\
&&
\Delta{m}^2_{31}
\sim
10^{-3} \, \mathrm{eV}^2
\,.
\label{dm2atm}
\end{equationarrayzero}%

Let us consider the oscillations of atmospheric and LBL neutrinos
and assume that
\begin{equation}
\frac{ \Delta{m}^2_{21} \, L }{ E } \ll 1
\,.
\label{a01}
\end{equation}
The distance $L$ travelled by atmospheric neutrinos
ranges from $\sim 10 - 20 \, \mathrm{km}$ for downward-going neutrinos
to $\sim 1.3 \times 10^4 \, \mathrm{km}$ for upward-going neutrinos
and the energy $E$ relevant for the oscillations of
atmospheric neutrino experiments
ranges from $\sim 100 \, \mathrm{MeV}$
to $\sim 100 \, \mathrm{GeV}$.
Therefore,
if the solar neutrino problem
is due to MSW transitions,
the condition (\ref{a01})
is not satisfied by atmospheric neutrinos with energies below
$ \sim 1 \, \mathrm{GeV} $.
In this case,
the following discussion applies only to high-energy
atmospheric neutrinos.
On the other hand,
the inequality (\ref{a01}) is valid in any case for
reactor LBL experiments with a source -- detector distance
$ L \lesssim 1 \, \mathrm{km} $
and
for accelerator LBL experiments
with a source -- detector distance
$ L \lesssim 10^3 \, \mathrm{km} $
and neutrino energy
$ E \gtrsim 1 \, \mathrm{GeV} $.

If the inequality (\ref{a01}) is satisfied,
from Eq.(\ref{prob2}), we have
for the  probability
of $\nu_\alpha\to\nu_\beta$ transitions 
\begin{equation}
P_{\nu_\alpha\to\nu_\beta}^{(\mathrm{atm,LBL})}
=
\left|
\delta_{\alpha\beta}
+
U_{{\alpha}3}^* \, U_{{\beta}3}
\left[ \exp\left( - i \frac{ \Delta{m}^2_{31} L }{ 2 E } \right) - 1 \right]
\right|^2
\,.
\label{a02}
\end{equation}
Therefore,
a mass hierarchy of three neutrinos implies that
the transition
probabilities in atmospheric and LBL experiments
depend only on the largest mass squared difference $\Delta{m}^2_{31}$
and on the
elements $U_{{\alpha}3}$ that connect flavour neutrinos with
the heaviest neutrino
$\nu_3$.
From Eq.(\ref{a02}) we obtain
\cite{DeRujula80,Barger-Whisnant88,BFP92,BGK96-atm,GKM98-atm}
\begin{equationarrayzero}
&&
P_{\nu_\alpha\to\nu_\beta}^{(\mathrm{atm,LBL})}
=
\frac{1}{2}
\,
A_{\alpha;\beta}
\left( 1 - \cos \frac{ \Delta{m}^2_{31} \, L }{ 2 \, E } \right)
\qquad
(\alpha\neq\beta)
\,,
\label{trans3}
\\
&&
P_{\nu_\alpha\to\nu_\alpha}^{(\mathrm{atm,LBL})}
=
1
-
\frac{1}{2}
\,
B_{\alpha;\alpha}
\left( 1 - \cos \frac{ \Delta{m}^2_{31} \, L }{ 2 \, E } \right)
\,,
\label{surv3}
\end{equationarrayzero}%
with the oscillation amplitudes
\begin{equationarrayzero}
&&
A_{\alpha;\beta}
=
4 \, |U_{{\alpha}3}|^2 \, |U_{{\beta}3}|^2
\,,
\label{a3}
\\
&&
B_{\alpha;\alpha}
=
4 \, |U_{{\alpha}3}|^2 \left( 1 - |U_{{\alpha}3}|^2 \right)
\,.
\label{b3}
\end{equationarrayzero}%
The expressions (\ref{trans3}) and (\ref{surv3}) 
have the same dependence on the
quantity
$ \Delta{m}^2_{31} L / 2 E $
as the standard expressions
(\ref{trans2}) and (\ref{surv2}) 
that describe neutrino oscillations in the case of two flavours,
with $\Delta m^2$ replaced by $\Delta m^2_{31}$.
This is due to the fact that only one mass-squared difference is
relevant for the oscillations of atmospheric and LBL neutrinos in the case of
the mass hierarchy (\ref{hierarchy}).
The expressions (\ref{a3}) and (\ref{b3}) describe, however,
all possible oscillations between three flavour neutrinos:
$\nu_\mu\leftrightarrows\nu_e$,
$\nu_\mu\leftrightarrows\nu_\tau$
and
$\nu_e\leftrightarrows\nu_\tau$. 

The unitarity of the mixing matrix implies that
$ \sum_{\alpha} |U_{{\alpha}3}|^2 = 1 $
and
LBL and atmospheric neutrino oscillations in the scheme under consideration are
described by only three parameters:
$\Delta{m}^2_{31}$,
$|U_{e3}|^2$
and
$|U_{\mu3}|^2$.

Let us also stress that the transition probabilities (\ref{trans3})
are independent on the phases of the mixing matrix $U$.
Thus, in the case of a hierarchy of neutrino masses
the relation (\ref{pCP}) is automatically satisfied and effects of CP
violation in the lepton sector cannot be revealed 
in neutrino oscillation experiments.

As it is seen from Eq.(\ref{surv3}),
the $\bar\nu_e$ survival probability 
in reactor LBL experiments
depends on the parameters
$\Delta{m}^2_{31}$ and $|U_{e3}|^2$.
This fact allows to obtain information on the value of $|U_{e3}|^2$
in LBL reactor experiments
and, in particular,
from the results of the first of them,
CHOOZ \cite{CHOOZ98},
in which no indications in favour of neutrino oscillations were found.
The exclusion curve in the plane of the two-neutrino mixing parameters 
$\sin^22\vartheta$ and $\Delta{m}^2$ that
was obtained in the CHOOZ experiment
is shown in Fig.~\ref{chooz}.

Comparing Eqs.(\ref{surv2}) and (\ref{surv3}),
one can see that
in the three-neutrino scheme under consideration in this section
\begin{equation}
\sin^22\vartheta
=
B_{e;e}
=
4 \, |U_{e3}|^2 \left( 1 - |U_{e3}|^2 \right)
\,.
\label{bee}
\end{equation}
Therefore,
we can obtain information on the parameter $|U_{e3}|^2$
from the exclusion plot obtained in the CHOOZ experiment.
From Eq.(\ref{bee}) we have
\begin{equation}
|U_{e3}|^2
=
\frac{1}{2}
\left( 1 \pm \sqrt{ 1 - B_{e;e} } \right)
\,,
\label{ue3}
\end{equation}
and the exclusion curves shown in Fig.~\ref{chooz}
imply the constraint
\begin{equation}
B_{e;e} \leq B_{e;e}^0
\,,
\label{bee0}
\end{equation}
with a value of $B_{e;e}^0$ that depends on $\Delta{m}^2_{31}$.
From Eqs.(\ref{ue3}) and (\ref{bee0})
it follows that $|U_{e3}|^2$ must satisfy one of the two inequalities
\begin{equation}
|U_{e3}|^2 \leq a_e^0
\qquad \mbox{or} \qquad
|U_{e3}|^2 \geq 1- a_e^0
\,,
\label{ue3b1}
\end{equation}
with
\begin{equation}
a_e^0
=
\frac{1}{2}
\left( 1 - \sqrt{ 1 - B_{e;e}^0 } \right)
\,.
\label{ae0}
\end{equation}
From the CHOOZ exclusion curve in Fig.~\ref{chooz} (90\% CL) we find 
$B_{e;e} \leq 0.18$ for $\Delta m^2_{31} \gtrsim 2 \times 10^{-3}$
eV$^2$ and therefore
\begin{equation}
|U_{e3}|^2 \leq 5 \times 10^{-2}
\qquad \mbox{or} \qquad
|U_{e3}|^2 \geq 0.95
\,.
\label{ue3b2}
\end{equation}

\subsubsection{Solar neutrinos}
\label{Three: Solar neutrinos}

Large values of $|U_{e3}|^2$ as those implied by the second inequality
in Eq.(\ref{ue3b2}) are excluded by the data of solar
neutrino experiments.
In order to prove this statement,
let us derive the expression
for the averaged survival probability of solar
electron neutrinos in the case of vacuum oscillations.
From Eq.(\ref{prob2}) with $n=3$,
taking into account that for solar neutrinos
$ \Delta{m}^2_{31} L / E \gg 1 $,
the survival probability averaged over the unobservable
fast oscillations due to $\Delta{m}^2_{31}$
is given by
\begin{equation}
P_{\nu_e\to\nu_e}^{(\mathrm{sun})}
=
\left|
|U_{e1}|^2
+
|U_{e2}|^2 \, \exp\left( - i \, \frac{ \Delta{m}^2_{21} L }{ 2 E } \right)
\right|^2
+
|U_{e3}|^4
\,.
\label{sun31}
\end{equation}
This expression can be rewritten in the form
\begin{equation}
P_{\nu_e\to\nu_e}^{(\mathrm{sun})}
=
\left( 1 - |U_{e3}|^2 \right)^2
P_{\nu_e\to\nu_e}^{(1,2)}
+
|U_{e3}|^4
\,,
\label{sun32}
\end{equation}
where
$P_{\nu_e\to\nu_e}^{(1,2)}$
is the two-generation survival probability
\begin{equation}
P_{\nu_e\to\nu_e}^{(1,2)}
=
1
-
\frac{1}{2}
\,
\sin^22\vartheta_{\mathrm{sun}}
\left( 1 - \cos \frac{ \Delta{m}^2_{\mathrm{sun}} L }{ 2 E } \right)
\label{}
\end{equation}
(see Eq.(\ref{surv2}))
with
$ \Delta{m}^2_{\mathrm{sun}} = \Delta{m}^2_{21} $
and the mixing angle $\vartheta_{\mathrm{sun}}$ determined by
\begin{equation}
\cos^2\vartheta_{\mathrm{sun}}
=
\frac{ |U_{e1}|^2 }{ 1 - |U_{e3}|^2 }
\,,
\qquad
\sin^2\vartheta_{\mathrm{sun}}
=
\frac{ |U_{e2}|^2 }{ 1 - |U_{e3}|^2 }
\,.
\label{sun33}
\end{equation}
The
relation (\ref{sun32}) is also valid if matter effects are taken into
account (see Eq.(\ref{aaa}) and Ref.~\cite{Shi-Schramm92}).
In this case, $P_{\nu_e\to\nu_e}^{(1,2)}$ is the two-neutrino
survival probability in matter given in Eq.(\ref{surviv}),
with $\Delta{m}^2$ replaced by $\Delta{m}^2_{21}$
and $\vartheta$ by $\vartheta_{\mathrm{sun}}$.  

Considering now the constraints (\ref{ue3b2}) on $|U_{e3}|^2$,
if $|U_{e3}|^2$ is large ($ |U_{e3}|^2 \geq 0.95 $),
from Eq.(\ref{sun32})
one can see that
$ P_{\nu_e\to\nu_e}^{(\mathrm{sun})} \geq 0.90 $.
Such a large value of $P_{\nu_e\to\nu_e}$
is incompatible with the deficit of solar 
$\nu_e$'s observed in all solar neutrino experiments.
Therefore,
from the results of the CHOOZ experiment we reach the conclusion that
the element $|U_{e3}|^2$ is small:
\begin{equation}
|U_{e3}|^2
\leq
5 \times 10^{-2}
\quad
\mbox{for} \quad \Delta{m}^2_{31} \gtrsim 2 \times 10^{-3} \, \mathrm{eV}^2
\,.
\label{ue3small}
\end{equation}
This limit implies that
\begin{equation}
P_{\nu_e\to\nu_e}^{(\mathrm{sun})}
\simeq
P_{\nu_e\to\nu_e}^{(1,2)}
\qquad \mbox{and} \qquad
\sin^22\vartheta_{\mathrm{sun}}
\simeq
4 \, |U_{e2}|^2 \left( 1 - |U_{e2}|^2 \right)
\,.
\label{sun34}
\end{equation}
Hence,
the oscillations of solar neutrinos are practically independent
of the value of $|U_{e3}|^2$
and are decoupled from the oscillations of atmospheric
and LBL neutrinos,
which depend only on the elements $U_{{\alpha}3}$ of the mixing matrix
(see Eqs.(\ref{a3}) and (\ref{b3})).
This means that
the two-generation analyses of the solar neutrino data
are appropriate also
in the three-neutrino scheme
and they give information on the values of
$ \Delta{m}^2_{21} = \Delta{m}^2_{\mathrm{sun}} $,
$ |U_{e1}| \simeq \cos\vartheta_{\mathrm{sun}} $
and
$ |U_{e2}| \simeq \sin\vartheta_{\mathrm{sun}} $
\cite{BG98-dec}.

According to a recent analysis of the solar neutrino data
\cite{FLM98-sun-AP},
which include preliminary data from Super-Kamiokande
\cite{SK-sun-97-Inoue,SK-sun-97-Svoboda},
the ranges of the mixing parameters allowed at 90\% CL
for the small and large mixing angle MSW solutions
and for the vacuum oscillation solution are,
respectively,
\begin{eqnarray}
\null \hspace{-1cm} \null
4 \times 10^{-6} \, \mathrm{eV}^2
\lesssim
\Delta{m}^2_{\mathrm{sun}}
\lesssim
1.2 \times 10^{-5} \, \mathrm{eV}^2
\,,
\null & \null & \null
3 \times 10^{-3}
\lesssim
\sin^22\vartheta_{\mathrm{sun}}
\lesssim
1.1 \times 10^{-2}
\,,
\label{23}
\\
\null \hspace{-1cm} \null
8 \times 10^{-6} \, \mathrm{eV}^2
\lesssim
\Delta{m}^2_{\mathrm{sun}}
\lesssim
3.0 \times 10^{-5} \, \mathrm{eV}^2
\,,
\null & \null & \null
0.42
\lesssim
\sin^22\vartheta_{\mathrm{sun}}
\lesssim
0.74
\,,
\label{24}
\\
\null \hspace{-1cm} \null
6 \times 10^{-11} \, \mathrm{eV}^2
\lesssim
\Delta{m}^2_{\mathrm{sun}}
\lesssim
1.1 \times 10^{-10} \, \mathrm{eV}^2
\,,
\null & \null & \null
0.70
\lesssim
\sin^22\vartheta_{\mathrm{sun}}
\leq
1
\,.
\label{25}
\end{eqnarray}
Therefore,
taking into account that $|U_{e3}|^2$ is small
and assuming that
$|U_{e2}|\leq|U_{e1}|$
(this choice is necessary only for the MSW solutions in order to fulfill
the resonance condition (\ref{rescond})),
we have
\begin{eqnarray}
|U_{e1}|
\simeq
1
\,,
\quad
|U_{e2}|
\simeq
0.03 - 0.05
\null & \null \quad \null & \null
\mbox{(small mixing MSW)}
\,,
\label{31}
\\
|U_{e1}|
\simeq
0.87 - 0.94
\,,
\quad
|U_{e2}|
\simeq
0.35 - 0.49
\null & \null \quad \null & \null
\mbox{(large mixing MSW)}
\,,
\label{32}
\\
|U_{e1}|
\simeq
0.71 - 0.88
\,,
\quad
|U_{e2}|
\simeq
0.48 - 0.71
\null & \null \quad \null & \null
\mbox{(vacuum oscillations)}
\,.
\label{33}
\end{eqnarray}
Notice that these ranges are statistically rather stable.
For example,
the range of
$\sin^22\vartheta_{\mathrm{sun}}$
allowed at 99\% CL
in the case of the large mixing angle MSW solution is
$
0.36
\lesssim
\sin^22\vartheta_{\mathrm{sun}}
\lesssim
0.85
$
\cite{FLM98-sun-AP},
which implies
$
|U_{e1}|
\simeq
0.83 - 0.95
$,
$
|U_{e2}|
\simeq
0.32 - 0.55
$
(compare with Eq.(\ref{32})).

\subsubsection{Atmospheric neutrinos}
\label{Three: Atmospheric neutrinos}

The atmospheric neutrino data of the Super-Kamiokande experiment
give an additional indication that $|U_{e3}|^2$ is small.
Indeed, the up-down asymmetry of electron
events $A_e$ that could be
generated by $\nu_\mu\to\nu_e$
or
$\nu_e\to\nu_\beta$
(with $\beta{\neq}e$)
transitions is compatible with zero (see Section 5.1.3) \cite{SK-atm-98}:
\begin{equation}
A_e
=
- 0.036 \pm 0.067 \pm 0.02
\,.
\label{AeSK}
\end{equation}
In the three-neutrino scheme under consideration,
the Super-Kamiokande atmospheric neutrino data can be explained by
$\nu_\mu\to\nu_\tau$
oscillations \cite{SK-atm-98,Gonzalez-Garcia98a,FLMS98}.
The analysis of the data under the assumption of two-generation mixing
yields the following constraints:
\begin{equation}
A_{\mu;\tau} \equiv \sin^22\vartheta_{\mathrm{atm}} \gtrsim 0.82
\,,
\qquad
5 \times 10^{-4} \, \mathrm{eV}^2
\lesssim
\Delta{m}^2
\lesssim
6 \times 10^{-3} \, \mathrm{eV}^2
\,,
\label{atm2SK}
\end{equation}
from which it follows with Eqs.(\ref{trans3}) and (\ref{a3})
that $|U_{\mu 3}|^2$ and
$|U_{\tau 3}|^2$ must be close to 0.5 and therefore $|U_{e 3}|^2$ small.

Coming back to Eq.(\ref{ue3small}) which implies that
$ |U_{e3}| \leq 0.22 $ we see that
the CHOOZ upper bound for $|U_{e3}|$ is not very small.
It is suggestive, however, to assume that
\begin{equation}
|U_{e3}| \ll 1
\,,
\label{ue3ll1}
\end{equation}
\textit{i.e.} that the element $|U_{e3}|$ is so small 
that it is unimportant for neutrino oscillations
\cite{Vissani97,BG98-dec}.
Such an assumption has been recently adopted by several authors
\cite{bi-maximal}.

Let us infer some consequences of the assumption (\ref{ue3ll1}).
First of all,
it is easy to see that if the inequality (\ref{ue3ll1})
is satisfied,
matter effects are not essential for
$\nu_\mu\to\nu_\tau$
transitions of atmospheric and LBL neutrinos.
Indeed,
the evolution equation for neutrinos in matter in the flavour
representation is (see Section 4)
\begin{equation}
i
\,
\frac{ \mathrm{d} }{ \mathrm{d}x }
\,
\psi
=
\frac{ 1 }{ 2 \, E }
\left(
U
\,
\Delta {M}^2
\,
U^{\dagger}
+
\hat{A}
\right)
\psi
\,,
\label{13}
\end{equation}
with
\begin{equation}
\psi
\equiv
\left(
\begin{array}{l}
a_{e}
\\
a_{\mu}
\\
a_{\tau}
\end{array}
\right)
\,,
\quad
\Delta M^2
\equiv
\mathrm{diag}( 0 , \Delta{m}^2_{21} , \Delta{m}^2_{31} )
\,,
\quad
\hat{A}
\equiv
\mathrm{diag}( A_{CC} , 0 , 0 )
\,,
\label{14}
\end{equation}
and
\begin{equation}
A_{CC} \equiv 2 E V_{CC}
\,,
\qquad
V_{CC} = \sqrt{2} G_{F} N_{e}
\,.
\label{141}
\end{equation}
Remember that for anti-neutrinos the matter potential term
$ A_{CC} $
must be replaced by
$ \bar{A}_{CC} = - A_{CC} $.

The inequality (\ref{a01})
implies that the phase generated by
$\Delta{m}^2_{21}$
can be neglected for atmospheric neutrinos
and $\Delta{M}^2$
can be approximated by
\begin{equation}
\Delta{M}^2
\simeq
\mathrm{diag}(0,0,\Delta{m}^2_{31})
\,.
\label{16}
\end{equation}
In this case
(taking into account that
the phases of the matrix elements $U_{\alpha3}$
can be included in the charged lepton fields)
we have
\begin{equation}
(
U
\,
\Delta{M}^2
\,
U^{\dagger}
)_{\alpha\beta}
\simeq
\Delta{m}^2_{31}
\,
|U_{\alpha3}|
\,
|U_{\beta3}|
\,.
\label{17}
\end{equation}
If the condition (\ref{ue3ll1}) is satisfied,
the elements of the matrix
$ U \, \Delta{M}^2 \, U^{\dagger} $
with indices  $\alpha=e$ or/and  $\beta=e$
can be neglected.
This means that the equation for
the wave function $a_e$ is decoupled from those of $a_\mu$, $a_\tau$:
\begin{equationarrayzero}
&&
i \,
\frac{ \mathrm{d}a_e }{ \mathrm{d}x }
=
\frac{ A_{CC} }{ 2 E }
\,
a_e
\,,
\label{171}
\\
&&
i \,
\frac{ \mathrm{d}a_\alpha }{\mathrm{d}x}
=
\frac{ \Delta{m}^2_{31} }{ 2 E }
\,
|U_{\alpha3}|
\,
\sum_{\beta=\mu,\tau}
|U_{\beta3}|
\,
a_\beta
\qquad
(\alpha=\mu,\tau)
\,.
\label{172}
\end{equationarrayzero}%
These equations imply that
$ P_{\nu_e\to\nu_e}^{(\mathrm{atm,LBL})} = 1 $
and
$ P_{\nu_\mu\to\nu_\tau}^{(\mathrm{atm,LBL})} $
is given by the two-generation expression for neutrino oscillations in vacuum,
\begin{equation}
P_{\nu_\mu\to\nu_\tau}^{(\mathrm{atm,LBL})}
=
\frac{1}{2}
\,
\sin^22\vartheta_{\mathrm{atm}}
\left( 1 - \cos \frac{ \Delta{m}^2_{\mathrm{atm}} L }{ 2 E } \right)
\label{173}
\end{equation}
(see Eq.(\ref{trans2})),
with $\Delta{m}^2_{\mathrm{atm}} = \Delta{m}^2_{31}$
and
the mixing angle $\vartheta_{\mathrm{atm}}$ determined by
\begin{equation}
\sin^2\vartheta_{\mathrm{atm}} = |U_{\mu3}|^2
\,,
\qquad
\cos^2\vartheta_{\mathrm{atm}} = |U_{\tau3}|^2
\,.
\label{174}
\end{equation}
Therefore,
if the condition (\ref{ue3ll1}) is satisfied,
there are no $\nu_\mu\leftrightarrows\nu_e$ oscillations
in atmospheric and LBL neutrino experiments
and matter effects are not important for
$\nu_\mu\to\nu_\tau$
transitions.
This means that
the two-generation analyses of
the atmospheric neutrino data in terms of $\nu_\mu\to\nu_\tau$
are appropriate
in the three-neutrino scheme under consideration and
yield information on the values of the parameters
$ \Delta{m}^2_{31} = \Delta{m}^2_{\mathrm{atm}} $
and
$ |U_{\mu3}| = \sin\vartheta_{\mathrm{atm}} $.

According to a recent analysis \cite{Gonzalez-Garcia98a}
of the atmospheric neutrino data,
the ranges of
$\Delta{m}^2_{\mathrm{atm}}$
and
$\sin^22\vartheta_{\mathrm{atm}}$
for $\nu_\mu\to\nu_\tau$ oscillations
allowed at 90\% CL
by the Super-Kamiokande multi-GeV data
and by the data of atmospheric neutrino oscillation experiments 
are, respectively,
\begin{eqnarray}
4 \times 10^{-4} \, \mathrm{eV}^2
\lesssim
\Delta{m}^2_{\mathrm{atm}}
\lesssim
8 \times 10^{-3} \, \mathrm{eV}^2
\,,
\null & \null \quad \null & \null
0.72
\lesssim
\sin^22\vartheta_{\mathrm{atm}}
\leq
1
\,,
\label{61}
\\
4 \times 10^{-4} \, \mathrm{eV}^2
\lesssim
\Delta{m}^2_{\mathrm{atm}}
\lesssim
6 \times 10^{-3} \, \mathrm{eV}^2
\,,
\null & \null \quad \null & \null
0.76
\lesssim
\sin^22\vartheta_{\mathrm{atm}}
\leq
1
\,.
\label{62}
\end{eqnarray}
Taking into account the comments after Eq.(\ref{a01}),
the ranges (\ref{62}) of the two-generation mixing parameters
can be used in order to constrain the values of
$|U_{\mu3}|$ and $|U_{\tau3}|$
only in the case of the vacuum oscillation solution
of the solar neutrino problem.
If the solar neutrino problem is due to MSW transitions,
we must use the less restrictive ranges (\ref{61}).
Thus, assuming that
$ |U_{\mu3}| \leq |U_{\tau3}| $,
we have
\begin{eqnarray}
|U_{\mu3}|
\simeq
0.49 - 0.71
\,,
\quad
|U_{\tau3}|
\simeq
0.71 - 0.87
\null & \null \quad \null & \null
\mbox{(solar MSW)}
\,,
\label{63}
\\
|U_{\mu3}|
\simeq
0.51 - 0.71
\,,
\quad
|U_{\tau3}|
\simeq
0.71 - 0.86
\null & \null \quad \null & \null
\mbox{(solar vacuum oscillations)}
\,.
\label{64}
\end{eqnarray}
As in the case of the ranges (\ref{31})--(\ref{33}),
also the ranges (\ref{63})--(\ref{64})
are statistically rather stable.
For example,
the range of
$\sin^22\vartheta_{\mathrm{sun}}$
allowed at 99\% CL
by all atmospheric neutrino data is
$
0.66
\lesssim
\sin^22\vartheta_{\mathrm{sun}}
\leq
1
$
\cite{Gonzalez-Garcia98a},
which implies
$
|U_{\mu 3}|
\simeq
0.46 - 0.71
$,
$
|U_{\tau 3}|
\simeq
0.71 - 0.89
$
(compare with Eq.(\ref{64})).

\subsubsection{The mixing matrix in the case $|U_{e3}| \ll 1$}
\label{Bilenky: The mixing matrix}

We have seen that,
if we assume that $|U_{e3}|$ is very small,
the results of the two-generation
analyses of the data of solar and atmospheric neutrino experiments allow to
determine the moduli of the elements of the first line and of the third
column of the mixing matrix $U$.
The moduli of the other elements can be
determined from  the unitarity of $U$.
The simplest way to do it is
to start from the parameterization (\ref{par3})
of the $3\times3$ mixing matrix $U$.

A very small $|U_{e3}|$ implies that $|s_{13}|\ll1$.
Since in the parameterization (\ref{par3}) the CP-violating phase 
$\delta_{13}$ is associated with $s_{13}$,
it follows that CP violation
is negligible in the lepton sector (see also footnote \ref{jarlskog})
and we have
\begin{equation}
U
\simeq
\left(
\begin{array}{ccc}
c_{12}
&
s_{12}
&
\ll 1
\\
-
s_{12}
c_{23}
&
c_{12}
c_{23}
&
s_{23}
\\
s_{12}
s_{23}
&
-
c_{12}
s_{23}
&
c_{23}
\end{array}
\right)
\,.
\label{dec:47}
\end{equation}
Using the information on
$|s_{12}|\simeq|U_{e2}|$
and
$|s_{23}|\simeq|U_{\mu3}|$
given by the two-generation analyses of the results of solar
and atmospheric neutrino experiments, we obtain
for the moduli of the elements of the mixing matrix \cite{BG98-dec}:
\begin{eqnarray}
\mbox{Small mixing MSW:}
\null & \null \quad \null & \null
\left(
\begin{array}{ccc}
\simeq 1
&
0.03 - 0.05
&
\ll 1
\\
0.02 - 0.05
&
0.71 - 0.87
&
0.49 - 0.71
\\
0.01 - 0.04
&
0.48 - 0.71
&
0.71 - 0.87
\end{array}
\right)
\,,
\label{dec:44}
\\
\mbox{Large mixing MSW:}
\null & \null \quad \null & \null
\left(
\begin{array}{ccc}
0.87 - 0.94
&
0.35 - 0.49
&
\ll 1
\\
0.25 - 0.43
&
0.61 - 0.82
&
0.49 - 0.71
\\
0.17 - 0.35
&
0.42 - 0.66
&
0.71 - 0.87
\end{array}
\right)
\,,
\label{dec:45}
\\
\mbox{Vacuum oscillations:}
\null & \null \quad \null & \null
\left(
\begin{array}{ccc}
0.71 - 0.88
&
0.48 - 0.71
&
\ll 1
\\
0.34 - 0.61
&
0.50 - 0.76
&
0.51 - 0.71
\\
0.24 - 0.50
&
0.36 - 0.62
&
0.71 - 0.86
\end{array}
\right)
\,.
\label{dec:46}
\end{eqnarray}
In the case of the small mixing angle MSW solution
of the solar neutrino problem
$|U_{e3}|\ll1$
could be of the same order of magnitude as $|U_{e2}|$.

The best fit of Super-Kamiokande data corresponds to maximal mixing,
$\vartheta_{23} = \pi/4 $
and the vacuum oscillation solution of the solar neutrino problem
includes the maximal mixing case
$\vartheta_{12} = \pi/4 $.
The combination of these two possibilities gives the
bi-maximal mixing matrix
\begin{equation}
U
=
\left(
\begin{array}{rrr}
\frac{1}{\sqrt{2}} & \frac{1}{\sqrt{2}} & 0
\\
-\frac{1}{2} &  \frac{1}{2} & \frac{1}{\sqrt{2}}
\\
\frac{1}{2} &  -\frac{1}{2} & \frac{1}{\sqrt{2}}
\end{array}
\right)
\label{001}
\end{equation}
that has been assumed recently by several authors \cite{bi-maximal}.
Notice that bi-maximal mixing
is not compatible with the large mixing angle MSW solution
of the solar neutrino problem
\cite{Giunti98-bimax}.

It is interesting to note that,
because of the large mixing of
$\nu_\mu$ and $\nu_\tau$
with $\nu_2$,
the transitions of solar $\nu_e$'s
in
$\nu_\mu$'s and $\nu_\tau$'s
are of comparable magnitude.
However,
this phenomenon
and the values of the entries
in the
$(\nu_\mu,\nu_\tau)$--$(\nu_1,\nu_2)$
sector of the mixing matrix
cannot be checked with solar neutrino experiments
because the low-energy
$\nu_\mu$'s and $\nu_\tau$'s
coming from the sun can be detected only with neutral-current
interactions,
which are flavour-blind.
Moreover,
it will be very difficult to check
the values of
$|U_{\mu1}|$, $|U_{\mu2}|$, $|U_{\tau1}|$ and $|U_{\tau2}|$
in laboratory experiments
because of the smallness of $\Delta m^2_{21}$.

In the derivation of Eqs.(\ref{dec:44})--(\ref{dec:46})
we have assumed that
$|U_{e2}|\leq|U_{e1}|$
and
$|U_{\mu3}|\leq|U_{\tau3}|$.
The other possibilities,
$|U_{e2}|\geq|U_{e1}|$
and
$|U_{\mu3}|\geq|U_{\tau3}|$,
are equivalent, respectively,
to an exchange of the first and second columns
and
to an exchange of the second and third rows
in the matrices (\ref{dec:44})--(\ref{dec:46}).
Unfortunately,
these alternatives are hard to distinguish experimentally
because of the above mentioned difficulty
to measure directly the values of
$|U_{\mu1}|$, $|U_{\mu2}|$, $|U_{\tau1}|$ and $|U_{\tau2}|$.
Only the choice $|U_{e2}|\leq|U_{e1}|$,
which is necessary for the MSW solutions
of the solar neutrino problem,
could be confirmed by the results of the new generation
of solar neutrino experiments
(Super-Kamiokande,
SNO,
ICARUS,
Borexino,
GNO
and others
\cite{future-sun})
if they will allow to exclude the vacuum oscillation solution.

\subsubsection{Accelerator long-baseline experiments}
\label{Three: Accelerator long-baseline experiments}

Future results
from reactor long-baseline neutrino oscillation experiments
(CHOOZ \cite{CHOOZ},
Palo Verde \cite{PaloVerde},
Kam-Land \cite{Kam-Land})
could allow to improve the upper bound (\ref{ue3small})
on $|U_{e3}|^2$.
In this section we discuss how
an improvement of this upper bound
could be obtained by future
accelerator long-baseline neutrino oscillation experiments
that are sensitive to $\nu_\mu\to\nu_e$ transitions
(K2K \cite{K2K},
MINOS \cite{MINOS},
ICARUS \cite{ICARUS}
and others \cite{NOE,AQUA-RICH,OPERA,NICE};
for a review see Ref.~\cite{Zuber98}).

If matter effects are not important,
in the scheme under consideration
the parameter
$\sin^22\vartheta_{\mu{e}}$
measured in
$\nu_\mu\to\nu_e$
long-baseline experiments is given by
(see Eq.(\ref{a3}))
\begin{equation}
\sin^22\vartheta_{\mu{e}}
=
4 |U_{e3}|^2 |U_{\mu3}|^2
\,.
\label{56}
\end{equation}
If accelerator long-baseline neutrino oscillation experiments
will not observe
$\nu_\mu\to\nu_e$ transitions
and will place an upper bound
$
\sin^22\vartheta_{\mu{e}}
\leq
\sin^22\vartheta_{\mu{e}}^{\mathrm{(max)}}
$,
it will be possible to obtain the limit
\begin{equation}
|U_{e3}|^2
\leq
\frac
{ \sin^22\vartheta_{\mu{e}}^{\mathrm{(max)}} }
{ 4 |U_{\mu3}|^2_{\mathrm{(min)}} }
\,,
\label{57}
\end{equation}
where
$|U_{\mu3}|^2_{\mathrm{(min)}}$
is the minimum value of
$|U_{\mu3}|^2$
allowed by the solution of the atmospheric neutrino anomaly
and
by the observation of
$\nu_\mu\to\nu_\tau$
long-baseline transitions.
For example,
taking 
$|U_{\mu3}|^2_{\mathrm{(min)}}=0.25$
(see Eq.(\ref{64}))
we have
$
|U_{e3}|^2
\leq
\sin^22\vartheta_{\mu{e}}^{\mathrm{(max)}}
$.
If a value of
$ \sin^22\vartheta_{\mu{e}}^{\mathrm{(max)}} \simeq 10^{-3} $,
corresponding to the sensitivity of the ICARUS experiment
for one year of running \cite{ICARUS},
will be reached,
it will be possible to
put the upper bound
$ |U_{e3}| \lesssim 3 \times 10^{-2} $.

The observation of
$\nu_\mu\to\nu_\tau$
transitions in long-baseline experiments
will allow to establish a lower bound for
$|U_{\mu3}|^2$
because
the parameter
$\sin^22\vartheta_{\mu\tau}$
is given in the scheme under consideration by
(see Eq.(\ref{a3}))
\begin{equation}
\sin^22\vartheta_{\mu\tau}
=
4 |U_{\mu3}|^2 |U_{\tau3}|^2
\,.
\label{81}
\end{equation}
From the unitarity relation
$|U_{e3}|^2+|U_{\mu3}|^2+|U_{\tau3}|^2=1$
it follows that
an experimental lower bound
$\sin^22\vartheta_{\mu\tau}\geq\sin^22\vartheta_{\mu\tau}^{\mathrm{(min)}}$
allows to constrain
$|U_{\mu3}|^2$
in the range
\begin{equation}
\frac{1}{2}
\left( 1 - \sqrt{ 1 - \sin^22\vartheta_{\mu\tau}^{\mathrm{(min)}} } \right)
\leq
|U_{\mu3}|^2
\leq
\frac{1}{2}
\left( 1 + \sqrt{ 1 - \sin^22\vartheta_{\mu\tau}^{\mathrm{(min)}} } \right)
\,.
\label{82}
\end{equation}
If
$\sin^22\vartheta_{\mu\tau}^{\mathrm{(min)}}$
is found to be close to one,
as suggested by the solution
of the atmospheric neutrino problem
(see Eqs.(\ref{61}) and (\ref{62})),
the lower bound
$
|U_{\mu3}|^2_{\mathrm{(min)}}
=
\frac{1}{2}
\left( 1 - \sqrt{ 1 - \sin^22\vartheta_{\mu\tau}^{\mathrm{(min)}} } \right)
$
is close to $1/2$.

If matter effects are important,
the extraction of an upper bound for
$|U_{e3}|^2$
from the data of
$\nu_\mu\to\nu_e$
accelerator long-baseline experiments
is more complicated.
In this case the probability
of
$\nu_\mu\to\nu_e$
oscillations
is given by \cite{GKM98-atm}
\begin{equation}
P_{\nu_\mu\to\nu_e}
=
\frac{ 4 |U_{e3}|^2 |U_{\mu3}|^2 }
{ \left( 1 - \frac{ A_{CC} }{ \Delta{m}^2_{31} } \right)^2
+ 4 |U_{e3}|^2 \, \frac{ A_{CC} }{ \Delta{m}^2_{31} } }
\,
\sin^2\left(
\frac{ \Delta{m}^2_{31} L }{ 4 E }
\,
\sqrt{ \textstyle \left( 1 - \frac{ A_{CC} }{ \Delta{m}^2_{31} } \right)^2
+ 4 |U_{e3}|^2 \, \frac{ A_{CC} }{ \Delta{m}^2_{31} } }
\right)
\,,
\label{83}
\end{equation}
where $E$ is the neutrino energy and $L$ is the distance of propagation.
This probability depends on the neutrino energy
not only through the explicit $E$
in the denominator of the phase,
but also through the energy dependence of
$ A_{CC} \equiv 2 E V_{CC} $.
For long-baseline neutrino beams
propagating in the lithosphere of the earth
the charged-current effective potential 
$ V_{CC} = \sqrt{2} G_{F} N_{e} $
is practically constant:
$ N_{e} \simeq 1.5 \, N_{\mathrm{avo}} \: \mathrm{cm}^{-3} $
and
$ V_{CC} \simeq 1.1 \times 10^{-13} \, \mathrm{eV} $ (see Eq.(\ref{litho})).

If long-baseline experiments will not observe
$\nu_\mu\to\nu_e$
transitions
(or will find that they have an extremely small probability)
for neutrino energies such that
$A_{CC}\lesssim\Delta{m}^2_{31}$,
it will mean that
$|U_{e3}|^2$
is small
and a fit of the experimental data with the formula (\ref{83})
will yield a stringent upper limit for
$|U_{e3}|^2$
(taking into account
the lower limit $|U_{\mu3}|^2\geq|U_{\mu3}|^2_{\mathrm{(min)}}$
obtained from the solution of the atmospheric neutrino anomaly
and
from the observation of
$\nu_\mu\to\nu_\tau$
long-baseline transitions).
On the other hand,
the non-observation of
$\nu_\mu\to\nu_e$
transitions
for neutrino energies such that
$A_{CC}\gg\Delta{m}^2_{31}$
does not provide any information on
$|U_{e3}|^2$
because in this case the transition probability (\ref{83})
is suppressed by the matter effect.
Hence,
in order to check the hypothesis
$|U_{e3}|\ll1$,
as well as to have some possibility to observe
$\nu_\mu\to\nu_e$
transitions if this hypothesis is wrong,
it is necessary that a
substantial part of the energy spectrum of the neutrino beam
lies below
\begin{equation}
\frac{ \Delta{m}^2_{31} }{ 2 V_{CC} }
\simeq
30 \, \mathrm{GeV}
\left( \frac{ \Delta{m}^2_{31} }{ 10^{-2} \, \mathrm{eV}^2 } \right)
\,.
\label{84}
\end{equation}
This requirement will be satisfied in the
accelerator long-baseline experiments under preparation
(K2K \cite{K2K},
MINOS \cite{MINOS},
ICARUS \cite{ICARUS}
and others \cite{NOE,AQUA-RICH,OPERA,NICE};
see \cite{Zuber98})
if
$\Delta{m}^2_{31}$
is not much smaller than
$ 10^{-2} \, \mathrm{eV}^2 $.

\subsection{Mixing of four massive neutrinos}
\label{Mixing of four massive neutrinos}

Let us consider now the case of
mixing of four neutrinos \cite{four-review,BGG98-EPJ,Okada-Yasuda97},
\textit{i.e.}, $n=4$ in Eqs.(\ref{state}) and (\ref{anti-state}),
\begin{equation}
|\nu_\alpha\rangle
=
\sum_{k=1}^{4} U_{{\alpha}k}^* \, |\nu_k\rangle
\,,
\qquad
|\bar\nu_\alpha\rangle
=
\sum_{k=1}^{4} U_{{\alpha}k} \, |\bar\nu_k\rangle
\qquad
(\alpha=e,\mu,\tau,s)
\,,
\label{4mix}
\end{equation} 
which allows to accommodate all the three existing indications
in favour of neutrino mixing.
Here $|\nu_s\rangle$ is a sterile
neutrino state (see Section~\ref{Dirac-Majorana mass term}).

\begin{figure}[t!]
\begin{center}
\setlength{\unitlength}{1cm}
\begin{tabular*}{0.9\textwidth}{@{\extracolsep{\fill}}cccccc}
\begin{picture}(1.4,4) % I
\thicklines
\put(0.5,0.2){\vector(0,1){3.8}}
\put(0.4,0.2){\line(1,0){0.2}}
\put(0.8,0.15){\makebox(0,0)[l]{$m_1$}}
\put(0.4,0.4){\line(1,0){0.2}}
\put(0.8,0.45){\makebox(0,0)[l]{$m_2$}}
\put(0.4,0.8){\line(1,0){0.2}}
\put(0.8,0.8){\makebox(0,0)[l]{$m_3$}}
\put(0.4,3.5){\line(1,0){0.2}}
\put(0.8,3.5){\makebox(0,0)[l]{$m_4$}}
\end{picture}
&
\begin{picture}(1.4,4) % II
\thicklines
\put(0.5,0.2){\vector(0,1){3.8}}
\put(0.4,0.2){\line(1,0){0.2}}
\put(0.8,0.2){\makebox(0,0)[l]{$m_1$}}
\put(0.4,0.6){\line(1,0){0.2}}
\put(0.8,0.55){\makebox(0,0)[l]{$m_2$}}
\put(0.4,0.8){\line(1,0){0.2}}
\put(0.8,0.85){\makebox(0,0)[l]{$m_3$}}
\put(0.4,3.5){\line(1,0){0.2}}
\put(0.8,3.5){\makebox(0,0)[l]{$m_4$}}
\end{picture}
&
\begin{picture}(1.4,4) % III
\thicklines
\put(0.5,0.2){\vector(0,1){3.8}}
\put(0.4,0.2){\line(1,0){0.2}}
\put(0.8,0.2){\makebox(0,0)[l]{$m_1$}}
\put(0.4,2.9){\line(1,0){0.2}}
\put(0.8,2.9){\makebox(0,0)[l]{$m_2$}}
\put(0.4,3.3){\line(1,0){0.2}}
\put(0.8,3.25){\makebox(0,0)[l]{$m_3$}}
\put(0.4,3.5){\line(1,0){0.2}}
\put(0.8,3.55){\makebox(0,0)[l]{$m_4$}}
\end{picture}
&
\begin{picture}(1.4,4) % IV
\thicklines
\put(0.5,0.2){\vector(0,1){3.8}}
\put(0.4,0.2){\line(1,0){0.2}}
\put(0.8,0.2){\makebox(0,0)[l]{$m_1$}}
\put(0.4,2.9){\line(1,0){0.2}}
\put(0.8,2.85){\makebox(0,0)[l]{$m_2$}}
\put(0.4,3.1){\line(1,0){0.2}}
\put(0.8,3.15){\makebox(0,0)[l]{$m_3$}}
\put(0.4,3.5){\line(1,0){0.2}}
\put(0.8,3.5){\makebox(0,0)[l]{$m_4$}}
\end{picture}
&
\begin{picture}(1.4,4) % A
\thicklines
\put(0.5,0.2){\vector(0,1){3.8}}
\put(0.4,0.2){\line(1,0){0.2}}
\put(0.8,0.2){\makebox(0,0)[l]{$m_1$}}
\put(0.4,0.6){\line(1,0){0.2}}
\put(0.8,0.6){\makebox(0,0)[l]{$m_2$}}
\put(0.4,3.3){\line(1,0){0.2}}
\put(0.8,3.25){\makebox(0,0)[l]{$m_3$}}
\put(0.4,3.5){\line(1,0){0.2}}
\put(0.8,3.55){\makebox(0,0)[l]{$m_4$}}
\end{picture}
&
\begin{picture}(1.4,4) % B
\thicklines
\put(0.5,0.2){\vector(0,1){3.8}}
\put(0.4,0.2){\line(1,0){0.2}}
\put(0.8,0.15){\makebox(0,0)[l]{$m_1$}}
\put(0.4,0.4){\line(1,0){0.2}}
\put(0.8,0.45){\makebox(0,0)[l]{$m_2$}}
\put(0.4,3.1){\line(1,0){0.2}}
\put(0.8,3.1){\makebox(0,0)[l]{$m_3$}}
\put(0.4,3.5){\line(1,0){0.2}}
\put(0.8,3.5){\makebox(0,0)[l]{$m_4$}}
\end{picture}
\\
(I) & (II) & (III) & (IV) & (A) & (B)
\end{tabular*}
\end{center}
\refstepcounter{figures}
\label{4spectra}
\footnotesize
Figure \ref{4spectra}.
The six types of neutrino mass spectra that can accommodate 
the solar, atmospheric and LSND scales of $\Delta{m}^2$.
\end{figure}

We do not make any assumptions
about the mass spectrum of four neutrinos and will consider all possibilities.
The six possible four-neutrino mass spectra
with three different scales of neutrino mass-squared differences
(atmospheric, solar and LSND)
are shown in Fig.~\ref{4spectra}.
In all these mass spectra there are two groups
of close masses separated by the ``LSND gap'' of the order of 1 eV.
The small mass-squared
differences correspond to
$\Delta{m}^2_{\mathrm{sun}}$
(the smallest one,
$\Delta{m}^2_{21}$ in schemes I and B,
$\Delta{m}^2_{32}$ in schemes II and IV,
$\Delta{m}^2_{43}$ in schemes III and A)
and
$\Delta{m}^2_{\mathrm{atm}}$
($\Delta{m}^2_{31}$ in schemes I and II,
$\Delta{m}^2_{42}$ in schemes III and IV,
$\Delta{m}^2_{21}$ in scheme A,
$\Delta{m}^2_{43}$ in scheme B)
and the largest mass squared difference
$ \Delta{m}^2_{41} = \Delta{m}^2_{\mathrm{LSND}} $
is relevant for the oscillations observed in the LSND experiment.

Let us consider SBL oscillations in the case of neutrino mass spectra
presented in Fig.~\ref{4spectra}.
Taking into account that
\begin{equation}
\frac{ \Delta{m}^2_{\mathrm{sun}} L }{ E } \ll 1
\qquad \mbox{and} \qquad
\frac{ \Delta{m}^2_{\mathrm{atm}} L }{ E } \ll 1
\,,
\label{4n01}
\end{equation}
from the general formula (\ref{prob2})
we obtain that the probabilities of $\nu_\alpha\to\nu_\beta$ transitions
and of $\nu_\alpha$ survival
in SBL experiments are given by
\begin{equationarrayzero}
&&
P_{\nu_\alpha\to\nu_\beta}^{(\mathrm{SBL})}
=
\frac{1}{2}
\,
A_{\alpha;\beta}
\left( 1 - \cos \frac{ \Delta{m}^2_{41} \, L }{ 2 \, E } \right)
\qquad
(\alpha\neq\beta)
\,,
\label{trans4}
\\
&&
P_{\nu_\alpha\to\nu_\alpha}^{(\mathrm{SBL})}
=
1
-
\frac{1}{2}
\,
B_{\alpha;\alpha}
\left( 1 - \cos \frac{ \Delta{m}^2_{41} \, L }{ 2 \, E } \right)
\,.
\label{surv4}
\end{equationarrayzero}%
The oscillation amplitudes
$A_{\alpha;\beta}$ and $B_{\alpha;\alpha}$ depend on the elements of
the mixing matrix and on the type of neutrino mass spectrum:
\begin{equationarrayzero}
&&
A_{\alpha;\beta}
=
4 \left| \sum_k U_{{\alpha}k}^* \, U_{{\beta}k} \right|^2
\,,
\label{a4}
\\
&&
B_{\alpha;\alpha}
=
4
\left( \sum_k |U_{{\alpha}k}|^2 \right)
\left( 1 - \sum_k |U_{{\alpha}k}|^2 \right)
\,,
\label{b4}
\end{equationarrayzero}%
where the index $k$ runs over the indices
of the first or (because of the unitarity of the mixing matrix)
of the second group
of neutrino masses (see Fig.~\ref{4spectra}).
The expressions (\ref{trans4}) and (\ref{surv4})
have the same form as the corresponding two-neutrino formulas
(\ref{trans2}) and (\ref{surv2})
and are a direct
generalization of the expressions
(\ref{trans3}) and (\ref{surv3})
presented in Section~\ref{Mixing of three massive neutrinos}
for the case of three massive neutrinos and
a neutrino mass hierarchy.
From Eq.(\ref{a4}) it is obvious that the amplitude of
$\bar\nu_\alpha\to\bar\nu_\beta$
oscillations
is equal to $A_{\alpha;\beta}$
and
\begin{equation}
P_{\bar\nu_\alpha\to\bar\nu_\beta}^{(\mathrm{SBL})}
=
P_{\nu_\alpha\to\nu_\beta}^{(\mathrm{SBL})}
\,.
\label{4n02}
\end{equation}
Thus,
CP violation in the lepton sector cannot be revealed in SBL
neutrino oscillation experiments if only one mass-squared difference is
relevant for SBL oscillations.

The expression (\ref{b4})
allows to obtain information on the quantities
$ \sum_k |U_{ek}|^2 $
and
$ \sum_k |U_{{\mu}k}|^2 $
from the results of SBL
$\bar\nu_e$ and $\nu_\mu$ disappearance experiments,
whose exclusion curves imply the constraint
\begin{equation}
B_{\alpha;\alpha}
\leq
B_{\alpha;\alpha}^0
\qquad
(\alpha=e,\mu)
\,.
\label{baa0}
\end{equation}
The upper bound $B_{\alpha;\alpha}^0$
depends on the value of $\Delta{m}^2_{41}$.
Equations (\ref{surv4}) and (\ref{baa0}) imply that each
$ \sum_k |U_{{\alpha}k}|^2 $
for $\alpha=e,\mu$
must satisfy one of the two inequalities
\begin{equation}
\sum_k |U_{{\alpha}k}|^2
\leq
a_\alpha^0
\qquad \mbox{or} \qquad
\sum_k |U_{{\alpha}k}|^2
\geq
1 - a_\alpha^0
\,,
\label{const4}
\end{equation}
with
\begin{equation}
a_\alpha^0
=
\frac{1}{2}
\left( 1 - \sqrt{ 1 - B_{\alpha;\alpha}^0 } \, \right)
\,.
\label{aa0}
\end{equation}

\begin{figure}[t!]
\begin{tabular*}{\linewidth}{@{\extracolsep{\fill}}cc}
\begin{minipage}{0.47\linewidth}
\begin{center}
\mbox{\epsfig{file=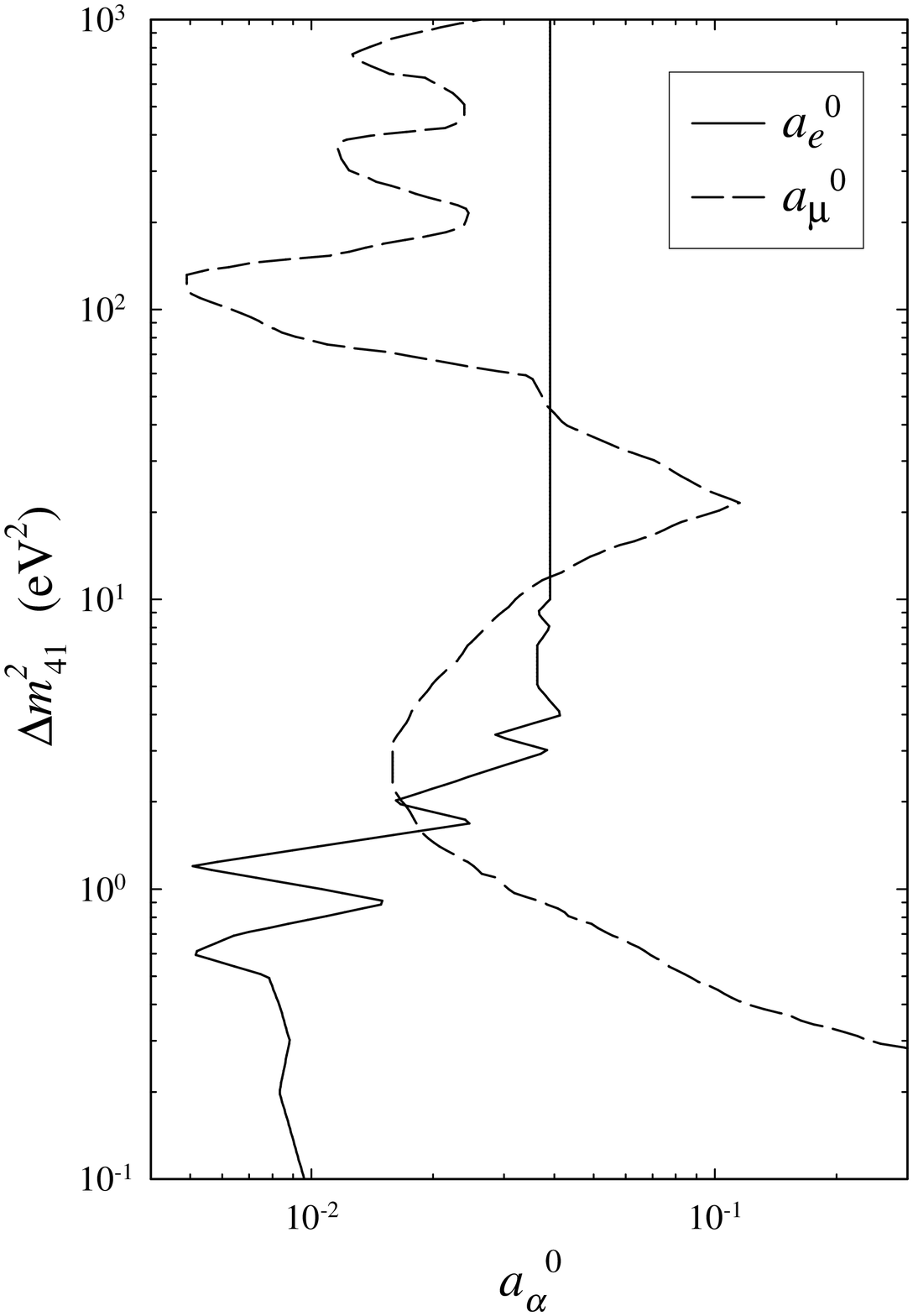,width=0.95\linewidth}}
\end{center}
\end{minipage}
&
\begin{minipage}{0.47\linewidth}
\begin{center}
\mbox{\epsfig{file=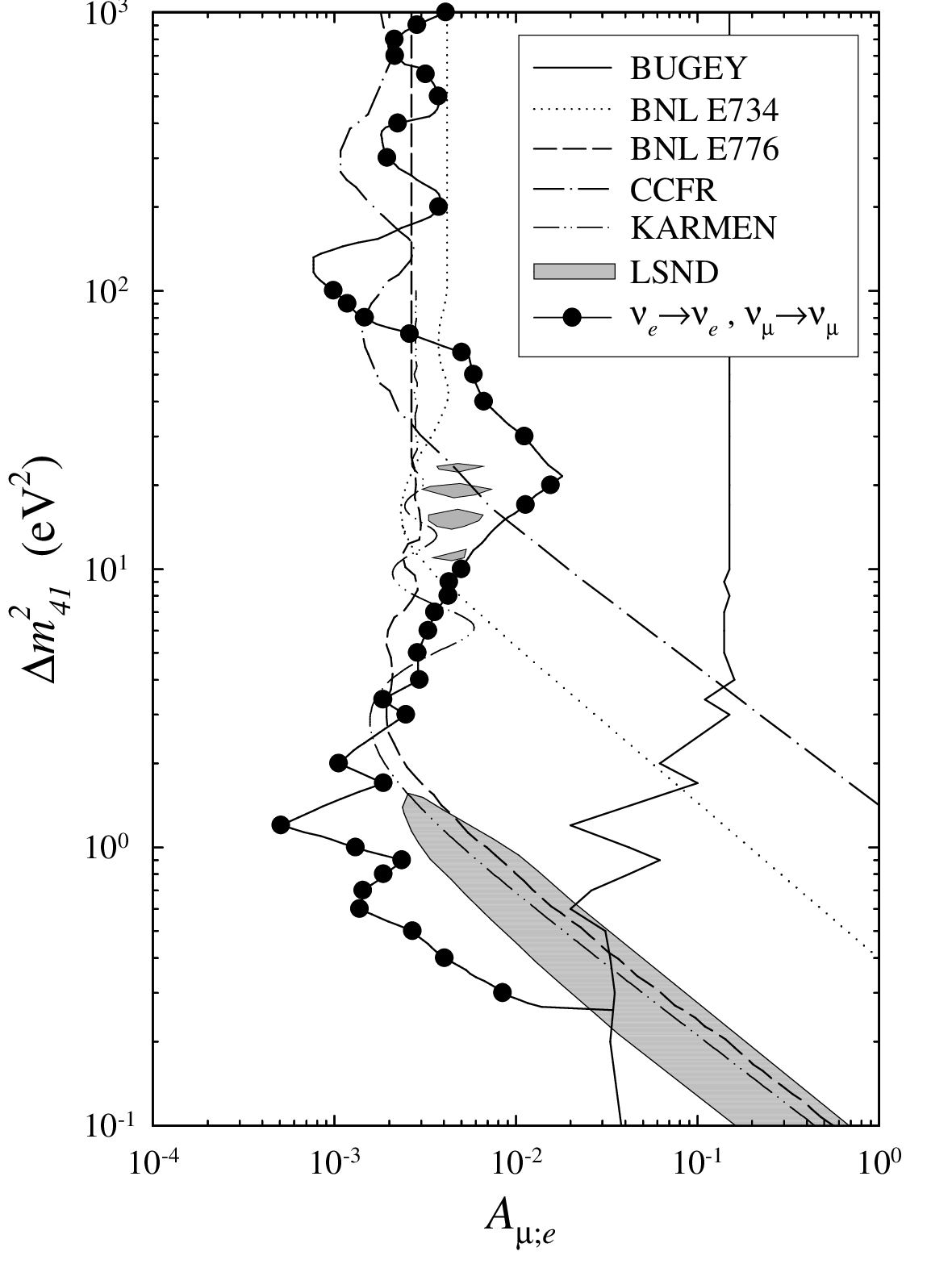,width=0.95\linewidth}}
\end{center}
\end{minipage}
\\
\begin{minipage}{0.47\linewidth}
\refstepcounter{figures}
\label{a0}
\footnotesize
Figure \ref{a0}.
Values of $a_e^0$ and $a_\mu^0$
(see Eq.(\ref{aa0}))
as functions of
$\Delta{m}^2_{41}$
obtained from the exclusion curves of the
Bugey \protect\cite{Bugey95},
CDHS \protect\cite{CDHS84} and CCFR \protect\cite{CCFR84}
SBL disappearance experiments.
\end{minipage}
&
\begin{minipage}{0.47\linewidth}
\refstepcounter{figures}
\label{amuel}
\footnotesize
Figure \ref{amuel}.
The region on the right of the solid line passing through the black circles
is excluded at 90\% CL
by the results of SBL disappearance experiments
for small
$ \left| U_{e4} \right|^2 $
and
$ \left| U_{\mu4} \right|^2 $
in the model with mixing of four neutrinos
and a mass hierarchy.
The shadowed regions are allowed at 90\% CL
by the results of the LSND experiment.
Also shown are the exclusion curves
of the
Bugey \protect\cite{Bugey95},
BNL E734 \protect\cite{BNLE734},
BNL E776 \protect\cite{BNLE776},
CCFR \protect\cite{CCFR97}
and
KARMEN (Bayesian analysis) \protect\cite{KARMEN98-nu98}
experiments.
\end{minipage}
\end{tabular*}
\end{figure}

The most stringent limits on
$B_{e;e}$
and
$B_{\mu;\mu}$
are given by the exclusion curves of the
Bugey \cite{Bugey95} reactor $\bar\nu_e\to\bar\nu_e$ experiment
and
of the CDHS \cite{CDHS84} and CCFR \cite{CCFR84}
accelerator $\nu_\mu\to\nu_\mu$ experiments.
The values of $a_e^0$ and $a_\mu^0$
corresponding to the 90\% CL
exclusion curves of these experiments
are shown in Fig.~\ref{a0}.
One can see that
\begin{equationarrayzero}
&&
a_e^0 \leq 4 \times 10^{-2}
\qquad \mbox{for} \qquad
\Delta{m}^2_{41} \gtrsim 0.1 \, \mathrm{eV}^2
\,,
\label{ae0u}
\\
&&
a_\mu^0 \leq 2 \times 10^{-1}
\qquad \mbox{for} \qquad
\Delta{m}^2_{41} \gtrsim 0.3 \, \mathrm{eV}^2
\,.
\label{am0u}
\end{equationarrayzero}%
Hence,
from Eq.(\ref{const4})
it follows that the quantities
$ \sum_k |U_{ek}|^2 $
and
$ \sum_k |U_{{\mu}k}|^2 $
are either small or large
(close to one).
We will show now that from the four possibilities for the two
quantities
$ \sum_k |U_{ek}|^2 $
and
$ \sum_k |U_{{\mu}k}|^2 $
(small-small, small-large, large-small, large-large)
for each neutrino mass spectrum in Fig.~\ref{4spectra}
only one possibility is compatible with the results of
solar and atmospheric neutrino experiments
\cite{BGG96-Nu96,BGG98-EPJ,BGG98-Pune}.

Let us consider, for example,
the hierarchical mass spectrum
\begin{equation}
m_1 \ll m_2 \ll m_3 \ll m_4
\,,
\label{hie4}
\end{equation}
which corresponds to type I in Fig.~\ref{4spectra}.
In this case,
the inequalities (\ref{const4}) become
\begin{equation}
|U_{{\alpha}4}|^2
\leq
a_\alpha^0
\qquad \mbox{or} \qquad
|U_{{\alpha}4}|^2
\geq
1 - a_\alpha^0
\qquad
(\alpha=e,\mu)
\,.
\label{ua4u}
\end{equation}

The survival probability of solar $\nu_e$'s
averaged over the fast, unobservable oscillations due to
$\Delta{m}^2_{41}$ and $\Delta{m}^2_{31}$
is given by \cite{BGKP96}
\begin{equation}
P_{\nu_e\to\nu_e}^{(\mathrm{sun})}
=
\left( 1 - \sum_{k=3,4} |U_{ek}|^2 \right)^2
P_{\nu_e\to\nu_e}^{(1,2)}
+
\sum_{k=3,4} |U_{ek}|^4
\,,
 \label{Psun4}
\end{equation}
where $P_{\nu_e\to\nu_e}^{(1,2)}$ is the $\nu_e$
survival probability
due to the mixing of $\nu_e$ with $\nu_1$ and $\nu_2$.
This probability has the two-generation form
(\ref{surv2}) and depends on the neutrino energy,
$ \Delta{m}^2 = \Delta{m}^2_{\mathrm{sun}} = \Delta{m}^2_{21} $ and 
\begin{equation}
\sin^22\vartheta
=
\sin^22\vartheta_{\mathrm{sun}}
=
\frac
{ 4 \, |U_{e1}|^2 \, |U_{e2}|^2 }
{ (|U_{e1}|^2 + |U_{e2}|^2)^2 }
\,.
\label{sun41}
\end{equation}
Though Eqs.(\ref{Psun4}) and (\ref{sun41}) were derived here for
vacuum oscillations, the considerations in Section 4.4 show that they
are also valid including matter effects.
If
$ |U_{e4}|^2 \geq 1-a^{0}_e $,
from Eqs.(\ref{ae0u}) and (\ref{Psun4})
one can see that
$P_{\nu_e\to\nu_e}^{(\mathrm{sun})}$
practically does not depend on the neutrino energy
and
$ P_{\nu_e\to\nu_e}^{(\mathrm{sun})} \gtrsim 0.92 $
for all solar neutrino energies.
This is not compatible with the solar neutrino
data \cite{Bahcall-Krastev-Smirnov98,pet-solar-vac,Vetrano98}.
Therefore,
we reach the conclusion that from the two possibilities
(\ref{ua4u}) for $|U_{e4}|^2$
only
\begin{equation}
|U_{e4}|^2 \leq a_e^0
\label{ue4u}
\end{equation}
is compatible with the results of solar neutrino experiments.

Let us now consider the
survival probability of atmospheric $\nu_\mu$'s.
From the general formula (\ref{prob2})
we obtain
\begin{equation}
P_{\nu_\mu\to\nu_\mu}^{(\mathrm{atm})}
=
\left|
\sum_{k=1,2}
|U_{{\mu}k}|^2
+
|U_{\mu3}|^2
\,
\exp \left( - i \frac{ \Delta{m}^2_{31} L }{ 2 E } \right)
+
|U_{\mu4}|^2
\,
\exp \left( - i \frac{ \Delta{m}^2_{41} L }{ 2 E } \right)
\right|^2
\,,
\label{atm41}
\end{equation}
where we have taken into account that for atmospheric
neutrinos\footnote{As explained after Eq.(\ref{a01}),
in the case of the MSW solutions of the solar neutrino problem
this inequality and the following equations are
valid only for multi-GeV neutrinos.}
$ \Delta{m}^2_{21} L / E \ll 1 $.
The probability averaged over the fast unobservable oscillations
due to $\Delta{m}^2_{41}$ is given by
\begin{equation}
P_{\nu_\mu\to\nu_\mu}^{(\mathrm{atm})}
=
\left(
\sum_{k=1,2}
|U_{{\mu}k}|^2
\right)^2
+
|U_{\mu3}|^4
+
2
\left(
\sum_{k=1,2}
|U_{{\mu}k}|^2
\right)
|U_{\mu3}|^2
\,
\cos \left( \frac{ \Delta{m}^2_{31} L }{ 2 E } \right)
+
|U_{\mu4}|^4
\,.
\label{atm42}
\end{equation}
Using the unitarity relation
\begin{equation}
\sum_{k=1}^{3}
|U_{{\mu}k}|^2
=
1 - |U_{\mu4}|^2
\,,
\label{atm43}
\end{equation}
the survival probability (\ref{atm42})
can be written as \cite{BGG98-EPJ,BGG98-Pune} (see also Section 4.4)
\begin{equation}
P_{\nu_\mu\to\nu_\mu}^{(\mathrm{atm})}
=
\left( 1 - |U_{\mu4}|^2 \right)^2
P_{\nu_\mu\to\nu_\mu}^{(1,2;3)}
+
|U_{\mu4}|^4
\,,
\label{atm44}
\end{equation}
where $P_{\nu_\mu\to\nu_\mu}^{(1,2;3)}$
is the survival probability of atmospheric $\nu_\mu$'s
due to the mixing of $\nu_\mu$ with
$\nu_3$ and $\nu_{2}$, $\nu_{1}$.
This probability has the two-generation form
\begin{equation}
P_{\nu_\mu\to\nu_\mu}^{(1,2;3)}
=
1
-
\frac{1}{2}
\,
\sin^22\vartheta_{\mathrm{atm}}
\left( 1 - \cos \frac{ \Delta{m}^2_{31} L }{ 2 E } \right)
\label{atm45}
\end{equation}
(see Eq.(\ref{surv2})),
with
\begin{equation}
\sin^22\vartheta_{\mathrm{atm}}
=
\frac
{ 4 \, |U_{\mu3}|^2 \left( |U_{\mu1}|^2 + |U_{\mu2}|^2 \right) }
{ \left( |U_{\mu1}|^2 + |U_{\mu2}|^2 + |U_{\mu3}|^2 \right)^2 }
\,.
\label{atm46}
\end{equation}
The zenith angle dependence of $\mu$-like events
in atmospheric neutrino experiments
is due to $P_{\nu_\mu\to\nu_\mu}^{(1,2;3)}$, which is multiplied by
$\left( 1 - |U_{\mu4}|^2 \right)^2$
in the expression (\ref{atm44}) for $P_{\nu_\mu\to\nu_\mu}^{(\mathrm{atm})}$.
If
$|U_{\mu4}|^2$
is large,
as given by the second inequality in Eq.(\ref{ua4u}),
$ |U_{\mu4}|^2 \geq 1 - a_\mu^0 $,
we have
$ \left( 1 - |U_{\mu4}|^2 \right)^2 \leq 4 \times 10^{-2} $
for $ \Delta{m}^2_{41} \gtrsim 0.3 \, \mathrm{eV}^2 $
(see Eq.(\ref{am0u})).
A zenith-angle variation of $\mu$-like events smaller than
$ 4 \times 10^{-2} $
is incompatible with the up-down asymmetry
of multi-GeV $\mu$-like events observed by the Super-Kamiokande
experiment (see Section~\ref{Indications: atmospheric neutrino experiments}).
Therefore,
from the two inequalities (\ref{ua4u}) for $|U_{\mu4}|^2$
only the first one is compatible
with the results of the atmospheric neutrino experiments:
\begin{equation}
|U_{\mu4}|^2 \leq a_\mu^0
\,.
\label{atm47}
\end{equation}
The same conclusion can be reached for all the neutrino mass spectra
in Fig.~\ref{4spectra}
in which one mass is separated by the other three
by the LSND gap,
\textit{i.e.},
for the spectra I, II, III, IV.
Hence,
in the framework of these spectra the results of SBL disappearance experiments
and of solar and atmospheric neutrino experiments
imply that
\begin{equation}
|U_{{e}j}|^2 \leq a_e^0
\,,
\qquad
|U_{{\mu}j}|^2 \leq a_\mu^0
\,,
\label{4nu01}
\end{equation}
with
$j=4$ in the schemes I and II
and
$j=1$ in the schemes III and IV.
These inequalities imply that
$
\stackrel{\makebox[0pt][l]
{$\hskip-3pt\scriptscriptstyle(-)$}}{\nu_{\mu}}
\to\stackrel{\makebox[0pt][l]
{$\hskip-3pt\scriptscriptstyle(-)$}}{\nu_{e}}
$
transitions in SBL experiments
are strongly suppressed.
Indeed,
from Eqs.(\ref{a4}) and (\ref{4nu01})
we have
\begin{equation}
A_{\mu;e}
=
4 \, |U_{ej}|^2 \, |U_{{\mu}j}|^2
\leq
4 \, a_e^0 \, a_\mu^0
\,.
\label{4nu02}
\end{equation}
Thus,
the amplitude of SBL
$
\stackrel{\makebox[0pt][l]
{$\hskip-3pt\scriptscriptstyle(-)$}}{\nu_{\mu}}
\to\stackrel{\makebox[0pt][l]
{$\hskip-3pt\scriptscriptstyle(-)$}}{\nu_{e}}
$
oscillations is quadratic in the small quantities
$|U_{ej}|^2$ and $|U_{{\mu}j}|^2$.
The upper bound (\ref{4nu02}) obtained from the values of
$a_e^0$ and $a_\mu^0$
presented in Fig.~\ref{a0}
is shown in Fig.~\ref{amuel}
by the solid line passing through the black circles.
The solid, dotted, dashed, dash-dotted and dash-dot-dotted
curves in Fig.~\ref{amuel}
are the 90\% CL
exclusion curves obtained in the
Bugey \cite{Bugey95},
BNL E734 \cite{BNLE734},
BNL E776 \cite{BNLE776},
CCFR \cite{CCFR97}
and
KARMEN (Bayesian analysis) \cite{KARMEN98-nu98}
experiments.
The shadowed regions are allowed by the results of the LSND
\cite{LSND-nu98,LSND-www}
experiments at 90\% CL.
From this figure one can see that the result of the LSND experiment,
taken together with the results of SBL disappearance experiments
and solar and atmospheric neutrino experiments,
disfavours the schemes I, II, III and IV of 
Fig.~\ref{4spectra}.\footnote{The presence of 
a small allowed region in Fig.~6.3 for
$ 0.2 \, \mathrm{eV}^2
\lesssim \Delta m^2_{41} \lesssim
0.3 \, \mathrm{eV}^2 $
is due to the fact that
for
$ \Delta m^2_{41} \lesssim 0.3 \, \mathrm{eV}^2 $
there are no constraints on $B^0_{\mu;\mu}$ from the results of
$\nu_\mu$ disappearance experiments.}

Let us consider now the spectra A and B in Fig.~\ref{4spectra}.
In this case the inequalities (\ref{const4}) apply
with
the index $k$ running over $1,2$ or $3,4$
(the two choices are equivalent because of the unitarity of the mixing matrix).
We consider explicitly only the spectrum A,
but all the following results are valid also for scheme B
and can be obtained with the index exchange $ 1,2 \leftrightarrows 3,4 $.

In the case of spectrum A,
the survival probability of solar $\nu_e$'s
is given by \cite{BGKP96}
\begin{equation}
P_{\nu_e\to\nu_e}^{(\mathrm{sun})}
=
\sum_{k=1,2} |U_{ek}|^4
+
\left( 1 - \sum_{k=1,2} |U_{ek}|^2 \right)^2
P_{\nu_e\to\nu_e}^{(3,4)}
\,,
\label{AB01}
\end{equation}
where $P_{\nu_e\to\nu_e}^{(3,4)}$
is the survival probability of solar $\nu_e$'s
due to the mixing of $\nu_e$ with
$\nu_3$ and $\nu_4$.
This probability has the two-generation form
(\ref{surv2}) and depends on the neutrino energy,
$ \Delta{m}^2 = \Delta{m}^2_{\mathrm{sun}} = \Delta{m}^2_{43} $
and
\begin{equation}
\sin^22\vartheta
=
\sin^22\vartheta_{\mathrm{sun}}
=
\frac
{ 4 \, |U_{e3}|^2 \, |U_{e4}|^2 }
{ \left( |U_{e3}|^2 + |U_{e4}|^2 \right)^2 }
\,.
\label{AB02}
\end{equation}
If
$ \sum_{k=1,2} |U_{ek}|^2 \geq 1 - a_e^0 $,
\textit{i.e.}, the second inequality (\ref{const4})
for $\alpha=e$ and $k=1,2$ is satisfied,
from Eq.(\ref{ae0u}) we have
$ \left( 1 - \sum_{k=1,2} |U_{ek}|^2 \right)^2 \leq 1.6 \times 10^{-3} $,
which implies that
$
P_{\nu_e\to\nu_e}^{(\mathrm{sun})}
\simeq
\sum_{k=1,2} |U_{ek}|^4
$
is constant
and
$
P_{\nu_e\to\nu_e}^{(\mathrm{sun})}
\gtrsim
0.5
$.
Since these constraints are incompatible with the solar neutrino data,
we reach the conclusion that only the first inequality (\ref{const4})
for $\alpha=e$ and $k=1,2$
is compatible with the results of solar neutrino experiments,
\textit{i.e.},
$ \sum_{k=1,2} |U_{ek}|^2 \leq a_e^0 $.

Now we consider the survival probability of atmospheric $\nu_\mu$'s.
In scheme A it is given by \cite{BGKP96}
\begin{equation}
P_{\nu_\mu\to\nu_\mu}^{(\mathrm{atm})}
=
\left( 1 - \sum_{k=3,4} |U_{{\mu}k}|^2 \right)^2
P_{\nu_\mu\to\nu_\mu}^{(1,2)}
+
\left( \sum_{k=3,4} |U_{{\mu}k}|^2 \right)^2
\,,
\label{AB04}
\end{equation}
where the probability $P_{\nu_\mu\to\nu_\mu}^{(1,2)}$
has the two-generation form (\ref{surv2})
with
$ \Delta{m}^2 = \Delta{m}^2_{\mathrm{atm}} = \Delta{m}^2_{21} $
and
\begin{equation}
\sin^22\vartheta
=
\sin^22\vartheta_{\mathrm{atm}}
=
\frac
{ 4 \, |U_{{\mu}1}|^2 \, |U_{{\mu}2}|^2 }
{ \left( |U_{{\mu}1}|^2 + |U_{{\mu}2}|^2 \right)^2 }
\,.
\label{AB05}
\end{equation}
If
$ \sum_{k=3,4} |U_{{\mu}k}|^2 \geq 1 - a_\mu^0 $,
which corresponds to the second of the two inequalities (\ref{const4})
for $\alpha=\mu$ and $k=3,4$,
the upper bound (\ref{am0u}) for $a_\mu^0$
implies that $P_{\nu_\mu\to\nu_\mu}^{(\mathrm{atm})}$
is practically constant
and incompatible with the zenith-angle dependence of
multi-GeV $\mu$-like events
observed in the Super-Kamiokande experiment \cite{SK-atm-98}.
Hence,
only the first inequality (\ref{const4})
for $\alpha=\mu$ and $k=3,4$
is compatible with the results of
the Super-Kamiokande atmospheric neutrino experiments
and we have
$ \sum_{k=3,4} |U_{{\mu}k}|^2 \leq a_\mu^0 $.

Summarizing,
in the framework of scheme A
the results of SBL disappearance experiments
and of solar and atmospheric neutrino experiments
imply that
\begin{equation}
\sum_{k=1,2} |U_{ek}|^2 \leq a_e^0
\qquad \mbox{and} \qquad
\sum_{k=3,4} |U_{{\mu}k}|^2 \leq a_\mu^0
\qquad
(\mathrm{A})
\,.
\label{AB06}
\end{equation}
The corresponding bounds in scheme B
are obtained with the exchange $ 1,2 \leftrightarrows 3,4 $:
\begin{equation}
\sum_{k=3,4} |U_{ek}|^2 \leq a_e^0
\qquad \mbox{and} \qquad
\sum_{k=1,2} |U_{{\mu}k}|^2 \leq a_\mu^0
\qquad
(\mathrm{B})
\,.
\label{AB07}
\end{equation}

The constraints (\ref{AB06}) and (\ref{AB07})
imply the following upper bound
for the amplitude of $\nu_\mu\to\nu_e$
transitions in SBL experiments:
\begin{equation}
A_{\mu;e}
=
4 \left| \sum_k U_{ek} U_{{\mu}k}^* \right|^2
\leq
4 \left( \sum_k |U_{ek}|^2 \right) \left( \sum_k |U_{{\mu}k}|^2 \right)
\leq
4 \, \mathrm{Min}( a_e^0 , a_\mu^0 )
\label{AB08}
\end{equation}
(here the index $k$ runs over $1,2$ or over $3,4$).
This bound is linear in the small quantities
$a_e^0$ and $a_\mu^0$
and turns out to be compatible with the results of the LSND experiment.
Actually,
this bound is worse than the unitarity bound
\begin{equation}
A_{\mu;e}
\leq
\mathrm{Min}( B_{e;e}^0 , B_{\mu;\mu}^0 )
\,,
\label{AB09}
\end{equation}
which causes the exclusion of the large $\sin^22\vartheta$
part of the LSND allowed region
(see Fig.~\ref{amuel}, where $B_{e;e}^0$ is represented by the Bugey curve).

We want to note that the bound $a^0_\mu$ ceases to exist shortly below
$\Delta m^2_{41} = 0.3$ eV$^2$ (see Fig.~\ref{a0}). On the other hand,
it follows from the LSND and Bugey experiments that 
$\Delta m^2_{41} \gtrsim 0.2$ eV$^2$ (see Fig.~\ref{amuel}). Thus, for
the small interval 
$0.2 \: \mathrm{eV}^2 \lesssim \Delta m^2_{41} \lesssim 0.3 \: 
\mathrm{eV}^2$ our arguments leading to schemes A and B have to be
considered with caution and it would be desirable to have
experimental information on $B_{\mu;\mu}$ in this interval.

Let us now discuss some consequences of
the schemes A and B,
taking into account the constraints (\ref{AB06}) and (\ref{AB07}).
First of all,
we will show that these bounds
lead to constraints on the neutrino mass measured
in tritium $\beta$-decay experiments
and on the effective Majorana mass that determines
the probability of neutrinoless double-$\beta$ decay.

The electron spectrum in the decay
$ {}^3\mathrm{H} \to
{}^3\mathrm{He} + e^{-} + \bar\nu_e $
is given by
\begin{equation}
{\displaystyle
\mathrm{d} N
\over\displaystyle
\mathrm{d} E
}
=
C \, p_e \, E_e
\left( Q - T \right)
F(E_e)
\sum_{i}
\left| U_{ei} \right|^2
\sqrt{
\left( Q - T \right)^2
-
m^2_i
}
\,,
\label{210}
\end{equation}
where $Q$ is the energy release,
$p_e$ and $E_e$
are the electron momentum and energy,
$ T = E_e - m_e $,
$F(E_e)$ is the Fermi function
and $C$ is a constant.
Taking into account that
$ m_1 \simeq m_2 \ll m_3 \simeq m_4 $,
the first inequality in Eq.(\ref{AB06}) implies that
in scheme A the effective neutrino mass
determined from the measurement of the high-energy part of the
$\beta$-spectrum of $^3\mathrm{H}$ is
practically equal to the ``LSND mass'' $m_4$:
\begin{equation}
m(^3\mathrm{H})
\simeq
\sum_{k=3,4} |U_{ek}|^2 \, m_4
=
\left( 1 - \sum_{k=1,2} |U_{ek}|^2 \right) m_4
\simeq
m_4
\,.
\label{AB10}
\end{equation}

If massive neutrinos are Majorana particles,
the matrix element of neutrinoless double-$\beta$ decay
is proportional to the effective Majorana mass
\cite{Doi81,Wolfenstein81-PL} (see Appendix A)
\begin{equation}
\langle{m}\rangle
=
\sum_{k=1}^{4} U_{ek}^2 \, m_k
\,.
\label{AB12}
\end{equation}
In the case of scheme A,
the first inequality in Eq.(\ref{AB06}) implies that
\begin{equation}
|\langle{m}\rangle|
\simeq
\left| \sum_{k=3,4} U_{ek}^2 \right| m_4
\simeq
m_4
\,
\sqrt{ 1 - 4 \, |U_{e4}|^2 \left( 1 - |U_{e4}|^2 \right) \sin^2\phi }
\,,
\label{AB13}
\end{equation}
where $\phi$
is the difference of the phases of $U_{e3}$ and $U_{e4}$.
Since $\phi$ is unknown,
from Eq.(\ref{AB13}) we obtain the constraints
\begin{equation}
\left| 1 - 2 \, |U_{e4}|^2 \right| m_4
\lesssim
|\langle{m}\rangle|
\lesssim
m_4
\,,
\label{AB14}
\end{equation}
where the upper and lower bounds
correspond to $\phi=0$ or $\pi$ and $\phi=\pm\pi/2$,
respectively.\footnote{If
CP is conserved, we have
$\phi=0$ or $\pi$
if the CP parities of $\nu_3$ and $\nu_4$
are equal
and $\phi=\pm\pi/2$
if they are opposite.
Indeed, if CP is conserved from Eq.(\ref{orth4})
we get $ U_{ek}^* = U_{ek} \rho_k $
where
$\rho_k = \pm 1$
and the CP parity of $\nu_k$ is
$ \eta^{\mathrm{CP}}_k = i \rho_k $
(see Eq.(\ref{etaCPk}))
\cite{Bilenky-Nedelcheva-Petcov84,Kayser84}.
Writing $ U_{ek} = |U_{ek}| e^{i\phi_k} $,
we obtain $\phi_k=0$ or $\pi$
if $\rho_k=+1$
and
$\phi_k=\pm\pi/2$
if $\rho_k=-1$.
Hence,
we have
$ \phi_3 - \phi_4 = 0 $ or $\pi$
if $\rho_3=\rho_4$
and
$ \phi_3 - \phi_4 = \pm \pi/2 $
if $\rho_3=-\rho_4$.}
Taking into account Eqs.(\ref{AB02}) and (\ref{AB06}),
we have
\begin{equation}
|U_{e3}| \simeq \cos\vartheta_{\mathrm{sun}}
\qquad \mbox{and} \qquad
|U_{e4}| \simeq \sin\vartheta_{\mathrm{sun}}
\,,
\label{AB15}
\end{equation}
and Eq.(\ref{AB14})
can be written as
\begin{equation}
m_4 \, \sqrt{ 1 - \sin^22\vartheta_{\mathrm{sun}} }
\lesssim
|\langle{m}\rangle|
\lesssim
m_4
\,.
\label{AB16}
\end{equation}
If
$ \sin^2 2\vartheta_{\mathrm{sun}} \ll 1 $,
which corresponds to the MSW solution
with a small mixing angle,
we have
$ |\langle{m}\rangle| \simeq m_4 $,
independent of the value of
$\phi$
and of the conservation of CP.
In the case of a large value of the parameter
$ \sin^2 2\vartheta_{\mathrm{sun}} $,
which correspond to the MSW solution
with a large mixing angle
or to the vacuum oscillation solution,
in the future
it will be possible to obtain information
about
the violation of
CP in the lepton sector
if both the
$^3\mathrm{H}$ $\beta$-decay
and
neutrinoless double-$\beta$ decay
experiments
will obtain positive results.
Indeed,
from the measurements of
$ m(^4\mathrm{H}) \simeq m_4 $,
$ |\langle{m}\rangle| $
and 
$ \sin^2 2\vartheta_{\mathrm{sun}} $
with the help of Eq.(\ref{AB16})
it will be possible to determine the value of
$ \sin^2 \phi $
(if $ \sin^2 2\vartheta_{\mathrm{sun}} $ is large).
Let us emphasize that
in the case of CP conservation
the relative CP parities of $\nu_3$ and $\nu_4$
can be determined:
the CP parities are equal
if $ \sin^2 \phi = 0 $
and opposite if
$ \sin^2 \phi = 1 $.

Considering now scheme B,
the first inequality in Eq.(\ref{AB07}) implies that
both $m(^3\mathrm{H})$ and $|\langle{m}\rangle|$ are strongly suppressed:
\begin{equationarrayzero}
&&
m(^3\mathrm{H})
\simeq
\left( \sum_{k=3,4} |U_{ek}|^2 \right) m_4
\leq
a_e^0 \, m_4
\ll
m_4
\,.
\label{AB21}
\\
&&
|\langle{m}\rangle|
\simeq
\left| \sum_{k=3,4} U_{ek}^2 \right| m_4
\leq
\left( \sum_{k=3,4} |U_{ek}|^2 \right) m_4
\leq
a_e^0 \, m_4
\ll
m_4
\,.
\label{AB22}
\end{equationarrayzero}%
Therefore, only
if scheme A is realized in nature,
there is some possibility to measure the effect of the LSND mass
in future tritium decay experiments
and neutrinoless double-$\beta$ decay experiments.

We discuss now the implications of the schemes A and B for
LBL neutrino oscillation experiments \cite{BGG98-bounds}.
The smallness of
$ \sum_{k=1,2} |U_{ek}|^2 $ in scheme A
and of
$ \sum_{k=3,4} |U_{ek}|^2 $ in scheme B
implies that the electron neutrino has a
small mixing with the neutrinos whose mass-squared difference is
responsible for the oscillations of atmospheric neutrinos
(\textit{i.e.},
$\nu_1$, $\nu_2$ in scheme A and $\nu_3$, $\nu_4$ in scheme B).
Hence,
as shown in the following part of this section,
the transition probability of
electron neutrinos and antineutrinos
into other states
in atmospheric and LBL experiments
is suppressed.

Taking into account that in scheme A we have 
$ \Delta{m}^2_{43} L / E \ll 1 $
and
$ \Delta{m}^2_{41} L / E \gg 1 $
in LBL experiments,
we obtain from the general formula (\ref{prob2})
the following expression for the measurable
probability of $\nu_\alpha\to\nu_\beta$ transitions
averaged over the fast oscillations due to $\Delta{m}^2_{41}$:
\begin{equation}
P_{\nu_\alpha\to\nu_\beta}^{(\mathrm{LBL;A})}
=
\left|
U_{{\beta}1} \, U_{{\alpha}1}^*
+
U_{{\beta}2} \, U_{{\alpha}2}^*
\exp\left( - i \, \frac{ \Delta{m}^2_{21} L }{ 2 E } \right)
\right|^2
+
\left| \sum_{k=3,4} U_{{\beta}k} \, U_{{\alpha}k}^* \right|^2
\,.
\label{AB31}
\end{equation}
The corresponding probability in scheme B is given by the exchange
$ 1,2 \leftrightarrows 3,4 $
and the probability of
$\bar\nu_\alpha\to\bar\nu_\beta$ transitions
is obtained changing
$
U_{{\beta}k} \, U_{{\alpha}k}^*
\to
U_{{\beta}k}^* \, U_{{\alpha}k}
$.

\begin{figure}[t!]
\begin{tabular*}{\linewidth}{@{\extracolsep{\fill}}cc}
\begin{minipage}{0.47\linewidth}
\begin{center}
\mbox{\epsfig{file=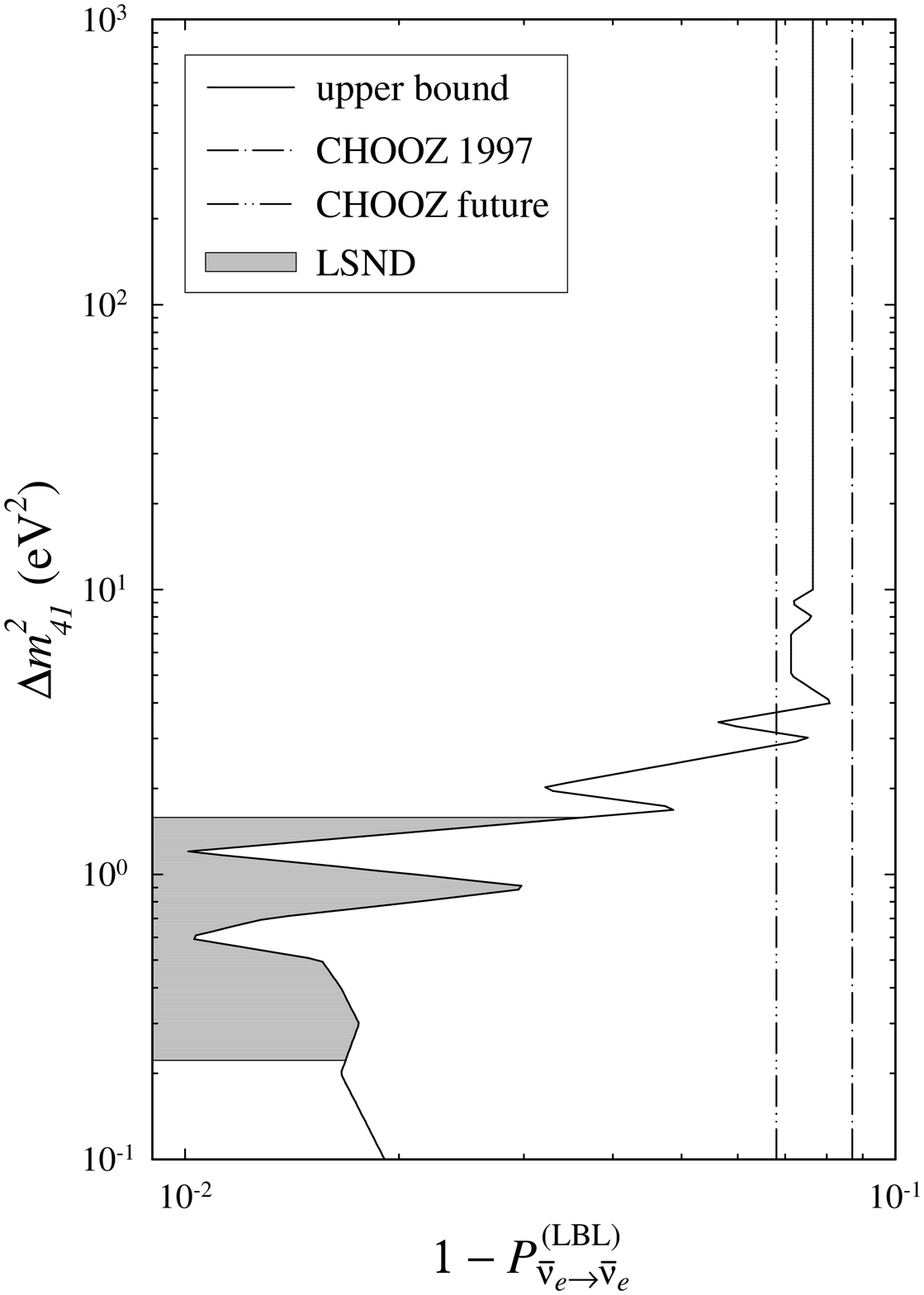,width=0.95\linewidth}}
\end{center}
\end{minipage}
&
\begin{minipage}{0.47\linewidth}
\begin{center}
\mbox{\epsfig{file=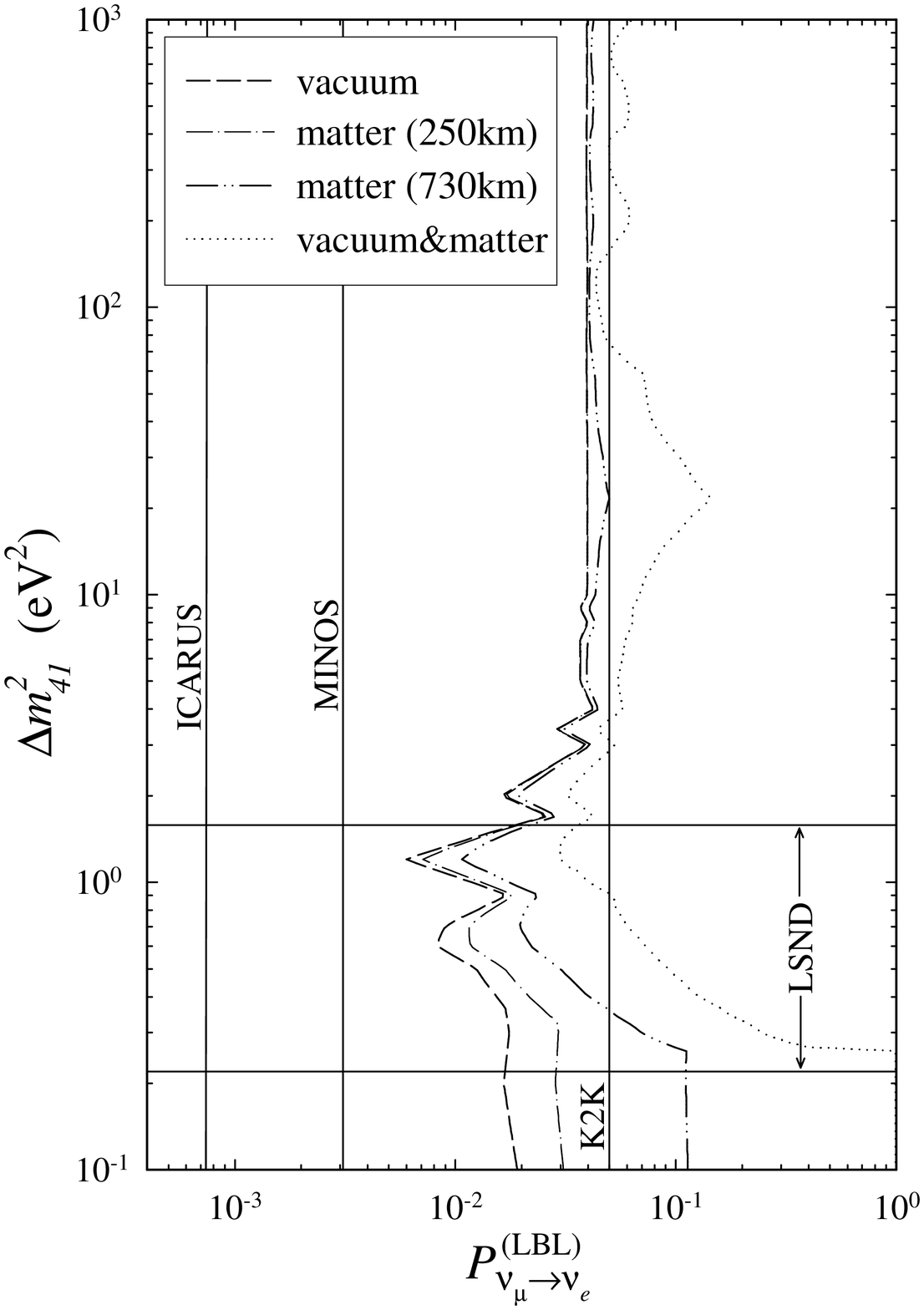,width=0.95\linewidth}}
\end{center}
\end{minipage}
\\
\begin{minipage}{0.47\linewidth}
\refstepcounter{figures}
\label{peelbl}
\footnotesize
Figure \ref{peelbl}.
The value of the bound (\ref{AB35}) on the transition
probability of $\bar \nu_e$ into all other neutrino types in LBL
experiments as a function of $\Delta m^2_{41}$ (solid line). 
The dash-dotted and dash-dot-dotted lines represent the upper bound 
obtained in the CHOOZ experiment \cite{CHOOZ98} and the final sensitivity
planned in this experiment, respectively. The shadowed region corresponds to
the range of $\Delta m^2_{41}$ allowed at 90\% CL by the results
of LSND and other experiments.
\end{minipage}
&
\begin{minipage}{0.47\linewidth}
\refstepcounter{figures}
\label{pmelbl}
\footnotesize
Figure \ref{pmelbl}.
Upper bounds on the probability of $\nu_\mu\to\nu_e$ transitions
in LBL experiments as functions of $\Delta m^2_{41}$. The dashed
line represents the bound (\ref{AB41}), whereas the dash-dotted and
dash-dot-dotted curves include matter corrections to this bound for the K2K
($L = 250$ km) and MINOS and ICARUS ($L = 730$ km) experiments, respectively.
The vertical lines show the planned sensitivities of these
experiments. The dotted curve depicts the bound (\ref{AB43}).
The range of $\Delta m^2_{41}$ allowed by LSND and
other experiments is also indicated.
\end{minipage}
\end{tabular*}
\end{figure}

Let us consider reactor LBL experiments
which measure the survival probability
$ P_{\bar\nu_e\to\bar\nu_e}^{(\mathrm{LBL})} $.
We will show that the negative results of SBL reactor experiments
and
the results of solar neutrino experiments
imply strong constraints on the transition probability
of LBL reactor $\bar\nu_e$'s into other states.
Indeed,
from Eq.(\ref{AB31}) and the corresponding one in scheme B,
for the survival probability of LBL reactor $\bar\nu_e$'s
we have the lower bounds
\begin{equationarrayzero}
&&
P_{\bar\nu_e\to\bar\nu_e}^{(\mathrm{LBL;A})}
\geq
\left( \sum_{k=3,4} |U_{ek}|^2 \right)^2
=
\left( 1 - \sum_{k=1,2} |U_{ek}|^2 \right)^2
\,,
\label{AB32}
\\
&&
P_{\bar\nu_e\to\bar\nu_e}^{(\mathrm{LBL;B})}
\geq
\left( \sum_{k=1,2} |U_{ek}|^2 \right)^2
=
\left( 1 - \sum_{k=3,4} |U_{ek}|^2 \right)^2
\,,
\label{AB33}
\end{equationarrayzero}%
in schemes A and B,
respectively.
From the inequalities (\ref{AB06}) and (\ref{AB07})
it follows that in both schemes
\begin{equation}
P_{\bar\nu_e\to\bar\nu_e}^{(\mathrm{LBL})}
\geq
\left( 1 - a_e^0 \right)^2
\,.
\label{AB34}
\end{equation}
Since $a_e^0$ is small
(see Fig.~\ref{a0} and Eq.(\ref{ae0u})),
the lower bound (\ref{AB34})
implies that the survival probability
$P_{\bar\nu_e\to\bar\nu_e}^{(\mathrm{LBL})}$
is close to one.
For the transition probability of $\nu_e$'s into any other state in LBL experiments,
\begin{equation}
1 - P_{\bar\nu_e\to\bar\nu_e}^{(\mathrm{LBL})}
=
\sum_{\beta{\neq}e}
P_{\bar\nu_e\to\bar\nu_\beta}^{(\mathrm{LBL})}
\,,
\label{AB38}
\end{equation}
we have the upper bound
\begin{equation}
1 - P_{\bar\nu_e\to\bar\nu_e}^{(\mathrm{LBL})}
\leq
a_e^0 \left( 2 - a_e^0 \right)
\,.
\label{AB35}
\end{equation}
The value of this bound,
which depends on $\Delta{m}^2_{41}$
through the dependence of $a_e^0$ on $\Delta{m}^2_{41}$
(see Fig.~\ref{a0}),
is shown in Fig~\ref{peelbl} (the solid line),
where it is compared with the upper bound for
$1 - P_{\bar\nu_e\to\bar\nu_e}^{(\mathrm{LBL})}$
obtained in the CHOOZ experiment \cite{CHOOZ98}
(dash-dotted line)
and with the final sensitivity of the CHOOZ experiment
(dash-dot-dotted line).
The shadowed region in Fig~\ref{peelbl}
corresponds to the range of $\Delta{m}^2_{41}$
allowed at 90\% CL
by the results of the LSND and other experiments.
One can see from Fig~\ref{peelbl}
that if $\Delta{m}^2_{41}$ lies in the LSND range,
the upper bound for
$1 - P_{\bar\nu_e\to\bar\nu_e}^{(\mathrm{LBL})}$
is much less than the sensitivity of the CHOOZ experiment
and it will be very difficult to measure
the disappearance of reactor $\bar\nu_e$
in future LBL experiments.

Let us consider now
$\nu_\mu\to\nu_e$
transitions, which will be investigated in the near future
by the
K2K \cite{K2K},
MINOS \cite{MINOS},
ICARUS \cite{ICARUS}
and/or other \cite{NOE,AQUA-RICH,OPERA,NICE}
experiments
(see \cite{Zuber98}).
With the help of the
Cauchy--Schwarz inequality,
from Eq.(\ref{AB31}) and Eq.(\ref{a4})
we obtain
\begin{equation}
P_{\nu_\mu\to\nu_e}^{(\mathrm{LBL})}
\leq
\left( \sum_k |U_{ek}|^2 \right)
\left( \sum_k |U_{{\mu}k}|^2 \right)
+
\frac{1}{4} \, A_{\mu;e}
\,,
\label{AB36}
\end{equation}
where the index $k$
runs over $1,2$ in the scheme A
and over $3,4$ in the scheme B.
From the constraints (\ref{AB06})and (\ref{AB07}),
in both schemes A and B we have
\begin{equation}
P_{\nu_\mu\to\nu_e}^{(\mathrm{LBL})}
\leq
a_e^0
+
\frac{1}{4} \, A_{\mu;e}^0
\,,
\label{AB37}
\end{equation}
where $A_{\mu;e}^0$ is the upper bound for the
oscillation amplitude $A_{\mu;e}$
measured in SBL experiments.

Since the CPT relation (\ref{pna})
and the conservation of probability, relation (\ref{AB38}),
imply that
\begin{equation}
\sum_{\alpha{\neq}e}
P_{\nu_\alpha\to\nu_e}^{(\mathrm{LBL})}
=
\sum_{\alpha{\neq}e}
P_{\bar\nu_e\to\bar\nu_\alpha}^{(\mathrm{LBL})}
=
1 - P_{\bar\nu_e\to\bar\nu_e}^{(\mathrm{LBL})}
\,,
\label{AB40}
\end{equation}
another upper limit for
$P_{\nu_\mu\to\nu_e}^{(\mathrm{LBL})}$
can be obtained from the inequality (\ref{AB35}):
\begin{equation}
P_{\nu_\mu\to\nu_e}^{(\mathrm{LBL})}
\leq
a_e^0 \left( 2 - a_e^0 \right)
\,.
\label{AB39}
\end{equation}
Combining the upper bounds (\ref{AB37}) and (\ref{AB39})
and taking into account that they are valid also for antineutrinos
\cite{BGG98-bounds},
we have
\begin{equation}
P^{(\mathrm{LBL})}_{\stackrel{\makebox[0pt][l]
{$\hskip-3pt\scriptscriptstyle(-)$}}{\nu_{\mu}}
\to\stackrel{\makebox[0pt][l]
{$\hskip-3pt\scriptscriptstyle(-)$}}{\nu_{e}}}
\leq
\mathrm{Min}\!\left(
a_e^0
+
\frac{1}{4} \, A_{\mu;e}^0
\, , \,
a_e^0 \left( 2 - a_e^0 \right)
\right)
\,.
\label{AB41}
\end{equation}
The curve corresponding
to this limit
obtained from the 90\% CL
exclusion plots of the Bugey
\cite{Bugey95}
experiment for
$a^{0}_{e}$
and
of the
BNL E734
\cite{BNLE734},
BNL E776
\cite{BNLE776},
CCFR
\cite{CCFR97}
and
KARMEN (Bayesian analysis)
\cite{KARMEN98-nu98}
experiments
for
$A_{\mu;e}^{0}$
is represented by the dashed line
in Fig.~\ref{pmelbl}.
For comparison,
the expected sensitivities
of the LBL accelerator neutrino experiments
K2K \cite{K2K},
MINOS \cite{MINOS}
and
ICARUS \cite{ICARUS} are also indicated
(the three straight vertical lines).
The dash-dotted and dash-dot-dotted
lines in Fig.~\ref{pmelbl} represent,
respectively,
the upper bound for
$P_{\nu_\mu\to\nu_e}^{(\mathrm{LBL})}$
in the K2K and MINOS, ICARUS experiments
corrected for the matter effects \cite{BGG98-bounds} due to the
propagation of the neutrino beam in the crust of the earth.
These effects are different for the K2K and MINOS, ICARUS experiments
because of the different value of $L$ in this experiments:
$ L \simeq 250 \, \mathrm{km} $
in K2K
and
$ L \simeq 730 \, \mathrm{km} $
in MINOS and ICARUS.
(It is obvious that the bound for other LBL experiments
with a neutrino beam from CERN to Gran Sasso
is the same as that for the ICARUS experiment.)

The region between the two horizontal solid lines in Fig.~\ref{pmelbl}
corresponds to the range of $\Delta{m}^2_{41}$
allowed at 90\% CL
by the results of the LSND and other experiments.
One can see that if $\Delta{m}^2_{41}$
is in the LSND region,
the sensitivity of the K2K experiment is not sufficient
for the observation of
$\nu_\mu\to\nu_e$
transitions,
whereas experiments with higher sensitivities
like MINOS and ICARUS could reveal these transitions.

Finally,
another upper bound on
$
P^{(\mathrm{LBL})}_{\stackrel{\makebox[0pt][l]
{$\hskip-3pt\scriptscriptstyle(-)$}}{\nu_{\mu}}
\to\stackrel{\makebox[0pt][l]
{$\hskip-3pt\scriptscriptstyle(-)$}}{\nu_{e}}}
$ 
can be obtained from Eq.(\ref{AB37}) noting that
\begin{equation}
A_{\mu;e}
\leq
4
\left( 1 - \sum_k |U_{{\mu}k}|^2 \right)
\left( 1 - \sum_k |U_{ek}|^2 \right)
\,,
\label{AB42}
\end{equation}
with $k=1,2$ in scheme A and $k=3,4$ in scheme B.
Taking into account the inequalities (\ref{AB06}) and (\ref{AB07}),
we obtain
\begin{equation}
P^{(\mathrm{LBL})}_{\stackrel{\makebox[0pt][l]
{$\hskip-3pt\scriptscriptstyle(-)$}}{\nu_{\mu}}
\to\stackrel{\makebox[0pt][l]
{$\hskip-3pt\scriptscriptstyle(-)$}}{\nu_{e}}}
\leq
a_e^0 + a_\mu^0 - 2 a_e^0 a_\mu^0
\,.
\label{AB43}
\end{equation}
This bound is stable against matter effects \cite{BGG98-bounds}
and its value is represented by the dotted curve in Fig.~\ref{pmelbl}.
For
$ a^0_\mu \ll a^0_e \ll 1 $
this bound is about half of that given by Eq.(\ref{AB39}).
However,
since $a^0_\mu$ is only small
in the same range of $\Delta{m}^2$ where
$A_{\mu;e}^{0}$ is small,
numerically the bound
(\ref{AB43})
turns out to be worse than
the bound (\ref{AB41})
(dashed line in Fig.~\ref{pmelbl})
and
the corresponding matter-corrected bound
(dash-dotted and dash-dot-dotted lines in Fig.~\ref{pmelbl}).

Summarizing the results presented in this section,
we have shown that only the two four-neutrino schemes A and B
are compatible with the results of all neutrino oscillation experiments.
If scheme A is realized in nature,
the effect of non-zero neutrino masses could be observed in the near future in
tritium $\beta$-decay experiments
and neutrinoless double-$\beta$ decay experiments,
whereas these effects are strongly suppressed in scheme B.
Furthermore,
we have shown that
the results of SBL $\bar\nu_e$ disappearance experiments
and solar neutrino experiments imply that
in both schemes A and B the electron neutrino has a
small mixing with the neutrinos whose mass-squared difference is
responsible for the oscillations of atmospheric neutrinos
(\textit{i.e.},
$\nu_1$, $\nu_2$ in scheme A and $\nu_3$, $\nu_4$ in scheme B).
As a consequence,
the transition probability of
electron neutrinos and antineutrinos
into other states
in atmospheric and LBL experiments
is strongly suppressed.

\begin{figure}[t!]
\begin{center}
\setlength{\unitlength}{1cm}
\begin{picture}(13.8,4.4)
%
%\put(0.0,0.0){\line(0,1){4.0}}
%\put(1.5,0.0){\line(0,1){4.0}}
%\put(3.3,0.0){\line(0,1){4.0}}
%\put(6.15,0.0){\line(0,1){4.0}}
%\put(7.65,0.0){\line(0,1){4.0}}
%\put(10.50,0.0){\line(0,1){4.0}}
%\put(12.30,0.0){\line(0,1){4.0}}
%\put(13.8,0.0){\line(0,1){4.0}}
%
\put(0.0,4.4){\makebox(0,0)[lt]{\underline{Scheme A}}}
\put(6.35,1.9){\makebox(1.1,0.0){$\Delta{m}^2_{{\rm SBL}}$}}
\put(13.8,4.4){\makebox(0,0)[rt]{\underline{Scheme B}}}
\put(1.5,0.0){\begin{picture}(1.8,4.4)
\thicklines
\put(1.0,0.2){\vector(0,1){3.8}}
\put(1.0,4.1){\makebox(0,0)[lb]{$m$}}
\put(0.9,0.2){\line(1,0){0.2}}
\put(1.2,0.2){\makebox(0,0)[l]{$\nu_1$}}
\put(0.9,0.6){\line(1,0){0.2}}
\put(1.2,0.6){\makebox(0,0)[l]{$\nu_2$}}
\put(0.9,3.3){\line(1,0){0.2}}
\put(1.2,3.25){\makebox(0,0)[l]{$\nu_3$}}
\put(0.9,3.5){\line(1,0){0.2}}
\put(1.2,3.55){\makebox(0,0)[l]{$\nu_4$}}
\put(0.8,0.4){\makebox(0,0)[r]{$\nu_\mu,\nu_\tau$}}
\put(0.8,3.4){\makebox(0,0)[r]{$\nu_e,\nu_s$}}
\end{picture}}
\put(3.3,0.0){\begin{picture}(2.85,4.0)
\thinlines
\put(0.0,0.6){\line(2,-1){0.4}}
\put(0.0,0.2){\line(2, 1){0.4}}
\put(0.6,0.4){\makebox(0,0)[l]{$\Delta{m}^2_{{\rm atm}}$}}
\put(0.0,3.6){\line(2,-1){0.4}}
\put(0.0,3.2){\line(2, 1){0.4}}
\put(0.6,3.4){\makebox(0,0)[l]{$\Delta{m}^2_{{\rm sun}}$}}
\put(2.0,3.6){\line(1,-2){0.85}}
\put(2.0,0.2){\line(1, 2){0.85}}
\end{picture}}
\put(7.65,0.0){\begin{picture}(2.85,4.0)
\thinlines
\put(2.85,0.5){\line(-2,-1){0.4}}
\put(2.85,0.1){\line(-2, 1){0.4}}
\put(2.25,0.3){\makebox(0,0)[r]{$\Delta{m}^2_{{\rm sun}}$}}
\put(2.85,3.6){\line(-2,-1){0.4}}
\put(2.85,3.2){\line(-2, 1){0.4}}
\put(2.25,3.4){\makebox(0,0)[r]{$\Delta{m}^2_{{\rm atm}}$}}
\put(0.85,3.6){\line(-1,-2){0.85}}
\put(0.85,0.2){\line(-1, 2){0.85}}
\end{picture}}
\put(10.50,0.0){\begin{picture}(1.8,4.4)
\thicklines
\put(0.8,0.2){\vector(0,1){3.8}}
\put(0.8,4.1){\makebox(0,0)[rb]{$m$}}
\put(0.7,0.2){\line(1,0){0.2}}
\put(0.6,0.15){\makebox(0,0)[r]{$\nu_1$}}
\put(0.7,0.4){\line(1,0){0.2}}
\put(0.6,0.45){\makebox(0,0)[r]{$\nu_2$}}
\put(0.7,3.1){\line(1,0){0.2}}
\put(0.6,3.1){\makebox(0,0)[r]{$\nu_3$}}
\put(0.7,3.5){\line(1,0){0.2}}
\put(0.6,3.5){\makebox(0,0)[r]{$\nu_4$}}
\put(1.0,0.3){\makebox(0,0)[l]{$\nu_e,\nu_s$}}
\put(1.0,3.3){\makebox(0,0)[l]{$\nu_\mu,\nu_\tau$}}
\end{picture}}
\end{picture}
\end{center}
\refstepcounter{figures}
\label{schemesAB}
\footnotesize
Figure \ref{schemesAB}.
The two types of neutrino mass spectra that can accommodate 
the solar, atmospheric and LSND scales of $\Delta{m}^2$ and the
mixing schemes that emerge if the Big-Bang Nucleosynthesis
constraint on the number of light neutrinos is less than 4.
\end{figure}

Furthermore,
it has been shown in Refs.~\cite{Okada-Yasuda97,BGGS98-BBN}
that in the schemes A and B, if
the standard Big-Bang Nucleosynthesis constraint
on the number of light neutrinos
(see \cite{BBN-standard} and references therein) is less than 4,
then one has a stringent limit on the mixing of the sterile neutrino
with the two massive neutrinos
that are responsible for the oscillations
of atmospheric neutrinos.
In this case,
$\nu_s$ is mainly mixed with the
two massive neutrinos that contribute to solar neutrino oscillations
($\nu_3$ and $\nu_4$ in scheme A
and
$\nu_1$ and $\nu_2$ in scheme B)
and the unitarity of the mixing matrix implies that
$\nu_\tau$ is mainly mixed with the
two massive neutrinos that contribute to the oscillations
of atmospheric neutrinos.
Hence,
the two schemes have the form shown in Fig.~\ref{schemesAB}
and have the following testable implications
for solar, atmospheric, long-baseline and
short-baseline neutrino oscillation experiments:
\begin{itemize}
\item
The solar neutrino problem
is due to
$\nu_e\to\nu_s$
oscillations.
This prediction will be checked by future solar neutrino
experiments
that can measure the ratio of
neutral-current and charged-current events
\cite{BG-sterile-sun}.
\item
The atmospheric neutrino anomaly is due to
$\nu_\mu\to\nu_\tau$
oscillations.
This prediction will be investigated by LBL experiments.
Furthermore,
the absence of
$\nu_\mu\to\nu_s$
atmospheric neutrino oscillations may be checked in the future
in the Super-Kamiokande atmospheric neutrino experiment
\cite{K2K,Vis98a,Bal98,Vis98b}.
\item
$\nu_\mu\to\nu_\tau$
and
$\nu_e\to\nu_s$
oscillations
are strongly suppressed in SBL experiments
\cite{BGGS98-BBN}.
\end{itemize}
If these prediction will be falsified by future experiments
it could mean that
some of the indications
given by the results of neutrino oscillations experiments
are wrong and neither of the two four neutrino schemes A and B
is realized in nature,
or that Big-Bang Nucleosynthesis occurs with
a non-standard mechanism
\cite{BBN-non-standard}.
On the other hand,
if the standard Big-Bang Nucleosynthesis constraint
on the number of light neutrinos is less than 4 and one of
the two four neutrino schemes depicted in Fig.~\ref{schemesAB}
is realized in nature,
at the leading order
the $4\times4$ neutrino mixing matrix has an extremely simple structure
in which the
$\nu_e,\nu_s$
and
$\nu_\mu,\nu_\tau$
sectors are decoupled:
in scheme A
\begin{equation}
U
\simeq
\left( \begin{array}{cccc}
0 & 0 & \cos\vartheta_{\mathrm{sun}} & \sin\vartheta_{\mathrm{sun}} \\
\cos\vartheta_{\mathrm{atm}} & \sin\vartheta_{\mathrm{atm}} & 0 & 0 \\
-\sin\vartheta_{\mathrm{atm}} & \cos\vartheta_{\mathrm{atm}} & 0 & 0 \\
0 & 0 & -\sin\vartheta_{\mathrm{sun}} & \cos\vartheta_{\mathrm{sun}}
\end{array} \right)
\label{mixA}
\end{equation}
and in scheme B
\begin{equation}
U
\simeq
\left( \begin{array}{cccc}
\cos\vartheta_{\mathrm{sun}} & \sin\vartheta_{\mathrm{sun}} & 0 & 0 \\
0 & 0 & \cos\vartheta_{\mathrm{atm}} & \sin\vartheta_{\mathrm{atm}} \\
0 & 0 & -\sin\vartheta_{\mathrm{atm}} & \cos\vartheta_{\mathrm{atm}} \\
-\sin\vartheta_{\mathrm{sun}} & \cos\vartheta_{\mathrm{sun}} & 0 & 0
\end{array} \right)
\,,
\label{mixB}
\end{equation}
where $\vartheta_{\mathrm{sun}}$ and $\vartheta_{\mathrm{atm}}$ are,
respectively,
the two-generation mixing angles relevant in
solar and atmospheric neutrino oscillations.
Therefore,
the oscillations of solar
and atmospheric neutrinos are independent
and
the two-generation analyses of solar and atmospheric neutrino oscillations
yield correct information on the mixing of four neutrinos.
The
$\stackrel{\scriptscriptstyle (-)}{\nu}_{\hskip-3pt \mu} \to
\stackrel{\scriptscriptstyle (-)}{\nu}_{\hskip-3pt e}$
transitions observed in the LSND experiment represent only a slight
distortion of this simple picture.

\section{Conclusions}
\label{Conclusions}
\setcounter{equation}{0}
\setcounter{figures}{0}
\setcounter{tables}{0}

In this review, we have discussed the phenomenological aspects
of neutrino oscillations. The phenomenon of neutrino oscillations
is possible if neutrinos are massive and mixed particles.
The problem of neutrino masses and mixing is the central problem of
today's neutrino physics and astrophysics.
More than 40 different experiments all over the
world are dedicated to the investigation of this problem
and several new
experiments are in preparation.

The investigation of the properties of neutrinos is considered as 
one of the most important direction for the search of new physics
(see Ref.~\cite{Mohapatra-Pal91}).
Massive neutrinos are among the plausible candidates
for dark matter particles and
the number of massive light neutrinos plays
crucial role in Big-Bang Nucleosynthesis \cite{BBN-standard}.

The Super-Kamiokande collaboration obtained recently a rather convincing
evidence in favour of transitions 
of atmospheric $\nu_{\mu}$ 
into other neutrino states ($\nu_{\tau}$ or  $\nu_{\mathrm{sterile}}$).
Other indications
in favour of neutrino mixing were obtained in 
all solar neutrino experiments and in the accelerator LSND experiment.

If all these indications will be confirmed by future experiments,
it will mean that the number of massive light neutrinos is larger
than the number of flavour neutrinos,
\textit{i.e.} that sterile neutrinos exist.
This would be an evidence that
neutrino masses and mixing are phenomena
due to physics beyond the Standard Model with right-handed
neutrino singlets.

If the indications in favour of a ``large''
$ \Delta{m}^2 \sim 1 \, \mathrm{eV}^2 $
obtained in the LSND experiment will not be confirmed by future
experiments,
the mixing of three massive neutrinos with a mass
hierarchy is a possible and plausible scenario.
In this case,
the Dirac or Majorana nature of massive neutrinos
can be determined only increasing
the sensitivity of the experiments on the search for
neutrinoless double-$\beta$ decay by almost two orders of magnitude. 
It has been shown by an
analysis of the existing data \cite{BGKM98-bb,BGG98-nu98} 
that the effective Majorana
mass in the case of three massive Majorana neutrinos and a mass hierarchy is
smaller than $ 10^{-2} \, \mathrm{eV} $,
whereas the present upper bound is about
$ 0.5 \, \mathrm{eV} $ \cite{Heidelberg-Moscow,Faessler-96-97}
and the sensitivity of the next generation of experiments 
will be of the order of $ 10^{-1} \, \mathrm{eV} $
\cite{next0bb}.

Future neutrino oscillation experiments will make possible to investigate in
detail the regions of $\Delta{m}^2$
and of the elements of the neutrino mixing matrix
allowed by the existing results of neutrino oscillation experiments.
Future solar neutrino experiments
(SNO,
ICARUS,
Borexino,
GNO
and others
\cite{future-sun})
will provide model-independent information on
the transition of high energy $^8\mathrm{B}$ neutrinos into other states
and
investigate in detail the fluxes of medium energy
$^7\mathrm{Be}$, $pep$  and CNO
neutrinos.
Future accelerator LBL experiments
(K2K \cite{K2K},
MINOS \cite{MINOS},
ICARUS \cite{ICARUS}
and others \cite{NOE,AQUA-RICH,OPERA,NICE,Zuber98})
will explore
the region of $\Delta{m}^2$ indicated by the atmospheric neutrino anomaly.
Besides $\nu_\mu\to\nu_\mu$ survival, also 
$\nu_\mu\to\nu_\tau$ and $\nu_\mu\to\nu_e$ transitions will be studied.
A very important task
for the neutrino oscillation experiments of the next generation
is the check of the indication in favour of neutrino oscillations
obtained in the LSND experiment.
The KARMEN experiment will reach the LSND sensitivity
in one or two years \cite{KARMEN98-Moriond}.
The proposed $\nu_\mu\to\nu_e$ SBL experiments
BooNE \cite{BooNE} at Fermilab,
I-216 \cite{I-216} at CERN,
ORLaND \cite{ORLaND} at Oak Ridge
and
NESS at the European Spallation Source \cite{NESS}
plan to achieve a sensitivity about two
orders of magnitude better than the one of the LSND experiment.

It is now widely believed that the recent measurement
of an up-down asymmetry of atmospheric $\mu$-like events
in the Super-Kamiokande detector represents a strong evidence in
favour of neutrino oscillations
generated by new physics beyond the Standard Model
(see, for example, Ref.~\cite{Wilczek-nu98}).
We think that in order to understand the
origin of neutrino masses and mixing
it is necessary to investigate in much more detail
all the phenomena generated by these properties:
neutrino oscillations, neutrinoless double-$\beta$ decay,
distortion of the high-energy part of the electron spectrum of 
tritium $\beta$ decay,
neutrino magnetic moments, etc.
There is no doubt that the next
generation of neutrino experiments will lead to a great
progress in the understanding the properties of neutrinos,
the most puzzling among the known particles.

\section*{Note added}

After this review was finished and presented to the editor (December 15,
1998), new Super-Kamiokande data were reported at WIN99 (Capetown,
January 24 -- 30, 1999)
by Y.\ Suzuki and at the VIIIth International Workshop on Neutrino
Telescopes (Venice, February 23 -- 26, 1999) by K. Scholberg and
K. Inoue. Also Soudan-2 published an updated result. 
Below we will shortly present the new development.

\subsection*{Atmospheric neutrinos}

For the double ratio $R$ (\ref{R}) the new values based on 736 days
for FC events and 685 days for PC events are
$$
R = \left\{
\begin{array}{ll}
0.67 \pm 0.02 \pm 0.05 & \quad \mbox{sub-GeV}, \\
0.66 \pm 0.04 \pm 0.08 & \quad \mbox{multi-GeV}.
\end{array}
\right.
$$
For the up-down asymmetry (\ref{asymm}) of $\mu$-like events in the
multi-GeV region the new value 
$$
A_\mu = -0.311 \pm 0.043 \pm 0.01
$$
is more than 7 standard deviations away from zero. Thus, the new data
confirm the Super-Kamiokande evidence in favour of oscillations of
atmospheric neutrinos. The updated dependence of the up-down asymmetry $A$ as
a function of the
lepton momentum for $e$-like and $\mu$-like events is given in Fig.~A.
The new allowed region in the plane of the oscillation parameters
under the assumption of
$\nu_\mu\to\nu_\tau$
oscillations is shown in Fig.~B.
The region allowed by Kamiokande and the allowed region found from the
analysis of previous Super-Kamiokande data
(see Fig.~\ref{kam-sk})
are also depicted.
As it is seen from Fig.~B,
the new Super-Kamiokande data favour \emph{higher} values of $\Delta{m}^2$
than the previous ones.
The best-fit values of the parameters are
$$
\sin^2 2\vartheta = 1
\,,
\qquad
\Delta m^2 = 3.5 \times 10^{-3} \: \mathrm{eV}^2
\,.
$$
This fit corresponds to $\chi^2_{\mathrm{min}}=62.1$
for 67 DOF
(the previous best-fit value of $\Delta m^2$
was
$ 2.2 \times 10^{-3} \: \mathrm{eV}^2 $).
The new allowed range for $\Delta m^2$ at 90\% CL
is
$ 10^{-3} \: \mathrm{eV}^2
\lesssim \Delta m^2 \lesssim
8 \times 10^{-3} \: \mathrm{eV}^2 $.

\begin{figure}[t!]
\begin{tabular*}{\linewidth}{@{\extracolsep{\fill}}cc}
\begin{minipage}{0.47\linewidth}
\begin{center}
\mbox{\epsfig{file=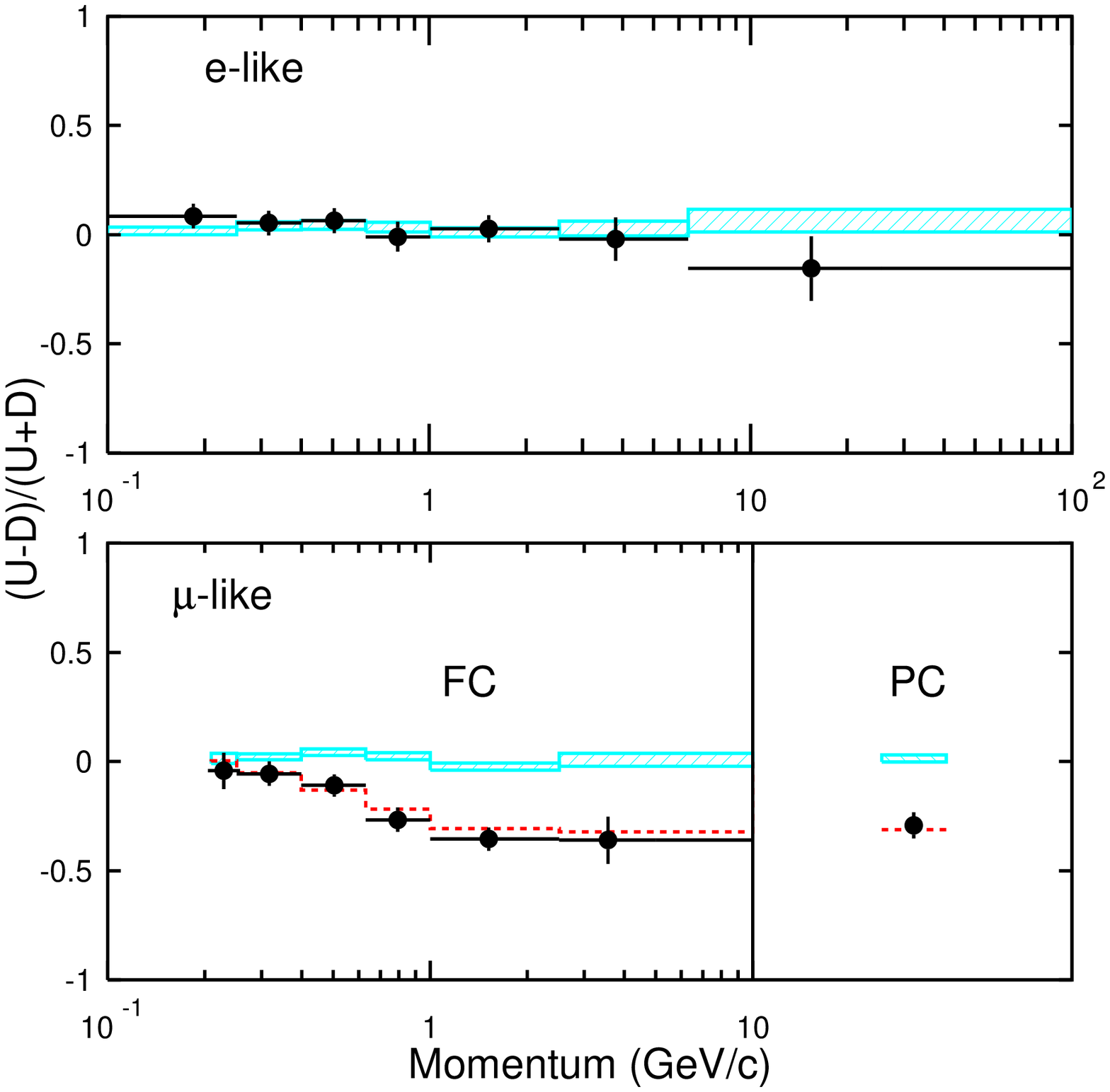,width=0.95\linewidth,clip=}}
\end{center}
\end{minipage}
&
\begin{minipage}{0.47\linewidth}
\begin{center}
\mbox{\epsfig{file=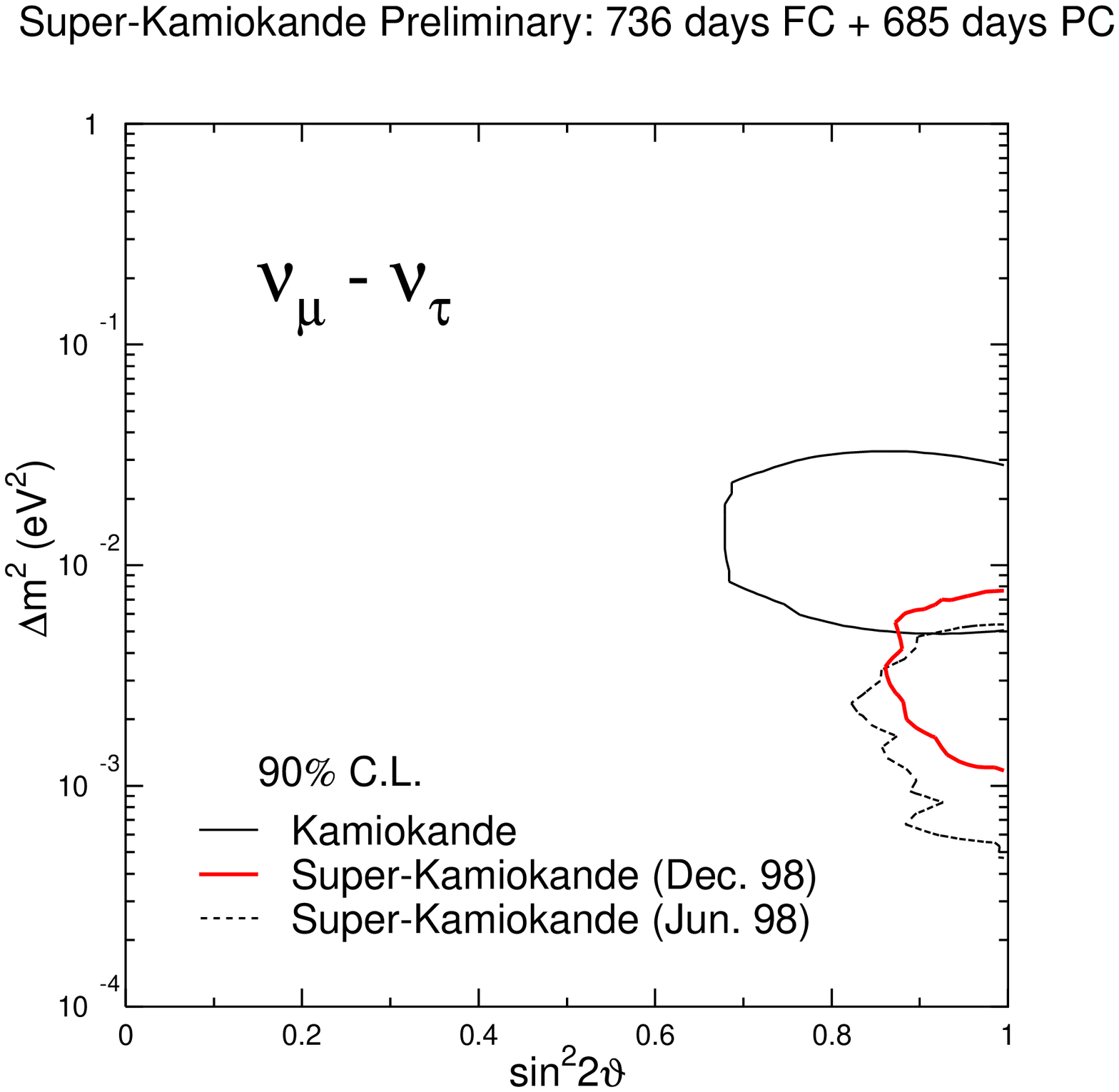,width=0.95\linewidth,clip=}}
\end{center}
\end{minipage}
\\
\begin{minipage}{0.47\linewidth}
\footnotesize
Figure A.
The up-down asymmetry $A$ of $e$-like and $\mu$-like events
as a function of the lepton momentum.
The hatched region shows the theoretical expectation
without neutrino oscillations.
The dashed line for $\mu$-like events
represents the fit of the data in the case of two-generation
$\nu_\mu\to\nu_\tau$ oscillations
with
$ \Delta{m}^2 = 3.5 \times 10^{-3} \, \mathrm{eV}^2 $
and
$ \sin^22\vartheta = 1.0$.
\end{minipage}
&
\begin{minipage}{0.47\linewidth}
\footnotesize
Figure B.
The region allowed at 90\% CL in the
$\sin^22\vartheta$--$\Delta{m}^2$
plane for
$\nu_\mu\to\nu_\tau$ oscillations 
obtained by Super-Kamiokande (December 98). For comparison,
the allowed regions found by the Kamiokande Collaboration and by the
Super-Kamiokande Collaboration (June 98) are also shown.
\end{minipage}
\end{tabular*}
\end{figure}

The Super-Kamiokande Collaboration also measured
the zenith angle dependence of the number of up-going muon events
(for a discussion see Section~5.1.2).
Also these data are compatible with the hypothesis of
$\nu_\mu\to\nu_\tau$ oscillations.
The best-fit values of the oscillation parameters
obtained from the data on the angular dependence
of through-going muon events are
$\sin^22\vartheta=1$
and
$\Delta m^2 = 3.2 \times 10^{-3} \: \mathrm{eV}^2$
(corresponding to $\chi^2_{\mathrm{min}}=7$ for 8 DOF).

The atmospheric neutrino FC and PC data can also be described if one
assumes that $\nu_\mu\to\nu_s$ transitions take place. 
In this case the best-fit
values of the oscillation parameters are given by 
$\sin^22\vartheta=1$
and
$\Delta m^2 = 4.5 \times 10^{-3}$ eV$^2$ with 
$\chi^2_{\mathrm{min}}=64.3$ for 67 DOF.

The Super-Kamiokande Collaboration reported also the observation
of 231.8 NC $\pi^0$ events (see Eq.(\ref{pi0})),
after subtraction of background,
with two $e$-like tracks with an invariant mass
in the range from 90 to 180 MeV.
As a preliminary result it was found that
$$
\left( \frac{\pi^0}{e} \right)_{\mathrm{data}}
\Big/
\left( \frac{\pi^0}{e} \right)_{\mathrm{MC}}
=
1.11 \pm 0.06 \, \mbox{(data stat)}
\pm 0.02 \, \mbox{(MC stat)}
\pm 0.26 \, \mbox{(syst)}
\,.
$$
In the case of
$\nu_\mu\to\nu_\tau$ transitions
this ratio should be equal to one,
whereas in the case of
$\nu_\mu\to\nu_s$ transitions
this ratio should be about 0.75.
The large systematic error in the present data does not allow to reach
a definite conclusion on the transition channel.

The updated value for the double ratio $R$ of the Soudan-2 experiment is
$R = 0.64 \pm 0.11 \pm 0.06$ (W.W.M. Allison \textit{et al.},
hep-ex/9901024). 

\subsection*{Solar neutrinos}

In the region of recoil electron energies between 6.5 and 20 MeV,
during 708 days, the Super-Kamiokande Collaboration has observed
$9531 \, {}^{+167}_{-155}$ events (about 13.5 events per day). For the
value of the $^8$B neutrino flux it was found
$\Phi^\mathrm{SK}_{^8\mathrm{B}} = (2.42 \pm 0.04 \pm 0.07) \times
10^6$ cm$^{-2}$s$^{-1}$. The ratio of this flux to the flux predicted
by the SSM \cite{BP98} is equal to $0.470 \pm 0.008 \pm 0.013$.

Now Super-Kamiokande has also data with an energy  
of the recoil electrons in the range from 5.5 to 6.5 MeV. This
lowering of the electron energy threshold is the 
most important difference of the new data with respect to the
previous ones.
The best fit to the recoil energy
spectrum is given by vacuum oscillations with the best-fit parameters
$\sin^2 2\vartheta = 0.80$ and $\Delta m^2 = 4.3 \times 10^{-10}$
eV$^2$. This is mainly because of the observed excess of events with
respect to the SSM prediction in the high energy region of the
recoil electron spectrum.

The contribution to the recoil spectrum in the region $E_e \gtrsim 14$
MeV comes from $hep$ neutrinos, which according to the SSM is small
(see Table 5.2). If the flux of $hep$ neutrinos is considered as a
free parameter, then from a fit of the Super-Kamiokande data under the
assumption of absence of neutrino oscillations it was found
$\Phi^\mathrm{fit}_{hep} = \left( 14.0 \,{}^{+7.6}_{-6.4} \right) 
\Phi^\mathrm{SSM}_{hep}$. The contribution of $hep$ neutrinos in the
Super-Kamiokande recoil energy spectrum was analysed in detail in the
papers of J. Bahcall and P. Krastev, Phys. Lett. B \textbf{436}, 243
(1998) and G. Fiorentini \textit{et al.}, Phys. Lett. B \textbf{444},
387 (1998).

Super-Kamiokande also presented a new measurement of the day-night
asymmetry with the result 
$N/D - 1 = 0.060 \pm 0.036 \,{}^{+0.028}_{-0.027}$.

\section*{Acknowledgements}

S.M.B. acknowledges
the support of the ``Sonderforschungsbereich 375-95 f\"ur
Astro-Teilchenphysik der Deutschen Forschungsgemeinschaft''.
C.G. would like to express his gratitude to the Korea Institute 
for Advanced Study (KIAS)
for the kind hospitality during part of this work.
Furthermore, we thank V.A. Naumov for reading a preliminary version of the
manuscript and providing several useful comments. We are particularly
grateful to K. Scholberg for providing us with the latest
Super-Kamiokande figures.

\appendix

\section{Properties of Majorana neutrinos and fields}

In this Appendix we summarize the main properties of Majorana
particles associated with Majorana fields. These
particles have spin 1/2 and all charges equal to zero.
A free Majorana field $\nu(x)$ satisfies the Dirac equation
\begin{equation}
\left( i \, \gamma^\mu \partial_\mu - m \right) \nu(x) = 0
\label{aa1}
\end{equation}
together with the Majorana condition
\begin{equation}
\nu(x) = \nu^c(x) \equiv \mathcal{C} \, \overline{\nu}^T(x)
\,,
\label{aa2}
\end{equation}
where $\nu^c$ is the charge-conjugate field and $\mathcal{C}$ is 
the charge-conjugation matrix which fulfills the relations
\begin{equation}
\mathcal{C} \gamma_\mu^T \mathcal{C}^{-1} = - \gamma_\mu
\,,
\qquad
\mathcal{C}^\dagger = \mathcal{C}^{-1}
\,,
\qquad
\mathcal{C}^T = - \mathcal{C}
\,.
\label{a03}
\end{equation}
From Eqs.(\ref{aa1}) and (\ref{aa2})
it follows that the Fourier expansion of the Majorana field is
\begin{equation}
\nu(x)
=
\int \frac{ \mathrm{d}^3p }{ \sqrt{(2\pi)^3 2 p_0} }
\sum_{r=\pm1}
\left[
a_r(p) \, u^r(p) \, e^{-ipx}
+
a_r^\dagger(p) \, v^r(p) \, e^{ipx}
\right]
\,,
\label{a04}
\end{equation}
where the spinors $u^r(p)$ and $v^r(p)$ are related by
\begin{equation}
v^r(p) = \mathcal{C} (\overline{u}^r(p))^T
\,.
\label{a05}
\end{equation}
The operators  $a_r(p)$ and $a_r^\dagger(p)$
annihilate and create
Majorana particles with momentum $p$ and helicity $r$
(particle $\equiv$ antiparticle for a Majorana field), respectively.

Let us consider the quantities
\begin{equation}
\overline{\nu_a} \, \mathcal{O}_j \, \nu_b
\,,
\label{a06}
\end{equation}
where $\nu_a$ and $\nu_b$ are two Majorana fields and
\begin{equation}
\mathcal{O}_j
=
1
\,, \
\gamma_\mu
\,, \
\sigma_{\mu\nu}
\,, \
\gamma_\mu \gamma_5
\,, \
\gamma_5
\,,
\qquad \mbox{for} \qquad
j = S \,,\, V \,,\, T \,,\, A \,,\, P
\,.
\label{a07}
\end{equation}
Taking into account the minus sign that
appears under interchange of two fermion operators
and the Majorana condition (\ref{aa2}),
we obtain
\begin{equation}
\overline{\nu_a} \, \mathcal{O}_j \, \nu_b
=
\overline{\nu_b} \, \mathcal{C} \, \mathcal{O}^T_j \, \mathcal{C}^{-1} \, \nu_a
\,.
\label{a08}
\end{equation}
From Eq.(\ref{a03}) we derive
\begin{equation}
\mathcal{C} \sigma^T_{\mu\nu} \mathcal{C}^{-1} = - \sigma_{\mu\nu}
\quad \mbox{and} \quad
\mathcal{C} \gamma^T_5 \mathcal{C}^{-1} = \gamma_5
\,.
\end{equation}
With the help of this equation and Eq.(\ref{a03}) it follows that
\begin{equation}
\overline{\nu_a} \, \mathcal{O}_j \, \nu_b
=
\eta_j \overline{\nu_b} \, \mathcal{O}_j \, \nu_a 
\end{equation} 
with
\begin{equation}
\eta_j = -1 \quad \mbox{for} \quad j=V,T \quad \mbox{and} \quad 
\eta_j = 1  \quad \mbox{for} \quad j=S,A,P \,.
\end{equation}
Thus, for a single Majorana field $\nu = \nu_a = \nu_b$ we obtain
\begin{equation}
\bar\nu \gamma_\mu \nu = 0 \quad \forall \mu \quad \mbox{and} \quad
\bar\nu \sigma_{\lambda\rho} \nu = 
\bar\nu \sigma_{\lambda\rho} \gamma_5 \nu = 0 \quad \forall \lambda,
\rho \,.
\end{equation}
This means that a Majorana particle has neither electric charge
nor magnetic moments or electric dipole moments.

For any fermion field there is a  relation between its
left-handed component and the right-handed component of its
charge-conjugate field (see Section 2.2):
\begin{equation}
\psi_R = ((\psi^c)_L)^c \,.
\label{aRL}
\end{equation}
In the case of a Dirac field, $\psi^c$ is different from $\psi$
and, therefore, the left and right-handed components are
independent, whereas from Eq.(\ref{aRL}) it follows for a Majorana
field $\nu$ that
\begin{equation}
\nu_R = (\nu_L)^c
\label{a21}
\end{equation}
and thus the left and right-handed components of a Majorana field
are \emph{not independent}
(in other words, a Majorana field is a two-component field).

Though Dirac and Majorana neutrinos differ in the number of
degrees of freedom, it is nevertheless difficult to
distinguish between the two in practice. The reason is that
in the standard weak interaction Lagrangians (\ref{CC}) and (\ref{NC})
only the left-handed components 
$\nu_{\ell L}$ ($\ell = e,\mu,\tau$) of the flavour neutrino fields
are present. These are connected with the left-handed components
$\nu_{kL}$ of the neutrino mass eigenfields by the relation
$\nu_{\ell L} = \sum_k U_{\ell k} \nu_{kL}$ (see Section 2). The
mass eigenfields could be of Dirac or Majorana type. In the
first case we have $\nu^D_k = \nu_{kL} + \nu_{kR}$, where the
$\nu_{kR}$ are independent degrees of freedom, whereas in the
second case we have $\nu^M_k = \nu_{kL} + (\nu_{kL})^c$. The only
difference between the Dirac and Majorana case with respect to
the standard weak interactions resides in the
mass term. In the
limit of vanishing neutrino masses the right-handed Dirac fields
decouple from the standard weak interaction Lagrangians and the
distinction between Dirac and Majorana gets lost. Then the
physical fields are simply given by the $\nu_{\ell L}$. In
summary, we have arrived at the following conclusion \cite{okubo}: 
With the
standard weak interactions one cannot distinguish between the
Dirac and Majorana nature of the neutrinos in the limit of
vanishing neutrino masses. In other words, for the
standard weak interactions, Dirac neutrinos with negative
helicity and Dirac antineutrinos with positive helicity are the
same as Majorana neutrinos with positive and negative
helicities, respectively, in this limit.

In neutrinoless double-$\beta$ decay, which can occur only with
Majorana neutrinos (see Section 6), one needs to calculate the
propagator 
\begin{equation}
\langle 0 | T \left( \nu_{eL}(x_1) \nu^T_{eL}(x_2) \right) | 0
\rangle \,. 
\end{equation}
With the Majorana mass eigenfields $\nu_k$ satisfying the relation
$\nu^T_k = -\bar \nu_k \mathcal{C}$ this propagator is obtained
by the following steps:
\begin{eqnarray}
\lefteqn{\langle 0 | T \left( \nu_{eL}(x_1) \nu^T_{eL}(x_2) \right) | 0
\rangle } \nonumber \\
& & = \frac{1-\gamma_5}{2} \sum_k U^2_{ek} 
\langle 0 | T \left( \nu_k(x_1) \nu^T_k(x_2) \right) | 0 \rangle 
\frac{1-\gamma^T_5}{2} \nonumber \\ 
& & = 
-\frac{1-\gamma_5}{2} \sum_k U^2_{ek} 
\langle 0 | T \left( \nu_k(x_1) \bar\nu_k(x_2) \right) | 0 \rangle 
\frac{1-\gamma_5}{2} \,\mathcal{C} 
\nonumber \\
& & =
-\sum_k U^2_{ek} m_k\, i\! \int\! \frac{d^4p}{(2\pi)^4} 
\frac{e^{-i p(x_1-x_2)}}{p^2 - m^2_k} \,
\frac{1-\gamma_5}{2} \,\mathcal{C} \,.
\label{Mprop}
\end{eqnarray}

Finally,
we want to prove that the CP parity of a Majorana field is $\pm i$.
The CP transformation of a Majorana field $\nu(x)$ is given by
\begin{equation}
U_{\mathrm{CP}} \, \nu(x) \, U_{\mathrm{CP}}^{-1}
=
\eta_{\mathrm{CP}} \, \gamma_0 \, \nu(x_{\mathrm{P}})
\,,
\label{a41}
\end{equation}
where the phase factor
$\eta_{\mathrm{CP}}$ is the CP parity of the field $\nu(x)$,
$U_{CP}$ is the unitary operator of CP conjugation,
$x\equiv(x^0,\vec{x})$
and
$x_{\mathrm{P}}\equiv(x^0,-\vec{x})$.
The relation (\ref{a41}) can be written as
\begin{equation}
U_{\mathrm{CP}} \, \mathcal{C} \, \overline{\nu}^T(x) \, U_{\mathrm{CP}}^{-1}
=
- \eta_{\mathrm{CP}}^* \, \gamma_0 \, \mathcal{C} \, \overline{\nu}^T(x_{\mathrm{P}})
\,.
\label{a42}
\end{equation}
Using the Majorana condition (\ref{aa2})
one obtains
\begin{equation}
U_{\mathrm{CP}} \, \nu(x) \, U_{\mathrm{CP}}^{-1}
=
- \eta_{\mathrm{CP}}^* \, \gamma_0 \, \nu(x_{\mathrm{P}})
\,.
\label{a43}
\end{equation}
From the comparison of the expressions (\ref{a41}) and (\ref{a43})
one finds that
$ \eta_{\mathrm{CP}} = - \eta_{\mathrm{CP}}^* $,
which implies that
\begin{equation}
\eta_{\mathrm{CP}}^2 = - 1
\label{a44}
\end{equation}
and
$ \eta_{\mathrm{CP}} = \pm i $.
As it can be seen from Eq.(\ref{a42}) written as
\begin{equation}
U_{\mathrm{CP}} \, \nu^c(x) \, U_{\mathrm{CP}}^{-1}
=
- \eta_{\mathrm{CP}}^* \, \gamma_0 \, \nu^c(x_{\mathrm{P}})
\,,
\label{a46}
\end{equation}
in general the product of the CP parities of particles and antiparticles is
equal to $-1$.
The relation (\ref{a44}) results from the fact that in the Majorana case
there is no difference between particles and antiparticles.

\end{document}